\documentclass[a4paper,fleqn,usenatbib]{mnras}

\usepackage{newtxtext,newtxmath}
\usepackage[T1]{fontenc}
\usepackage{ae,aecompl}

\usepackage{graphicx}	% Including figure files
\usepackage{amsmath}	% Advanced maths commands
\usepackage{amssymb}	% Extra maths symbols

\title[SF outside galaxies]{GASP XIII. Star formation in gas outside galaxies}

\author[B.M. Poggianti et al.]{Bianca M. Poggianti,$^1$\thanks{bianca.poggianti@oapd.inaf.it}
Marco Gullieuszik,$^1$ 
Stephanie Tonnesen, $^2$ 
Alessia Moretti,$^1$\newauthor
Benedetta Vulcani,$^1$ 
Mario Radovich,$^1$  
Yara Jaff\'e,$^3$
Jacopo Fritz,$^4$
Daniela Bettoni,$^1$\newauthor
Andrea Franchetto,$^{1,5}$
Giovanni Fasano,$^1$
Callum Bellhouse$^6$
\\
$^1$INAF-Astronomical Observatory of Padova, vicolo dell'Osservatorio 5, 35122 Padova, Italy \\ 
$^{2}$Center for Computational Astrophysics, Flatiron Institute, 162 5th Ave, New York, NY 10010, USA\\
$^{3}$Instituto de F\'isica y Astronom\'ia, Universidad de Valpara\'iso, Avda. Gran Breta\~na 1111 Valpara\'iso, Chile\\
$^4$Instituto de Radioastronomia y Astrofisica, UNAM, Campus Morelia, A.P. 3-72, C.P. 58089, Mexico \\ 
$^5$Dipartimento di Fisica e Astronomia, Universita' di Padova, vicolo dell'Osservatorio 5, 35122 Padova, Italy\\
$^{6}$University of Birmingham School of Physics and Astronomy, Edgbaston, Birmingham, UK\\
}

\date{Accepted XXX. Received YYY; in original form ZZZ}

\begin{document}
\label{firstpage}
\pagerange{\pageref{firstpage}--\pageref{lastpage}}
\maketitle

% Abstract of the paper
\begin{abstract}
Based on MUSE data from the GASP survey, we study 
the $\rm H\alpha$-emitting 
extraplanar tails of 16 cluster galaxies at $z
\sim 0.05$ undergoing ram pressure stripping. We demonstrate that 
the dominating ionization mechanism of this gas (between 64\%
and 94\% of the $\rm H\alpha$ emission in the tails depending on the
diagnostic diagram used) is photoionization by
young massive stars due to ongoing star formation (SF) taking place 
in the stripped tails. 
This SF occurs 
 in dynamically quite cold HII clumps with
a median $\rm H\alpha$ velocity dispersion $\sigma = 27\rm
\, km\, s^{-1}$. We study the characteristics of
over 500 star-forming clumps in the tails and find median values of
 $\rm H\alpha$ luminosity $ L_{\rm H\alpha} = 4\times10^{38}
\rm erg\, s^{-1}$, dust extinction $A_V=0.5$ mag,  star formation
rate SFR$=0.003\, M_{\odot}\, yr^{-1}$, ionized gas density $n_e =52\rm \,cm^{-3}$,
ionized gas mass $ M_{gas} = 4\times10^4 M_{\odot}$, and stellar mass
$M_{*} = 3\times10^6\,M_{\odot}$. The tail clumps follow
scaling relations ($M_{gas}-M_{*}$, $ L_{\rm H\alpha} -\sigma$, SFR-$M_{gas} $) similar to disk clumps, and  
their stellar masses are comparable to
Ultra Compact Dwarfs and Globular Clusters.
The diffuse gas component in the tails is ionized by 
a combination of SF  and composite/LINER-like emission likely due to thermal
conduction or turbulence.
The stellar photoionization component of the diffuse gas can be due
either to leakage of ionizing photons 
from the HII clumps with an average escape fraction of 18\%, or
lower luminosity HII regions that we
cannot individually identify.

%We speculate that...[fate of these complexes]
%Our results show that 
\end{abstract}

% Select between one and six entries from the list of approved keywords.
% Don't make up new ones.
\begin{keywords}
galaxies: evolution -- galaxies: clusters: intracluster medium --
galaxies: peculiar
\end{keywords}

%%%%%%%%%%%%%%%%%%%%%%%%%%%%%%%%%%%%%%%%%%%%%%%%%%

%%%%%%%%%%%%%%%%% BODY OF PAPER %%%%%%%%%%%%%%%%%%

\section{Introduction}

Understanding how stars form would be the key to many of the most
pressing astrophysical questions across all fields of research, from
cosmology to galaxy evolution to planet formation. The star formation process is
responsible for the generation of energy and chemical elements in the
Universe, and is the root of galaxy formation.
% and the origin of life.
%evolution.
%, and drives what we can directly or
%indirectly observe to understand cosmic evolution.
Star formation is therefore at the heart of astrophysics, and yet it
is still a poorly understood phenomenon. The complexity of the physical
processes involved, over a wide range of physical scales, still makes a
comprehensive theory of star formation a challenging goal in spite of
progress \citep{Krumholz2014}.
Empirically, a relation between a galaxy's gas content and its SFR is
well established \citep{Schmidt1959, Kennicutt1998b}, and multiwavelength
observations of SF at different scales depict a
phenomenological description of the relation between interstellar
medium phases and SF activity \citep{KennicuttEvans2012}, but a
thorough understanding of what drives the star formation histories of galaxies
must still be developed.
%and provides the shining component that guide our
%investigation of the Universe.

One of the important questions to be answered is what controls
the clustering properties of SF. Massive stars form almost
exclusively in stellar-cluster-forming clumps \citep{Evans1999}. The clustered and
hierarchical structure of star-forming regions depends strongly on 
the physical conditions in which stars form. Turbulence appears to
play an important role \citep[e.g.][]{Elmegreen2014, Gouliermis2017}
and ambient conditions influence the formation of molecular clouds and
their properties as a consequence of radiative, thermal,
magneto-hydrodynamical and dynamical processes. 
These conditions can greatly vary between galaxies, within individual galaxies, and with redshift, leading to clumps of various sizes
and masses \citep{Elmegreen2013}.
In the last decade, the characteristics of star-forming clumps at
high- and low-z have received a surge of interest, due to the realization
% it has become clear 
that  the progenitors of local disk galaxies are high redshift galaxies
in which giant, massive star-forming clumps
($10^7-10^9 M_{\odot}$) are observed \citep{Elmegreen2007, Cava2018}. 

Probing SF in different regimes, ambient conditions and epochs is
fundamental to obtain an observational picture that can inform
our theoretical understanding. For example, the SF conditions in the
outskirts of galaxy disks can be substantially different from those in
the disks themselves, providing clues about non-linear star formation laws
and the distribution of stellar masses in the low SF regime \citep{Elmegreen2017}
 and starburst galaxies can help us understand how SF proceeds in
the high gas density regime \citep{KennicuttEvans2012}.

At the present epoch, most new stars are formed in galaxy disks, but SF is known to take place
also in more exotic environments, such as the tidal tails of merging systems \citep{Elmegreen1993, Boquien2009, Bournaud2004, 
Schweizer2006, Duc2012, Mullan2011, Mulia2015, Vulcani2017c}.

%In any environment and at any epoch, the availability of gas and 
%suitable conditions for the gas to cool to form new stars drive the star
%formation activity in galaxies.

In this paper we investigate the extra-galactic SF
occurring in the tails of gas that
is being stripped from galaxies, and study the properties of the
clumps that form within this gas.
% due to their interaction with the
%intergalactic medium in galaxy clusters. 
Gas can be removed from galaxies by various physical mechanisms, and 
in clusters the most efficient one is ram pressure stripping,
due to the pressure exerted by the hot ($10^8$ K) intracluster medium
(ICM) on the galaxy interstellar medium (ISM)  \citep{Gunn1972}. The
ram pressure stripped gas can produce tails up to
more than 100kpc long in which new stars can be formed.
A summary of both observational and theoretical evidence for this in the literature 
is deferred until \S8 to facilitate comparison with the results of this paper. 

GASP \citep[GAs Stripping Phenomena in galaxies with MUSE,][]{Poggianti2017a}
 is an ESO Large Program aimed at studying processes that
remove gas from galaxies.  Target galaxies were chosen to have
unilateral debris or tails in B-band images, suggestive of gas-only
removal, excluding clear mergers and tidal interactions (see
  \cite{Poggianti2017a}.
All targets are at redshift $0.04<z<0.07$. They are located in
different environments (galaxy clusters, groups, filaments and
isolated) and span a wide range of galaxy stellar masses, from $10^9$
to $10^{11.5} M_{\odot}$.
  
MUSE Integral Field spectroscopy of these galaxies allows us a
detailed investigation of the ionized gas phase and the stellar
component both in the disks and in the extraplanar tails. 
This program provides a direct observational window on galaxies in
various stages of ram pressure stripping in clusters \citep{Jaffe2018}, from
pre-stripping (undisturbed galaxies of a control sample), to initial
stripping, peak stripping \citep{Poggianti2017a, Bellhouse2017, Gullieuszik2017, Moretti2018}, and
post-stripping \citep{Fritz2017}, passive and devoid of
gas, as well as on a number of physical processes in groups and
filaments ranging from stripping to gas accretion, mergers and cosmic
web (\citealt{Vulcani2017c, Vulcani2018a,Vulcani2018b}, Vulcani et al. in prep.).

In this paper we focus on galaxies in clusters, and use the MUSE data to
investigate the origin of the ionization of the stripped gas and the star formation that takes place within it. After presenting
our sample and observations (\S2) and describing the methods
of analysis employed (\S3), we present our results on ionization
mechanisms (\S4.1) and discuss the location of ongoing star formation
in \S4.2. The physical properties of the star-forming clumps 
(velocity dispersion, $\rm H\alpha$ luminosity, dust extinction,
star formation rate, gas density and star formation rate density, gas
mass and stellar mass) and a few scaling relations
linking various properties of the clumps are presented in \S5. The diffuse
component of $\rm H\alpha$ emission is separately discussed in \S6,
and a summary of global SFR in the disks and tails is given in \S7.
The discussion (\S8) includes a summary of previous observational
results and theoretical expectations, and a summary of our results is
given in \S9.

This paper adopts a \cite{Chabrier2003} IMF and standard concordance
cosmology parameters $H_0 = 70 \, \rm km \, s^{-1} \, Mpc^{-1}$, ${\Omega}_M=0.3$
and ${\Omega}_{\Lambda}=0.7$.

\section{Sample and observations}

All the observations used in this paper have been obtained from the 
GASP survey. The GASP sample comprises 64 cluster galaxies that are
stripping candidates, plus another 30 candidates in groups, filaments 
and isolated and  20 undisturbed galaxies that represent a control sample.

For this work, we have selected galaxies
satisfying the following criteria: a) only cluster members; b) 
%with evidence for ongoing stripping or very strong stripping, 
with clear
%as testified by the presence 
tails of extraplanar $\rm H\alpha$ emitting gas
and c) without any nearby companion that
could affect the gas morphologies by tidal interactions.
In this way we are
excluding galaxies with $\rm H\alpha$ truncated disks that have gas left
only in the central regions of the disk, which are in an advanced phase
of stripping (Post-stripping galaxies in \citealt{Jaffe2018}). An example of such a truncated disk is JO36, studied in
detail in \cite{Fritz2017}.
We are also excluding face-on galaxies with unwinding spiral arms in
which, though the gas kinematics clearly indicates stripping, due to
the viewing angle no clear gas tails are visible.
%the gas tails are harder to identify.
%theidentification of the stripped gas requires 
%harder to see based simply on gas and
%stellar morphologies, and where 
%a detailed kinematical analysis. 
These are the subject of
a dedicated work (Bellhouse et al. in prep.). 
%e) galaxies with a
%nearby companion whose gas morphologies are most probably affected by
%tidal interactions.

In the following we will focus on the 16 galaxies with long
$\rm H\alpha$ tails (at least 20 kpc from the disk) complying to the selection
criteria described above and observed by GASP at the time of selection,
stressing that no additional selection criterion
(based for example on the gas ionization properties) was employed. 
%Thus the sample can be considered
%representative of galaxies undergoing ram pressure  
%The sample galaxies are listed in Table~1.
Table~1 lists their name, host cluster, redshift and cluster redshift,
cluster velocity dispersion,
coordinates and stellar masses. % and references.

\begin{table*}
\centering  
%\caption{Sample galaxies} % in the MUSE spectral range.} %\label{tab:decimal}
\caption{Sample galaxies. Columns are: 1) GASP ID number from \citet{Poggianti2016}; 
2) host cluster; 3) galaxy redshift; 4) cluster
  redshift; 5) cluster velocity dispersion; 6) and 7) RA and DEC; 8)
  galaxy stellar mass; 9) number of clumps in tail (total number of
  clumps); 10) references.
Refs: (1) \citet{Poggianti2017a}; (2) \citet{Moretti2018}; (3) \citet{Gullieuszik2017}; (4) \citet{Bellhouse2017};
(5) \citet{Poggianti2017b}; (6) \citet{Merluzzi2013}; (7)
\citet{George2018}; (8) \citet{Moretti2018b}. Cluster redshifts are taken from  \citet{Biviano2017},
\citet{Moretti2017} and \citet{Cava2009}.}
\begin{tabular}{lccccccccc}
\hline  
$ID_{P16}$ & cluster & $z_{gal}$ & $z_{clu}$ & $\sigma_{clu} \, (km/s)$ & RA(J2000) & DEC(J2000)  
& $M_*(M_{\odot})$ & $N_{tail \, clumps}$ & refs \\
&&&&&&&& $(N_{all \, clumps})$ &\\
\hline  
JO113 	&A3158  	& 0.0553 & 0.0594 & 948 &	03:41:49.225 &
-53:24:12.16  &$4.7^{5.5}_{3.2} \times 10^{9}$ & 4(20) & 	-- \\ 
JO135 	&A3530   & 0.0542 & 0.0548 &  674 &	12:57:04.322 & -30:22:30.19
&$1.1^{1.2}_{0.8}  \times 10^{11}$ &	15(77) &  (5)  \\
JO141 	&A3532  	& 0.0588 & 0.0555 & 662 &	12:58:38.371 &
-30:47:32.31  &$4.4^{5.2}_{2.4}  \times 10^{10}$ &	13(55) & -- \\ 
JO147	&A3558	& 0.0498 & 0.0486 &   910 &    13:26:49.731 &
-31:23:44.79 &$1.3^{1.4}_{0.7} \times 10^{11}$& 29(72) &	(6) \\
JO160 	&A3558   &  0.0483 & 0.0486 & 910 &	13:29:28.584 &	-31:39:25.46
&$1.1^{1.7}_{0.9}  \times 10^{10}$ &	6(39) & -- \\  
JO171 	&A3667   & 0.0520 & 0.0558 &   1031 &	20:10:14.753 &	-56:38:29.49
&$3.6^{4.3}_{2.8}  \times 10^{10}$ & 27(93) &	(2) \\ 
JO175 	&A3716   &  0.0468 & 0.0457 & 753 & 	20:51:17.593 & -52:49:22.34
&$3.4 ^{3.6}_{2.7} \times 10^{10}$ &  34(80) & (5) \\ 
JO194 	&A4059   & 0.0410 & 0.0490 &  744 &	23:57:00.740 &	-34:40:49.94
&$1.3^{1.8}_{1.2} \times 10^{11}$ & 84(223) &	(5) \\ 
JO201 	&A85       & 0.0446 & 0.0559 &  859 &	00:41:30.295 &
-09:15:45.98 &$4.4^{7.8}_{4.1}  \times 10^{10}$ & 57(148) &	(4,5,7,8) \\ 
JO204 	&A957   &  0.0424 & 0.0451 &  631 &	10:13:46.842 &	-00:54:51.27
&$5.5^{6.1}_{3.2}  \times 10^{10}$ & 41(121) &	(3,5,8) \\ 
JO206 	&IIZW108 & 0.0513 & 0.0486&   575 & 	21:13:47.410 & +02:28:35.50
&$7.8^{10.4}_{5.3}  \times 10^{10}$ & 68(139) & (1,5,8) \\ 
JO49  	&A168     & 0.0450 & 0.0453&  498 &	01:14:43.924 &	+00:17:10.07
&$5.9^{6.3}_{3.5}  \times 10^{10}$ & 6(75) &	-- \\ 
JO60  	&A1991   & 0.0623 &0.0584 &  570 &	14:53:51.567 &	+18:39:04.79
&$ 2.1^{2.9}_{1.5}  \times 10^{10}$ &	17(78) & --  \\ 
JO95  	&A2657  	& 0.0433 & 0.0400 &	829 & 23:44:26.659 &
+09:06:54.54  &$2.6^{3.3}_{1.4}  \times 10^{9}$ & 5(46) &	-- \\
JW100 	& A2626	 & 0.0602 & 0.0548 &   650 & 23:36:25.054 &	+21:09:02.64 &
$2.9^{3.1}_{1.2}  \times  10^{11}$ & 66(131) & (5,8) \\ 
JW39  	&A1668	 & 0.0650 & 0.0634&  654 &  13:04:07.719 &	+19:12:38.41
&$1.6^{1.8}_{1.0}  \times 10^{11}$ &	49(159) &-- \\
%JO28  	&A151   	&&&	01:10:09.355 &	-15:34:24.57 &$1.9 \times 10^{9}$ &	-- \\
%JO69	&A2399	&&&	7:19.291 &	-07:46:44.16 &$5.0 \times 10^{9}$ &	-- \\
%JO181	&A3880	&&&	8:03.757 &	-30:18:03.39 &$1.2 \times 10^{9}$ &	-- \\
%JO27  	&A151   	&&&	01:10:48.500 &	-15:04:42.92 &$1.7 \times 10^{9}$ &	-- \\
\hline  
\end{tabular}
\end{table*}
 
Observations were carried out in service mode with the MUSE
spectrograph mounted at the VLT in wide-field mode with natural
seeing.  MUSE \citep{Bacon2010} is an
integral-field spectrograph composed of 24 IFU modules with a $4k
\times 4k$ CCD each. It has  0.2''$\times$0.2'' pixels and covers
a 1'$\times$1' field-of-view. It covers the spectral range between 4800
and 9300 \AA $\,$ sampled at 1.25 \AA/pixel with a spectral resolution
FWHM=2.6 \AA.

Most of our target galaxies were observed with one MUSE pointing, and
some with two pointings in order to cover the length of the tail. On each
pointing, $4\times$675sec exposures were taken in clear, dark-time,
$<1$'' seeing conditions.
The data were reduced with the most recent available version of the MUSE pipeline
(\citealt{Bacon2010}; http://www.eso.org/sci/software/pipelines/muse),
as described in \cite{Poggianti2017a}.

\section{Methods}

The methods employed to analyze the MUSE data are described in detail in
\cite{Poggianti2017a} and are summarized below.

To derive emission line fluxes, velocities and velocity dispersions
with associated errors we make use of KUBEVIZ \citep{Fossati2016}, an IDL public software
that fits Gaussian line profiles using the MPFit package \citep{Markwardt2009}. The MUSE spectral range covers the $\rm H\beta$, [OIII]5007,
[OI]6300, $\rm H\alpha$, [NII]6583, [SII]6717, 6731 lines that are of
interest for this paper.
In our analysis, before performing the fits, we average filter the
data cube in the spatial direction with a $5 \times 5$ kernel, 
corresponding to our worst seeing conditions of $1" = 0.7-1.3$ kpc at
the redshifts of our galaxies. The velocity dispersions are corrected
for the instrumental line width at each wavelength \citep[see][]{Fumagalli2014}, which
at the observed $\rm H\alpha$ wavelengths of these galaxies is about 46 $\rm km \, s^{-1}$.

Before running KUBEVIZ, we correct the MUSE data cube
for Galactic extinction and subtract the stellar-only component of each spectrum
derived with our spectrophotometric code SINOPSIS \citep{Fritz2017}.
SINOPSIS uses the latest SSP models from S. Charlot \& G. Bruzual
(2018, in preparation) based on stellar evolutionary tracks from
\cite{Bressan2012} and stellar atmosphere spectra from a
compilation of different authors. SINOPSIS also includes the nebular
emission lines for the young SSPs computed with the Cloudy code \citep{Ferland2013}.
In addition to the best fit stellar-only model cube that is subtracted
from the observed cube, SINOPSIS provides for each MUSE spaxel stellar masses,
luminosity-weighted and mass-weighted ages and star formation
histories in four broad age bins.

All the SFRs in this paper are computed from the $\rm H\alpha$
luminosity corrected both for stellar absorption and for dust
extinction, adopting the \cite{Kennicutt1998a}'s relation: $\rm SFR (M_{\odot}
\, yr^{-1}) = 4.6 \times 10^{-42} L_{\rm H\alpha} (erg \, s^{-1})$. 
The extinction is estimated from the Balmer decrement
assuming a value $\rm H\alpha/H\beta = 2.86$ and the \cite{Cardelli1989} extinction law.

The ionized gas density $n$ is derived from the ratio R=[SII]6716/[SII]6731
adopting a gas
temperature T=10000 K and the calibration of \cite{Proxauf2014}
which is valid for the interval R=0.4-1.435.

The mass of the ionized gas is estimated as  $M_{gas} = N_{protons} \times m_H =
\frac{L_{H\alpha} \times m_p}{n \alpha_{H\alpha} h \nu_{H\alpha}}$ \citep{Poggianti2017a}
where $L_{\rm H\alpha}$ is the luminosity of the $\rm H\alpha$ line
corrected for stellar absorption and dust extinction,  $m_H =1.6737 \times
10^{-24} gr$ is the mass of the hydrogen atom, 
$\rm \alpha_{H\alpha}$ is the effective
$\rm H\alpha$ recombination coefficient ($1.17 \times 10^{-13} cm^{3}
\, s^{-1}$), $\rm h \nu_{H\alpha}$ is the energy
of the $\rm H\alpha$ photon ($0.3028 \times 10^{-11} erg$) 
%for a case B recombination, n=10000 $\rm cm^{-3}$, and T=10000 K, 
and $n$ is the gas density. 

\subsection{Definition of tails and $\rm H\alpha$ clumps}

The white-light MUSE images and $\rm H\alpha$-flux maps of our sample
galaxies are shown in Fig.~1. 
The black contour is the line we defined to have an
estimate of the "galaxy boundary", as described in Gullieuszik et
al. (in prep.). It is computed from the 
map of the stellar continuum in the H$\alpha$ region. As a starting 
point we used the isophote with a surface brightness 1$\sigma$ above the average sky background level.
This isophote does not have an elliptical symmetry
because of the (stellar and gaseous) emission from the stripped gas
tails. For this reason, 
we fit an ellipse to the undisturbed side of the isophote and we 
replaced the isophote on the disturbed side with the ellipse.
In the following we will refer to the galaxy emission outside of the
resulting contour as ``tail''.
%stripped gas or tails as those regions beyond the resulting line.
%Hereafter we will name ``tail'' all the galaxy emission 
%outside of these stellar contours. 

All our galaxies present bright $\rm H\alpha$ knots with logarithmic $\rm H\alpha$
surface brightness typically between -16.5 and -15 $\rm erg \, s^{-1}
\, cm^{-2} \, arcsec^{-2}$. As discussed throughout this paper, these are star-forming
clumps embedded in regions of more diffuse emission.
We identify these clumps as described in detail in \cite{Poggianti2017a} 
using a shell script including IRAF and FORTRAN routines,
searching the local minima of the laplace+median filtered $\rm
H\alpha$ MUSE image. % whose robustness index exceeds a given value.
The size of these clumps (i.e. their radius, having assumed circular symmetry) is estimated 
considering outgoing shells until the average counts reach a threshold
value that defines the underlying
diffuse emission. This radius is therefore an isophotal radius of the
kind derived with an isophote method used also in other works 
\citep{Wisnioski2012}. Isophotal radii are larger than so-called
``core-radii'', which are derived as 1$\sigma$ widths of Gaussian
profile fits (see Fig.~1 in \citealt{Wisnioski2012}).
For this work the $\rm H\alpha$ flux within each clump is measured 
including %the counts below the threshold set for 
the underlying diffuse emission in which the clump is embedded
and equally sharing the counts of spaxels belonging to overlapping clumps.
% Here we use L1, so we include the counts below the threshold set for
% the diffuse emission

\begin{figure*}
\centerline{\includegraphics[width=3.5in]{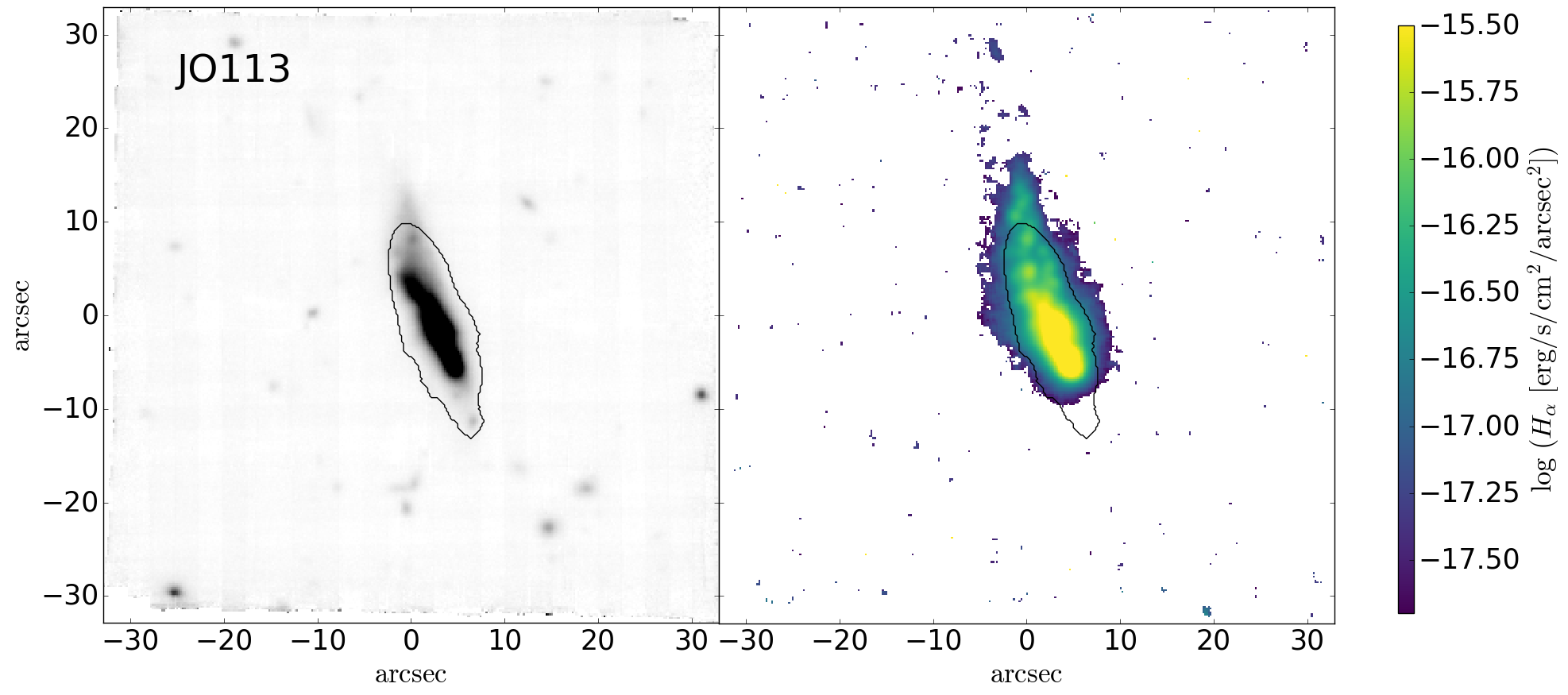}\includegraphics[width=3.5in]{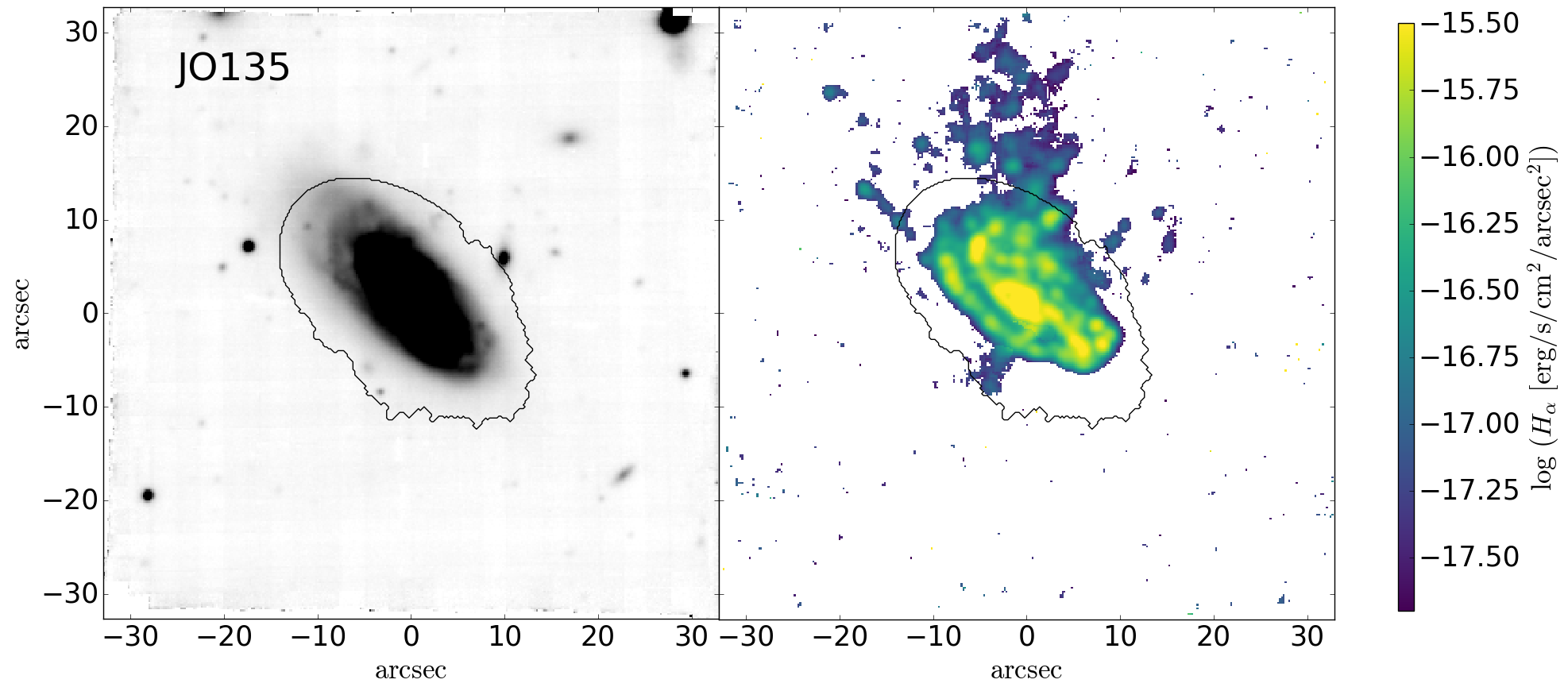}}
\centerline{\includegraphics[width=3.5in]{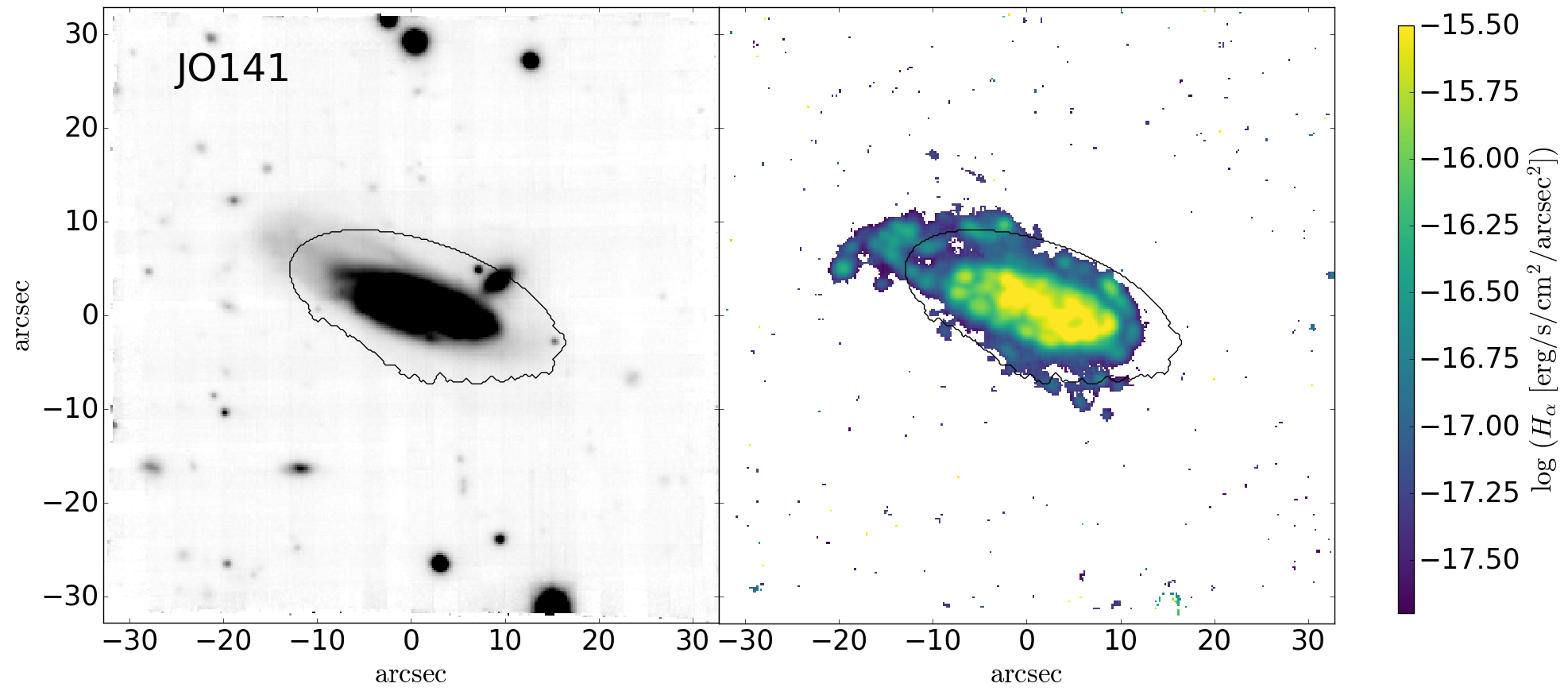}\includegraphics[width=3.5in]{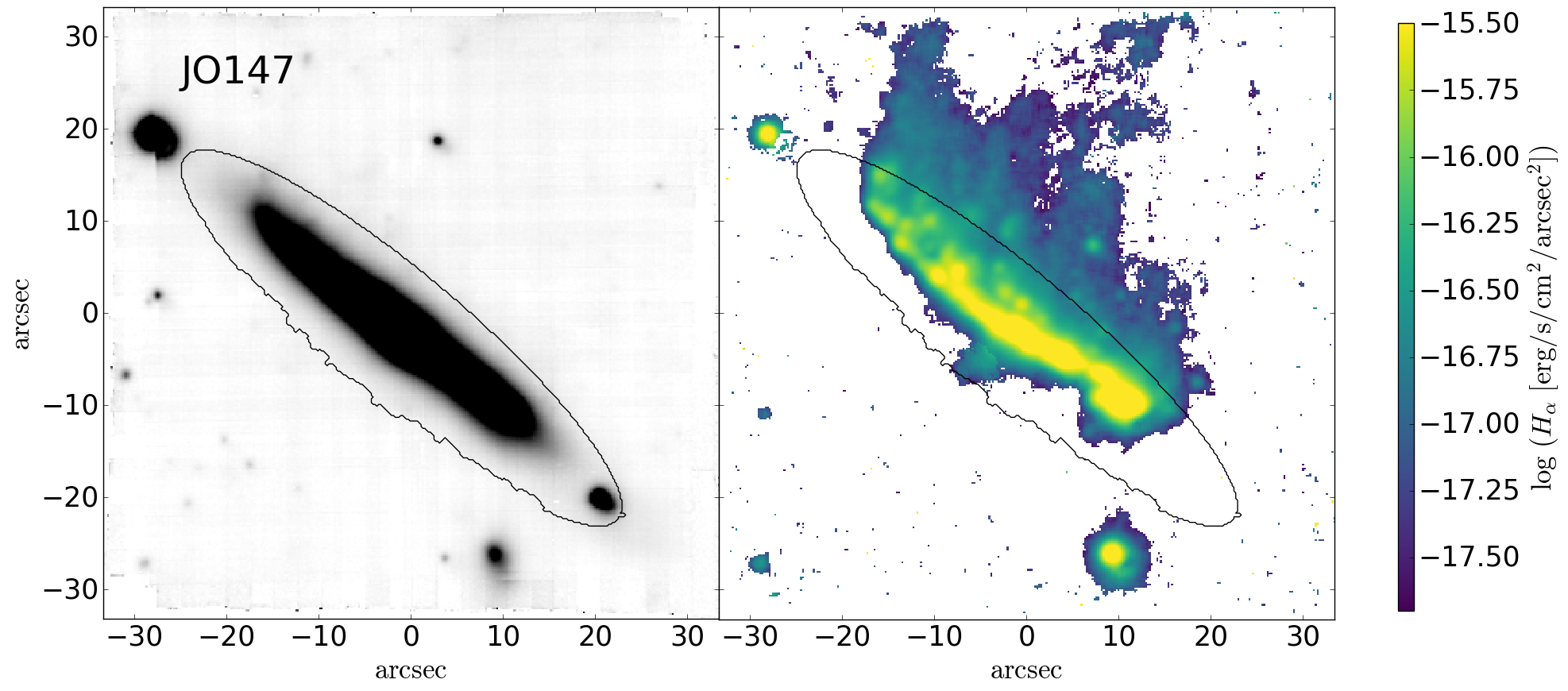}}
\centerline{\includegraphics[width=3.5in]{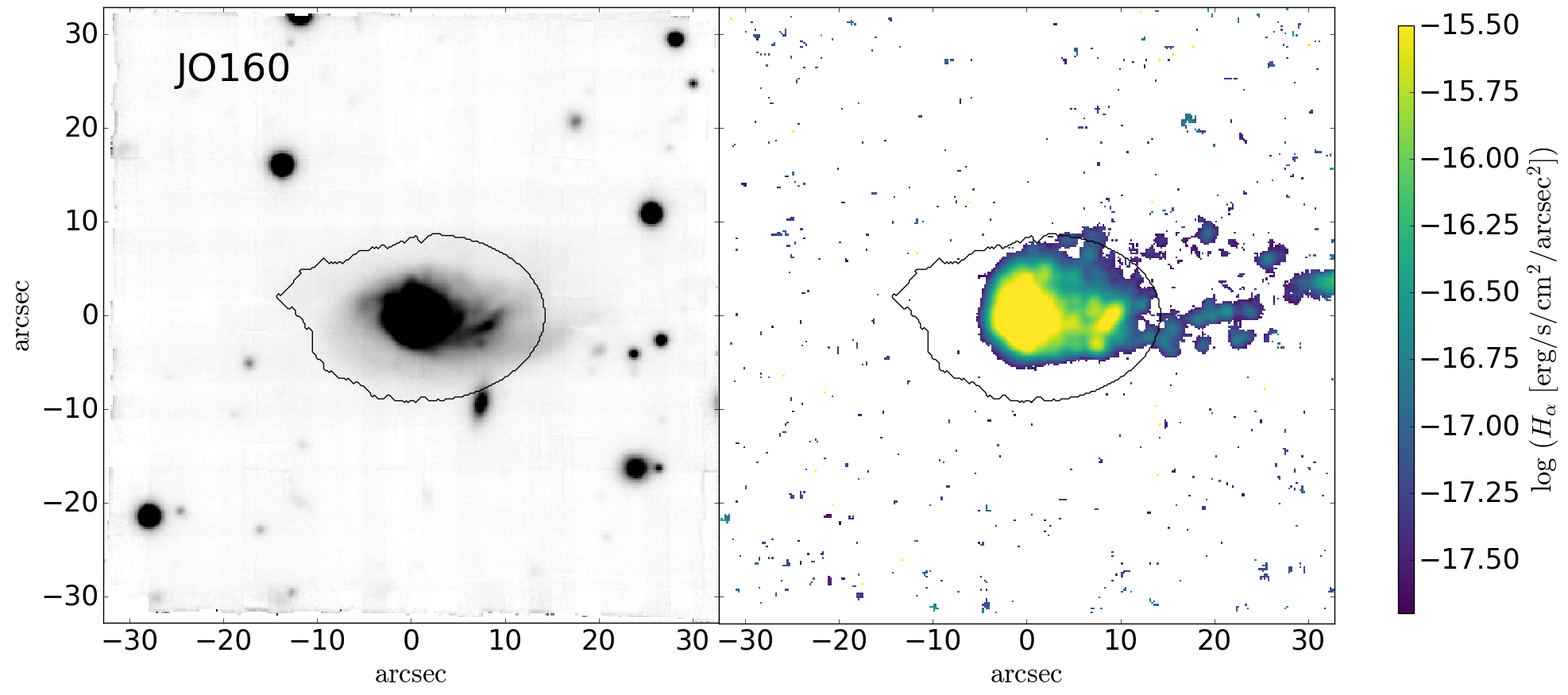}\includegraphics[width=3.5in]{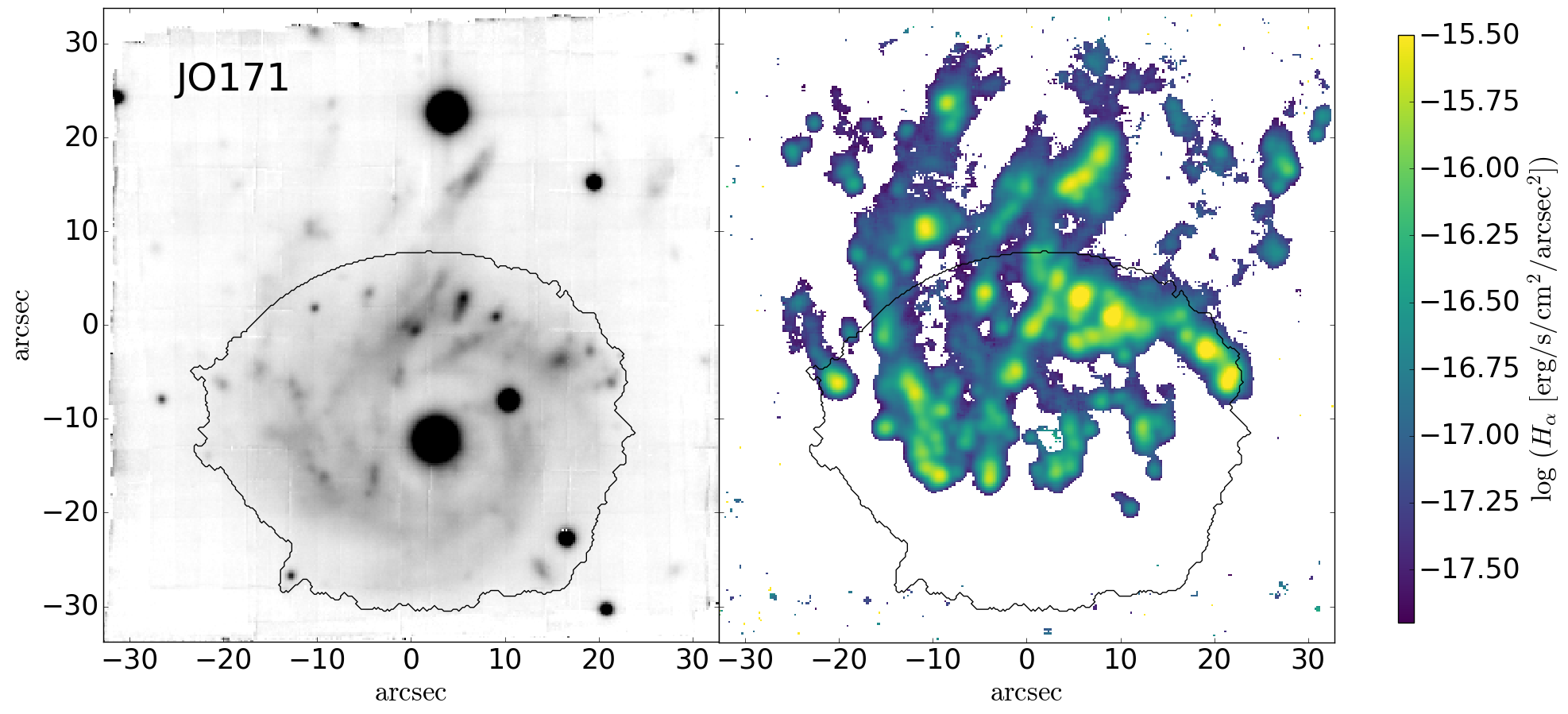}}
\centerline{\includegraphics[width=3.5in]{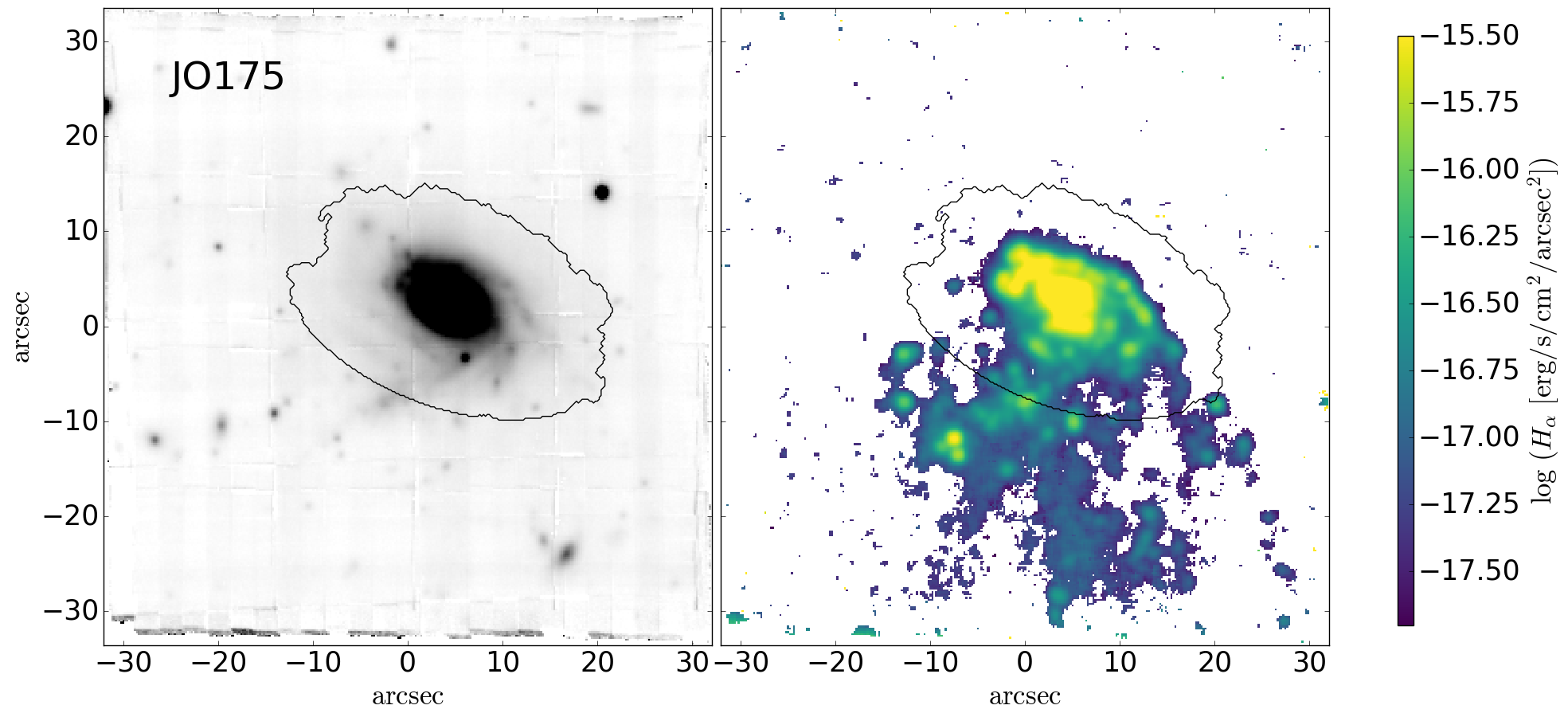}\includegraphics[width=3.5in]{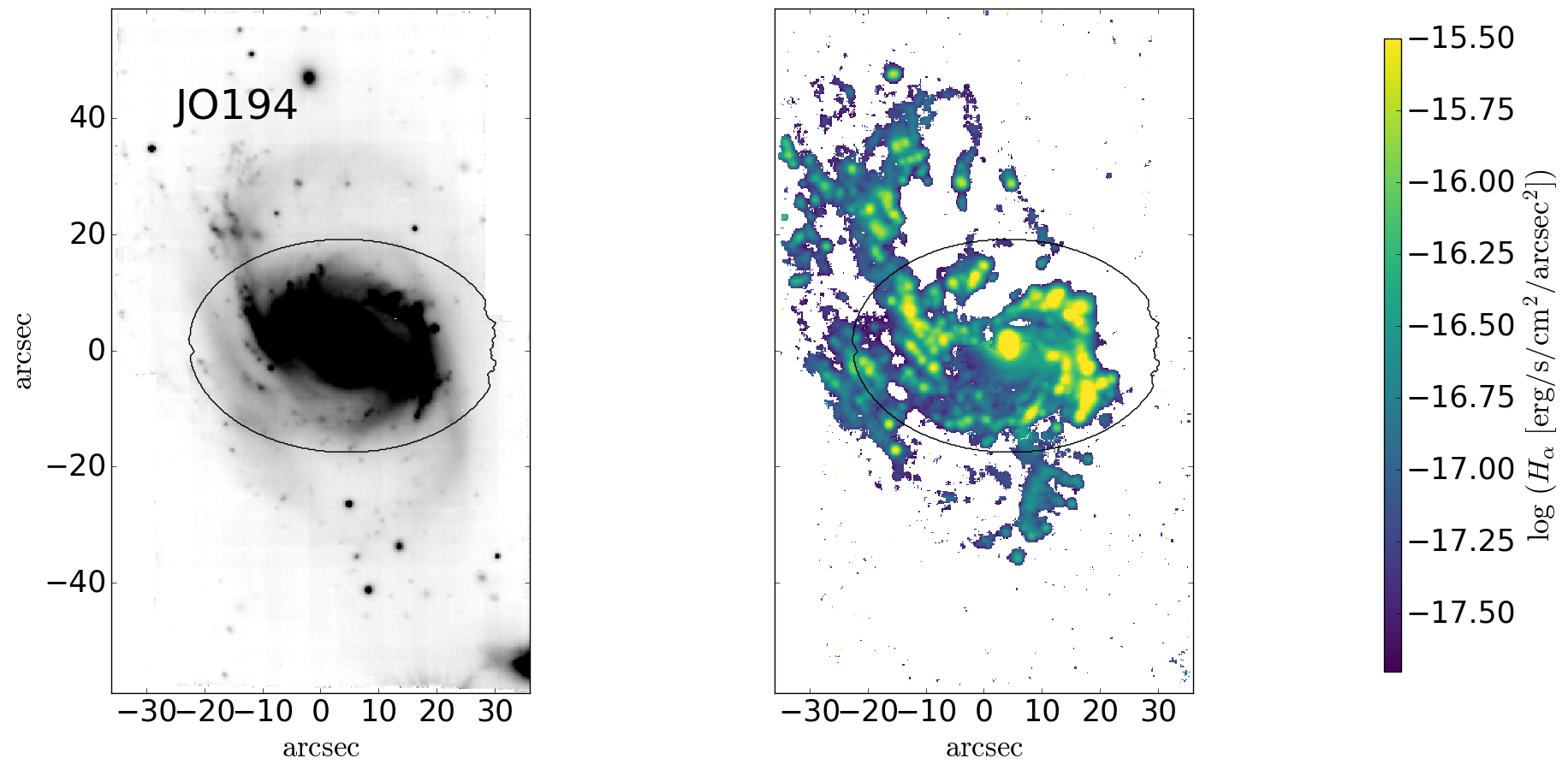}}
%\centerline{\includegraphics[width=2.9in]{JO135_white.png}\hspace{-1cm}\includegraphics[width=3.1in]{plotline_JO135_Ha_smo5_sn40_noflag.png}}
%\centerline{\includegraphics[width=2.9in]{JO141_white.png}\hspace{-1cm}\includegraphics[width=3.1in]{plotline_JO141_Ha_smo5_sn40_noflag.png}}
%\centerline{\includegraphics[width=2.9in]{JO147_white.png}\hspace{-1cm}\includegraphics[width=3.1in]{plotline_JO147_Ha_smo5_sn40_noflag.png}}
\caption{MUSE white image (left) and $\rm H\alpha$ flux map (right) for
each galaxy. Black contours are the stellar contours described in \S3.1}
\end{figure*}

\begin{figure*}
\centerline{\includegraphics[width=3.5in]{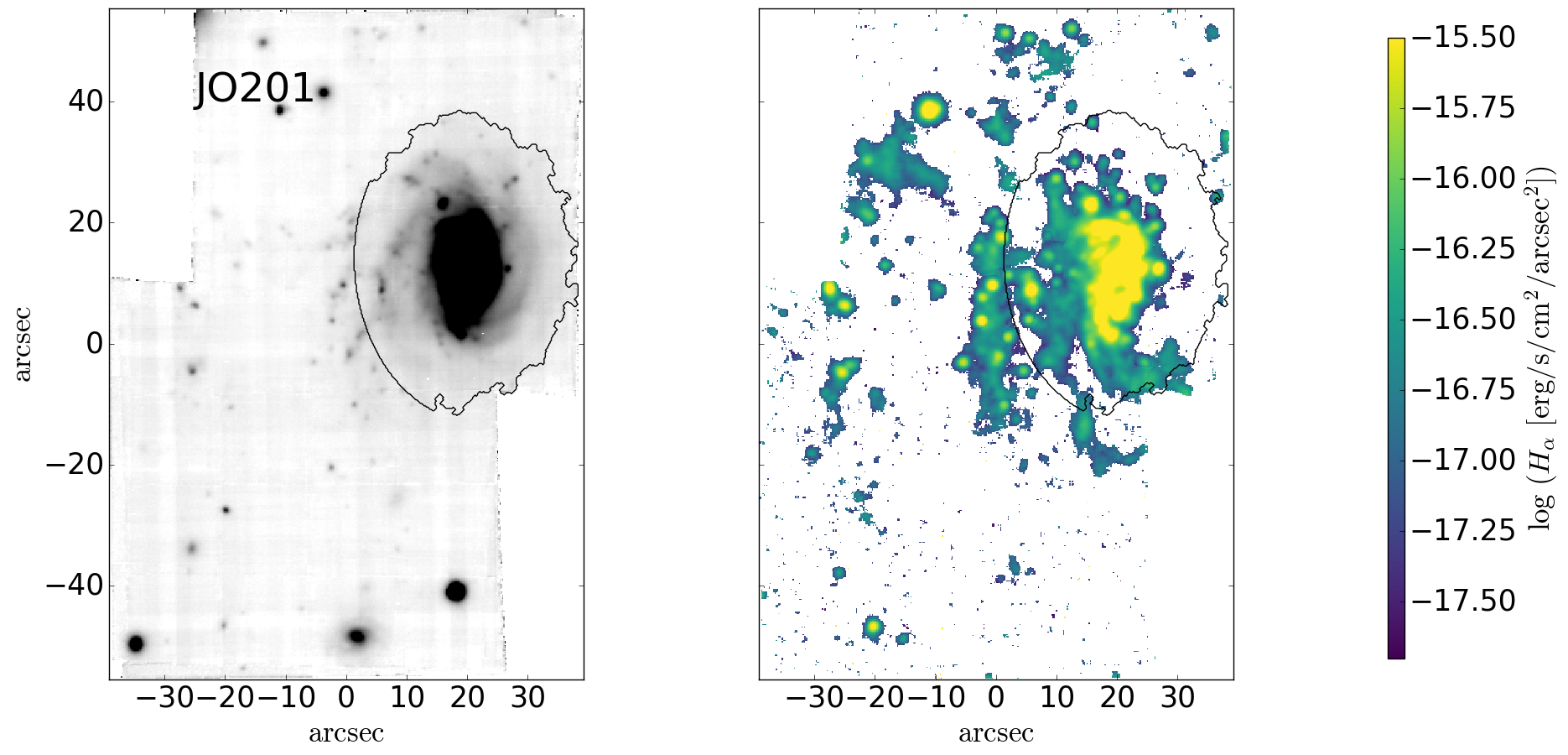}\includegraphics[width=3.5in]{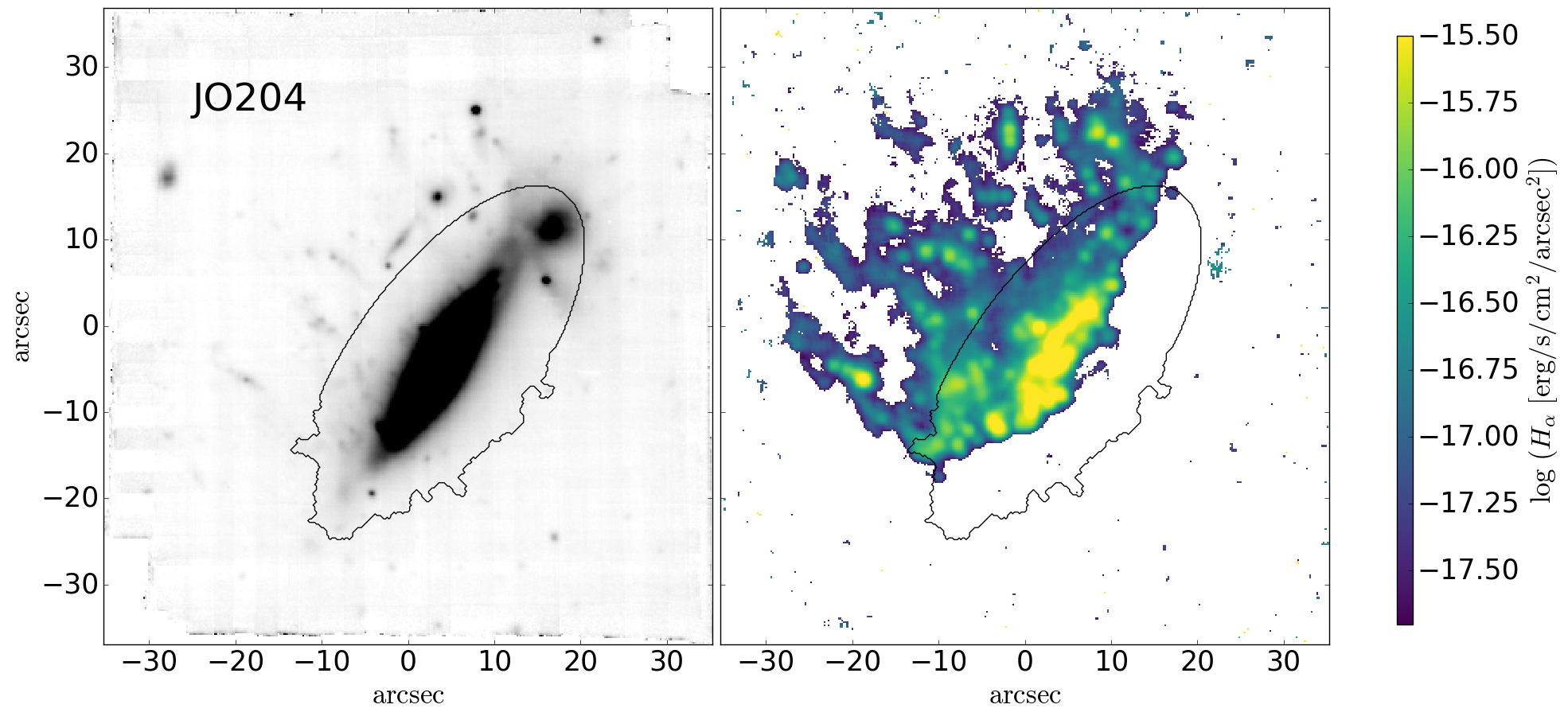}}
\centerline{\includegraphics[width=3.5in]{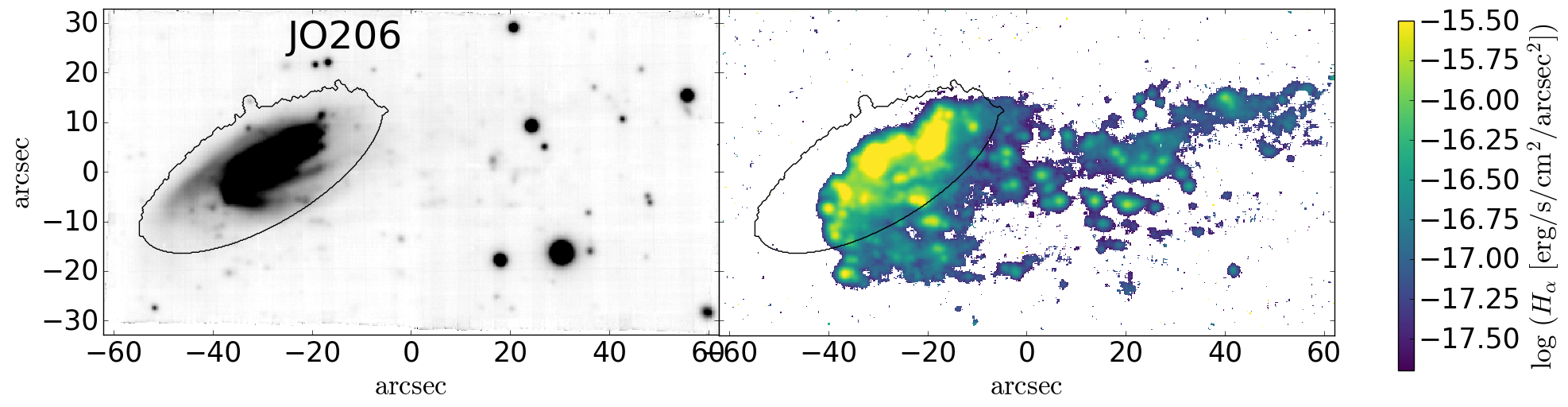}\includegraphics[width=3.5in]{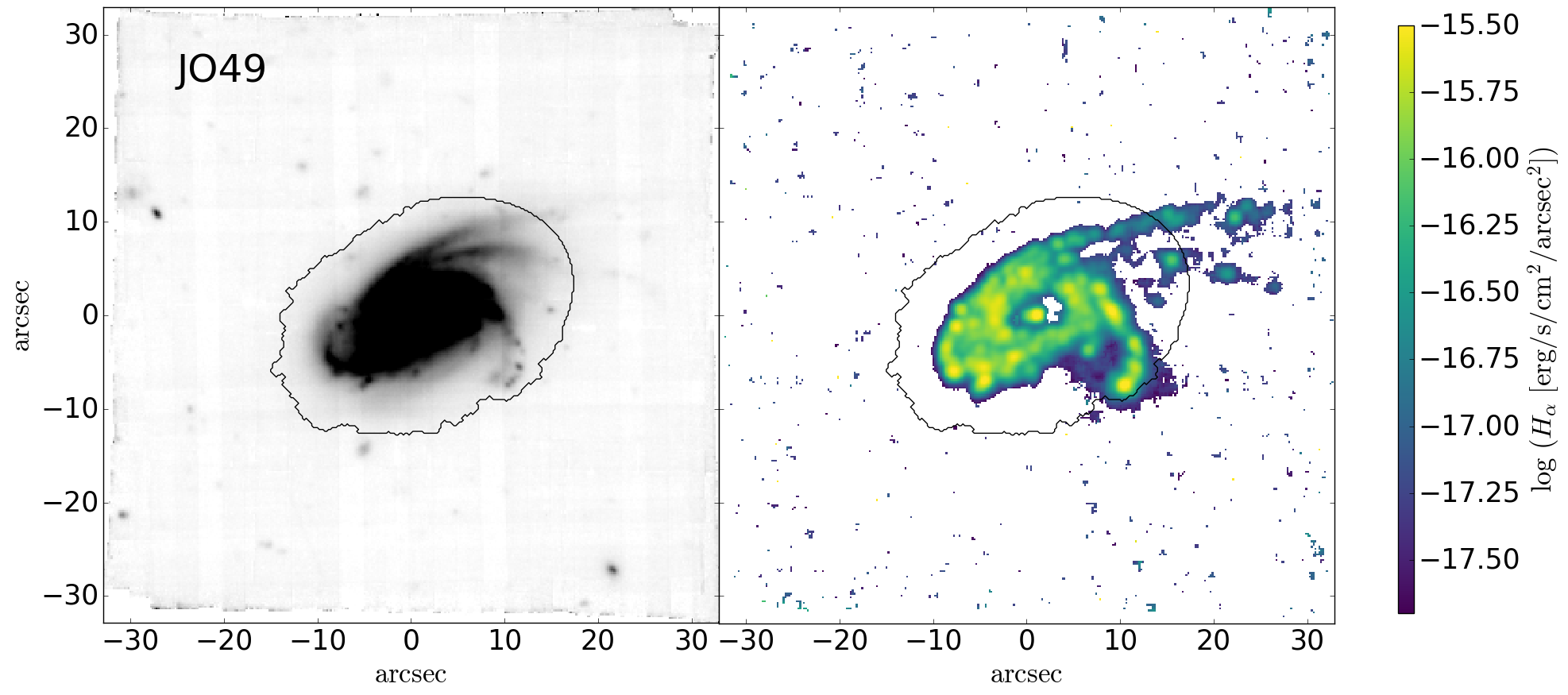}}
\centerline{\includegraphics[width=3.5in]{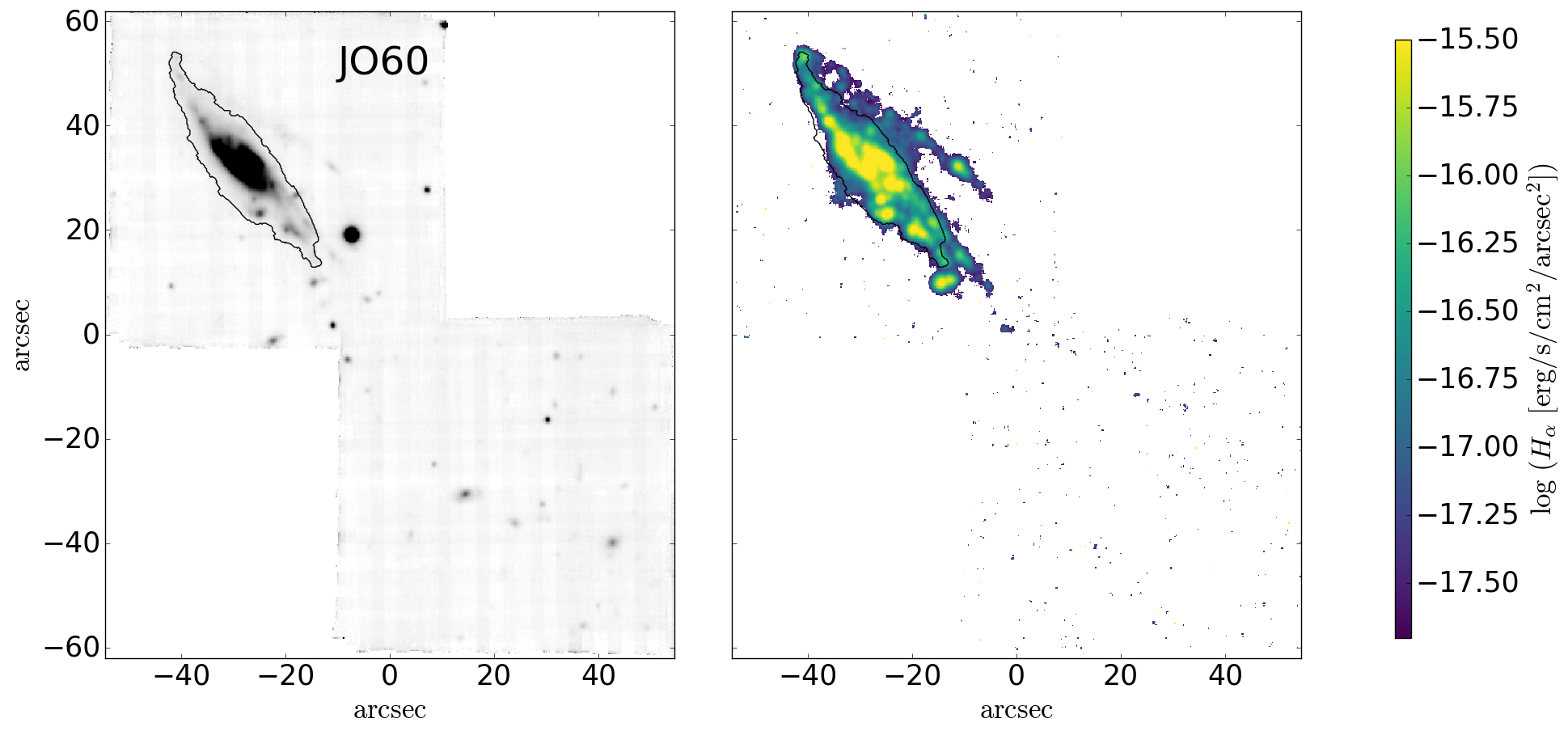}\includegraphics[width=3.5in]{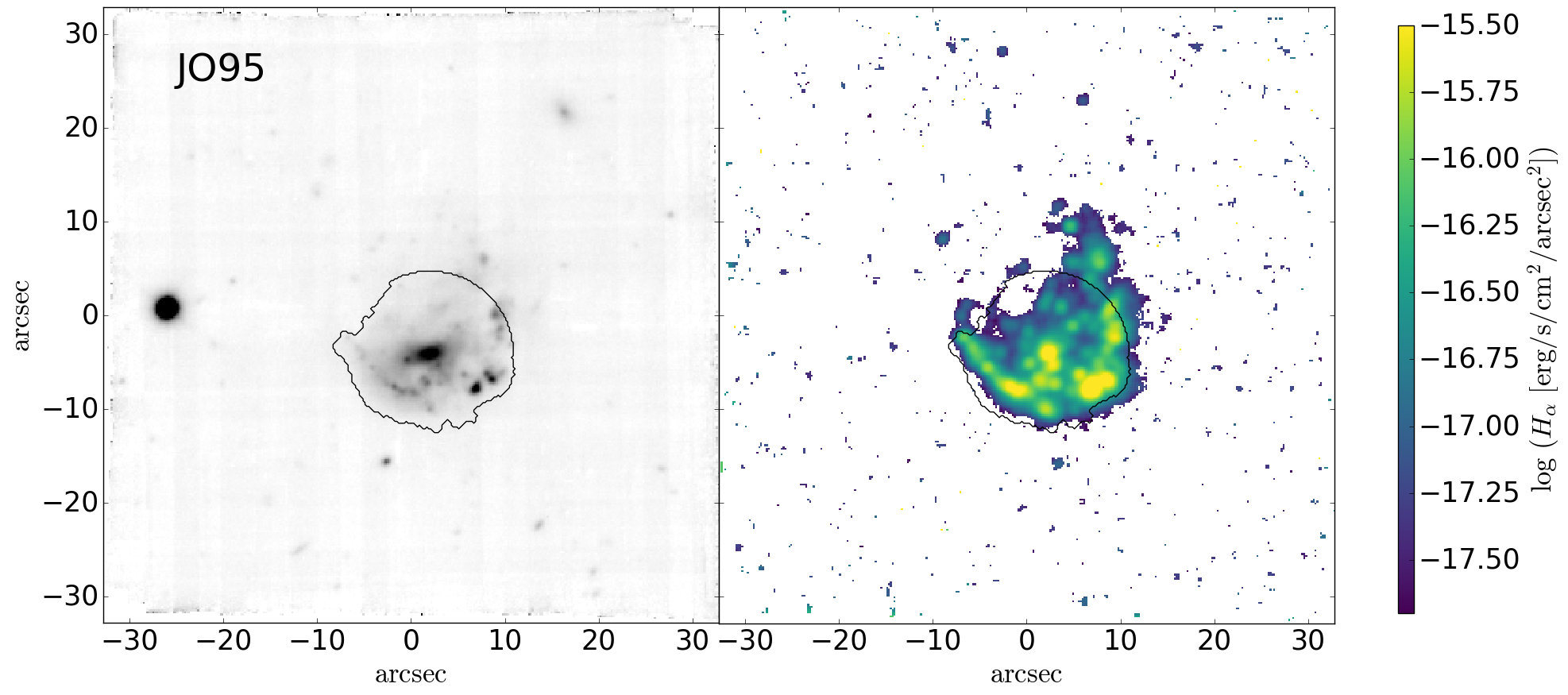}}
\centerline{\includegraphics[width=3.5in]{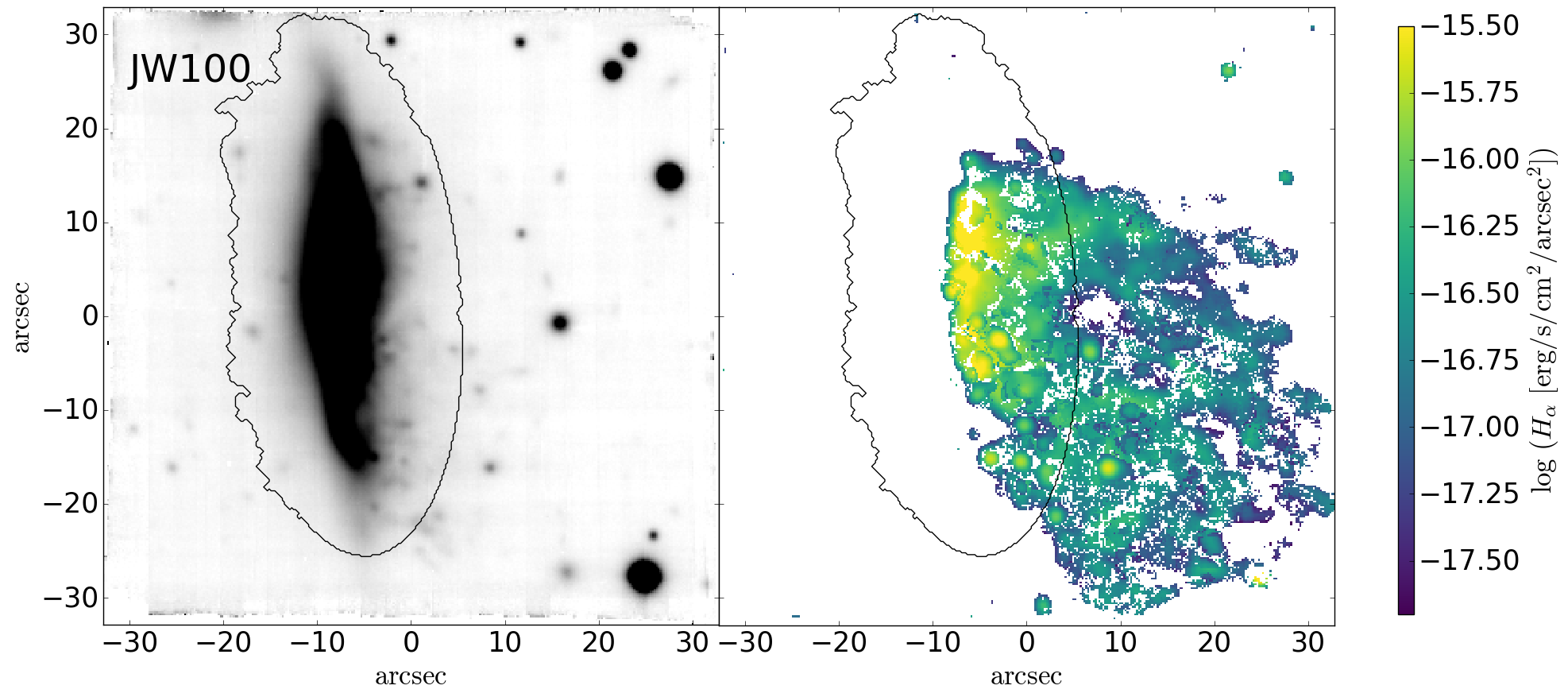}\includegraphics[width=3.5in]{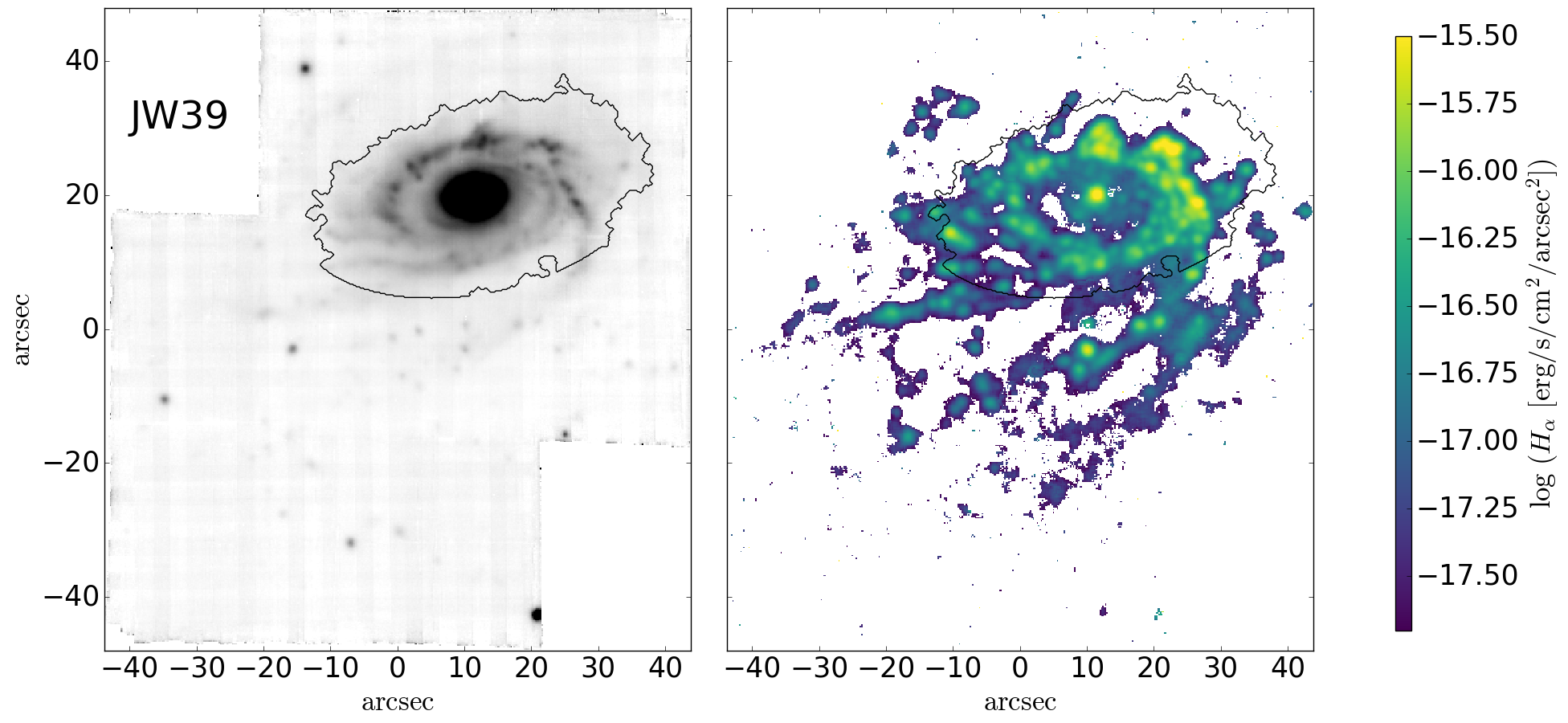}}
%\centerline{\includegraphics[width=2.9in]{JO135_white.png}\hspace{-1cm}\includegraphics[width=3.1in]{plotline_JO135_Ha_smo5_sn40_noflag.png}}
%\centerline{\includegraphics[width=2.9in]{JO141_white.png}\hspace{-1cm}\includegraphics[width=3.1in]{plotline_JO141_Ha_smo5_sn40_noflag.png}}
%\centerline{\includegraphics[width=2.9in]{JO147_white.png}\hspace{-1cm}\includegraphics[width=3.1in]{plotline_JO147_Ha_smo5_sn40_noflag.png}}
\contcaption{}
\end{figure*}

\section{Results}

\subsection{Ionization mechanisms}

\begin{table}
\centering  
%\caption{Sample galaxies} % in the MUSE spectral range.} %\label{tab:decimal}
\caption{Percentage of $\rm H\alpha$ tail emission due to SF,
  Composite, AGN and LINER according to the NII DD, and percentage of 
  SF according to OI and SII DDs.}
\begin{tabular}{lrrrrrr}
\hline  
$ID_{P16}$ & $\rm SF_{NII}$ & $\rm Comp._{NII}$ & $\rm AGN_{NII}$ &
$\rm LINER_{NII}$ & $\rm SF_{SII}$ & $\rm SF_{OI}$ \\
\hline  
   JO113  &  99.1 & 0.9  & 0  &  0       &   72.3   &   19.3   \\    
   JO135  &  52.6 & 17.7 &20.2& 9.5   &  68.1   &  35.7    \\     
   JO141  &  93.0 & 7.0  & 0  &  0       &   83.9   &   47.3  \\     
   JO147  &  59.6 & 37.6 & 0  & 2.8     &   92.6   &  62.7    \\    
   JO160  &  93.9 & 6.1  & 0  &  0       &   77.7   &   12.8   \\    
   JO171  &  93.8 & 6.2  & 0  &  0        &   96.7   &  63.0    \\    
   JO175  &  87.7 & 12.3 & 0  &  0       &   91.2   &  49.7    \\    
   JO194  &  66.1 & 33.8 & 0  &  0       &   97.8   &  86.1    \\    
   JO201  &  94.1 & 5.9  & 0  &  0        &   99.2   &  77.3    \\    
   JO204  &  77.2 & 13.8 & 8.9& 0.2    &  85.4   &   70.2    \\   
   JO206  &  95.6 & 4.4  & 0  & 0        &   97.1   &   56.6   \\    
   JO49   &  97.2 & 2.7  & 0  & 0.1       &   99.3   &  77.1    \\    
   JO60   &  96.1 & 3.8  & 0 & 0.1     &   87.8   &   66.9   \\    
   JO95   &  99.7 & 0.3  & 0  &  0        &   89.1   &   51.6   \\    
   JW100  &  --   &  --  &--  & --        &   84.4  & 11.4     \\    
   JW39   &  89.2 & 10.8 & 0  &  0        &   97.1   &  73.5    \\    
\hline  
\end{tabular}
\end{table}

To investigate the gas ionization mechanism we employ
three standard diagnostic diagrams (hereafter DD):
[OIII]5007/$\rm H\beta$ vs [NII]6583/$\rm H\alpha$ (NII DD), [OIII]5007/$\rm H\beta$ vs [SII]6717,6731/$\rm H\alpha$ (SII
DD) and [OIII]5007/$\rm 
H\beta$ vs [OI]6300/$\rm H\alpha$ (OI DD).\footnote{For JW100 we have excluded from our analysis the [NII]
  line which is affected by a sky line.} 
To separate in these diagrams the regions powered by Star-formation,
Composite (SF+LINER/AGN), AGN and LINER emission 
we adopt the division lines by \cite[K03]{Kauffmann2003}, \cite[K01]{Kewley2001}
\cite[K06]{Kewley2006} and \cite[SB10]{Sharp2010}.

Fig.~2 presents all three DDs of individual spaxels of each galaxy as well as the  galaxy
map color-coded by ionization mechanism. Note that only spaxels in the
tails (those outside of the stellar contours) are plotted in the DD,
while the map displays both tails and disks.

Only spaxels with a S/N$>3$ in all the four lines used in each diagram are
considered in Fig.~2. As a consequence, we can assess the ionization
source only for a fraction of the spaxels: the median fraction of $\rm
H\alpha$ luminosity in the tails for which the measurement is possible in our sample
is 58.0\% with a dispersion of 18.7\%
%$\sigma_{median}=18.7$\% 
for the NII DD , 63.6\%$\pm14.2$ for SII DD and 
57.2\%$\pm15.5$ for OI DD. Thus, the
reader should keep in mind that for
about 40\% of the  $\rm H\alpha$ luminosity in the tails, the
ionization mechanism cannot be determined from DDs because one or more
lines are too faint. 

We note that for a few of our galaxies there are regions where a
single component Gaussian does not provide a good fit to the observed
spectrum, due to the presence of gas at different velocities along the
line of sight, or emission around an AGN. The two galaxies for which
this effect is more important in the tails are JW100 and JO201, for
which detailed diagnostic diagrams based on two Gaussian component
fits were presented in \cite{Poggianti2017b} and Bellhouse et
al. (submitted), respectively. Since we have verified that in the
tails (that are the subject of the current paper)  the 
single component fit yields DD results very similar to
the double component, hereafter we only show the single component
results.

From the spaxel-by-spaxel analysis of Fig.~2, a number of conclusions can be drawn.

%First of all, different DDs can give different ionization mechanisms for the same
%spaxel. In particular, while the 
%NII DD and and SII DD generally show an excellent agreement, 
%the OI DD .

%This is a well known problem ({\bf MARIO, REFERENCES?}).
% in a
%region of the diagram that is believed to be occupied by shocks.
%{\bf MARIO, WHAT CAN WE SAY HERE?}

For the majority of galaxies, the 
NII DD and and SII DD generally show an excellent agreement, 
and they indicate star
formation as the predominant ionization mechanism in the 
tails (in JO113, JO141,  JO160, JO171, JO175, JO201, JO204,
JO206, JO49, JO60, JO95, JW100, JW39 plus
JO147 and JO194 from SII DD)
or a SF+Composite origin (NII DD for JO194, and especially for JO147).

In contrast, a non-negligible fraction of spaxels in 
the tails of all galaxies has an [OI]/$\rm H\alpha$ ratio that is too
high for being powered by SF, though also the [OI] DD indicates a
significant contribution from SF in the tails at the location of the
brightest  $\rm H\alpha$ clumps. The [OI]-LINER-like
emission dominates the tails of JO147 and JW100.
%, for which also the
%NII DD indicates a composite origin (INCLUDE NII JW100??)

Thus, the OI DD suggests a larger contribution from ``LINER/AGN''
emission than the other two DDs, as previously found by e.g.
\cite{Yoshida2008,Fossati2016}.  
This is expected in the presence of shocks 
\citep[see e.g.][]{Rich2011}, which are particularly effective in triggering the 
%[SII]/Ha and 
[OI]/Ha ratios, compared to [NII]/Ha \citep{Rich2015}. 
Interestingly, the LINER-like [OI] emission is mostly found 
in the regions \textit{surrounding} the bright $\rm H\alpha$ clumps,
and in those clumps with high gas velocity dispersion, as shown below.
The exact source of [OI] excitation in the LINER regions of the tails is
unknown. 
%Shocks might play a role, 
At their current resolution, simulations do not predict shocks in the stripped tails, and
we hypothesize that
thermal conduction at the boundaries where the stripped gas meets the hot ICM
might play an important role.

The dominant ionization mechanism in the tails, however, is clearly star formation.
According to the NII/SII/OI DDs, the median fraction of $\rm H\alpha$
luminosity in the tails powered by SF is 94(100 if
SF+Composite)/91/64\%, 
%(mean 97/55/89\%), 
ranging from galaxy to galaxy between 100/99/87\% and 66/68/12 (Table~2).
Thus, between $\sim$60\% and 100\% (depending on the DD employed)
of the overall tail $\rm H\alpha$ emission for which a
ionization mechanism can be identified is powered
by SF.

Finally, we note that all three DDs agree that
%Based on the same DDs, and as confirmed also by the OI DD, 
two of the galaxies (JO135 and JO204) present a ionization cone from
the central AGN that extends for several kpc and contributes to the
ionization of gas even in the tails \citep[see also][]{Poggianti2017b}. 
Nonetheless, also in these galaxies a large fraction of the tail $\rm
H\alpha$ emission appears to be powered by SF
%in these two galaxies still appears to be powered by SF 
(53\%/68\%/37\% in JO135 and
77\%/85\%/71\% in JO204 from the NII/SII/OI DDs, see Table~2).

%It is striking that in Fig.~2 the regions with the highest $\rm H\alpha$ surface
%brightness (the clumps) tend to be powered by SF
%according to all three diagrams with a few exceptions.
%In the following, we will 

\begin{figure*}
\centerline{\includegraphics[width=2.3in]{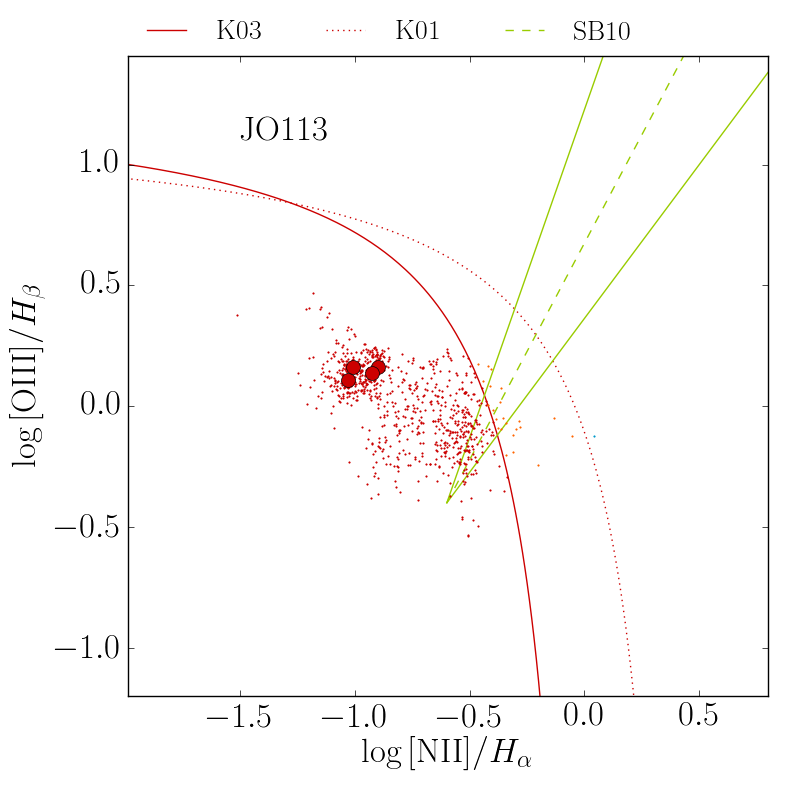}\includegraphics[width=2.3in]{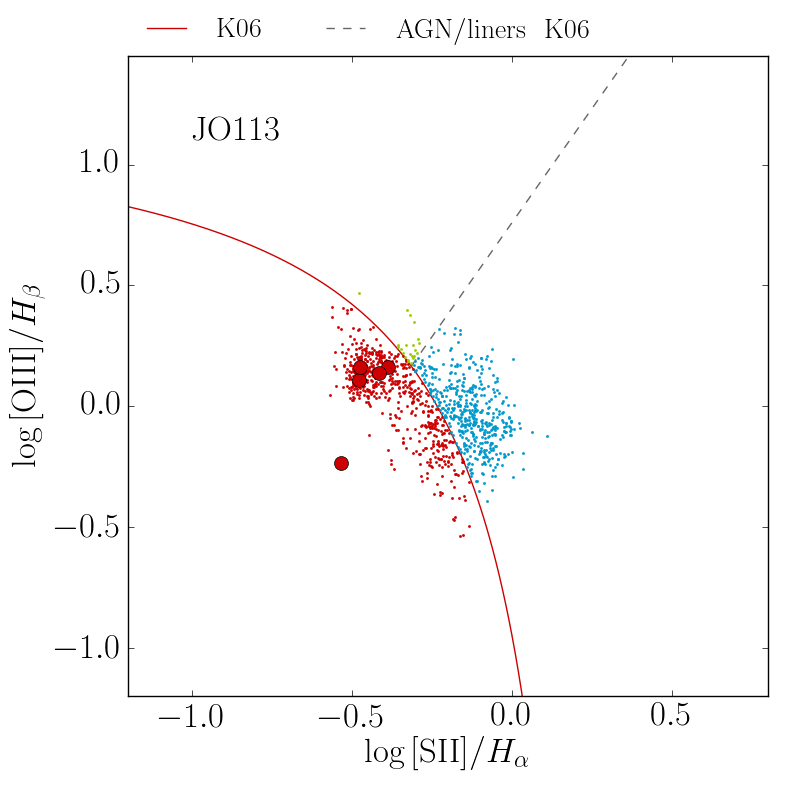}\includegraphics[width=2.3in]{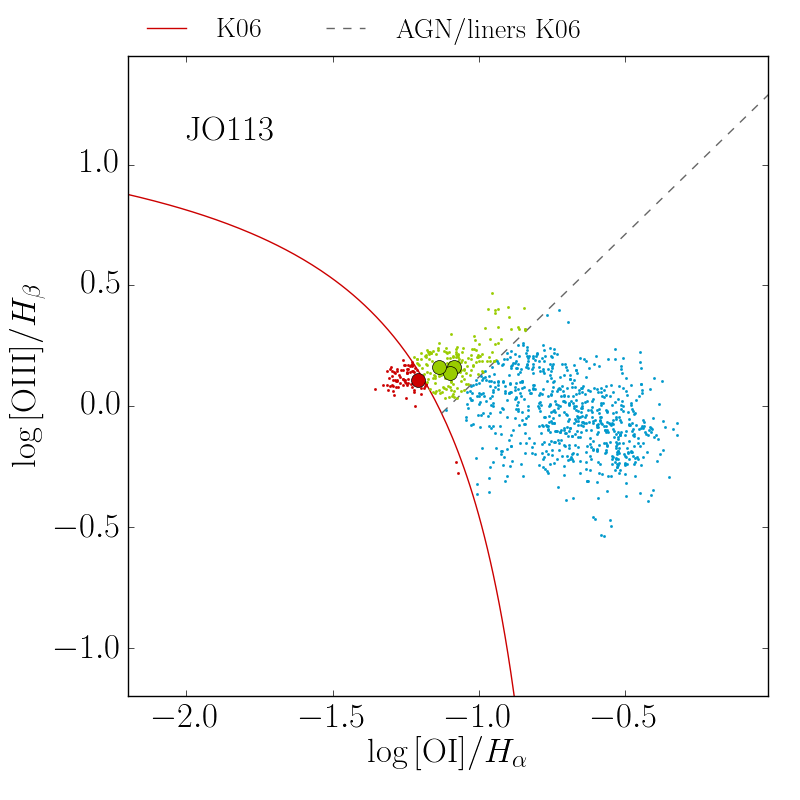}}
\centerline{\includegraphics[width=2.3in]{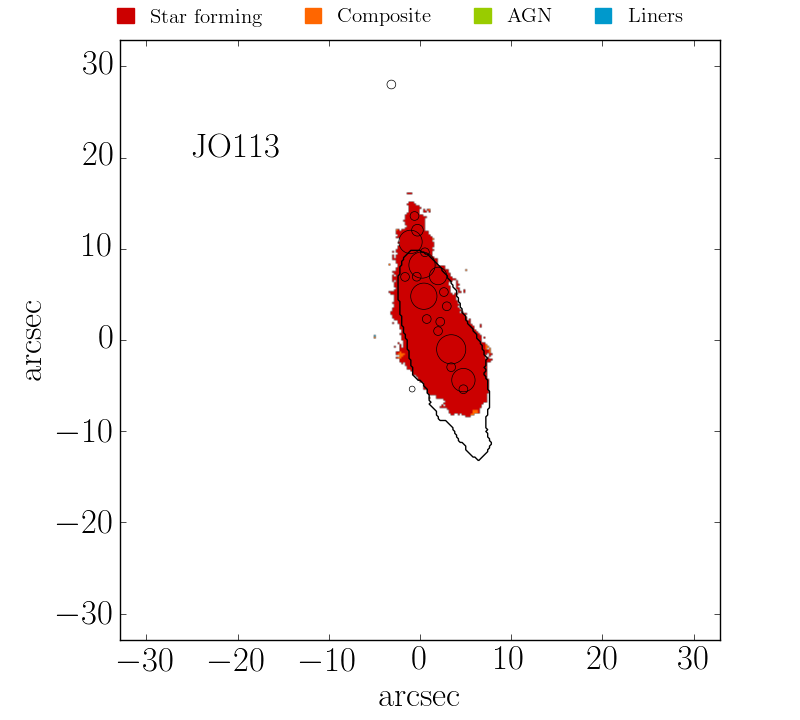}\includegraphics[width=2.3in]{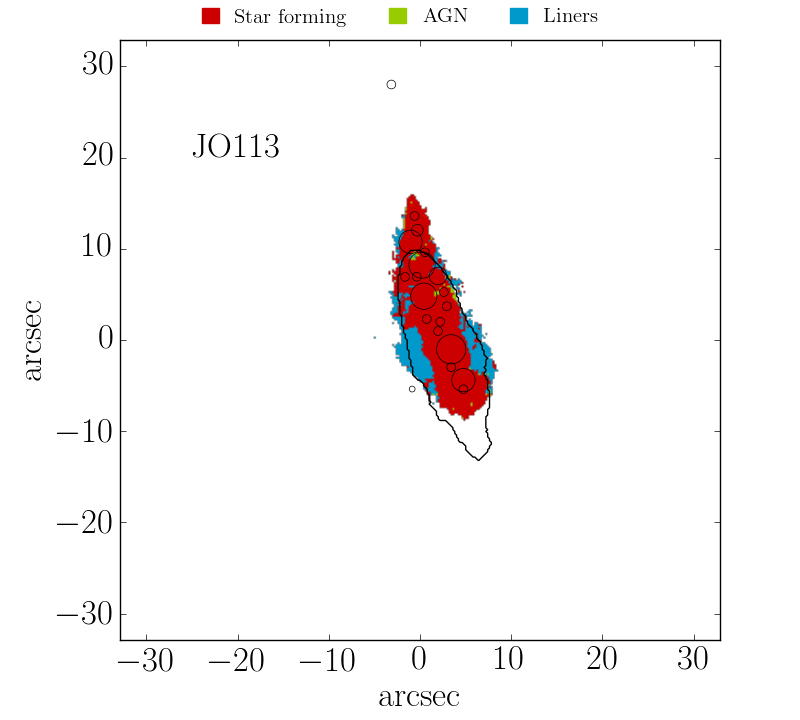}\includegraphics[width=2.3in]{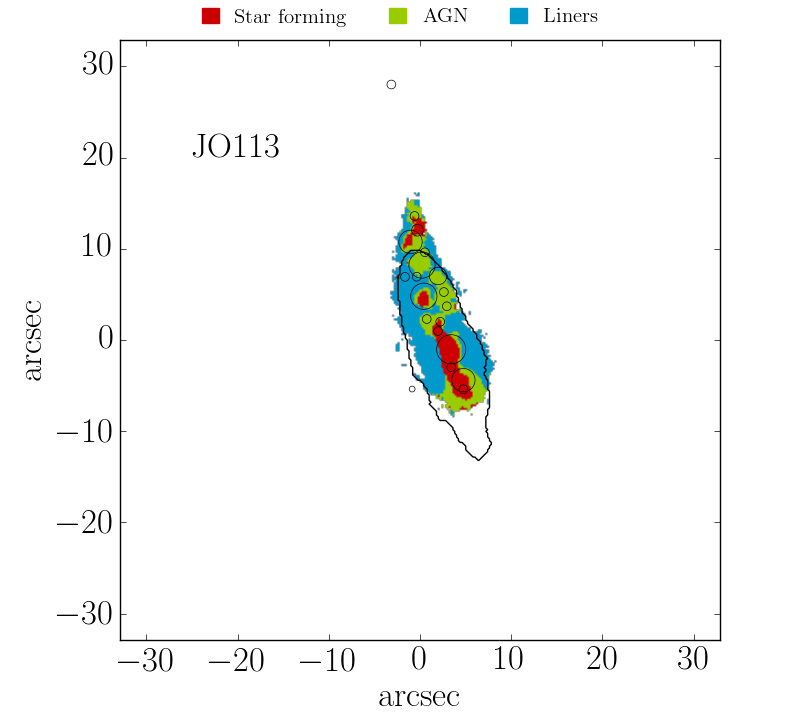}}
\centerline{\includegraphics[width=2.3in]{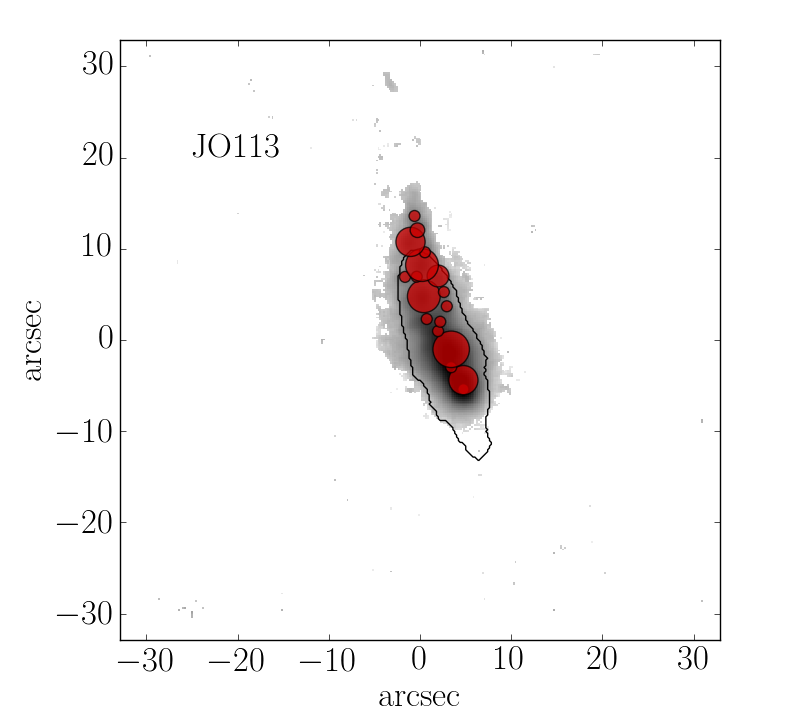}\includegraphics[width=2.3in]{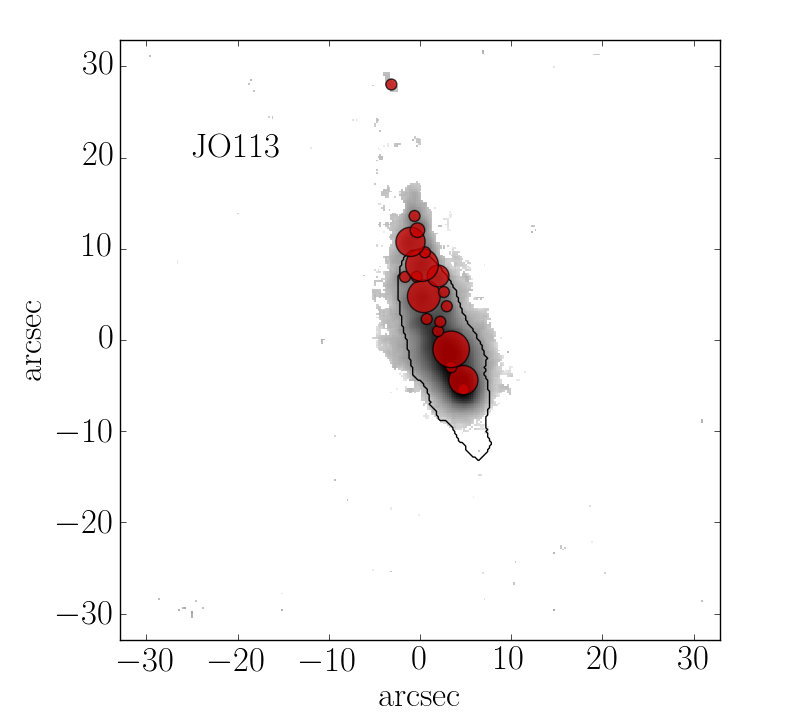}\includegraphics[width=2.3in]{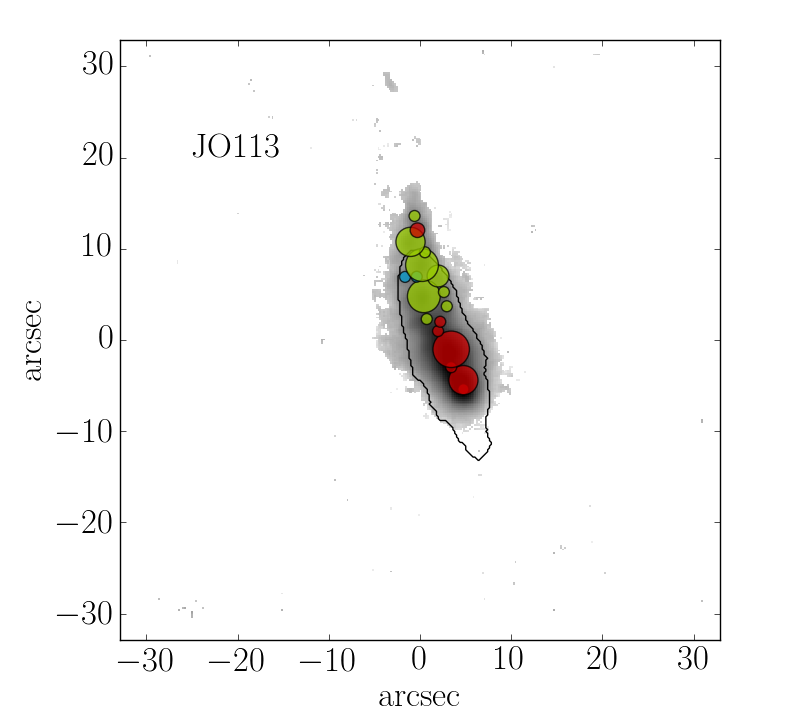}}
\caption{Diagnostic diagram results: NII DD (left), SII DD (center) 
  and OI DD (right). For each galaxy: 
Top panels: DDs for individual spaxels (small points) and 
  clumps (large circles); Middle panels: spaxel map color-coded for 
  ionization mechanism (see legend on top of middle panels) with clump 
  contours and stellar contours overplotted; Bottom panels:
  color-coded map of the clumps with stellar contour overplotted. The diagrams include only 
  spaxels and clumps in the tails, i.e. that are outside of the line contour showing the 
  stellar disk (see text for details). Only spaxels with a S/N$>3$ in 
  all the four lines used are plotted. See footnote 1) for
    JW100.}
\end{figure*}

\begin{figure*}
\centerline{\includegraphics[width=2.3in]{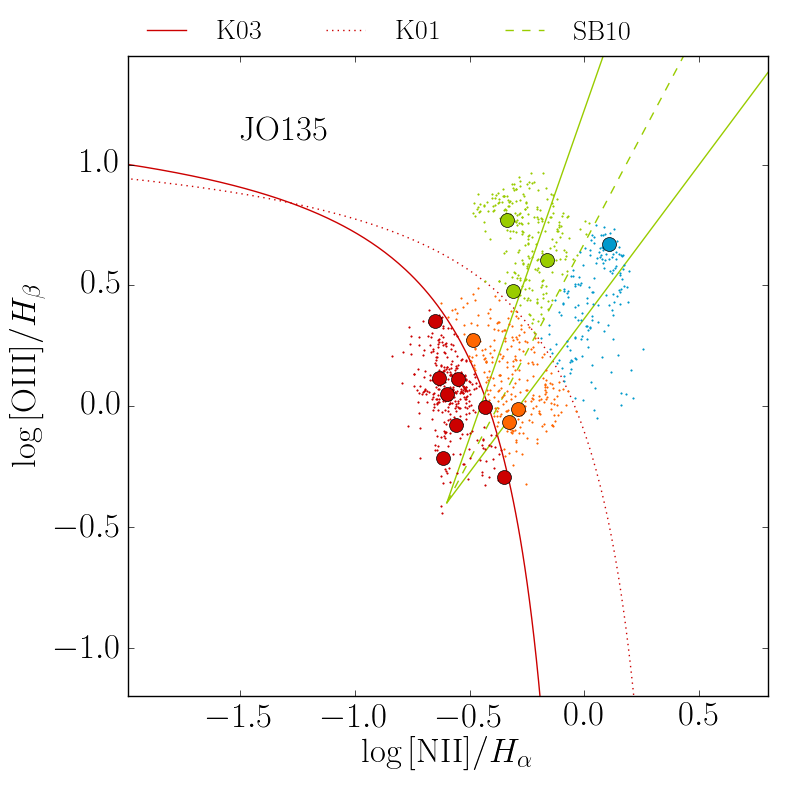}\includegraphics[width=2.3in]{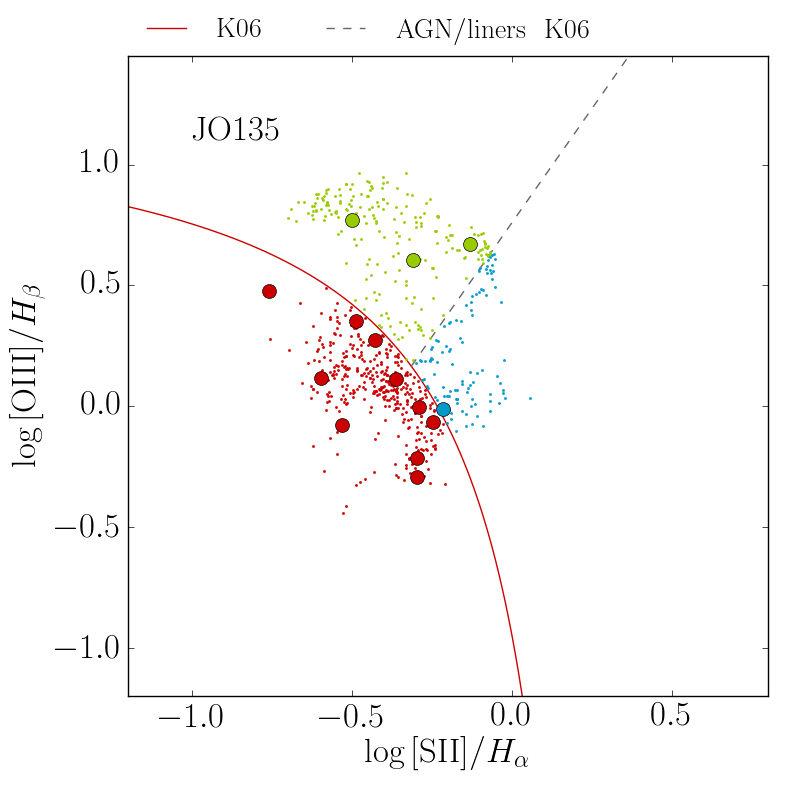}\includegraphics[width=2.3in]{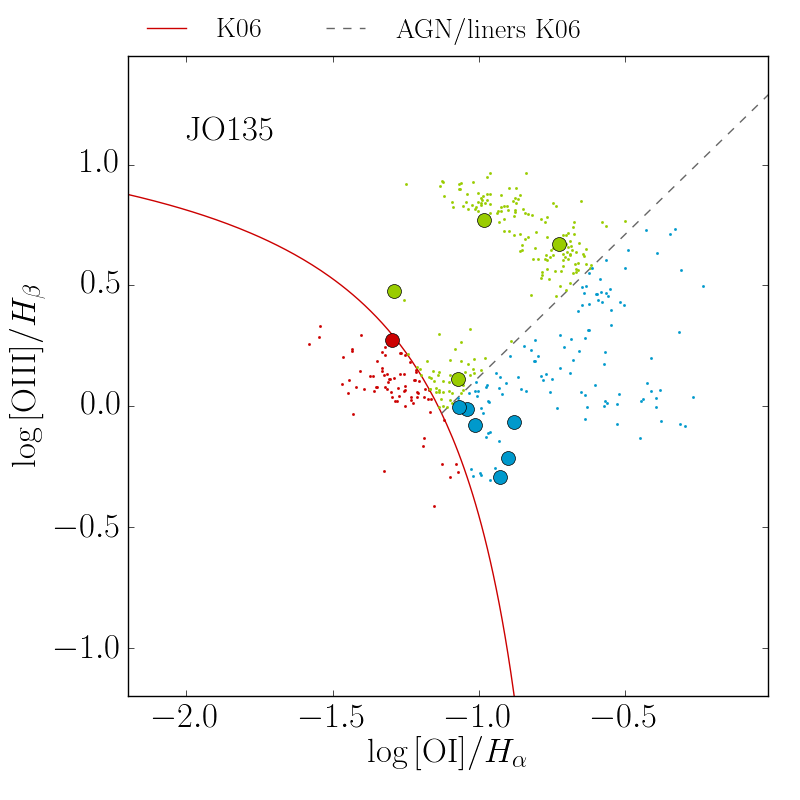}}
\centerline{\includegraphics[width=2.3in]{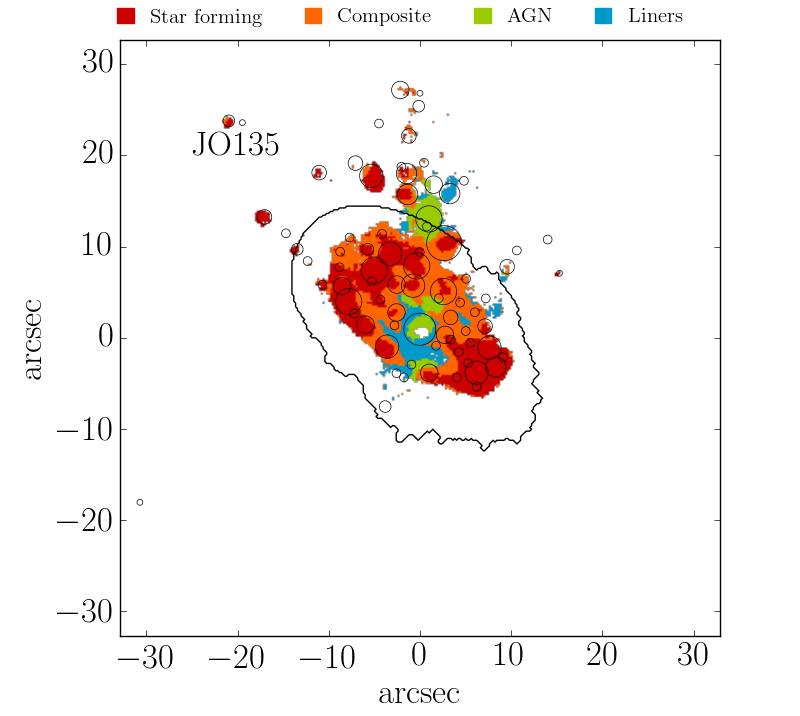}\includegraphics[width=2.3in]{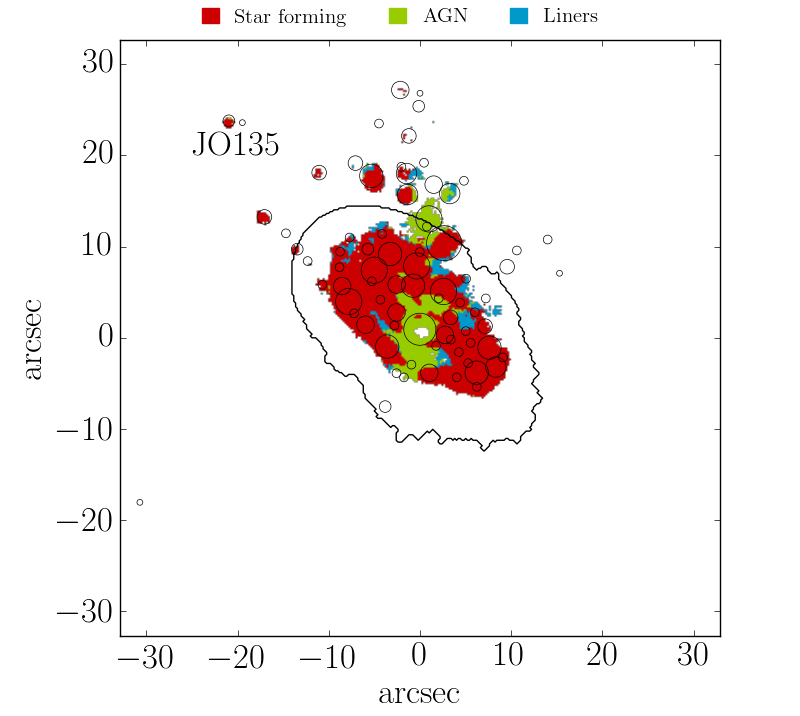}\includegraphics[width=2.3in]{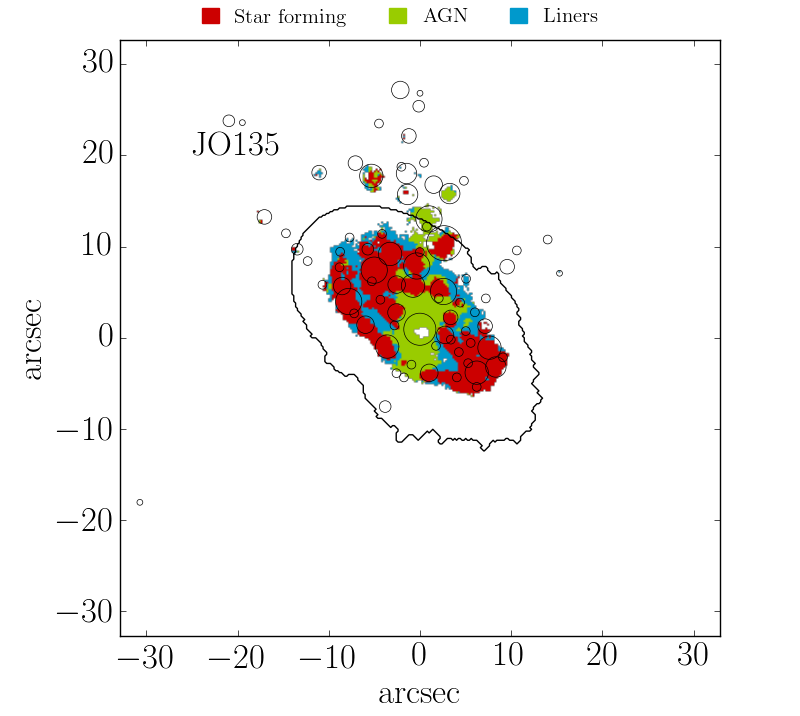}}
\centerline{\includegraphics[width=2.3in]{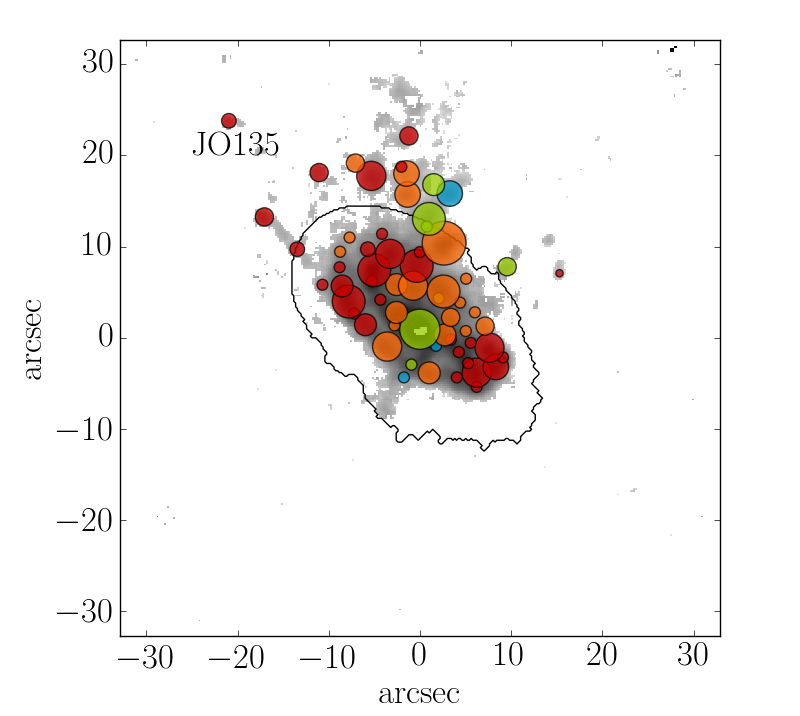}\includegraphics[width=2.3in]{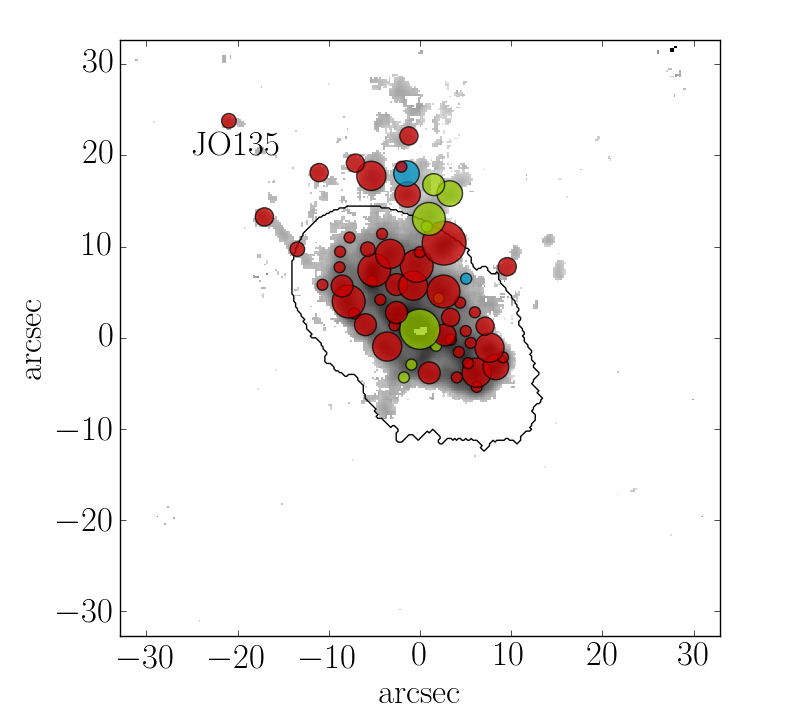}\includegraphics[width=2.3in]{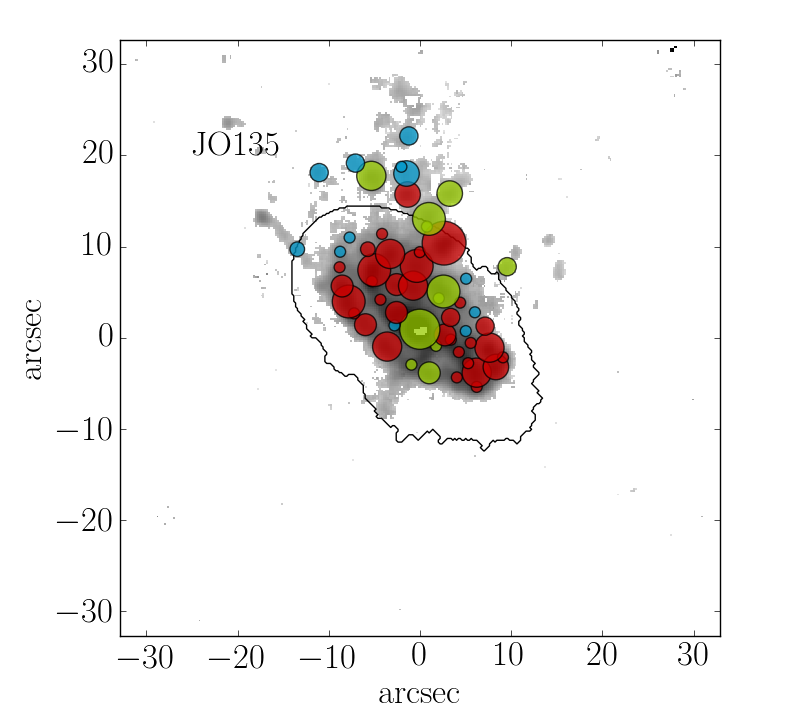}}
%\centerline{\includegraphics[width=1.7in]{JO113_bpt_blobs.png}\includegraphics[width=1.7in]{JO113_bpt_blobs_O.png}\includegraphics[width=1.7in]{JO113_bpt_blobs_S.png}}
\contcaption{}
%\caption{Diagnostic diagram results: NII DD (left), SII DD (center) 
%  and OI DD (right). For each galaxy: 
%Top panels: DDs for individual spaxels (small points) and 
%  clumps (large circles); Middle panels: spaxel map color-coded for 
%  ionization mechanism (see legend on top of middle panels) with clump 
%  contours and stellar contours overplotted; Bottom panels:
%  color-coded map of the clumps with stellar contour overplotted. The diagrams include only 
%  spaxels and clumps in the tails, i.e. that are outside of the line contour showing the 
%  stellar disk (see text for details). Only spaxels with a S/N$>3$ in 
%  all the four lines used are plotted. {\bf See footnote 1) for JW100.}}
\end{figure*}

\begin{figure*}
\centerline{\includegraphics[width=2.4in]{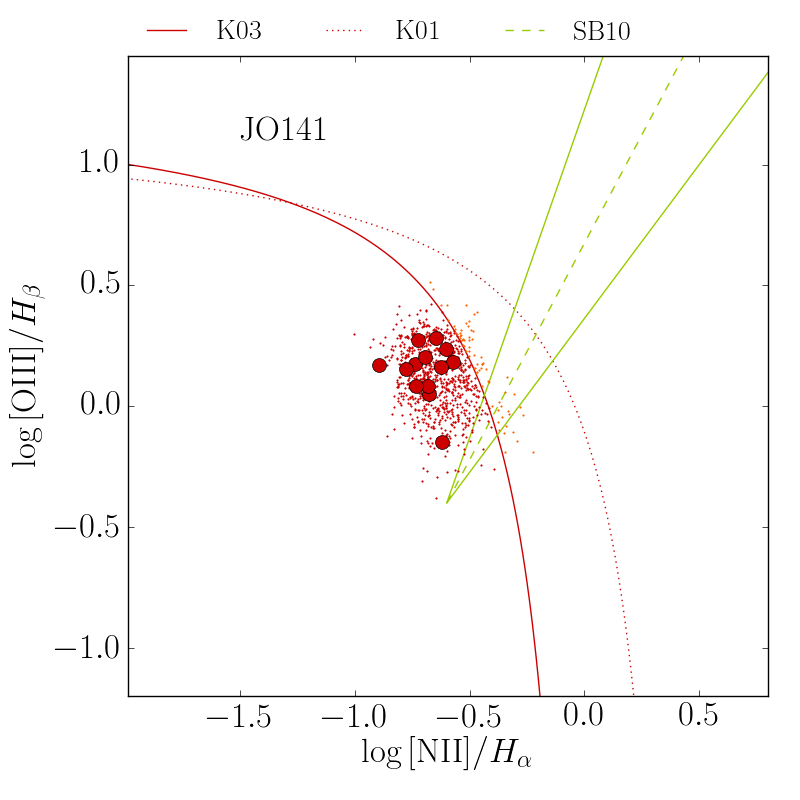}\includegraphics[width=2.4in]{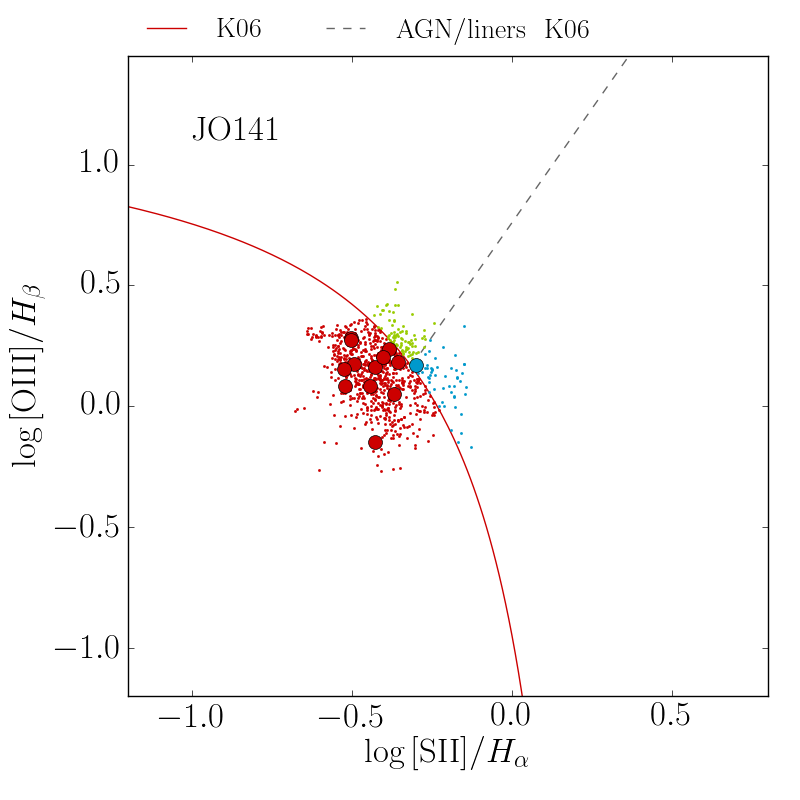}\includegraphics[width=2.4in]{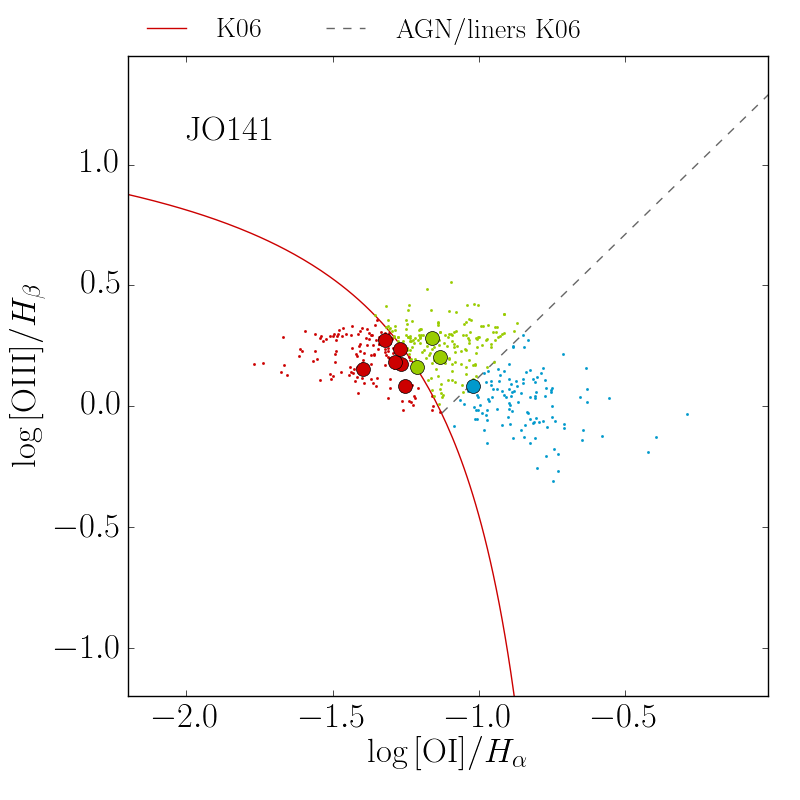}}
\centerline{\includegraphics[width=2.4in]{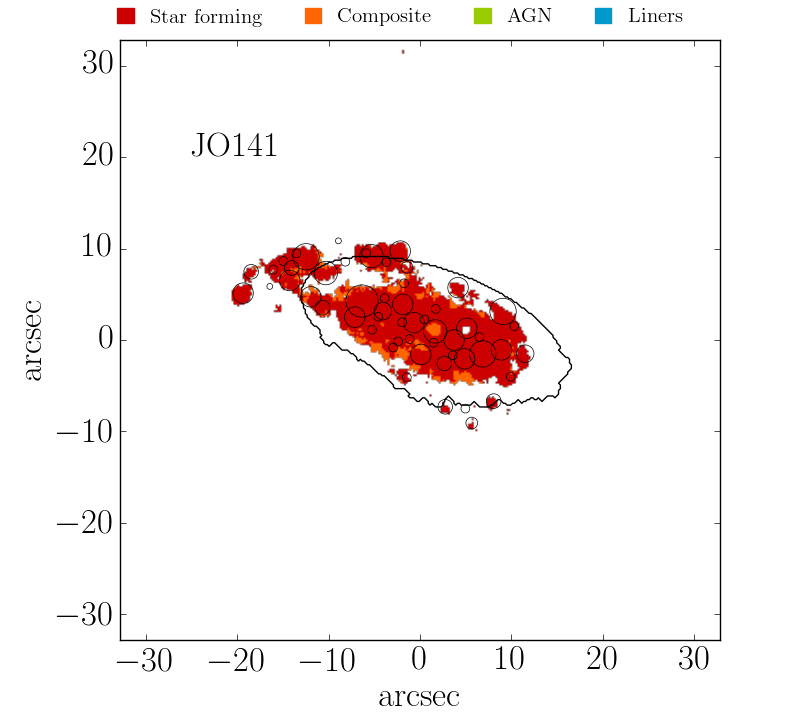}\includegraphics[width=2.4in]{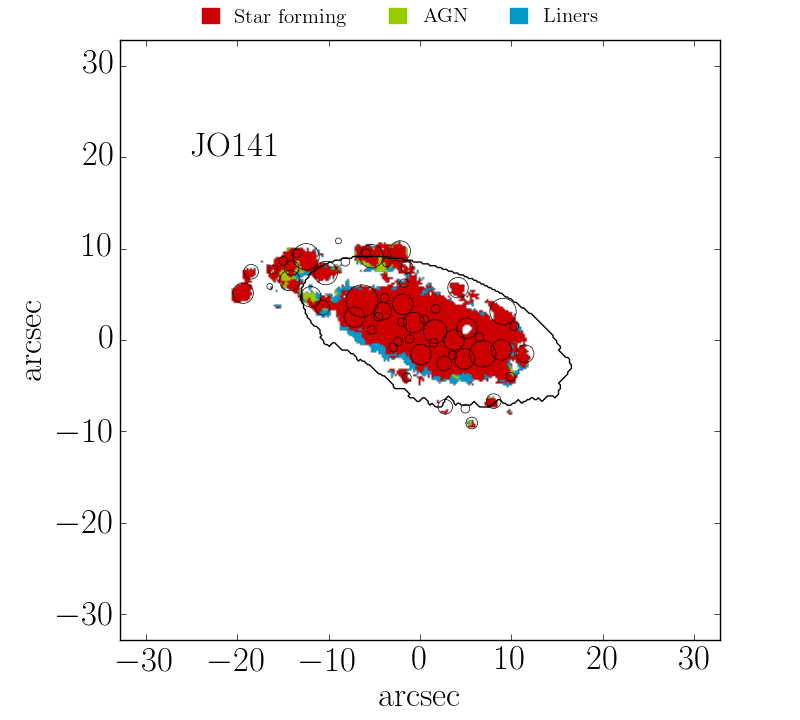}\includegraphics[width=2.4in]{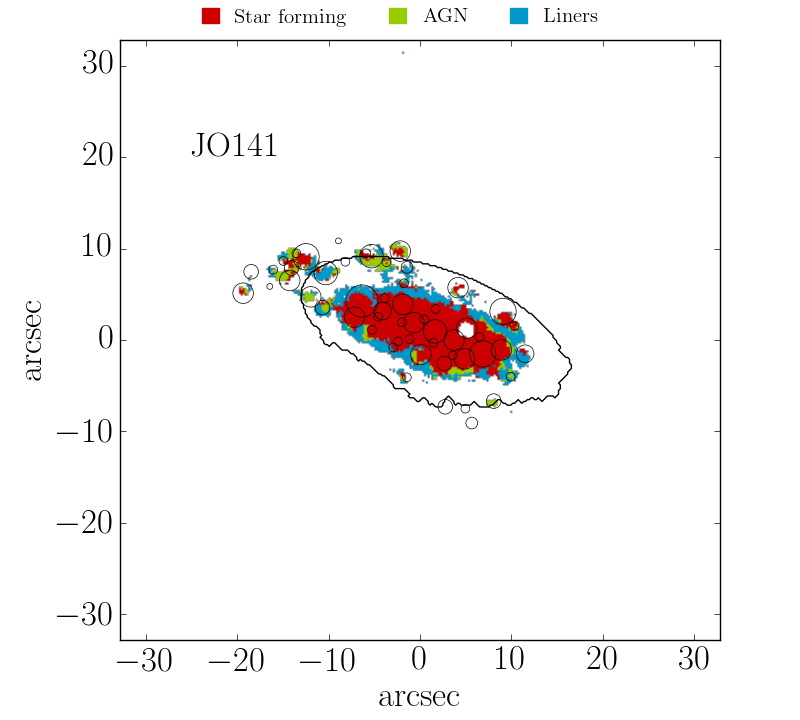}}
\centerline{\includegraphics[width=2.4in]{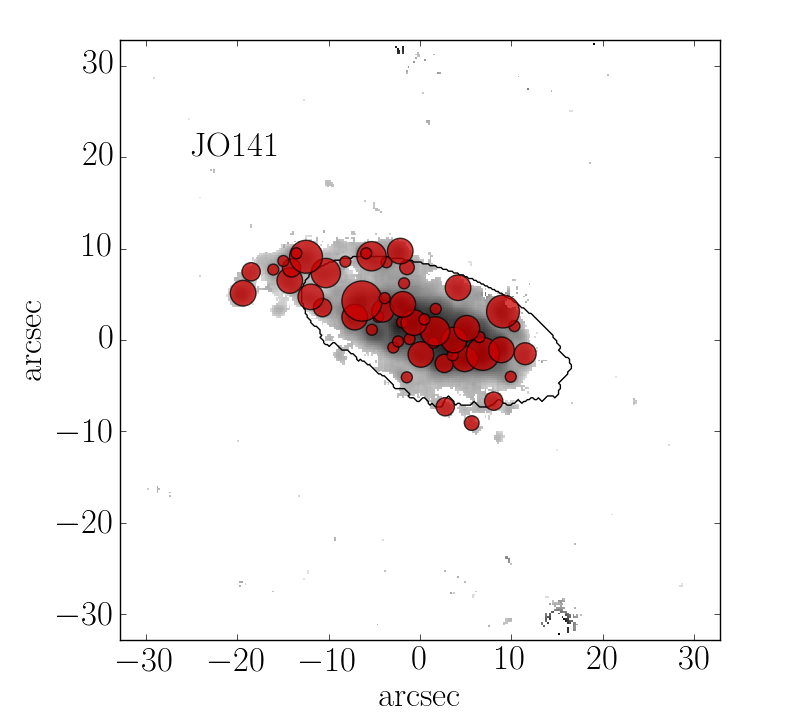}\includegraphics[width=2.4in]{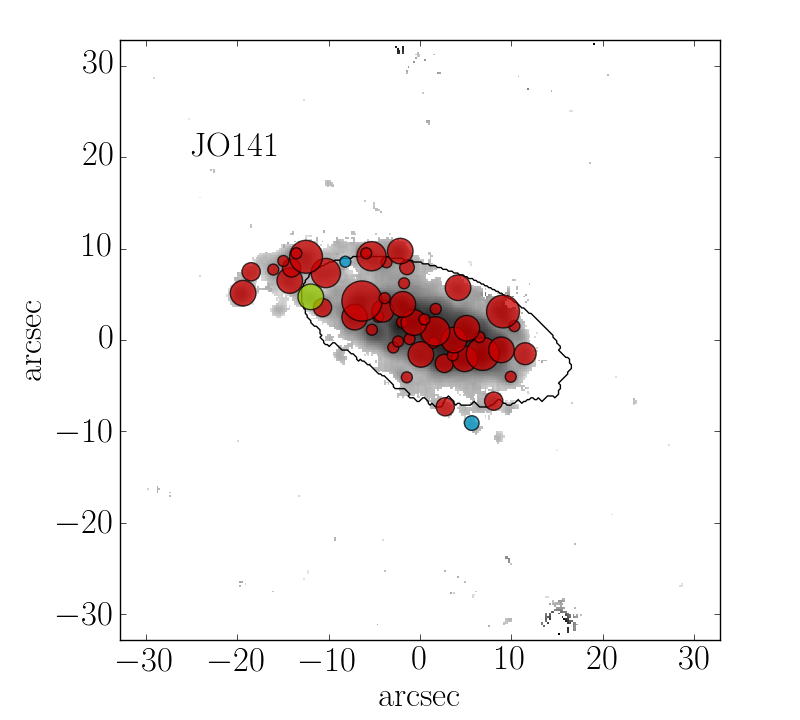}\includegraphics[width=2.4in]{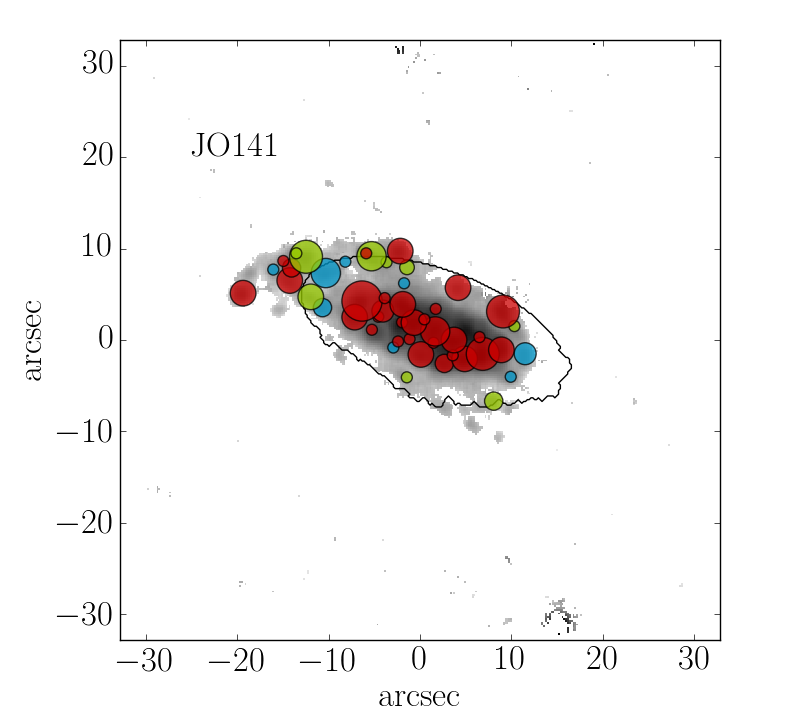}}
\contcaption{}
\end{figure*}

\begin{figure*}
\centerline{\includegraphics[width=2.4in]{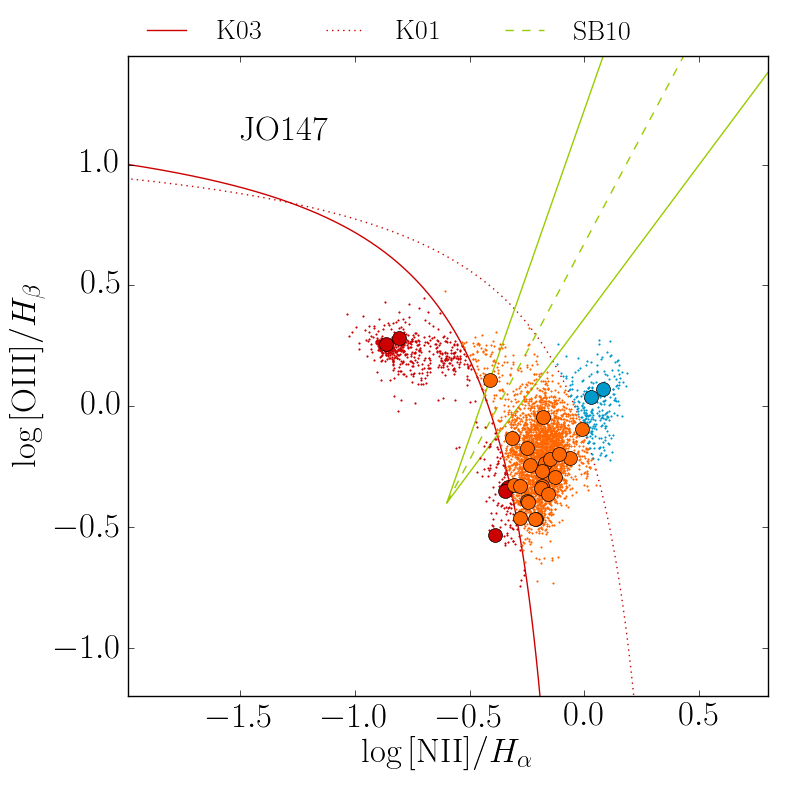}\includegraphics[width=2.4in]{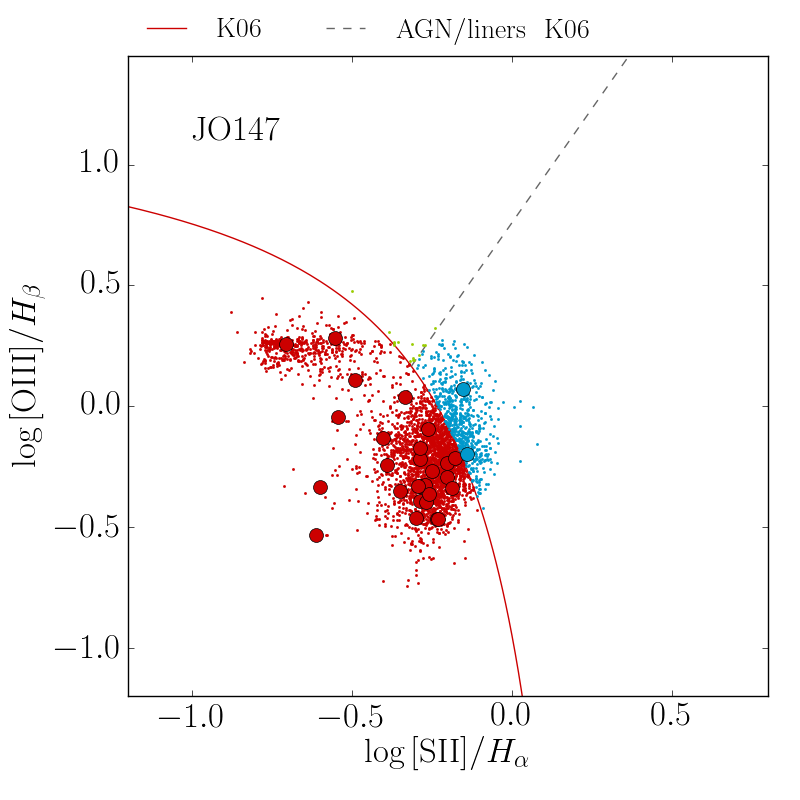}\includegraphics[width=2.4in]{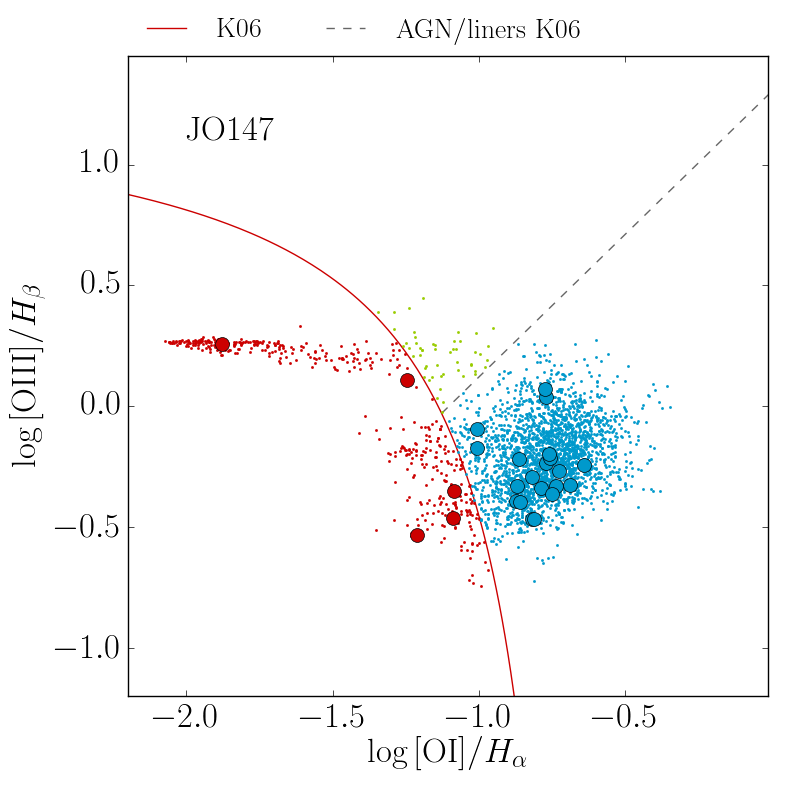}}
\centerline{\includegraphics[width=2.4in]{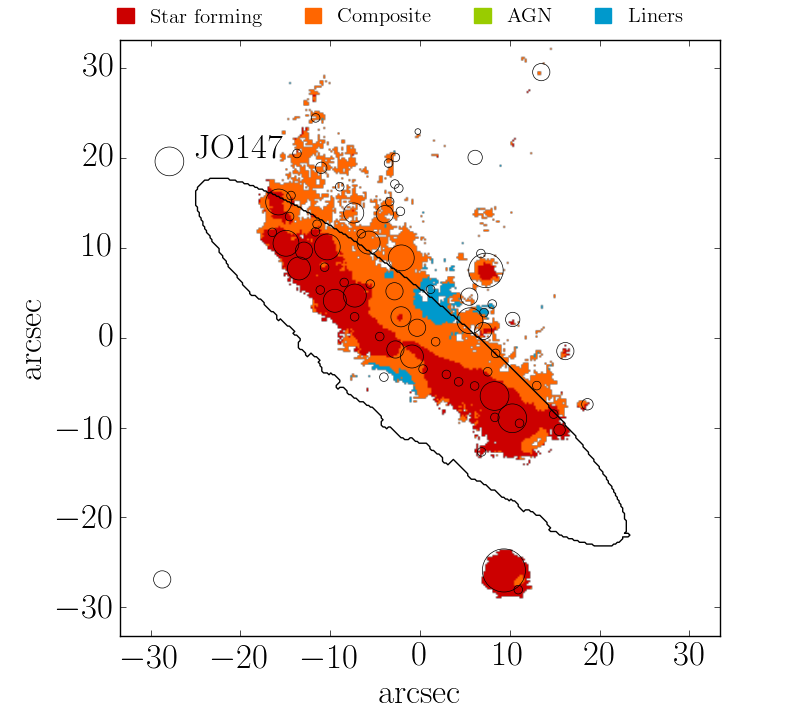}\includegraphics[width=2.4in]{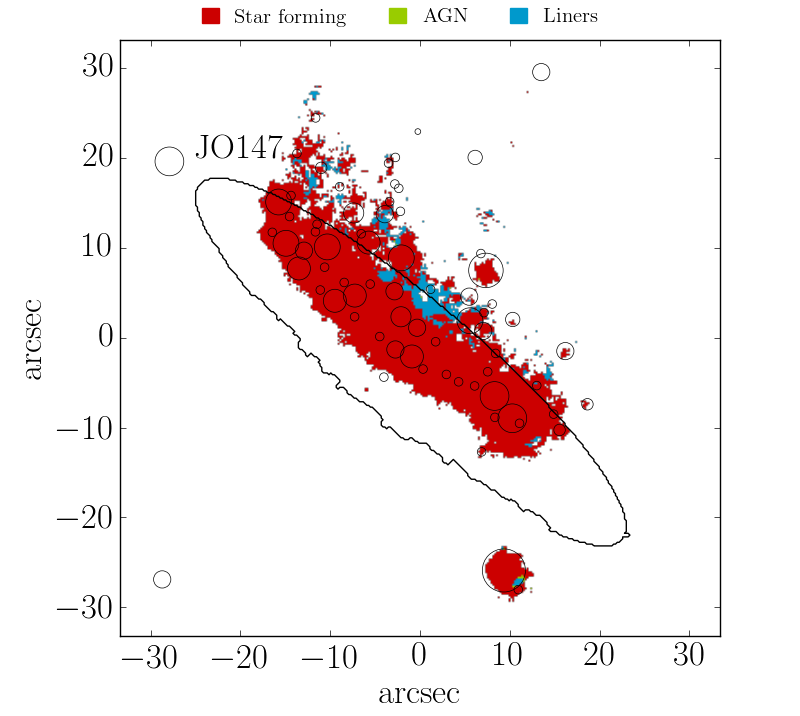}\includegraphics[width=2.4in]{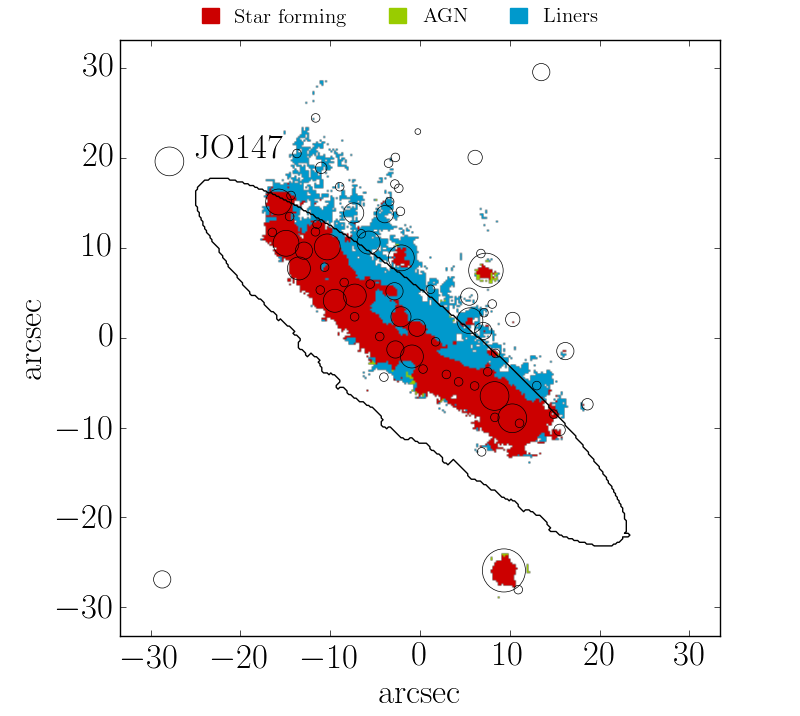}}
\centerline{\includegraphics[width=2.4in]{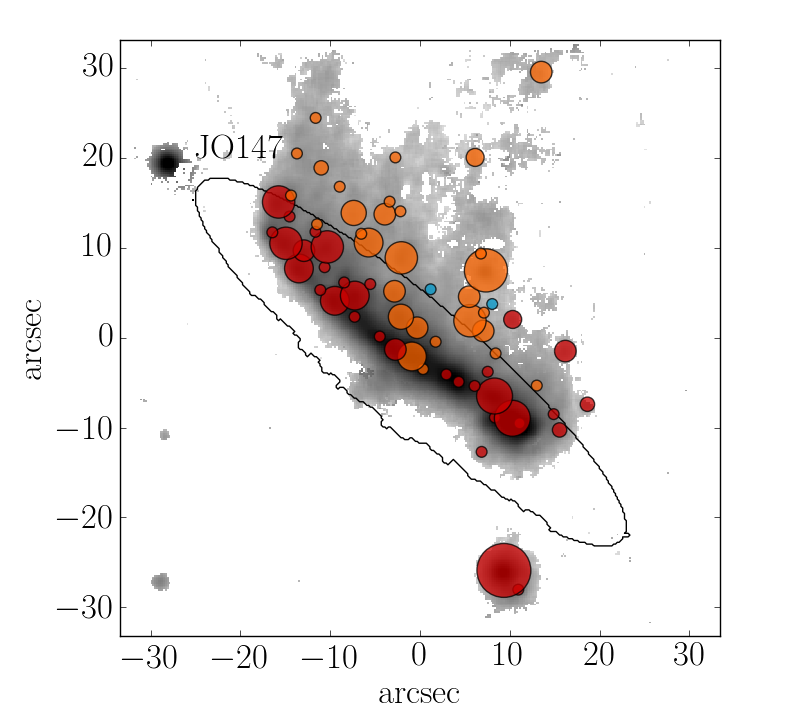}\includegraphics[width=2.4in]{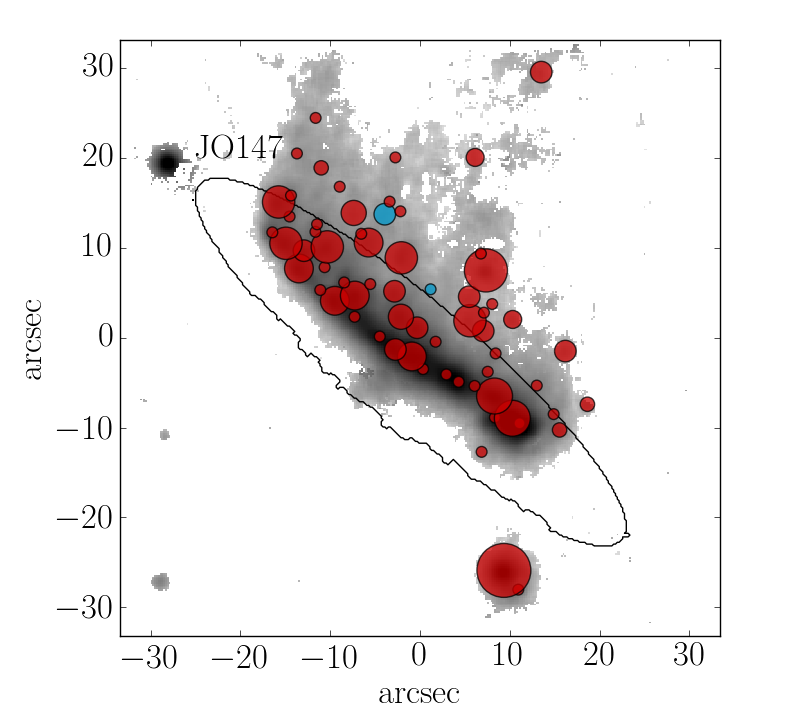}\includegraphics[width=2.4in]{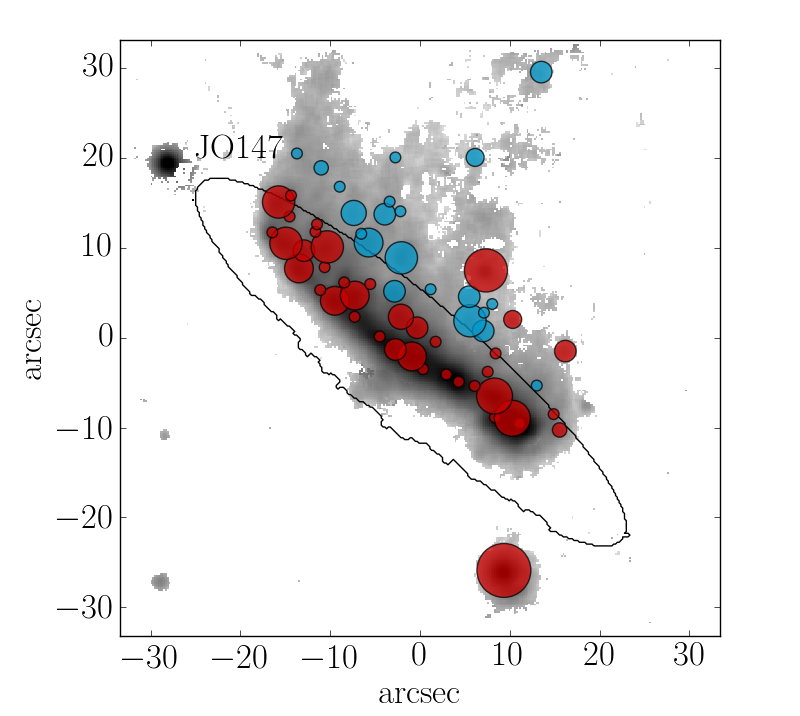}}
\contcaption{}
\end{figure*}

%\begin{figure*}
%\centerline{\includegraphics[width=2.1in]{JO135_bptmap_N.png}\includegraphics[width=2.1in]{JO135_bptmap_O.png}\includegraphics[width=2.1in]{JO135_bptmap_S.png}}
%\centerline{\includegraphics[width=2.1in]{JO135_bpt_N.png}\includegraphics[width=2.1in]{JO135_bpt_O.png}\includegraphics[width=2.1in]{JO135_bpt_S.png}}
%\centerline{\includegraphics[width=2.1in]{JO135_bptmap_blobs.png}\includegraphics[width=2.1in]{JO135_bptmap_blobs_O.png}\includegraphics[width=2.1in]{JO135_bptmap_blobs_S.png}}
%\centerline{\includegraphics[width=2.1in]{JO135_bpt_blobs.png}\includegraphics[width=2.1in]{JO135_bpt_blobs_O.png}\includegraphics[width=2.1in]{JO135_bpt_blobs_S.png}}
%\caption{Line-ratio diagnostic diagrams and maps for spaxels (two top 
 % panels) and knots (two bottom panels). The diagrams include only 
 % spaxels and knots that are outside of the line contour showing the 
 % stellar disk (see text for details).}
%\end{figure*}

\begin{figure*}
\centerline{\includegraphics[width=2.4in]{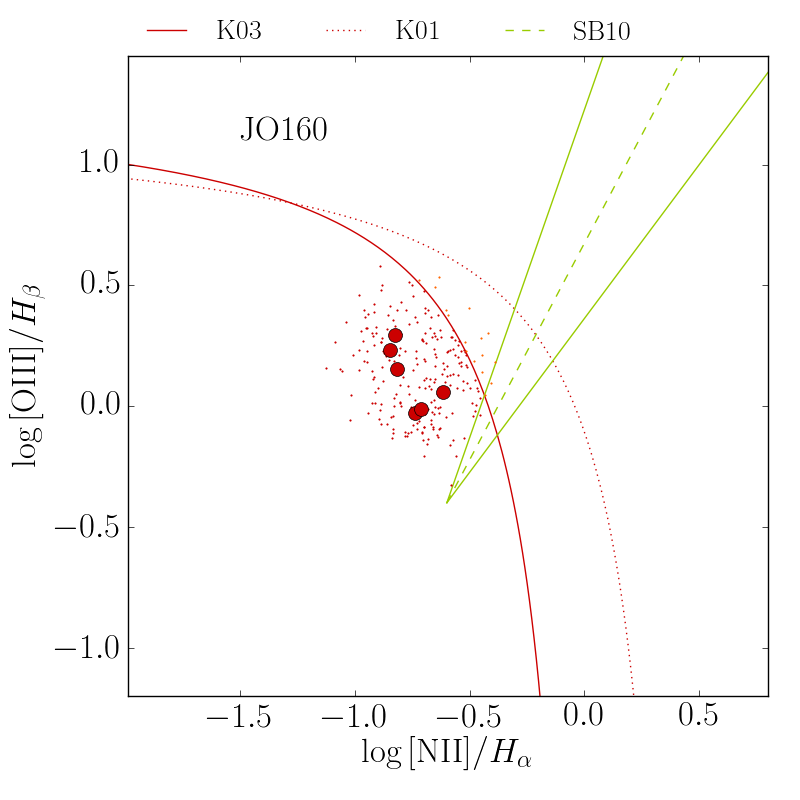}\includegraphics[width=2.4in]{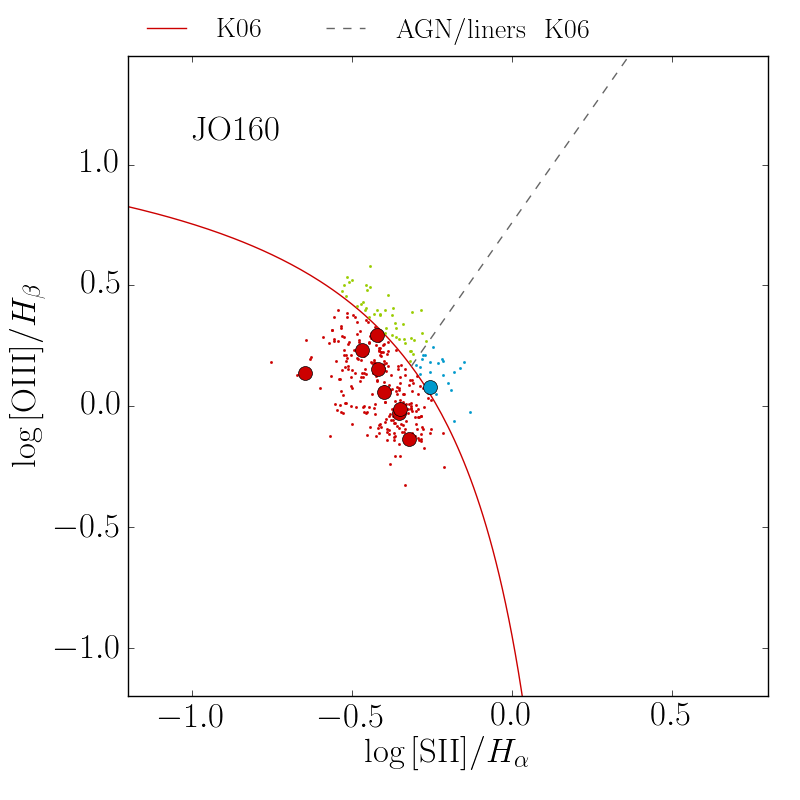}\includegraphics[width=2.4in]{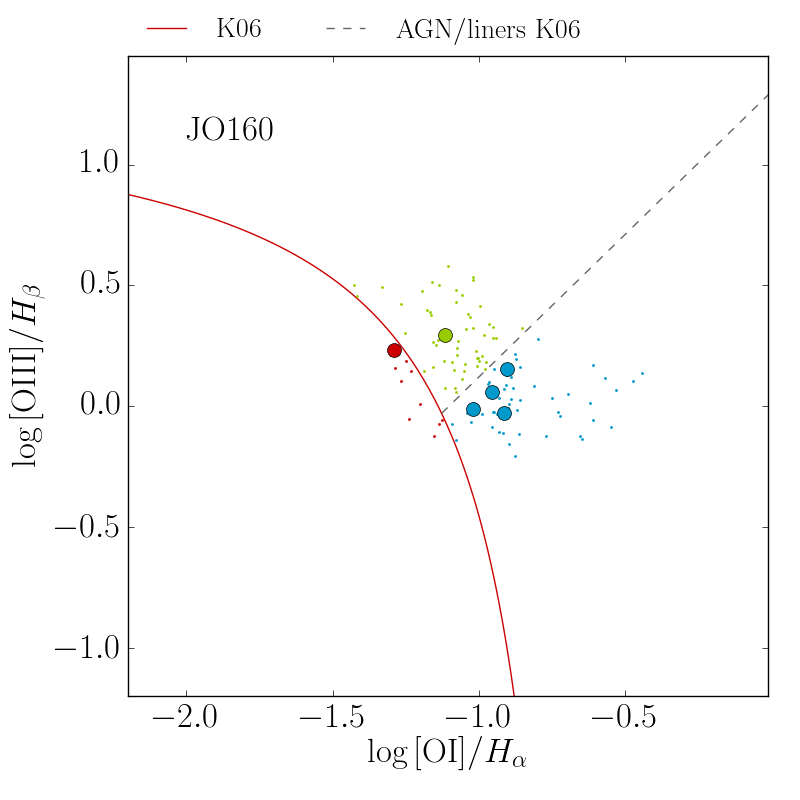}}
\centerline{\includegraphics[width=2.4in]{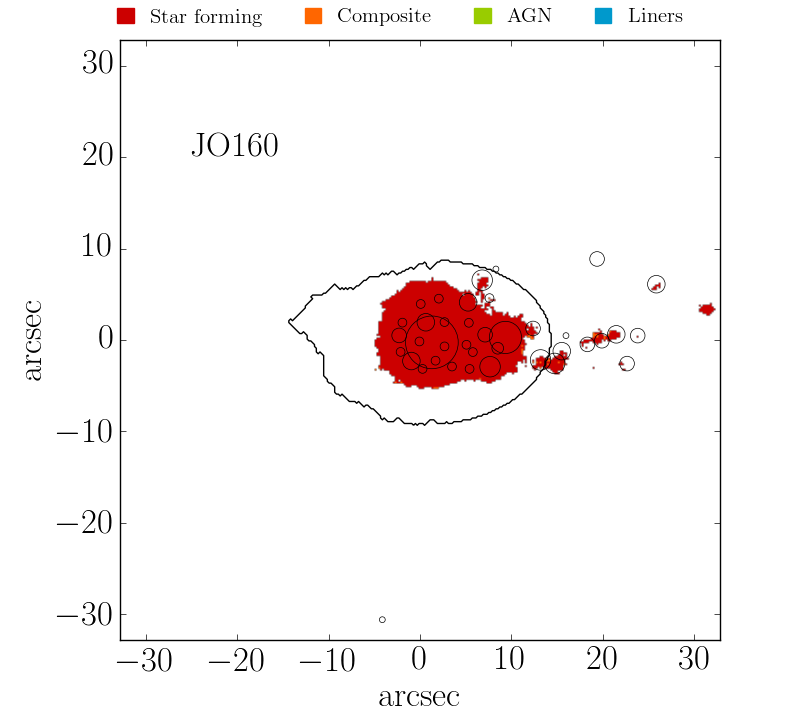}\includegraphics[width=2.4in]{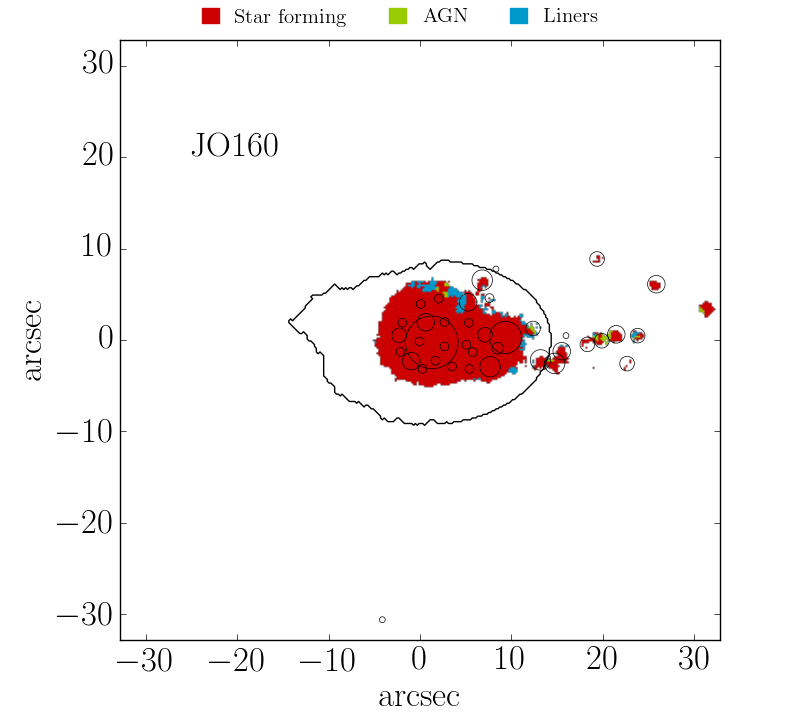}\includegraphics[width=2.4in]{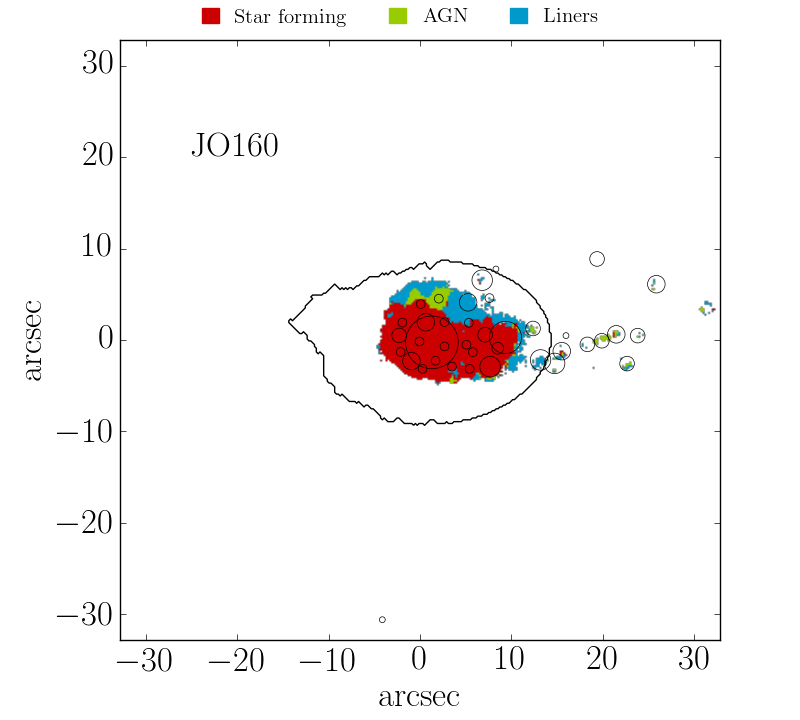}}
\centerline{\includegraphics[width=2.4in]{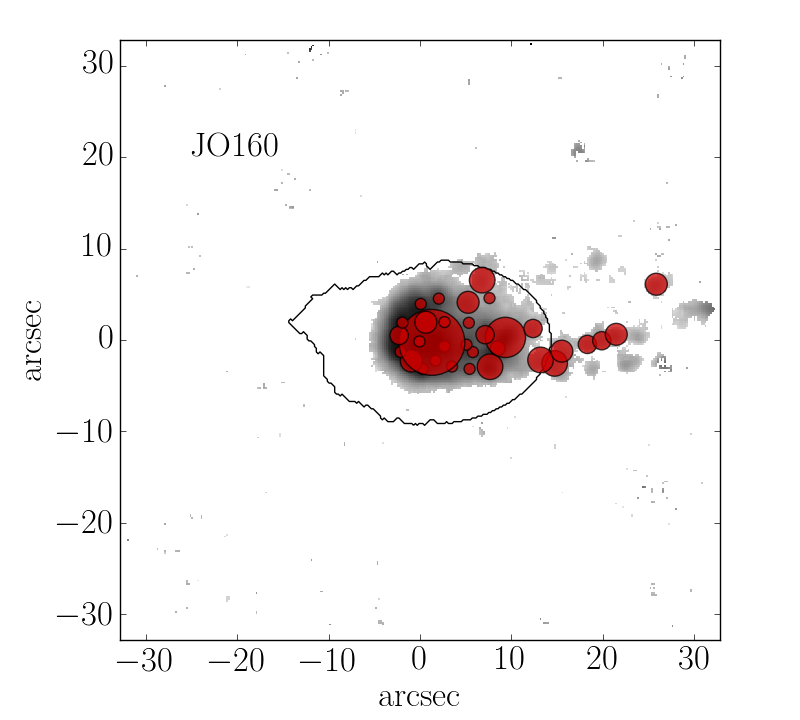}\includegraphics[width=2.4in]{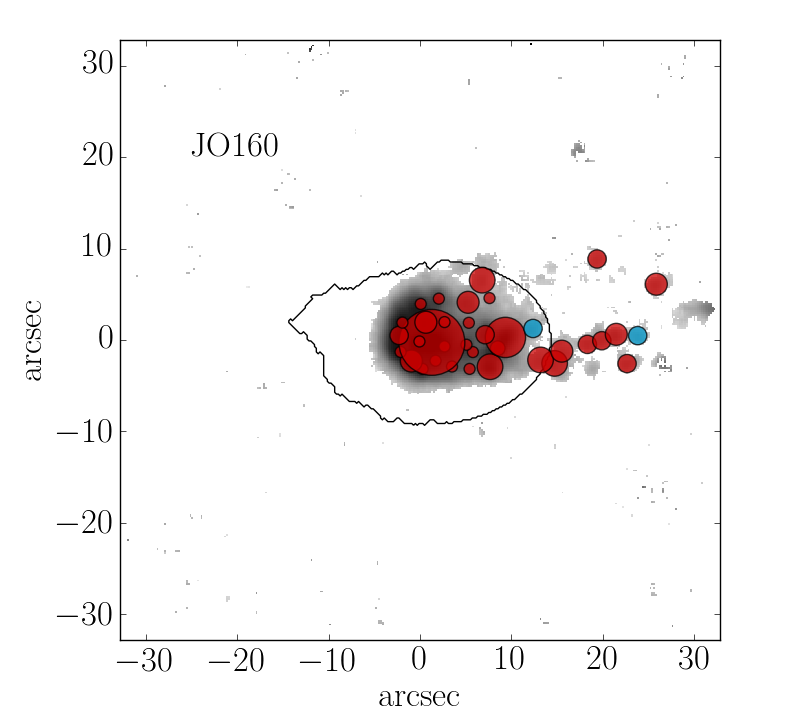}\includegraphics[width=2.4in]{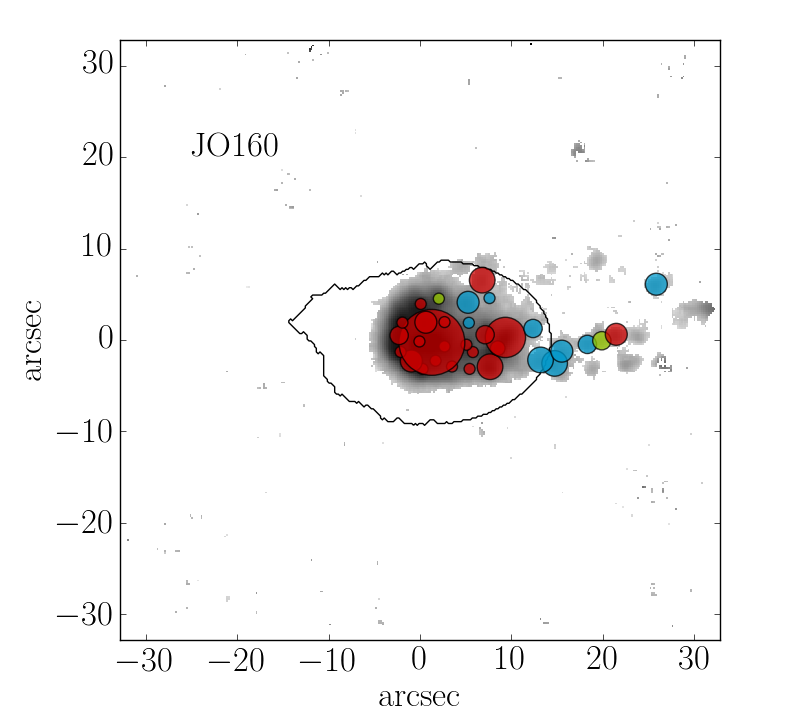}}
\contcaption{}
\end{figure*}

\begin{figure*}
\centerline{\includegraphics[width=2.4in]{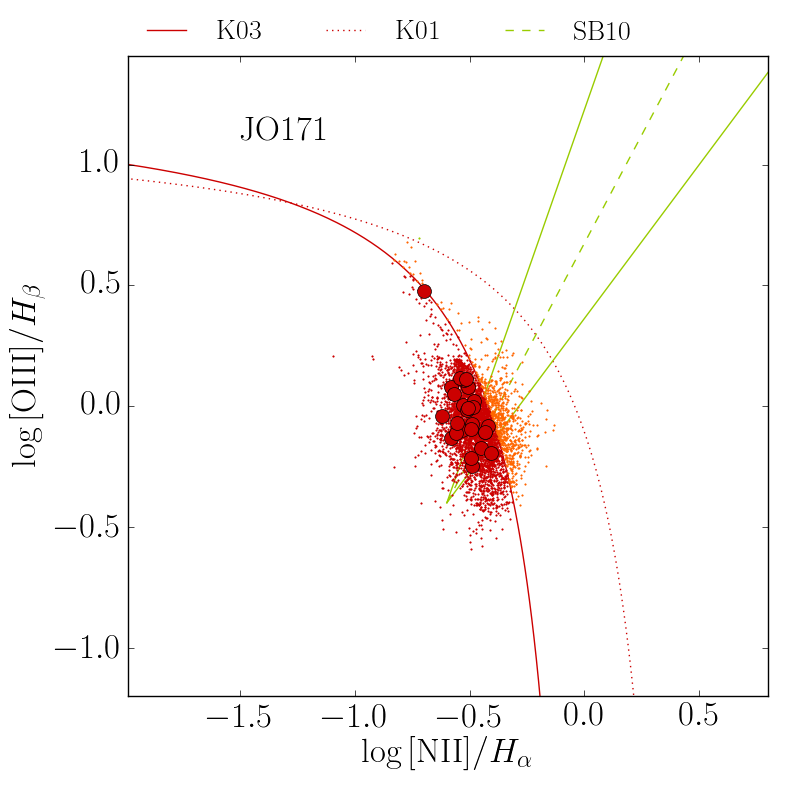}\includegraphics[width=2.4in]{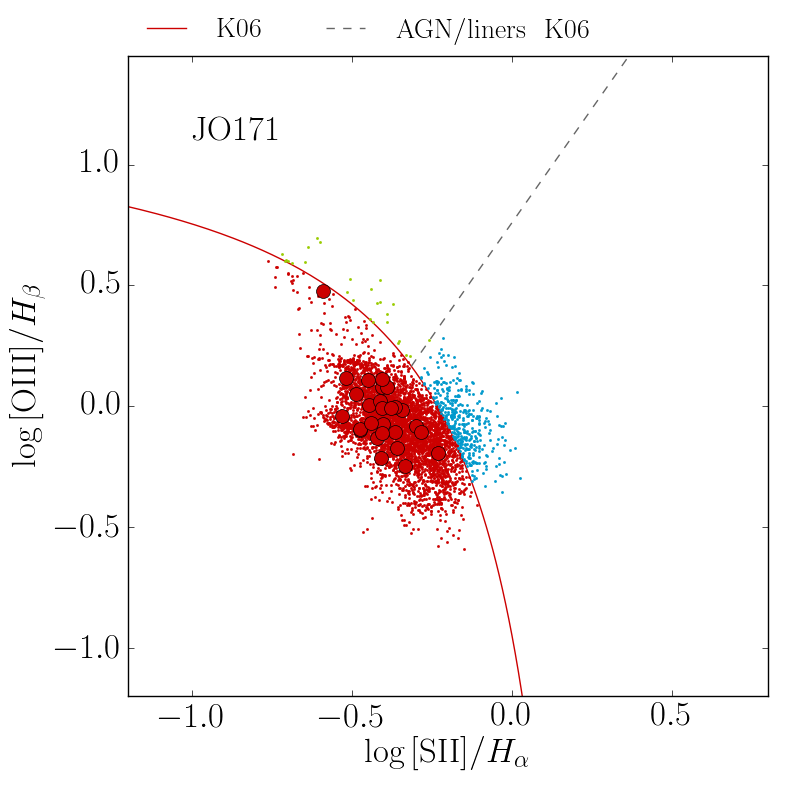}\includegraphics[width=2.4in]{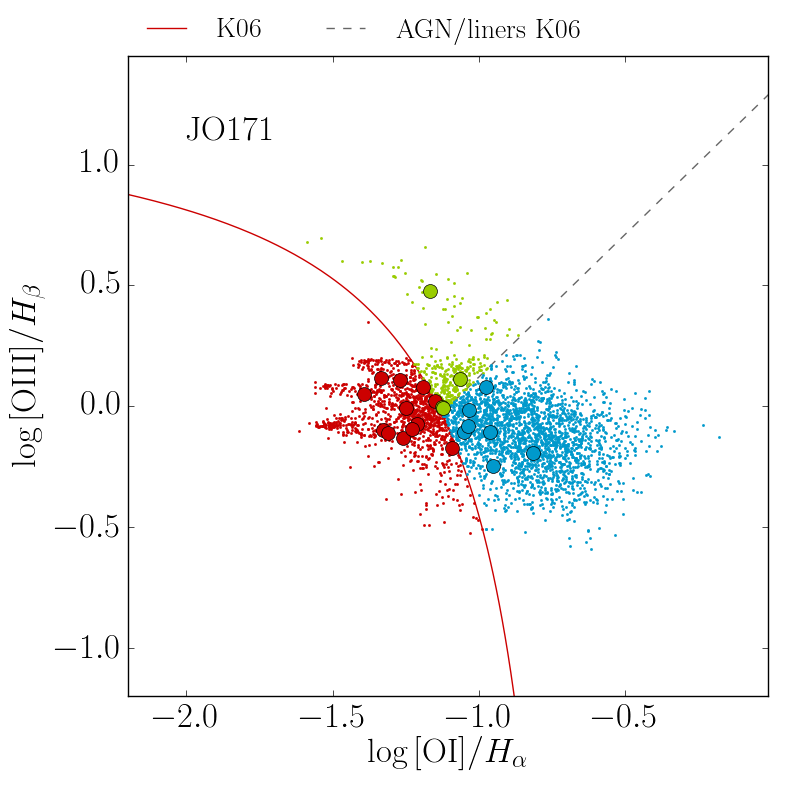}}
\centerline{\includegraphics[width=2.4in]{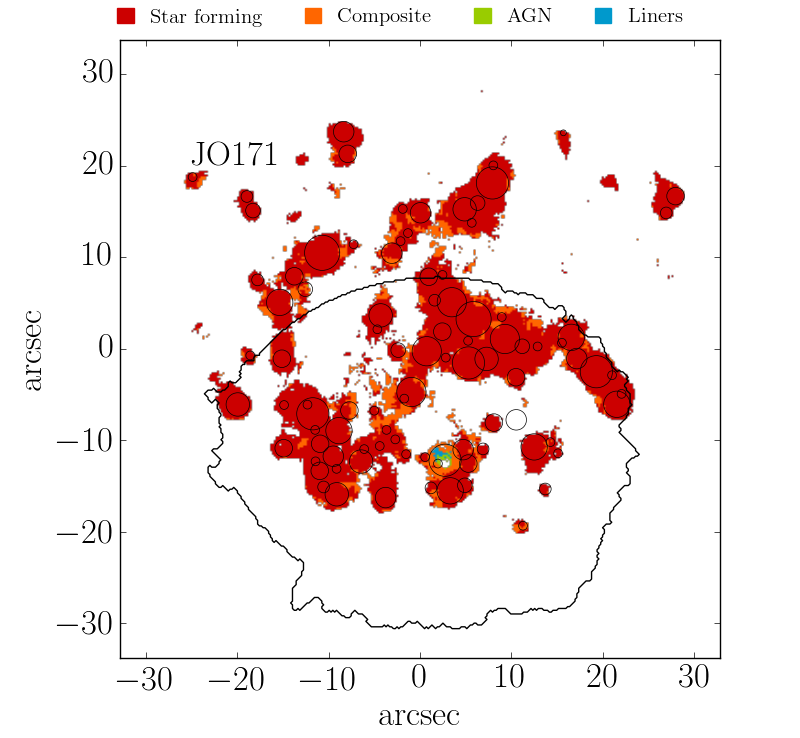}\includegraphics[width=2.4in]{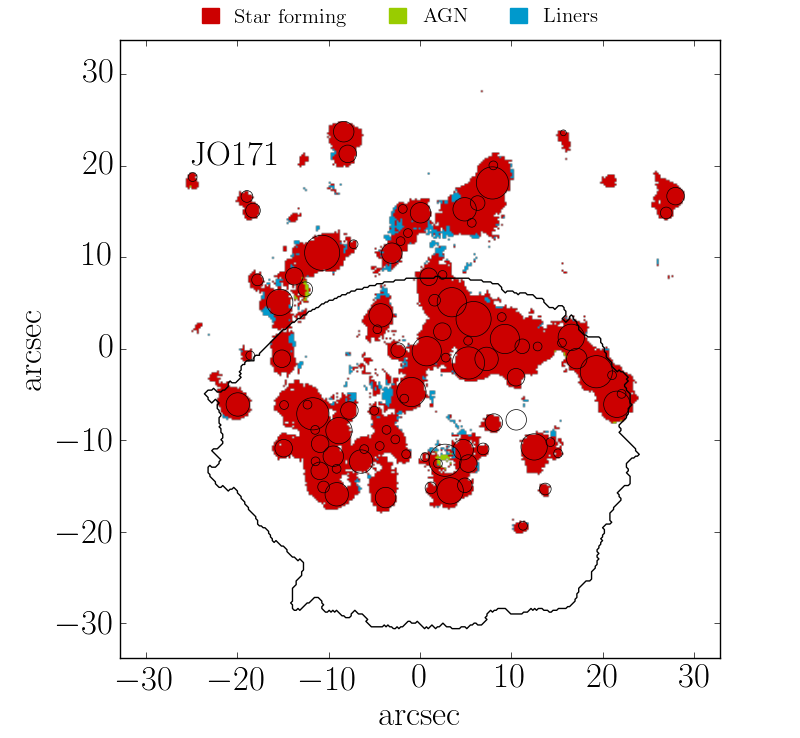}\includegraphics[width=2.4in]{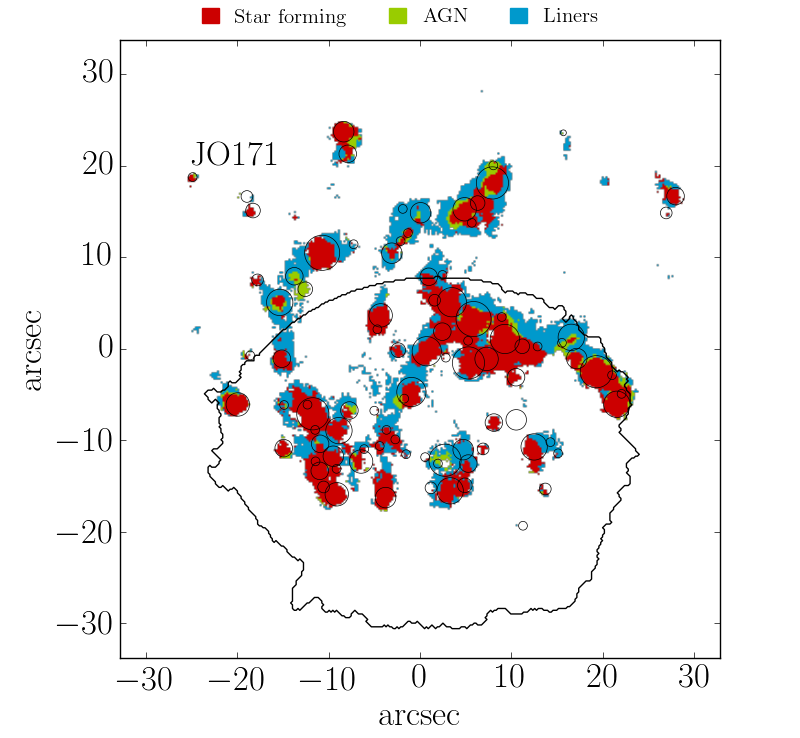}}
\centerline{\includegraphics[width=2.4in]{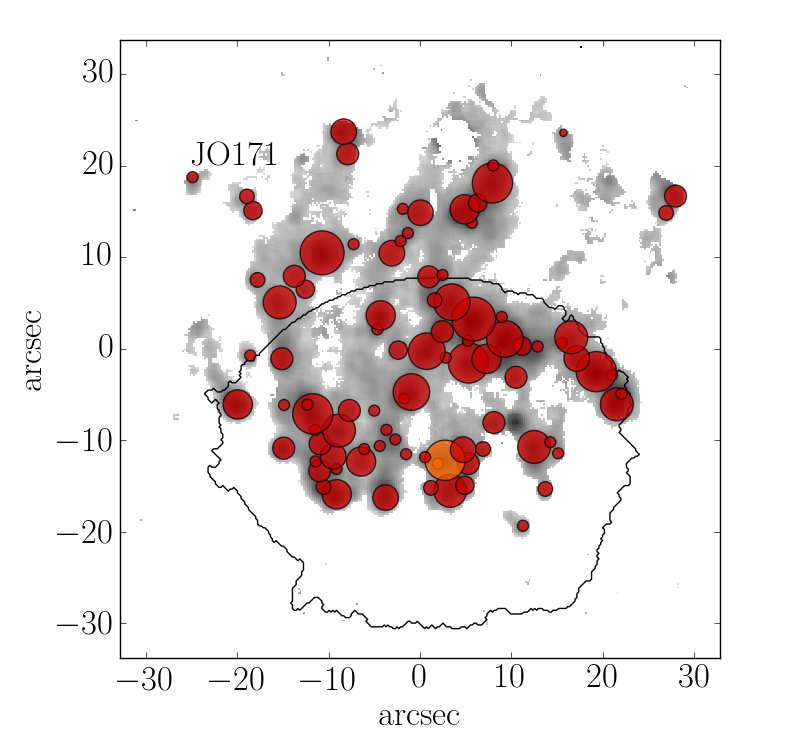}\includegraphics[width=2.4in]{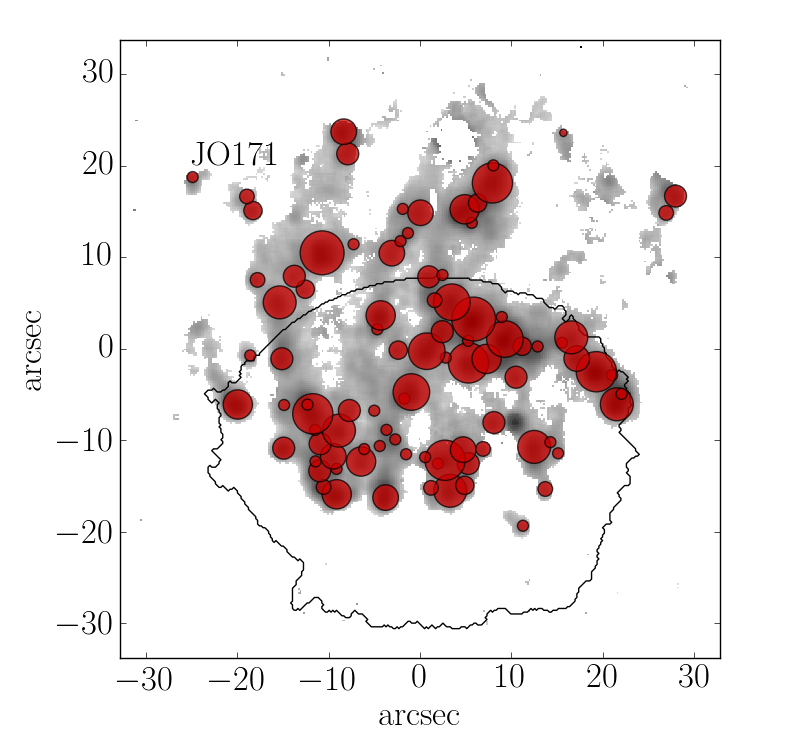}\includegraphics[width=2.4in]{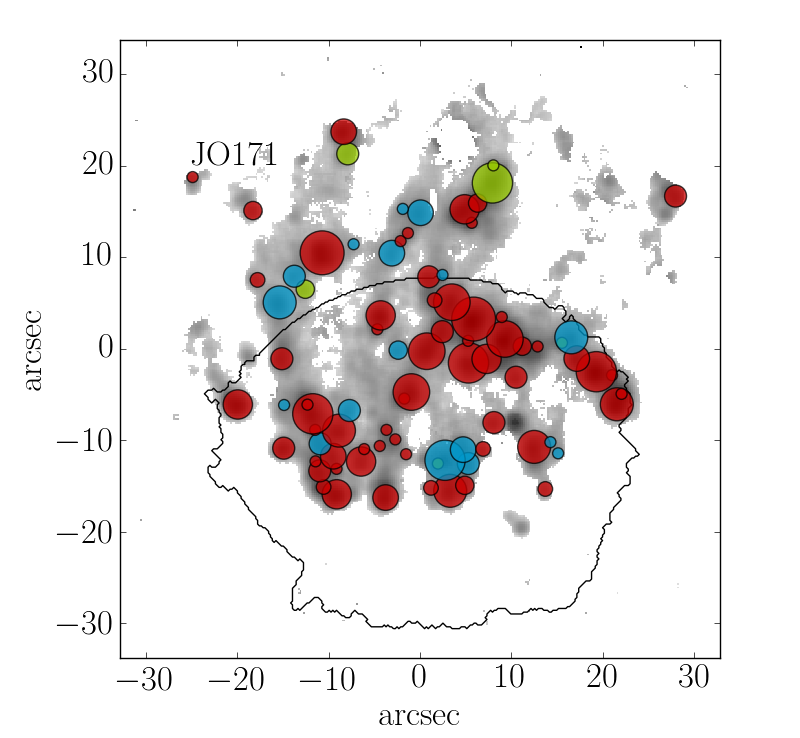}}
\contcaption{}
\end{figure*}

\begin{figure*}
\centerline{\includegraphics[width=2.4in]{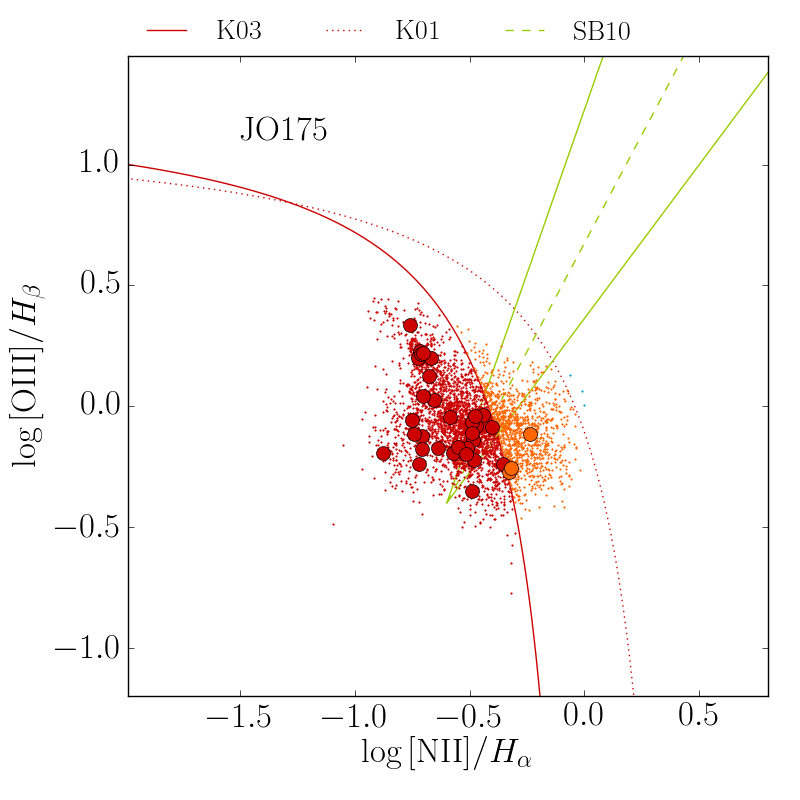}\includegraphics[width=2.4in]{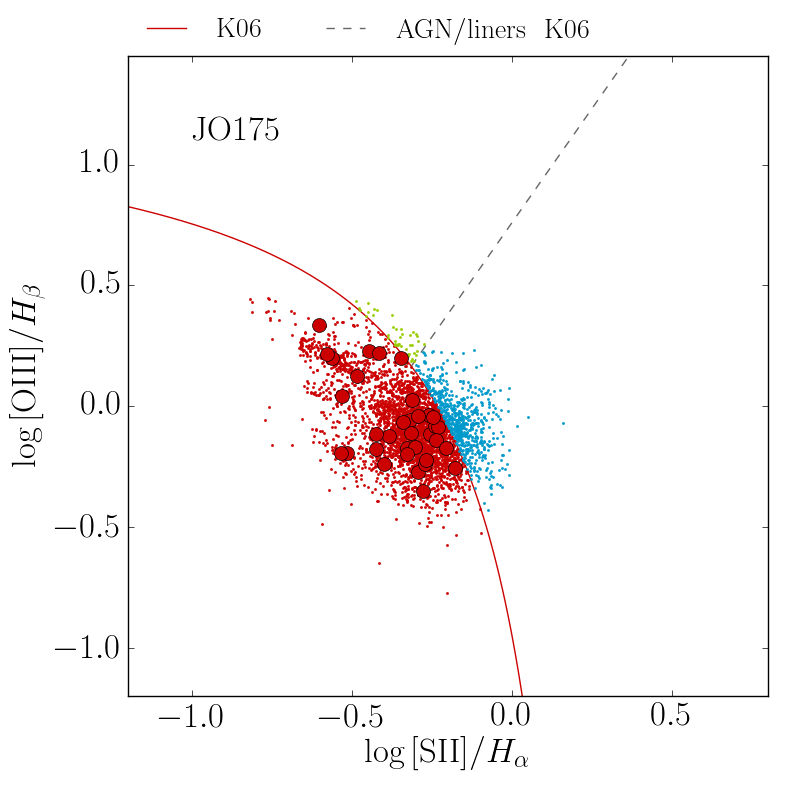}\includegraphics[width=2.4in]{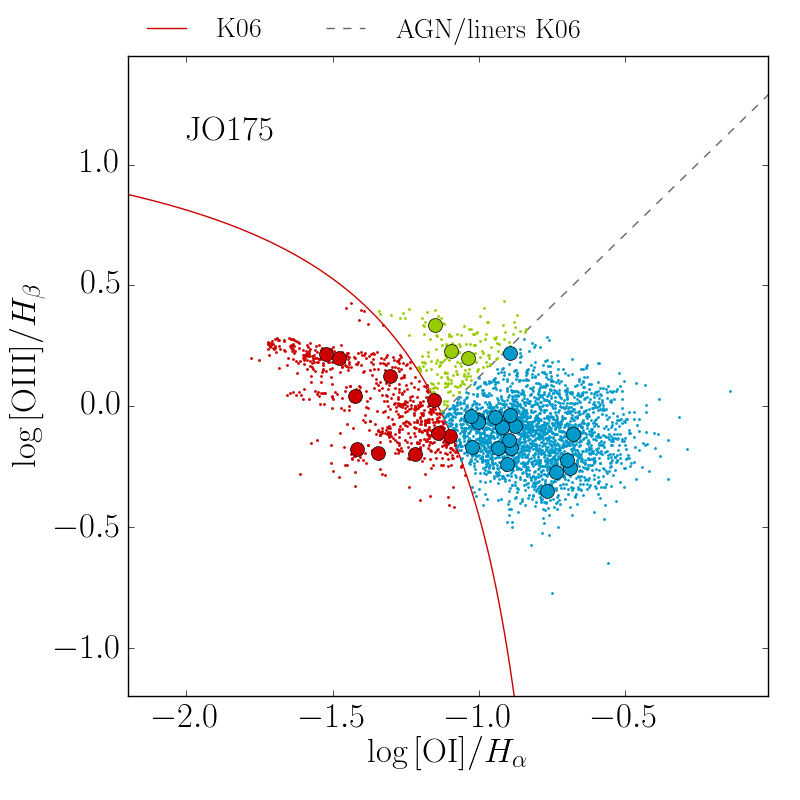}}
\centerline{\includegraphics[width=2.4in]{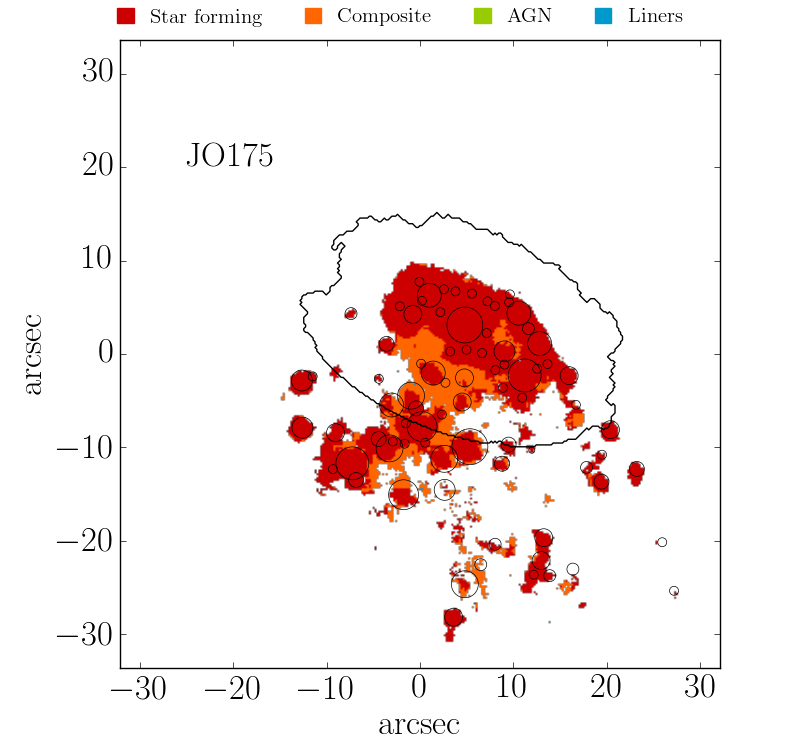}\includegraphics[width=2.4in]{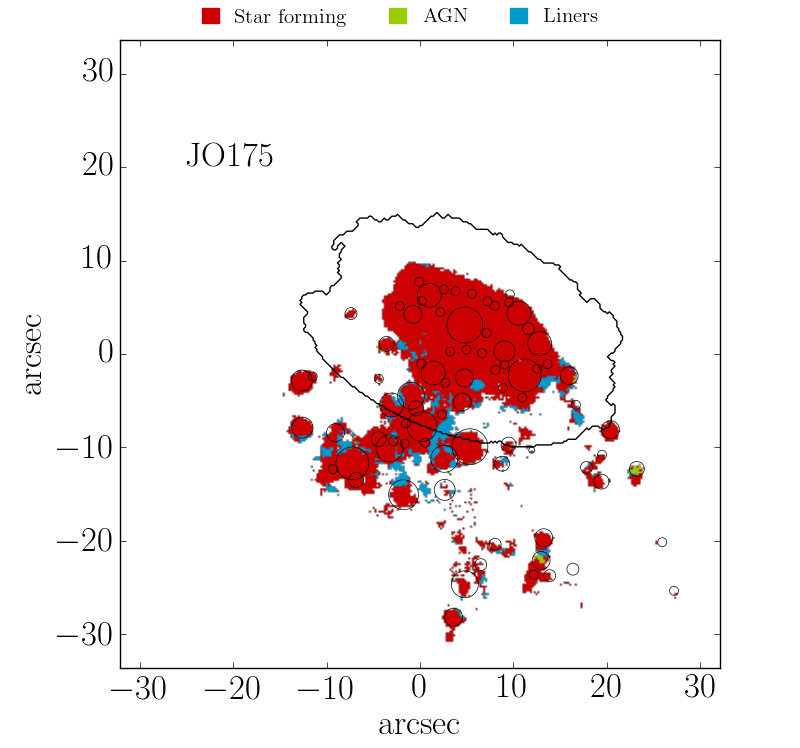}\includegraphics[width=2.4in]{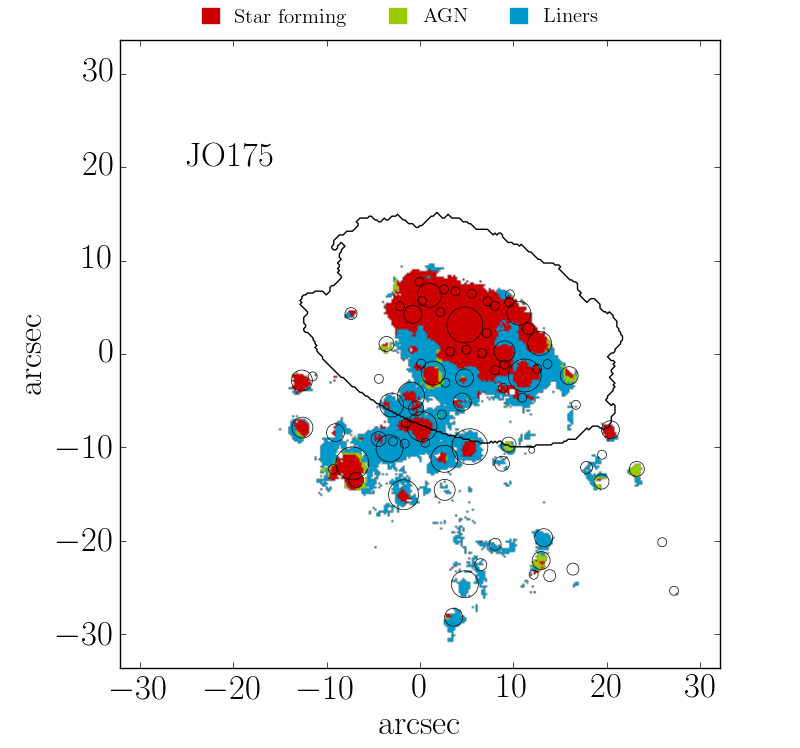}}
\centerline{\includegraphics[width=2.4in]{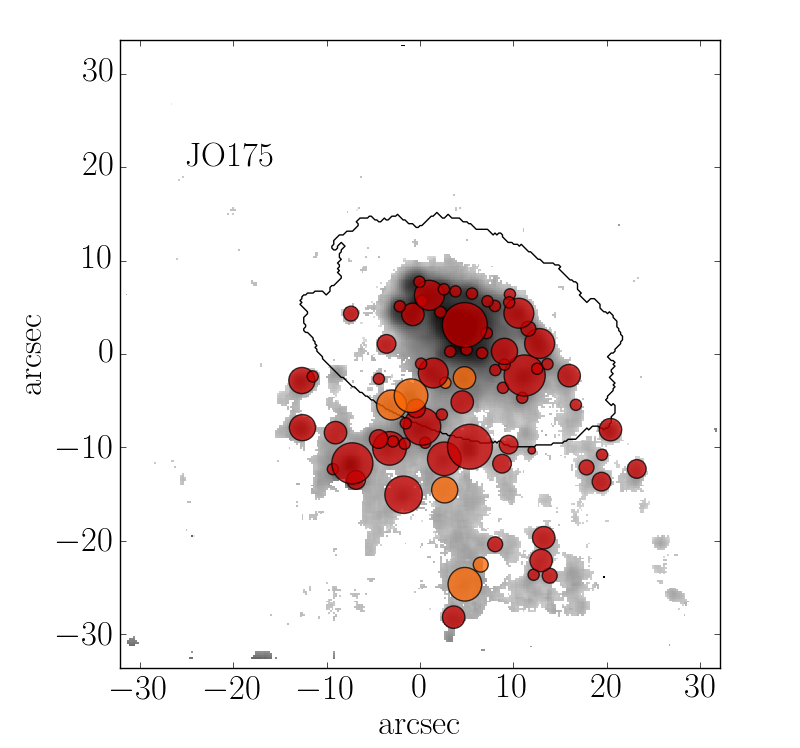}\includegraphics[width=2.4in]{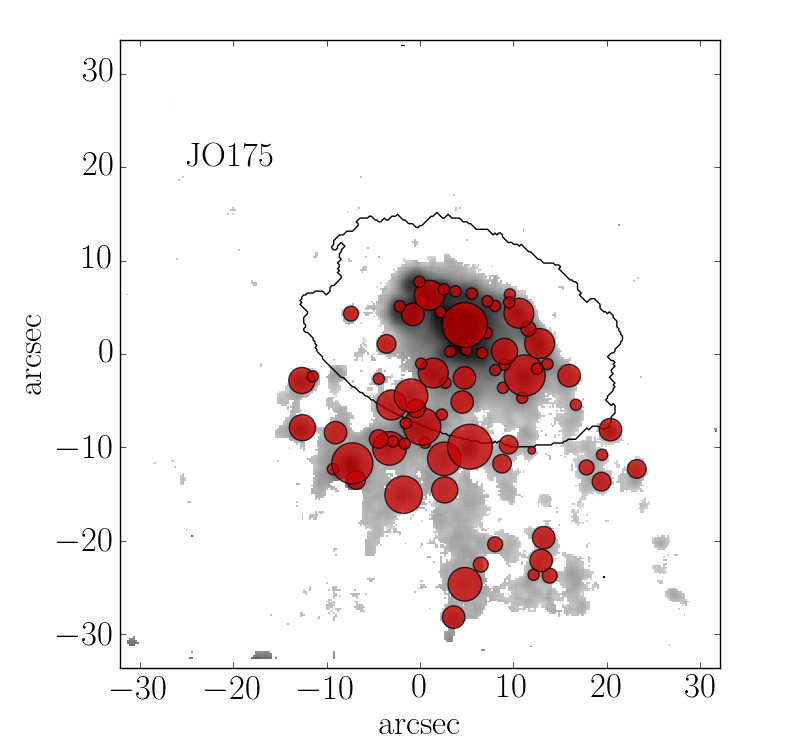}\includegraphics[width=2.4in]{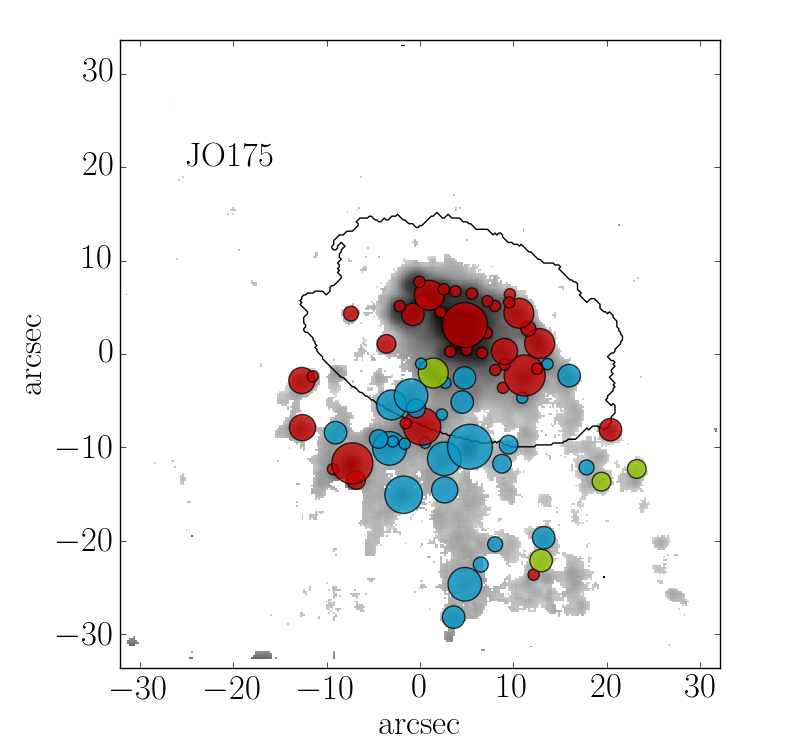}}
\contcaption{}
\end{figure*}

\begin{figure*}
\centerline{\includegraphics[width=2.2in]{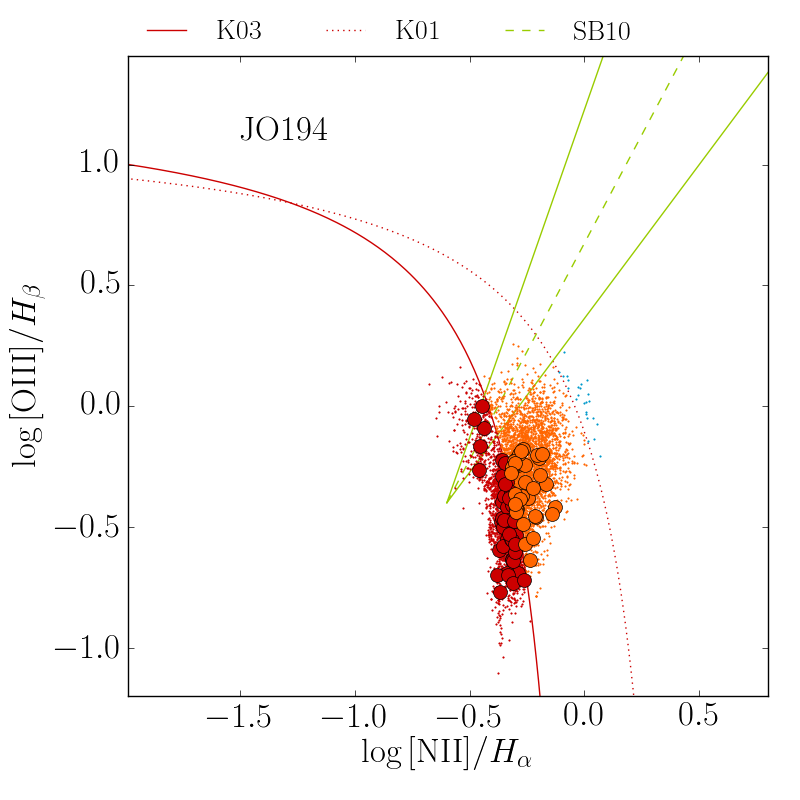}\includegraphics[width=2.2in]{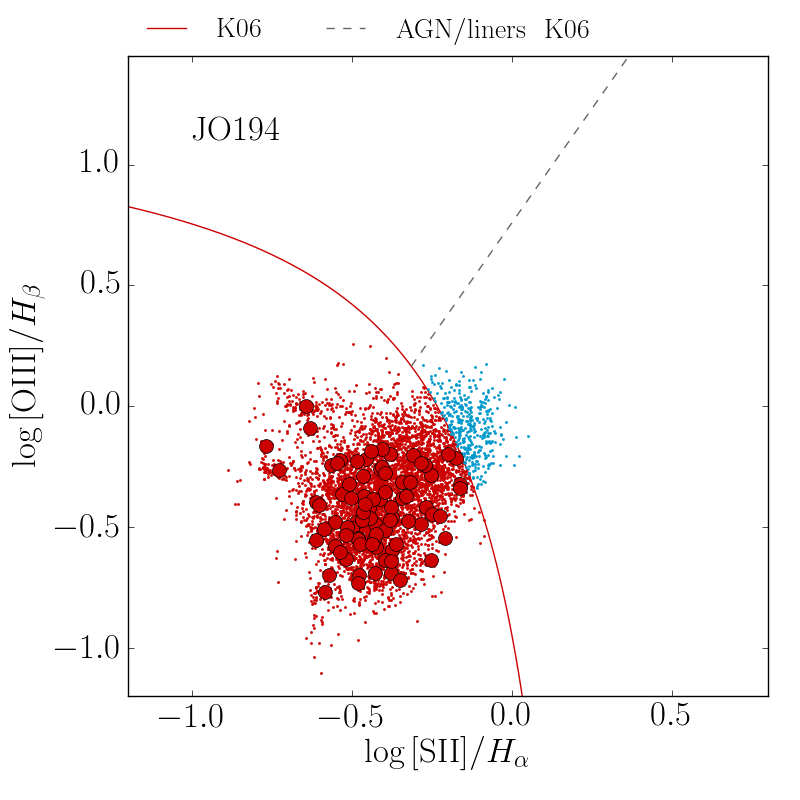}\includegraphics[width=2.2in]{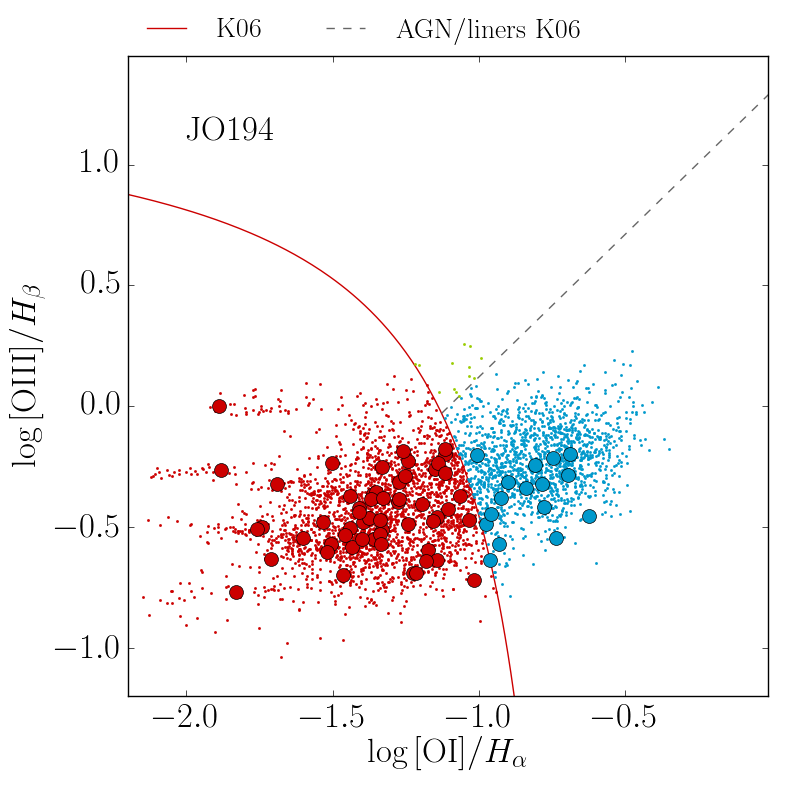}}
\centerline{\includegraphics[width=2.2in]{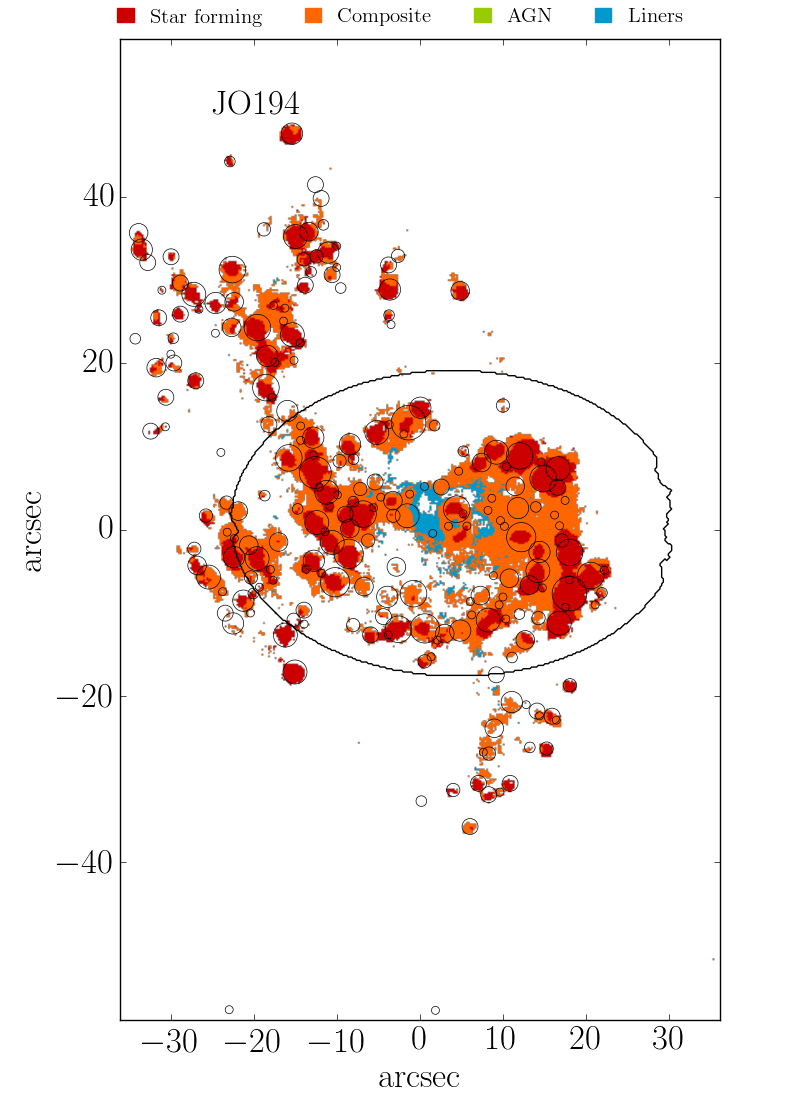}\includegraphics[width=2.2in]{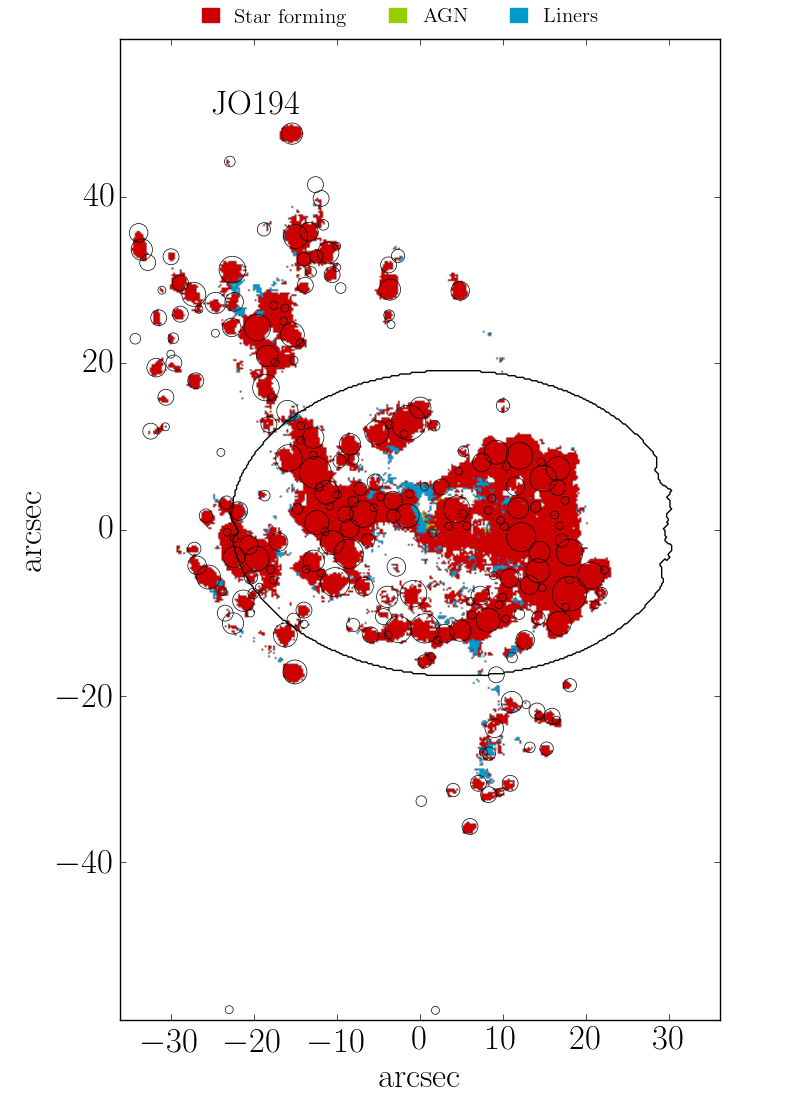}\includegraphics[width=2.2in]{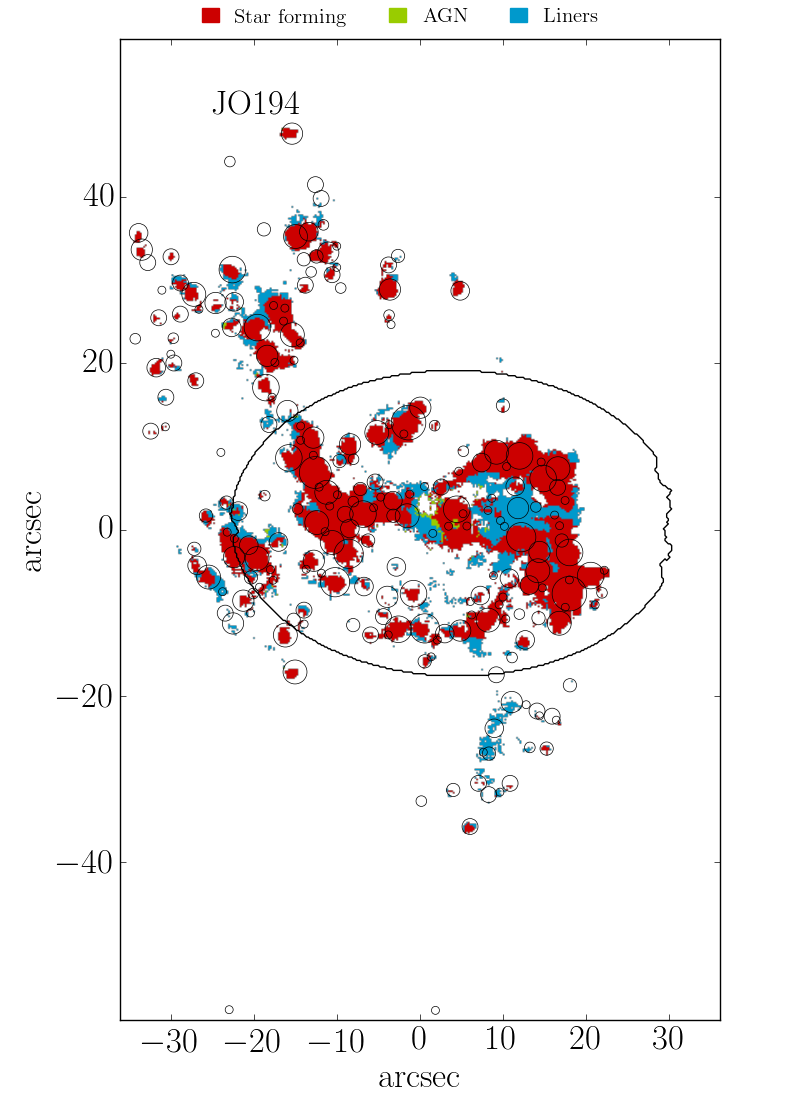}}
\centerline{\includegraphics[width=2.2in]{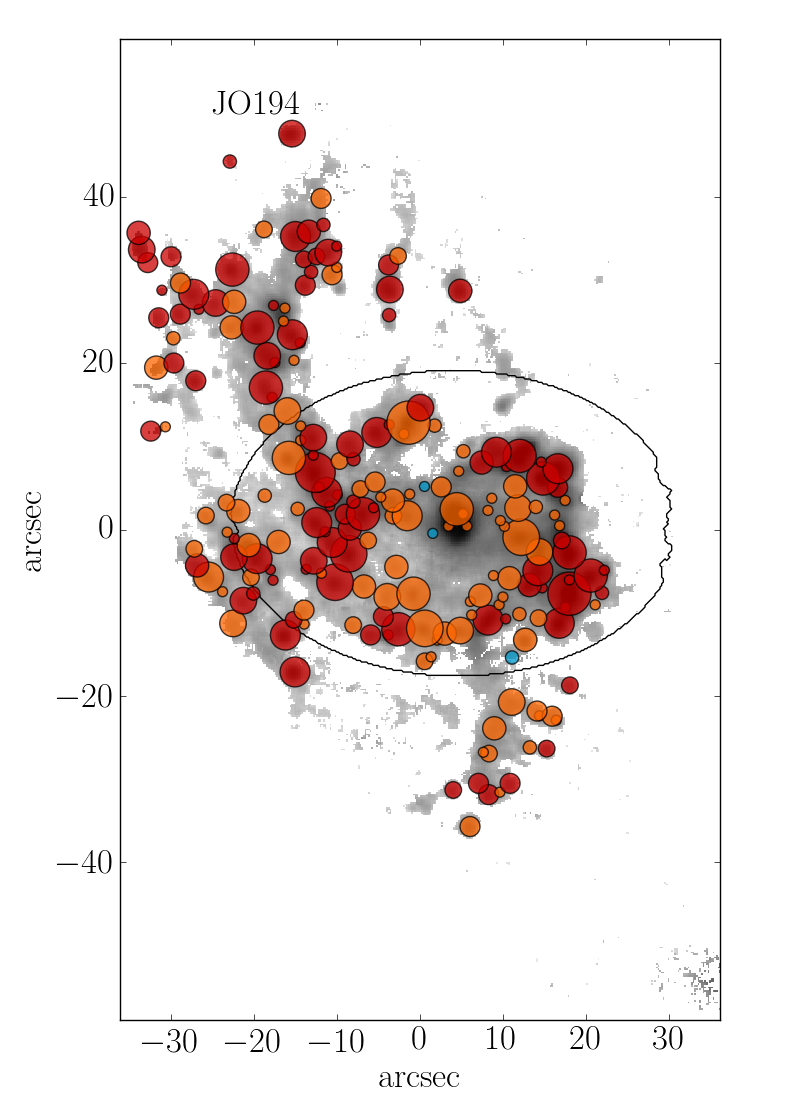}\includegraphics[width=2.2in]{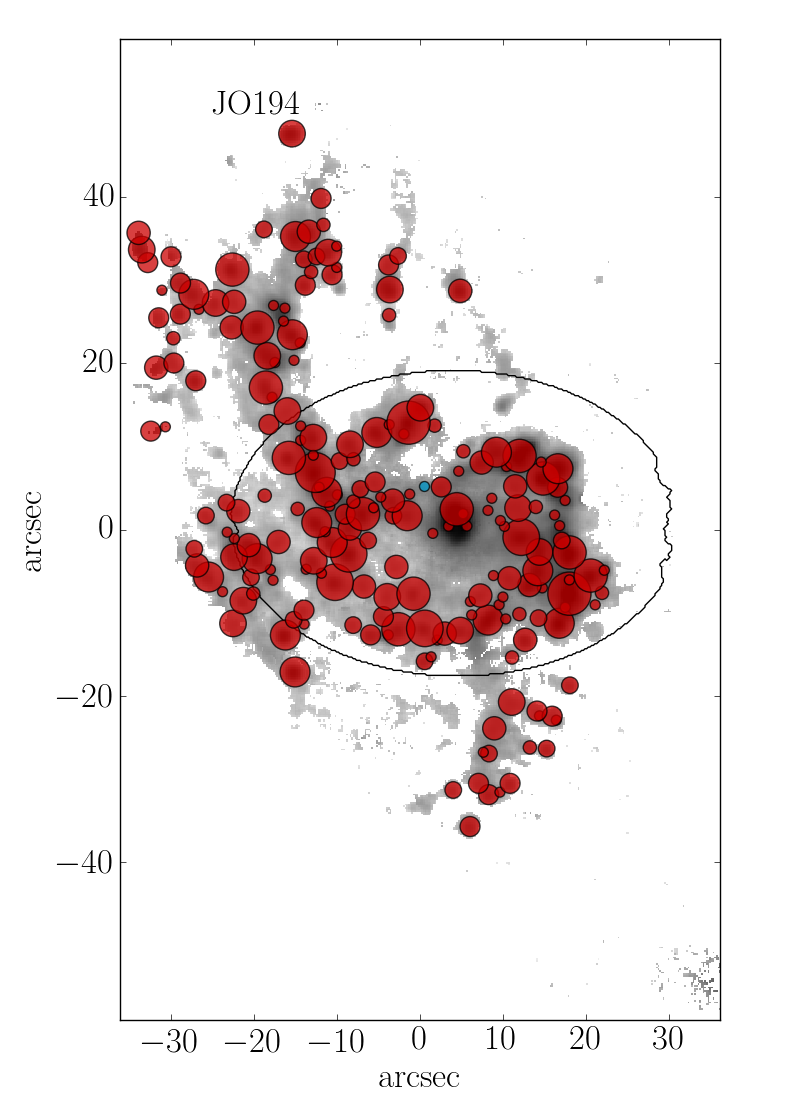}\includegraphics[width=2.2in]{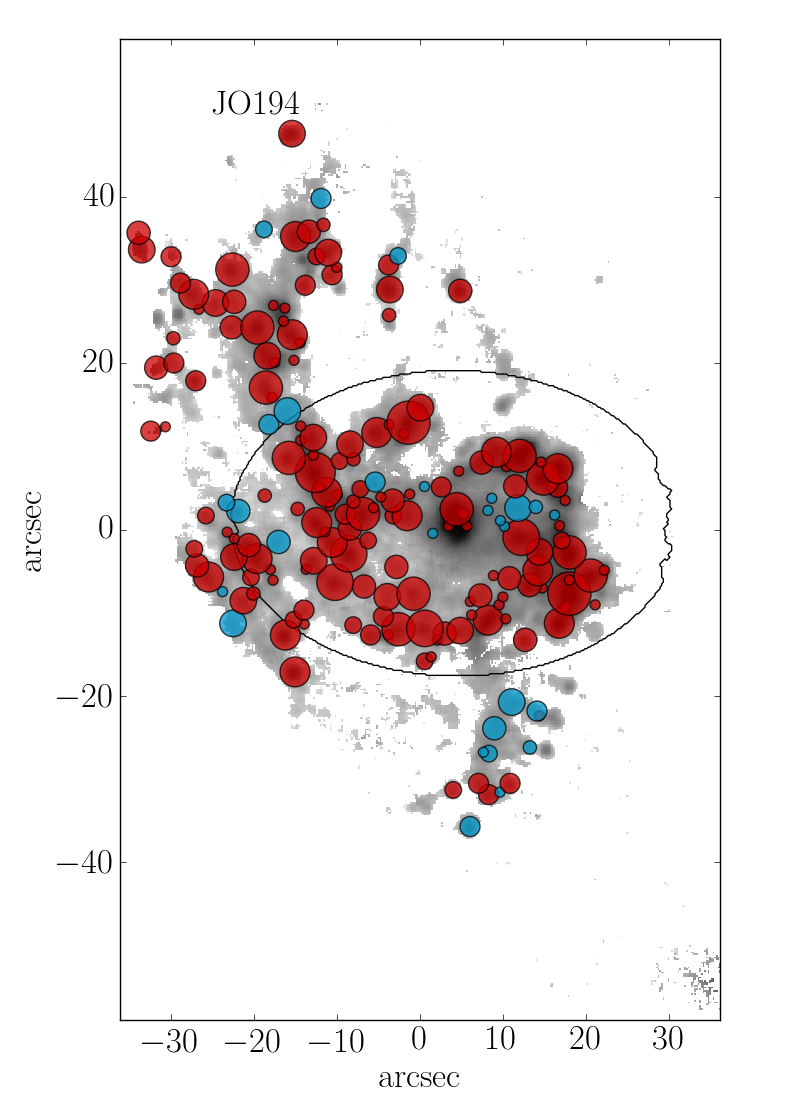}}
\contcaption{}
\end{figure*}

\clearpage 

\begin{figure*}
\centerline{\includegraphics[width=2.2in]{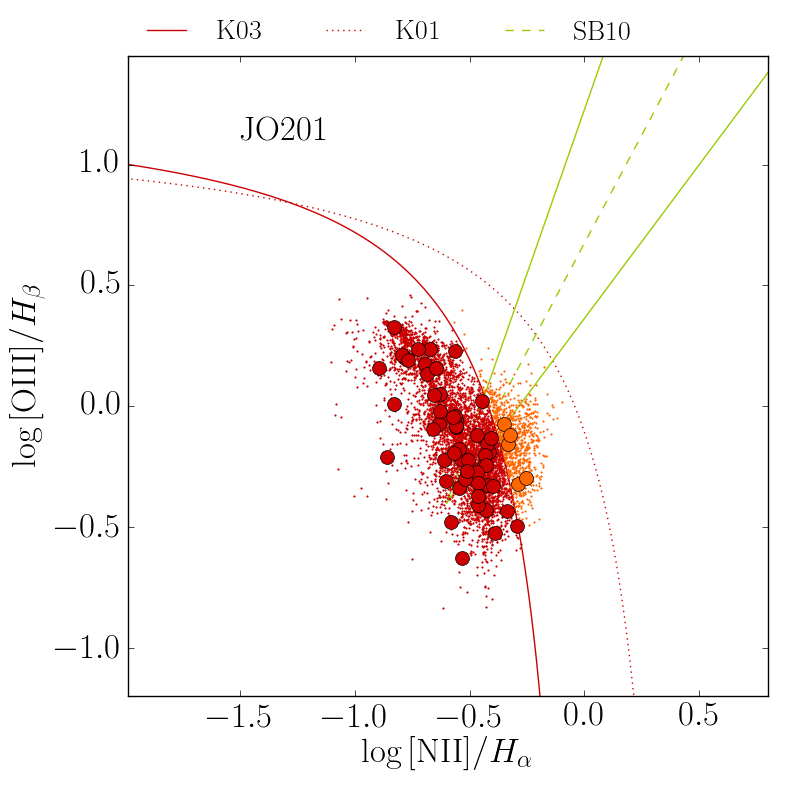}\includegraphics[width=2.2in]{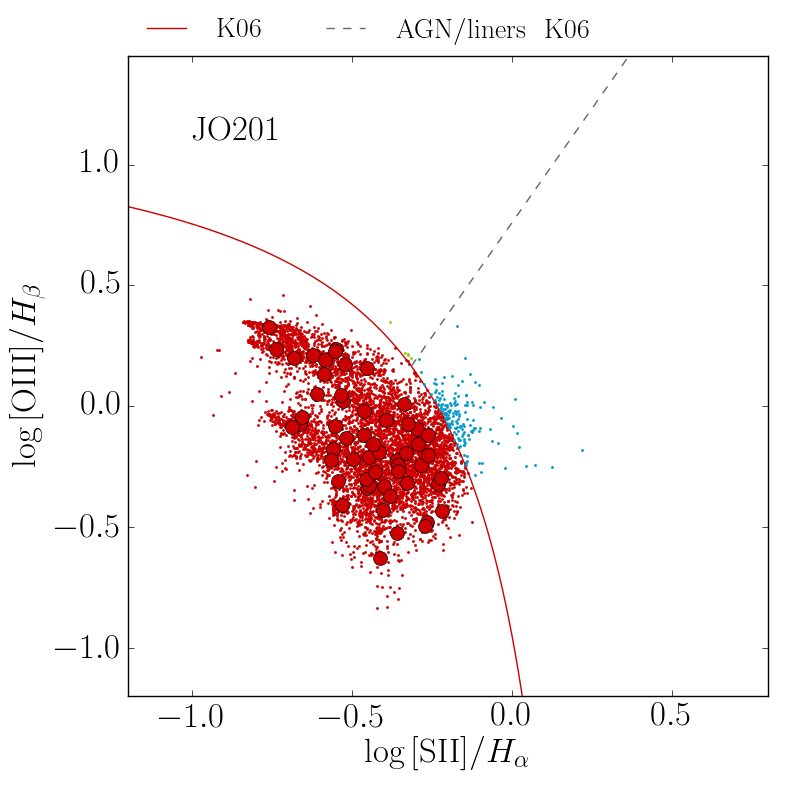}\includegraphics[width=2.2in]{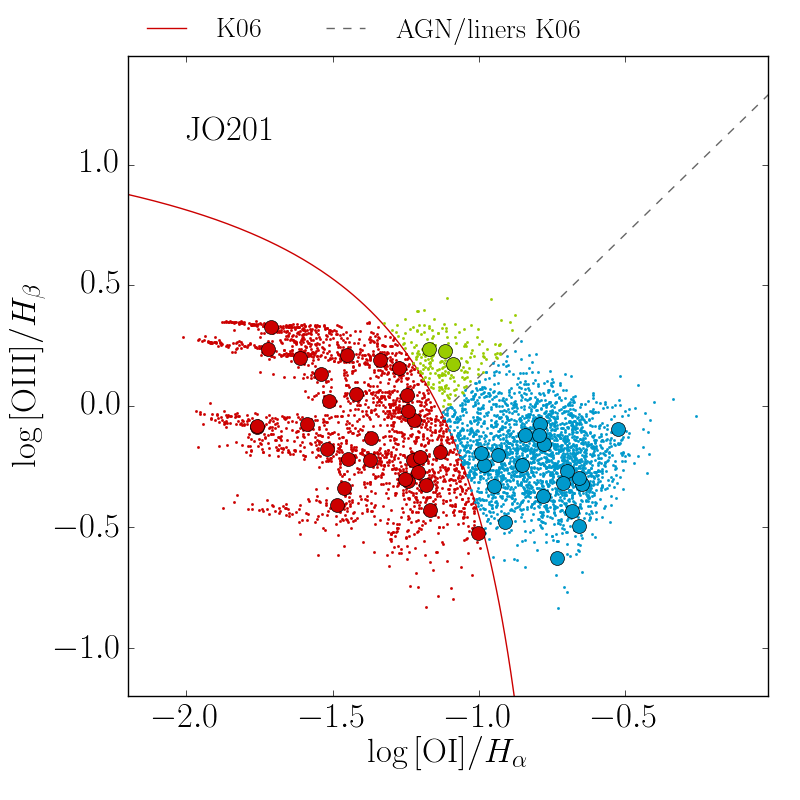}}
\centerline{\includegraphics[width=2.2in]{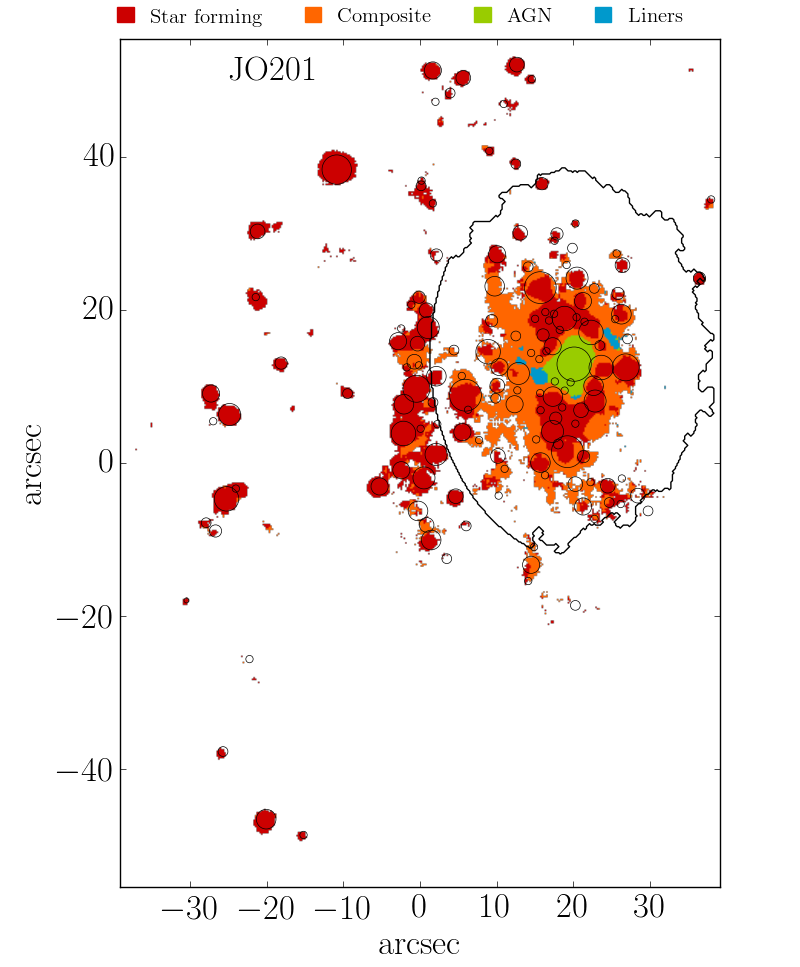}\includegraphics[width=2.2in]{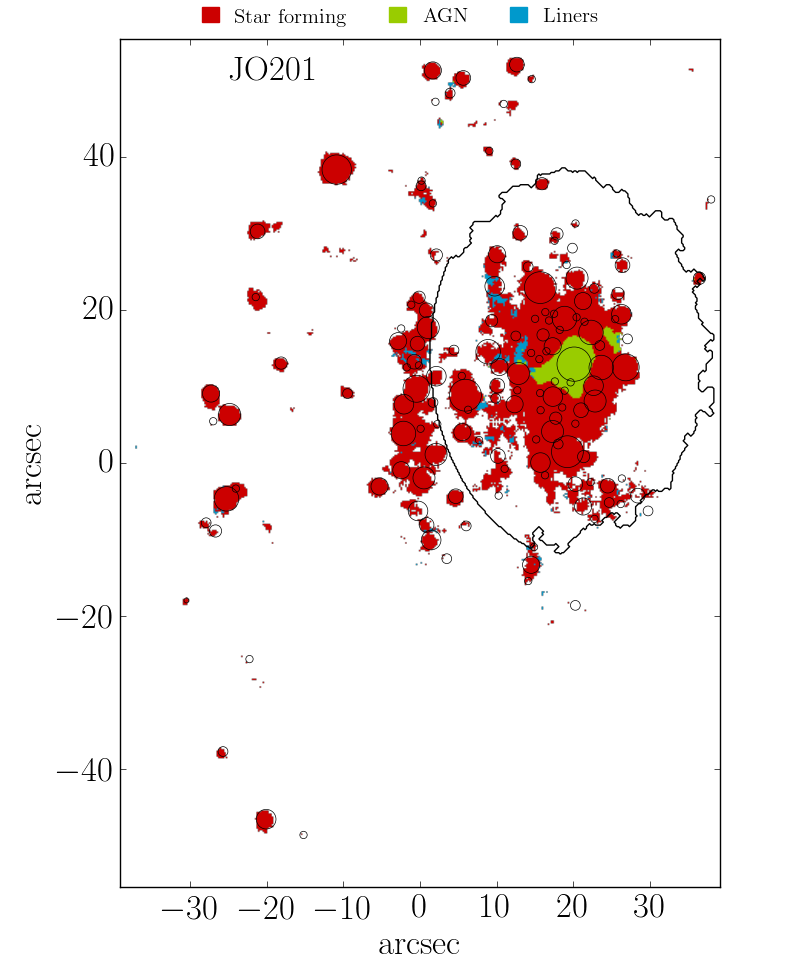}\includegraphics[width=2.2in]{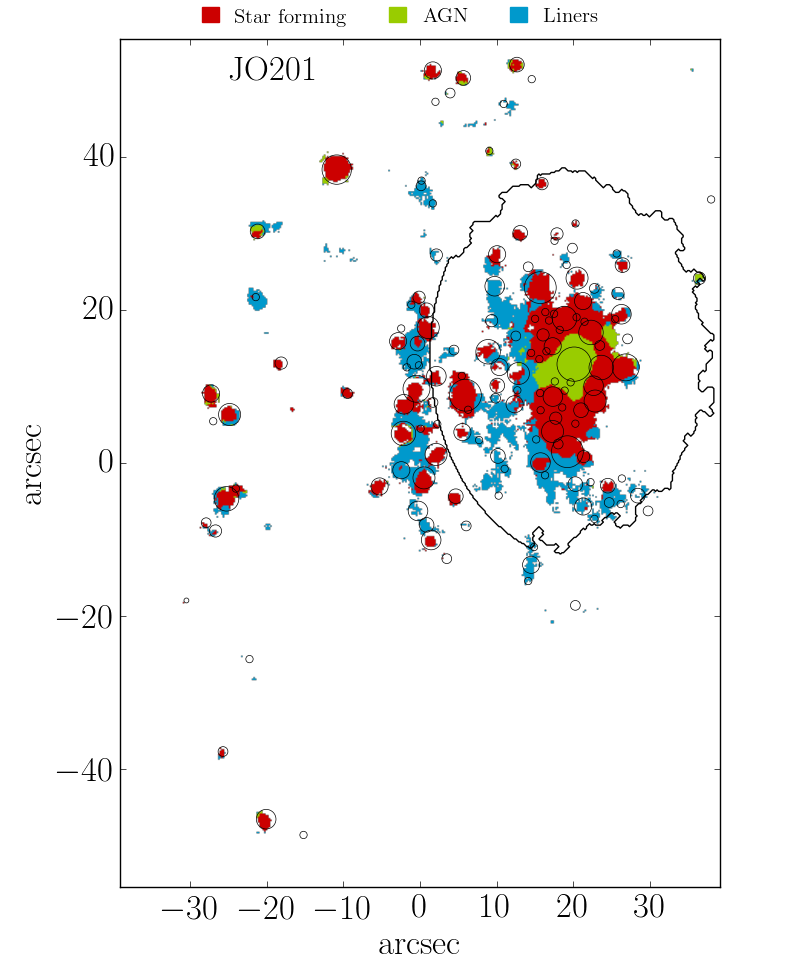}}
\centerline{\includegraphics[width=2.2in]{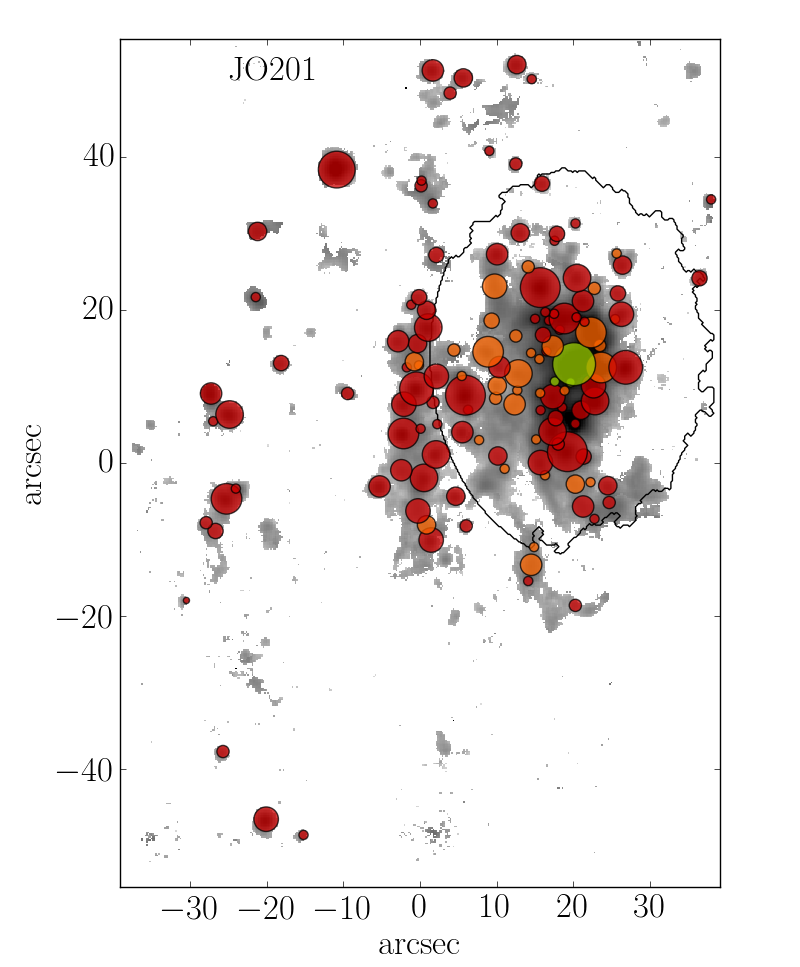}\includegraphics[width=2.2in]{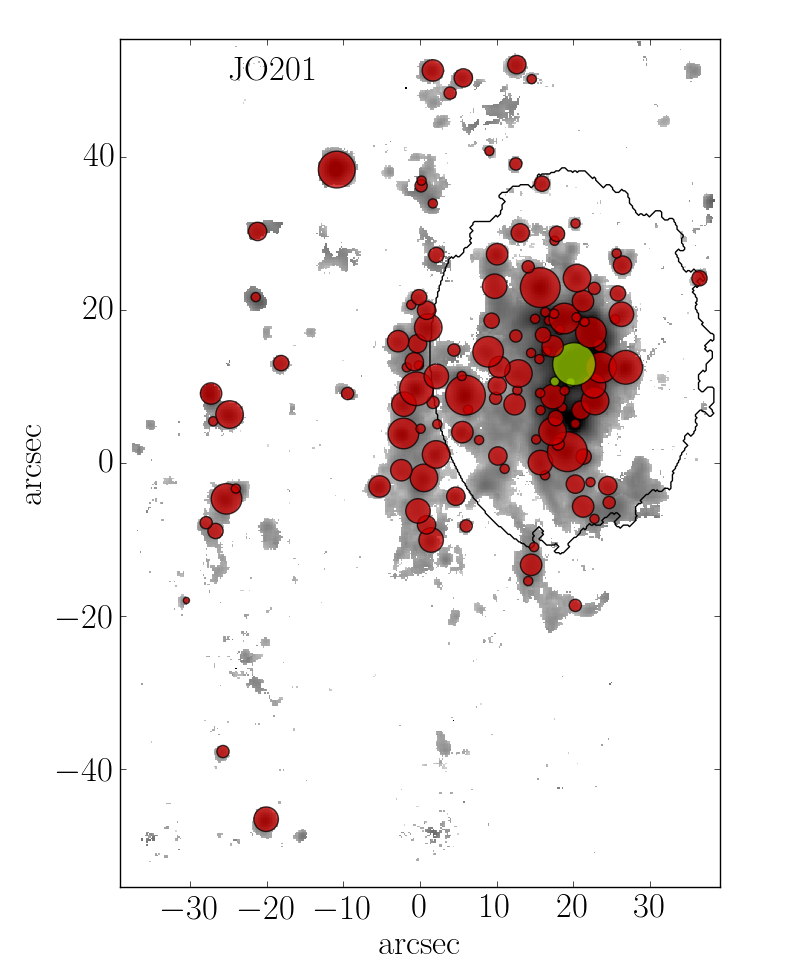}\includegraphics[width=2.2in]{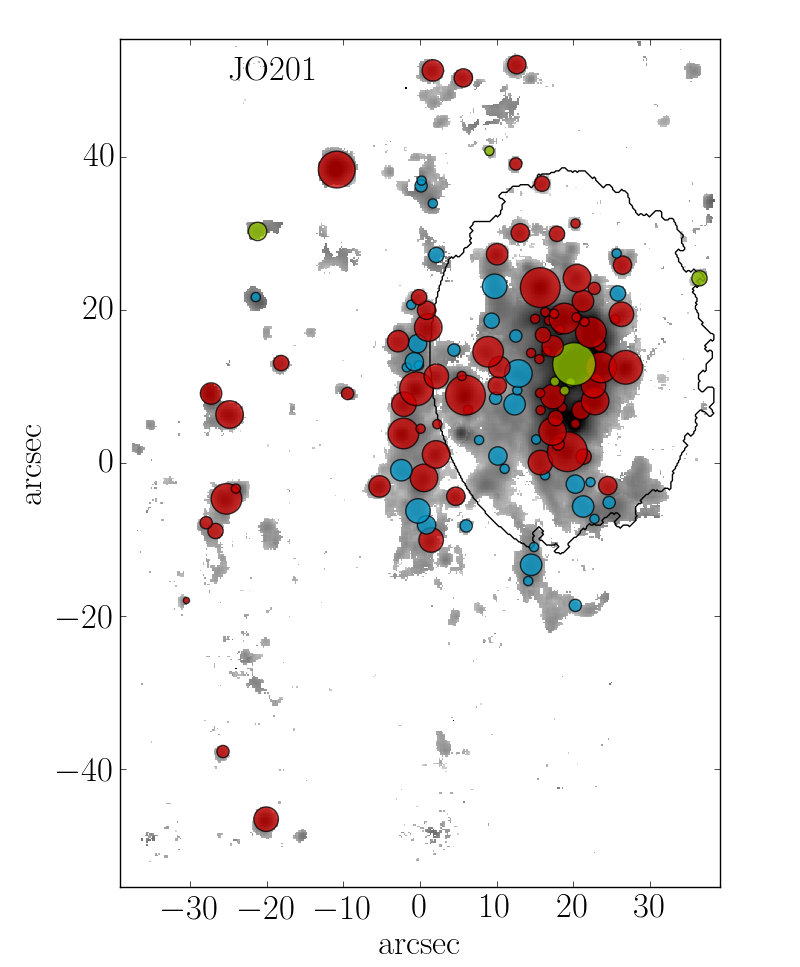}}
\contcaption{}
\end{figure*}

\begin{figure*}
\centerline{\includegraphics[width=2.4in]{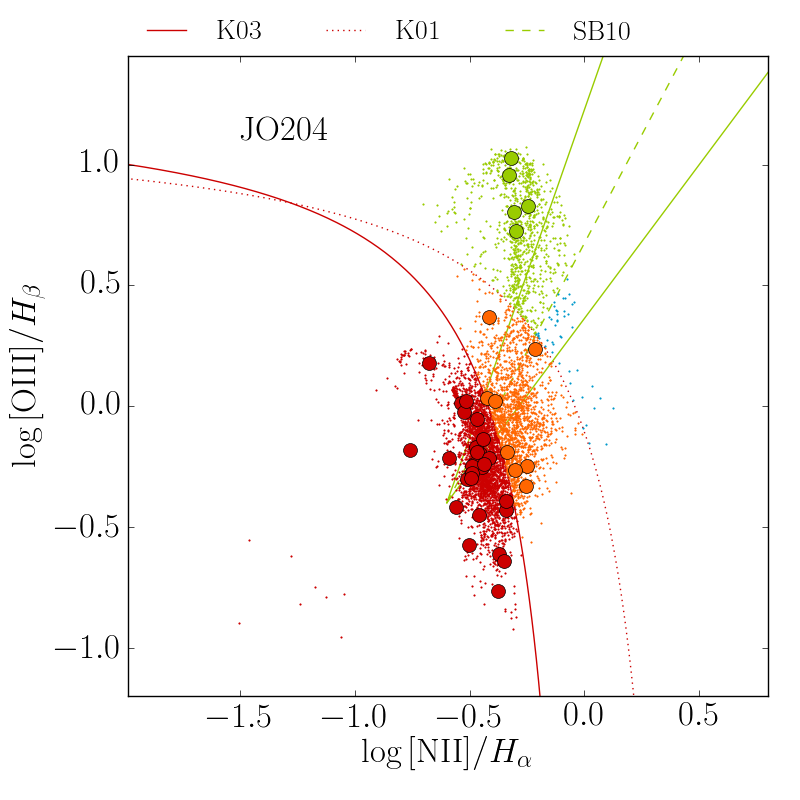}\includegraphics[width=2.4in]{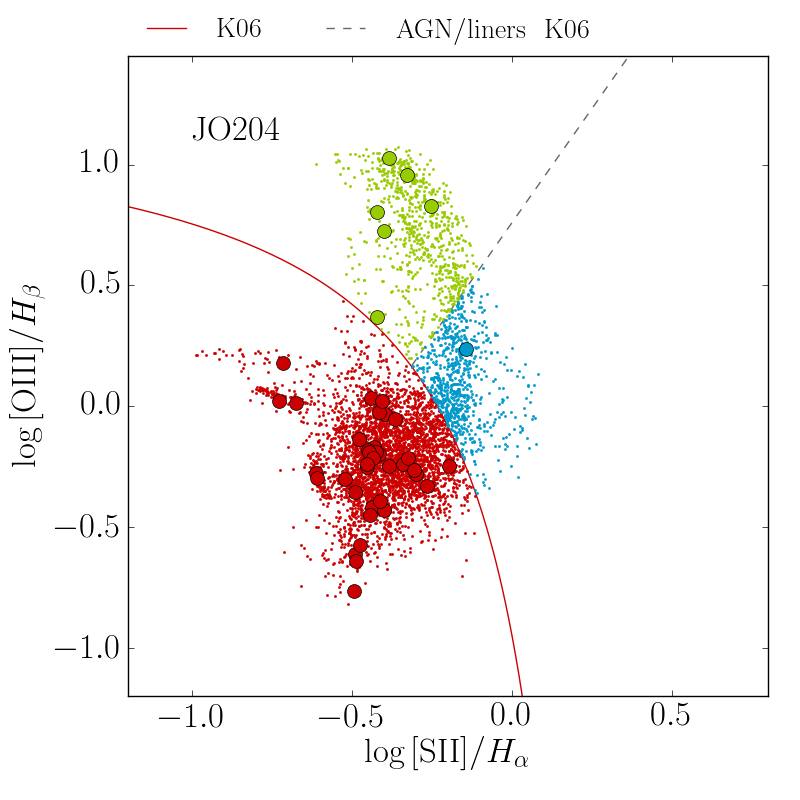}\includegraphics[width=2.4in]{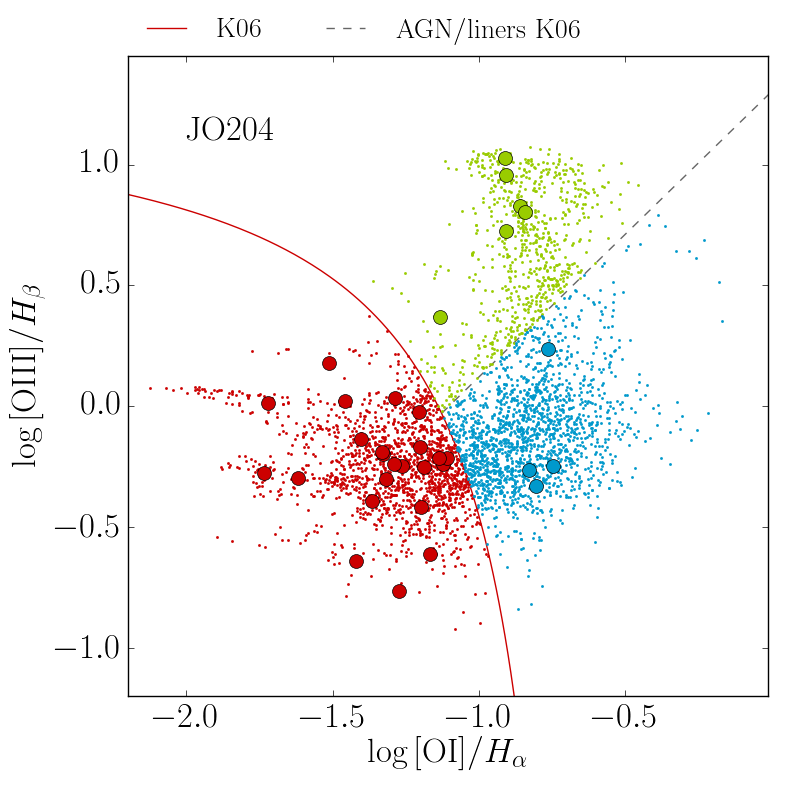}}
\centerline{\includegraphics[width=2.4in]{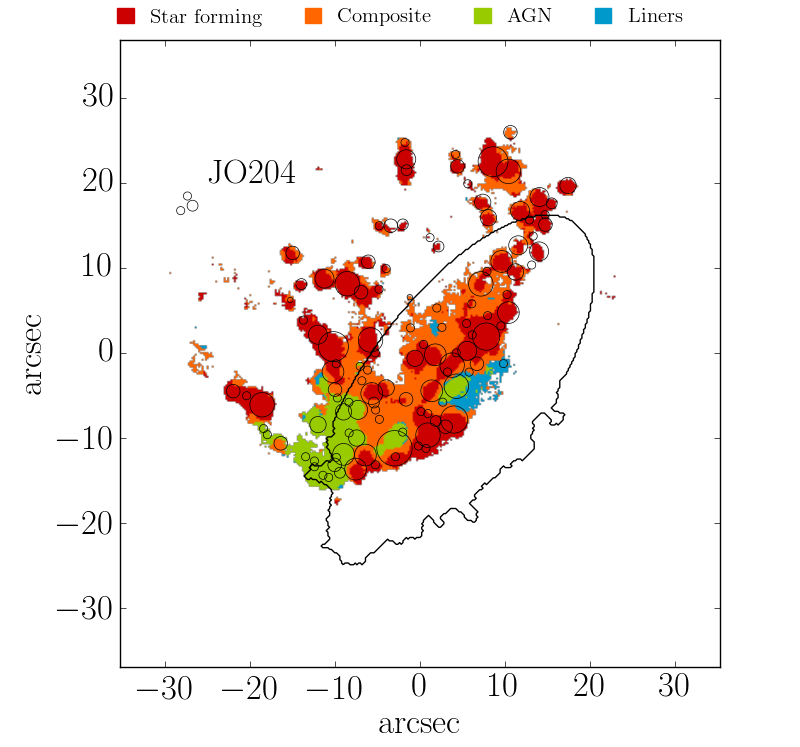}\includegraphics[width=2.4in]{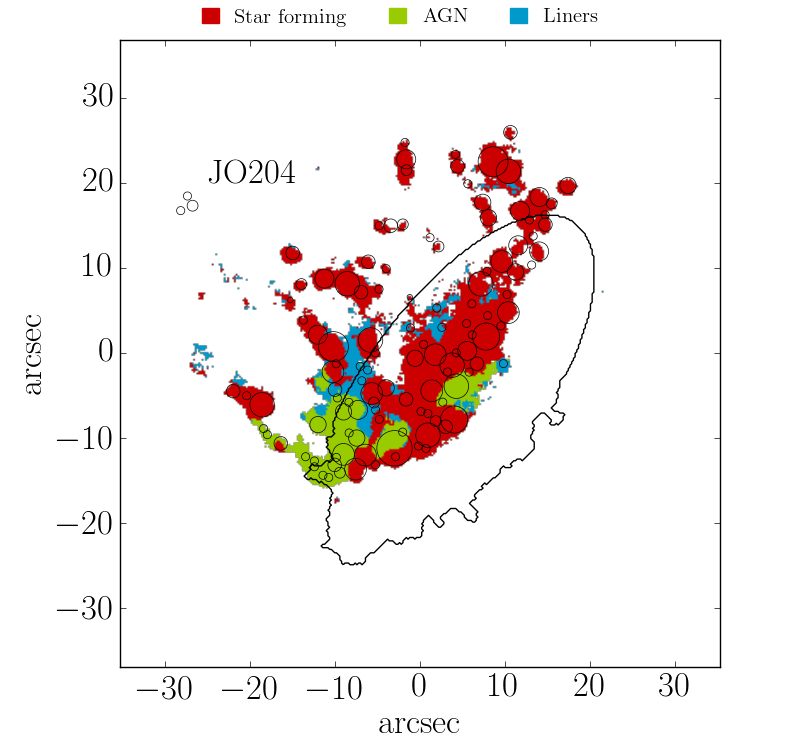}\includegraphics[width=2.4in]{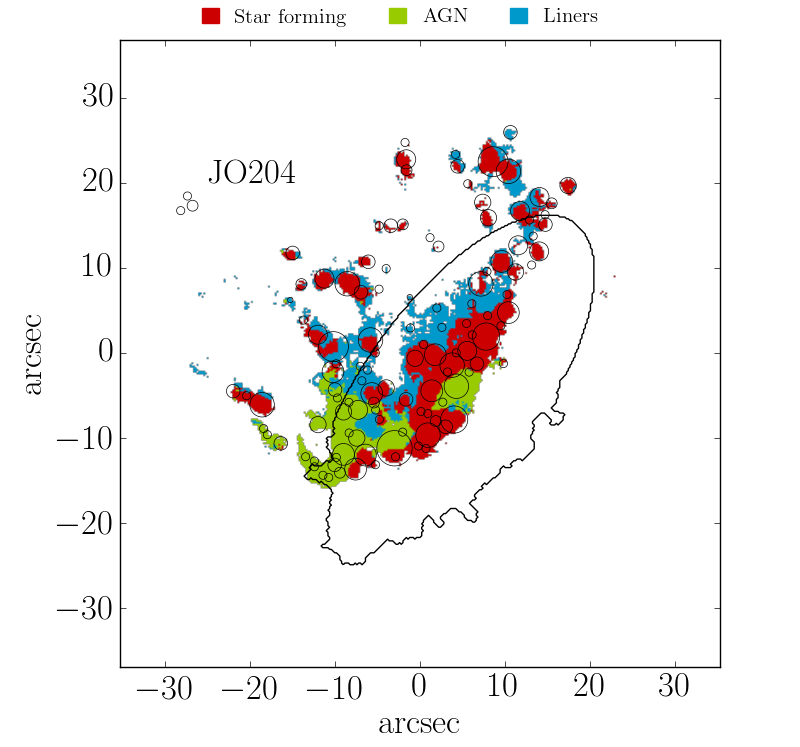}}
\centerline{\includegraphics[width=2.4in]{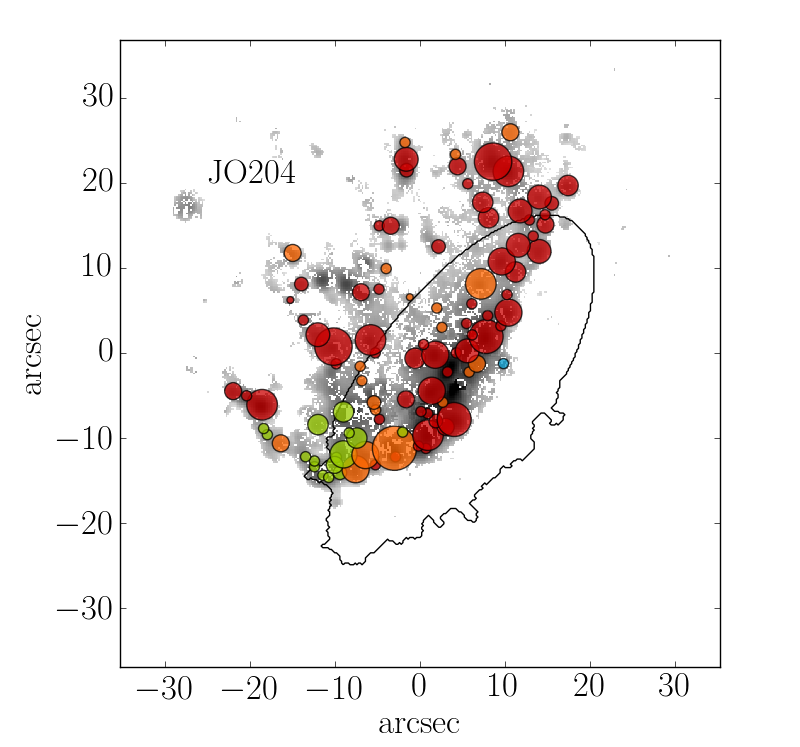}\includegraphics[width=2.4in]{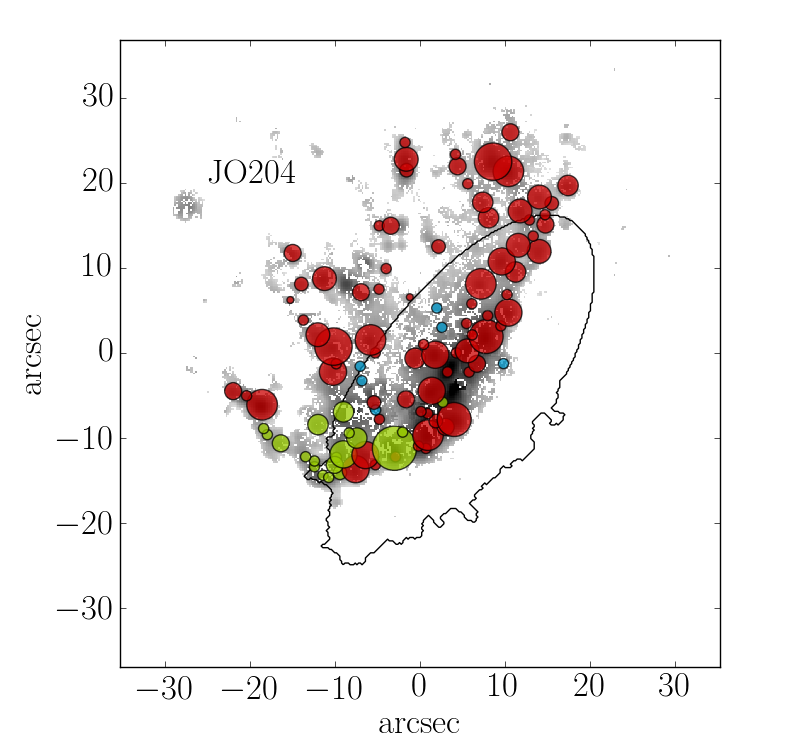}\includegraphics[width=2.4in]{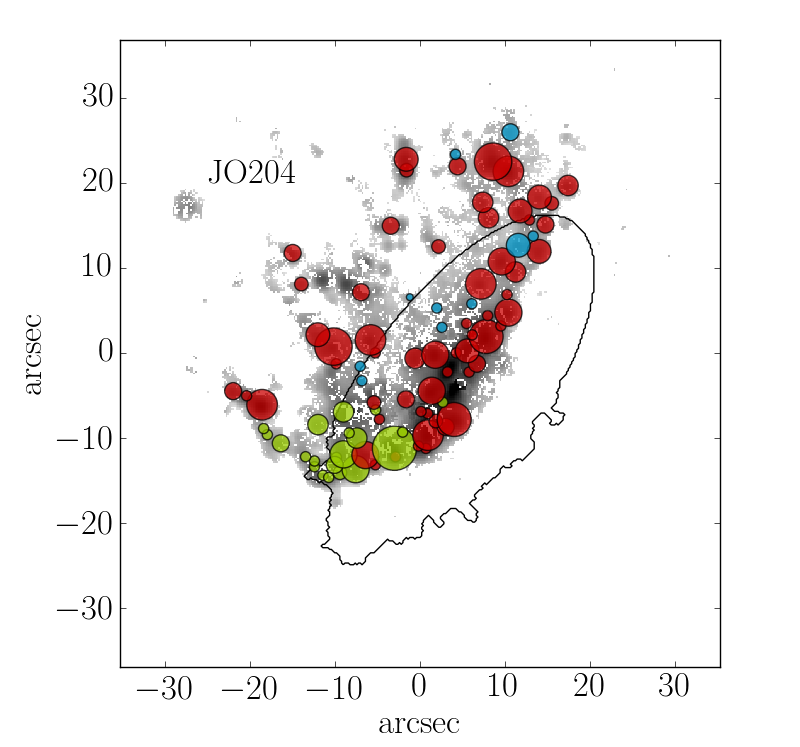}}
\contcaption{}
\end{figure*}

\begin{figure*}
\centerline{\includegraphics[width=2.4in]{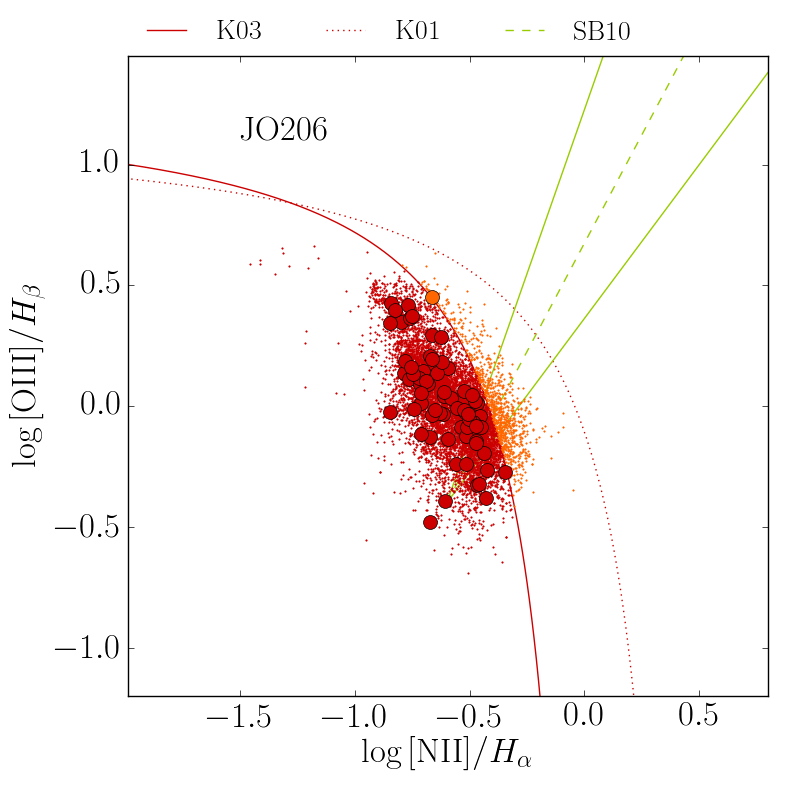}\includegraphics[width=2.4in]{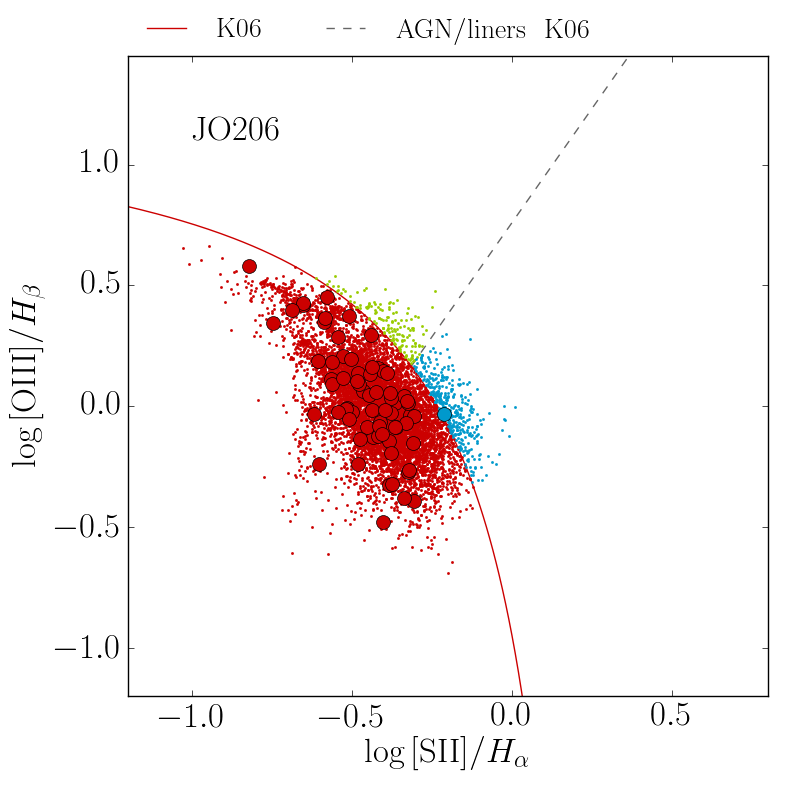}\includegraphics[width=2.4in]{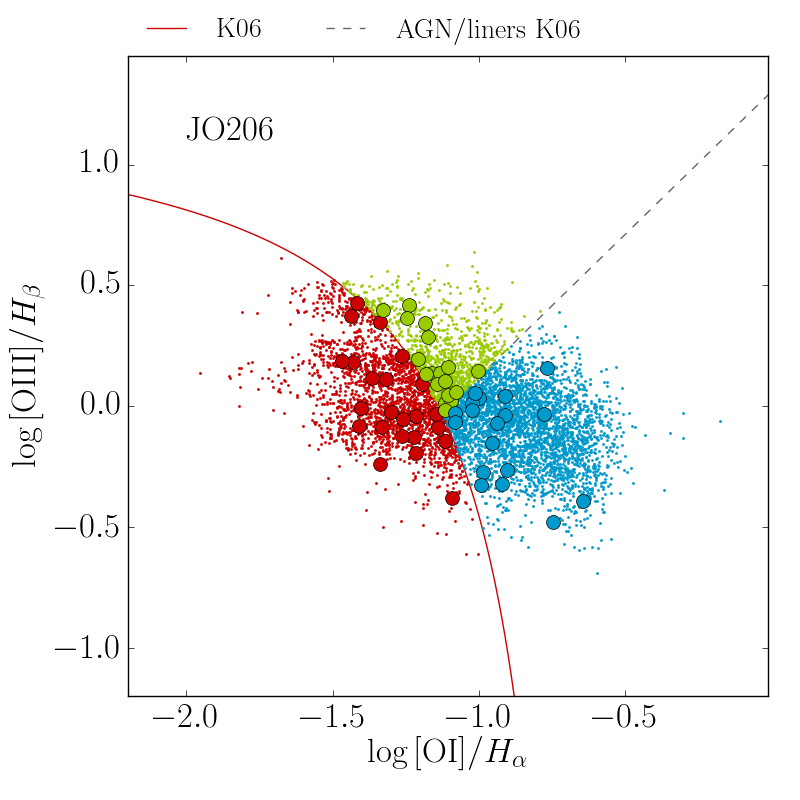}}
\centerline{\includegraphics[width=2.4in]{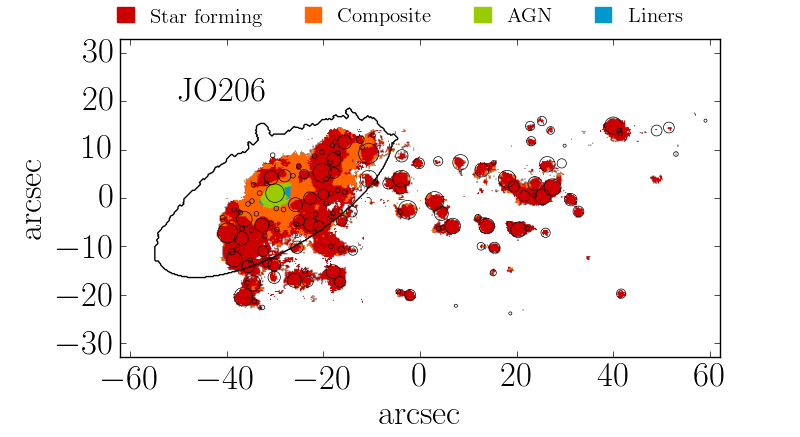}\includegraphics[width=2.4in]{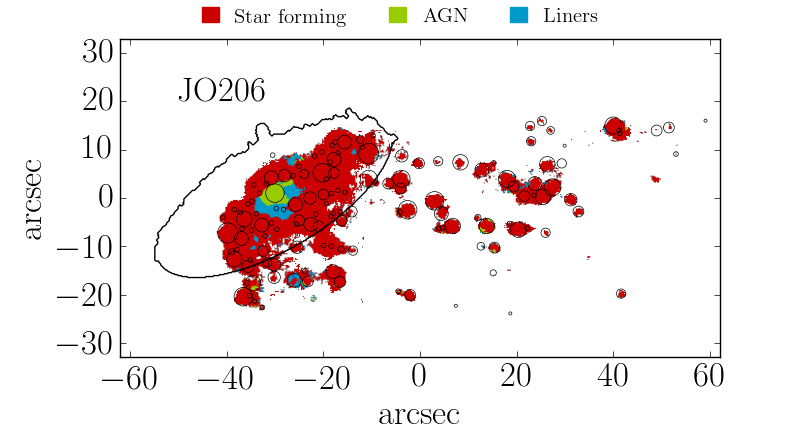}\includegraphics[width=2.4in]{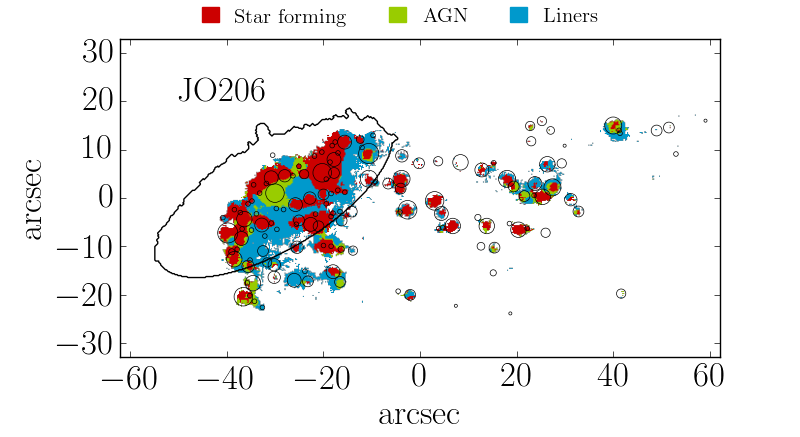}}
\centerline{\includegraphics[width=2.4in]{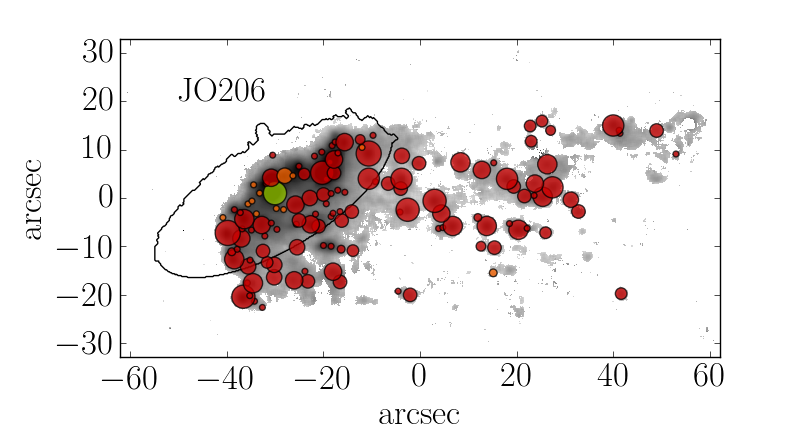}\includegraphics[width=2.4in]{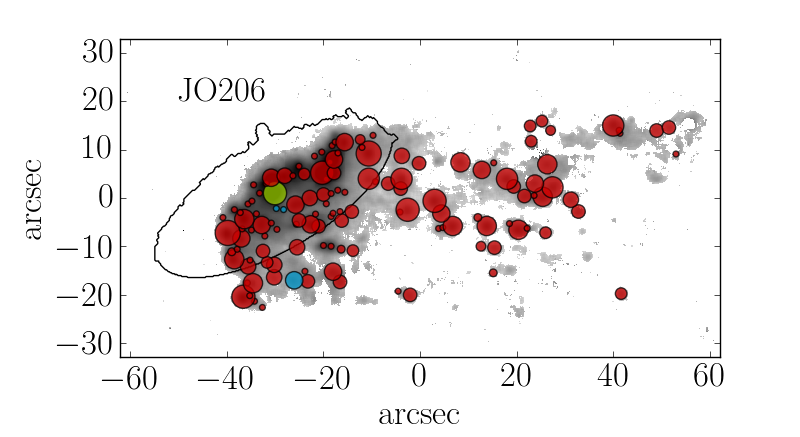}\includegraphics[width=2.4in]{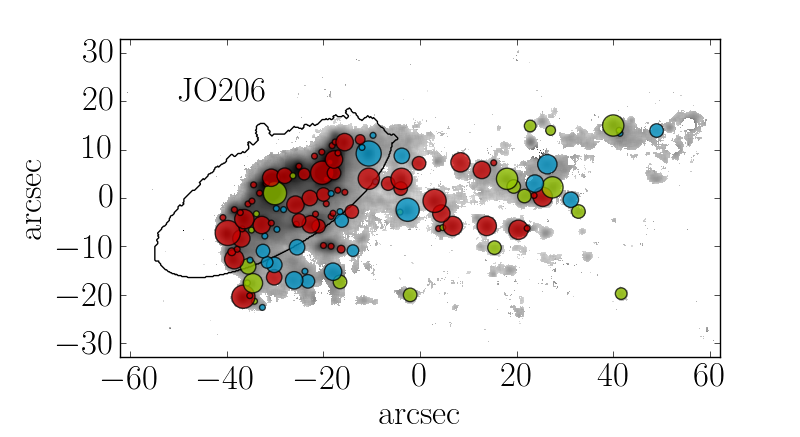}}
\contcaption{}
\end{figure*}

\begin{figure*}
\centerline{\includegraphics[width=2.4in]{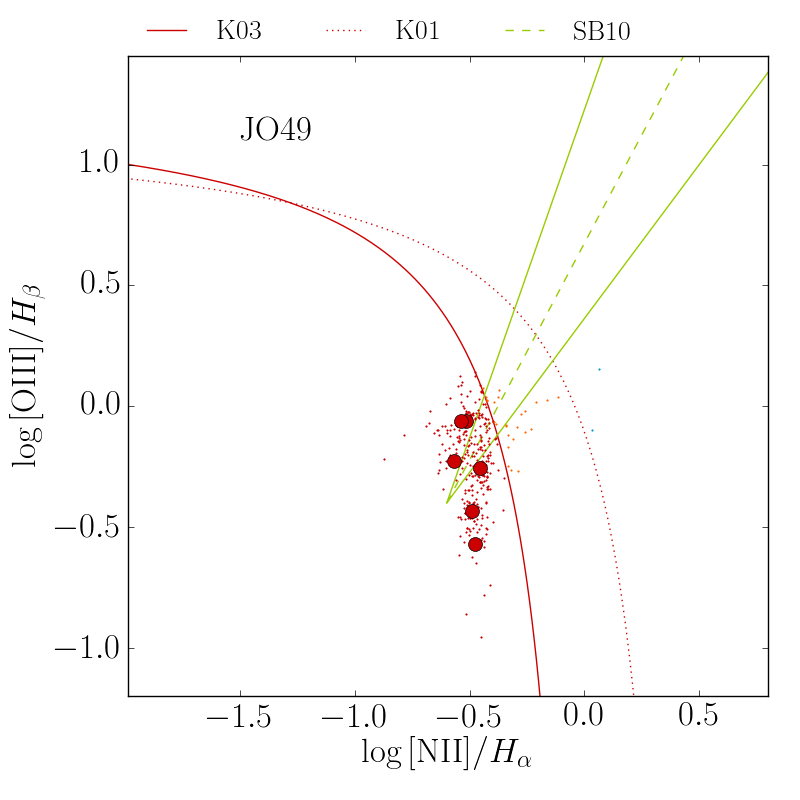}\includegraphics[width=2.4in]{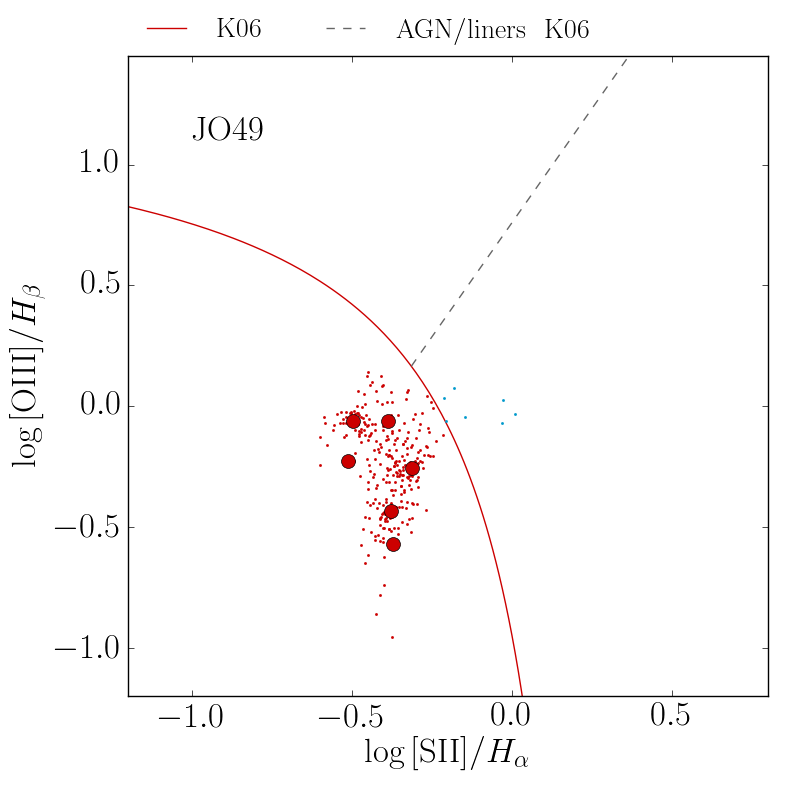}\includegraphics[width=2.4in]{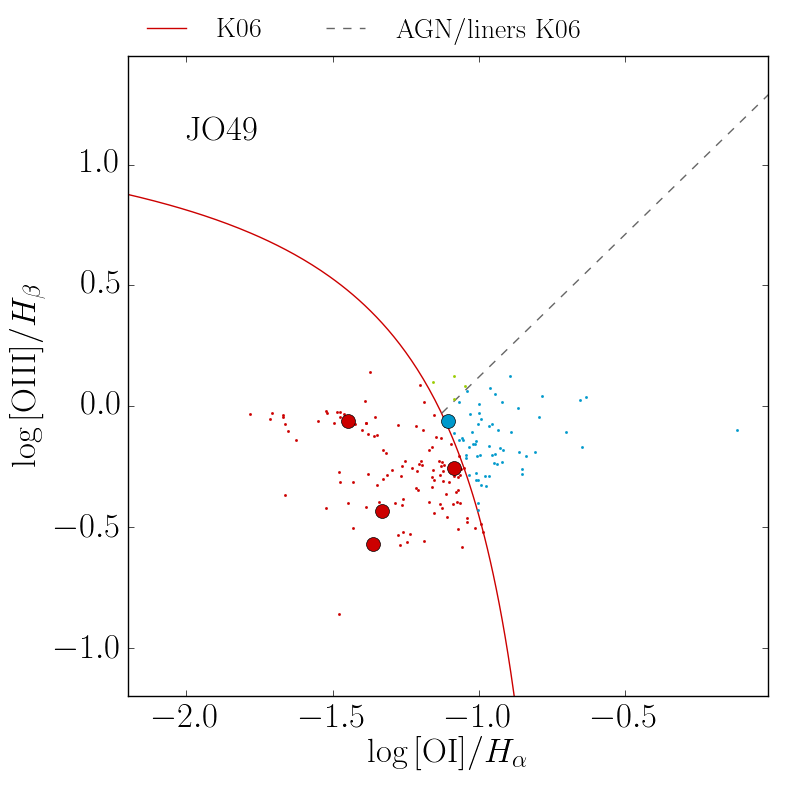}}
\centerline{\includegraphics[width=2.4in]{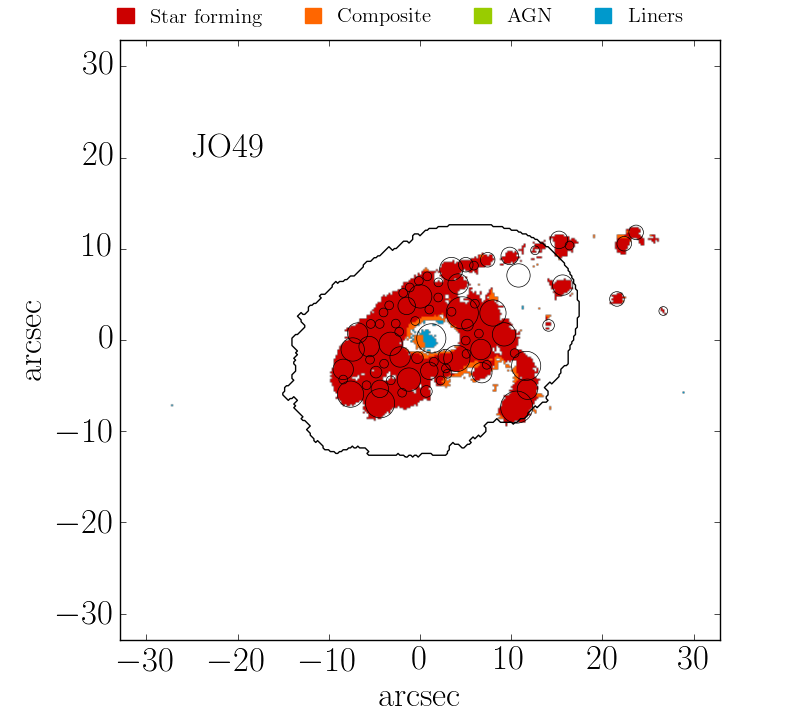}\includegraphics[width=2.4in]{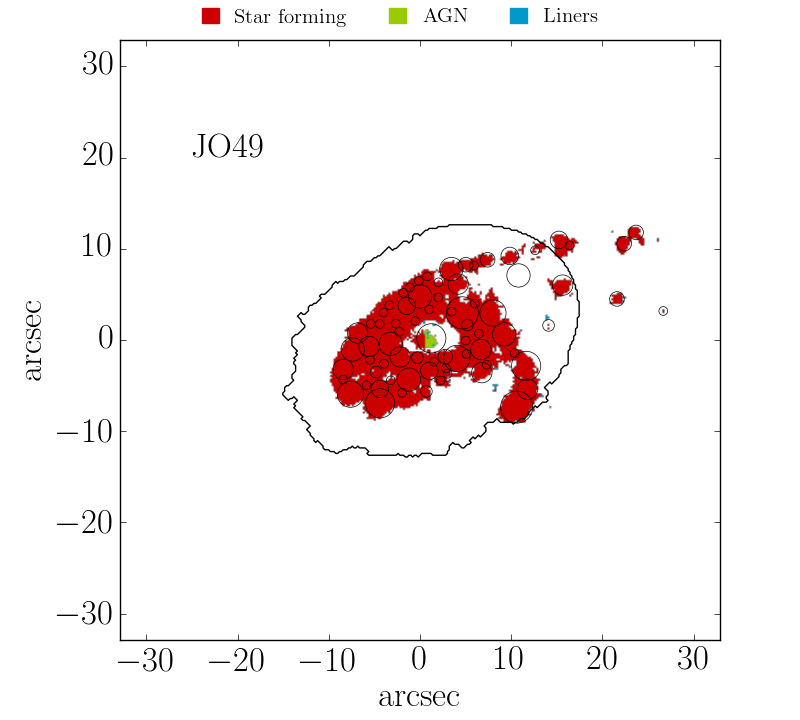}\includegraphics[width=2.4in]{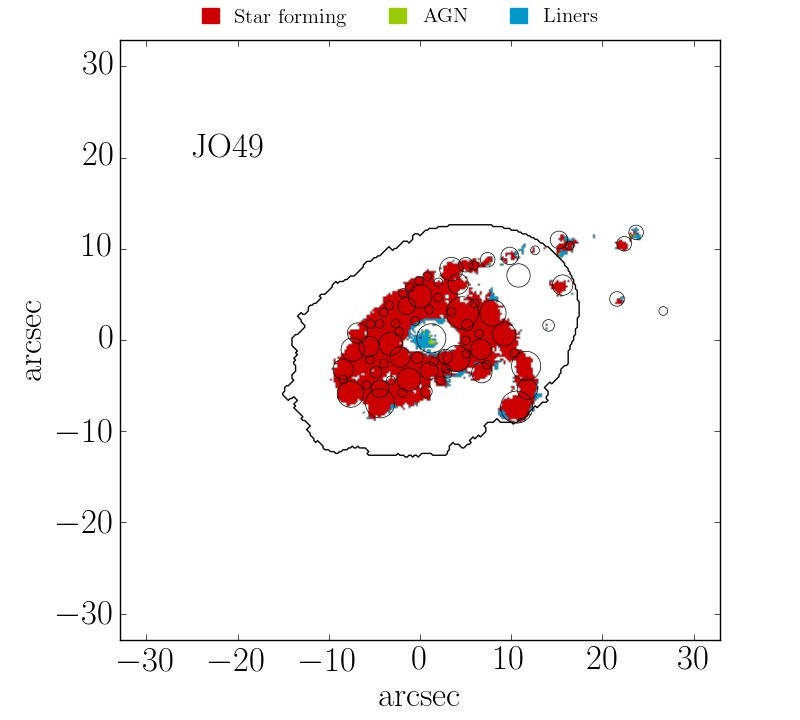}}
\centerline{\includegraphics[width=2.4in]{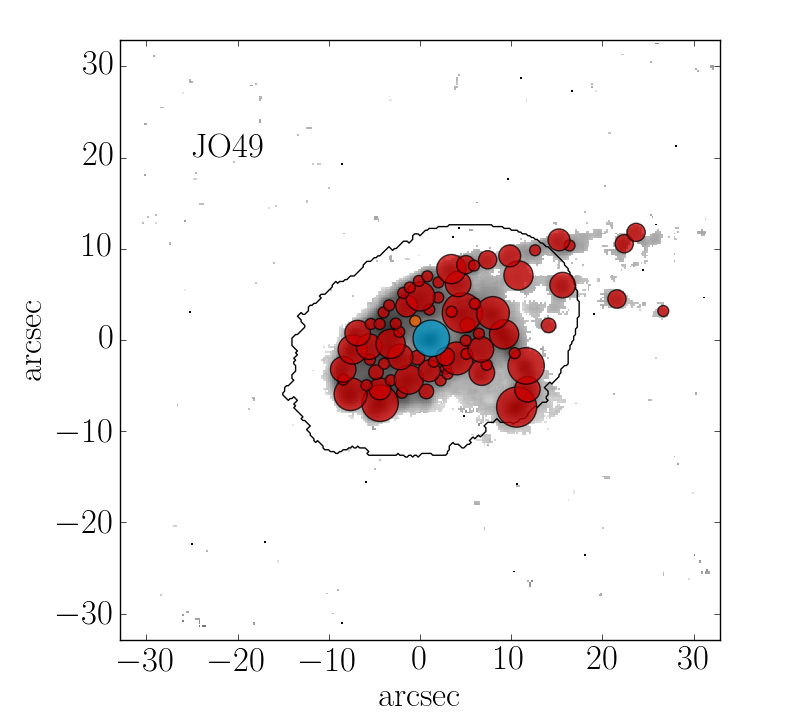}\includegraphics[width=2.4in]{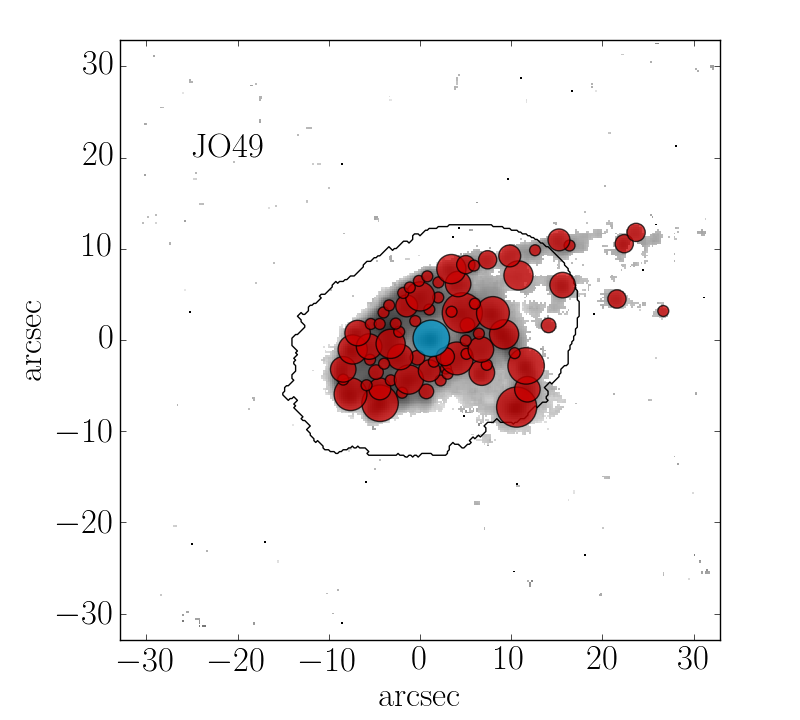}\includegraphics[width=2.4in]{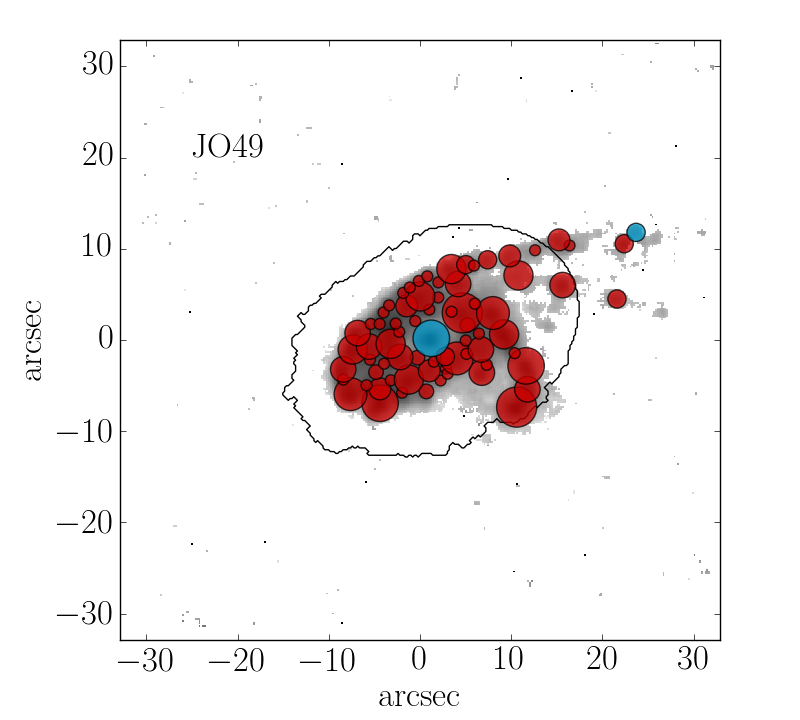}}
\contcaption{}
\end{figure*}

\begin{figure*}
\centerline{\includegraphics[width=2.4in]{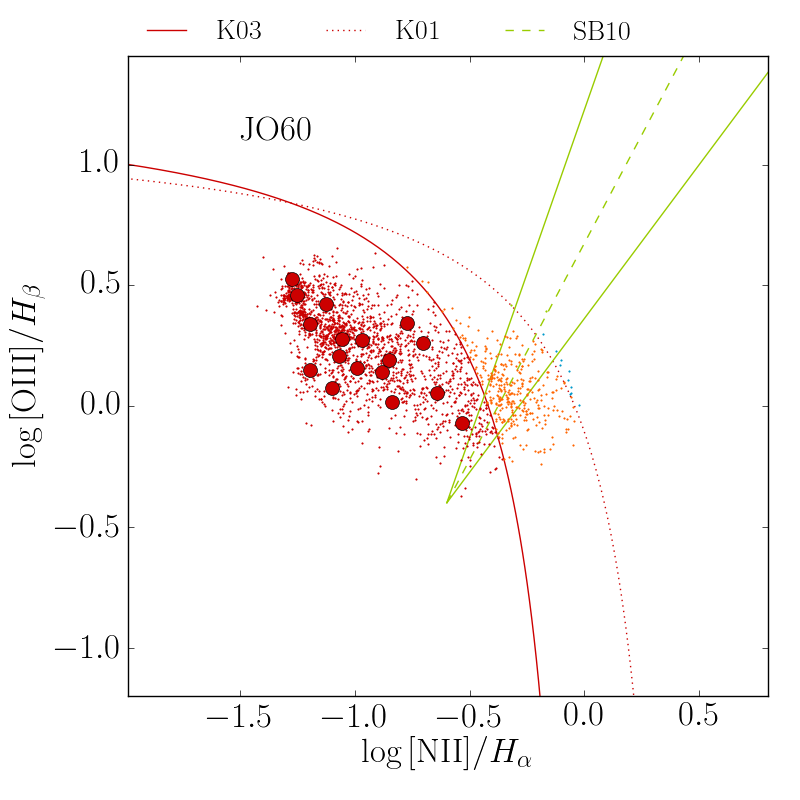}\includegraphics[width=2.4in]{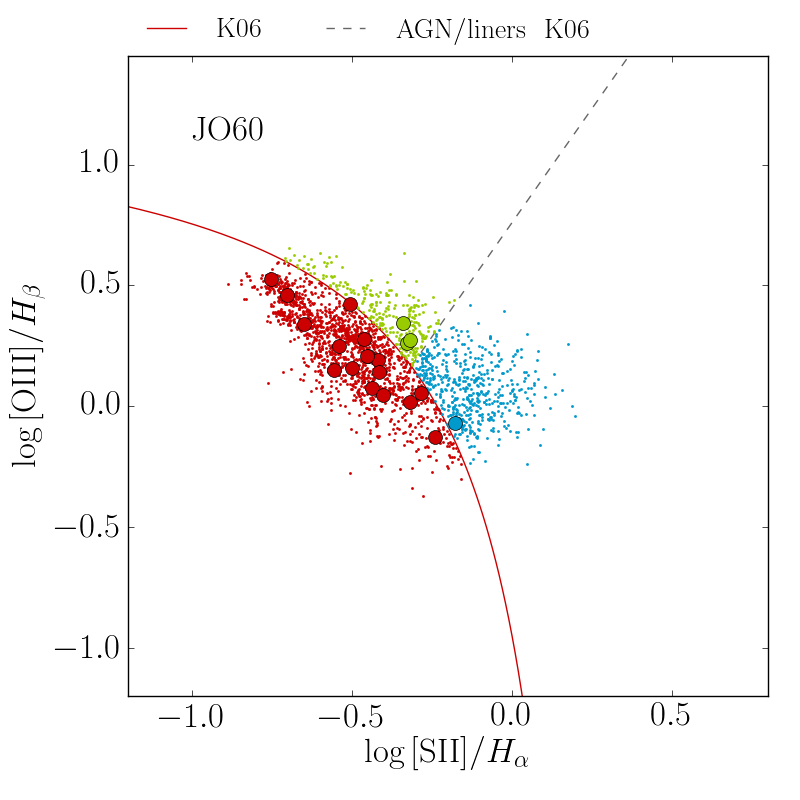}\includegraphics[width=2.4in]{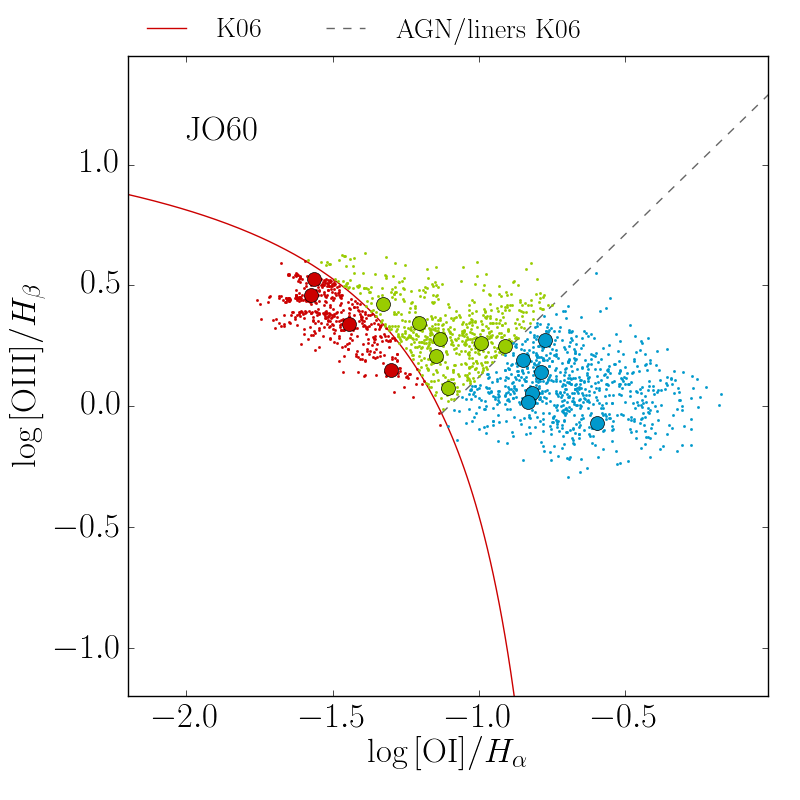}}
\centerline{\includegraphics[width=2.4in]{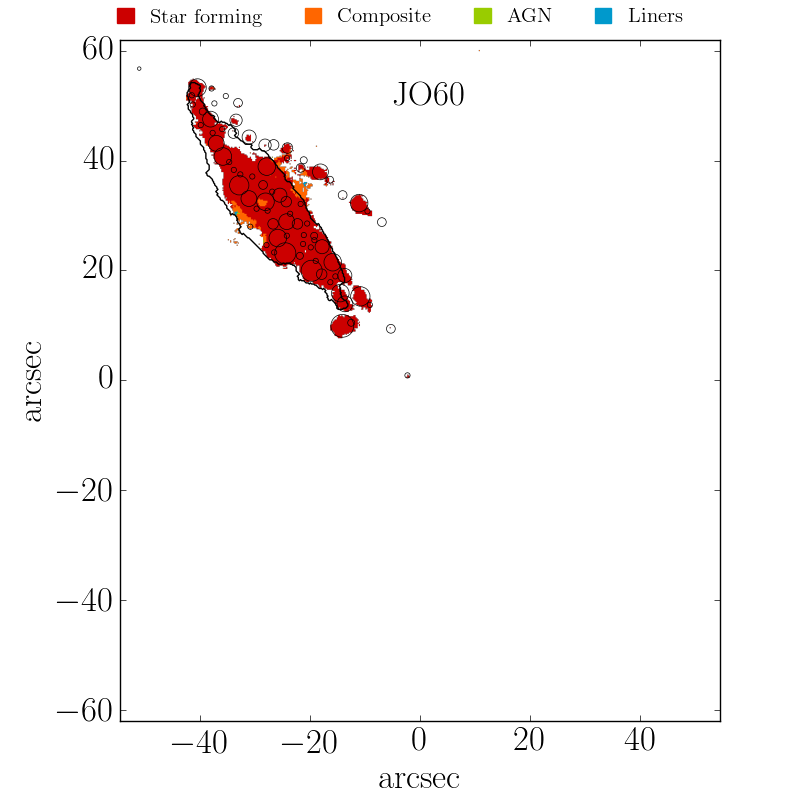}\includegraphics[width=2.4in]{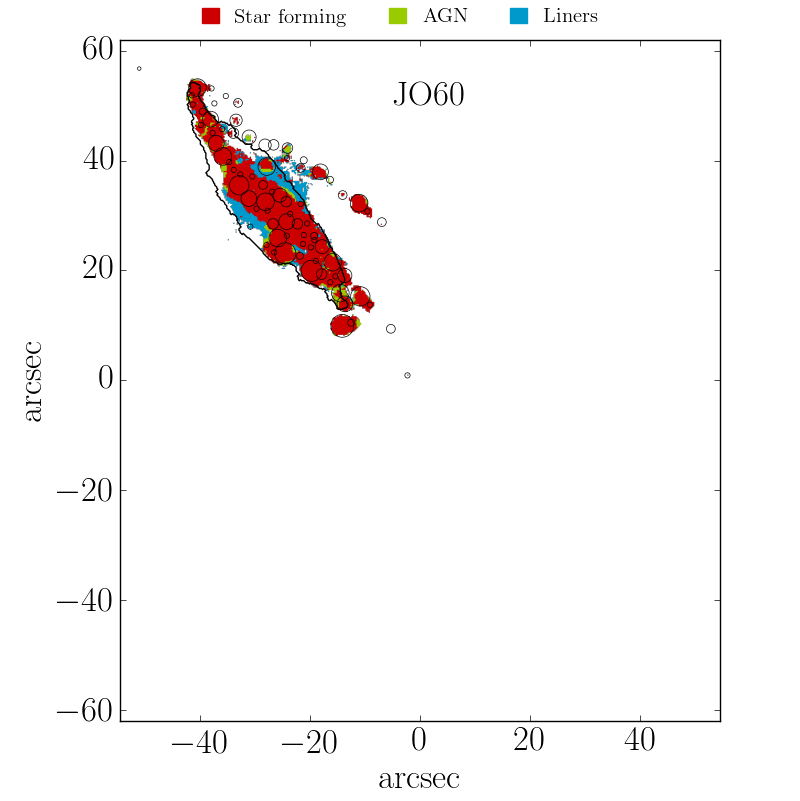}\includegraphics[width=2.4in]{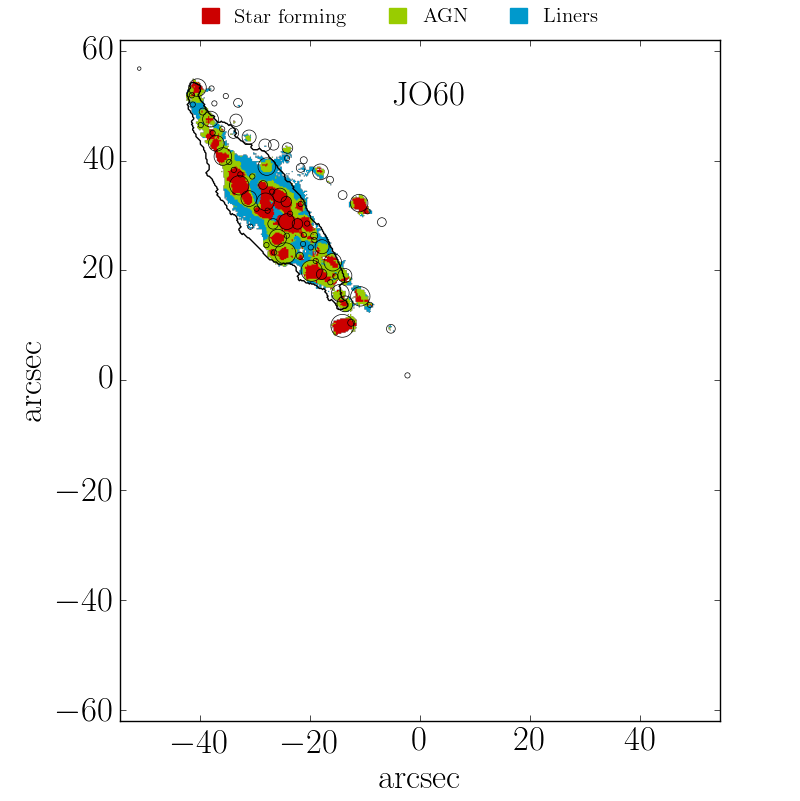}}
\centerline{\includegraphics[width=2.4in]{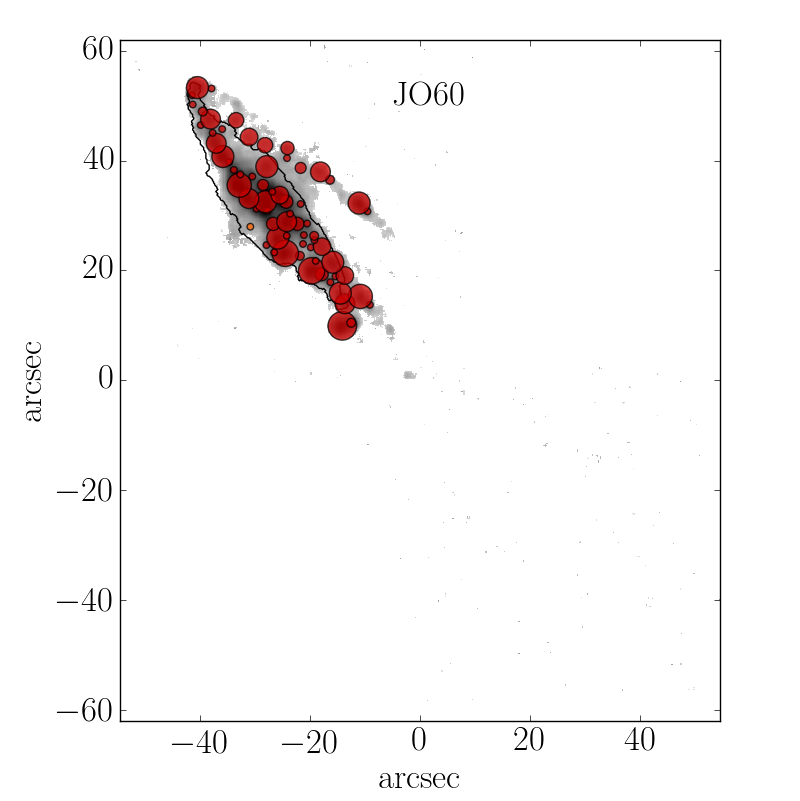}\includegraphics[width=2.4in]{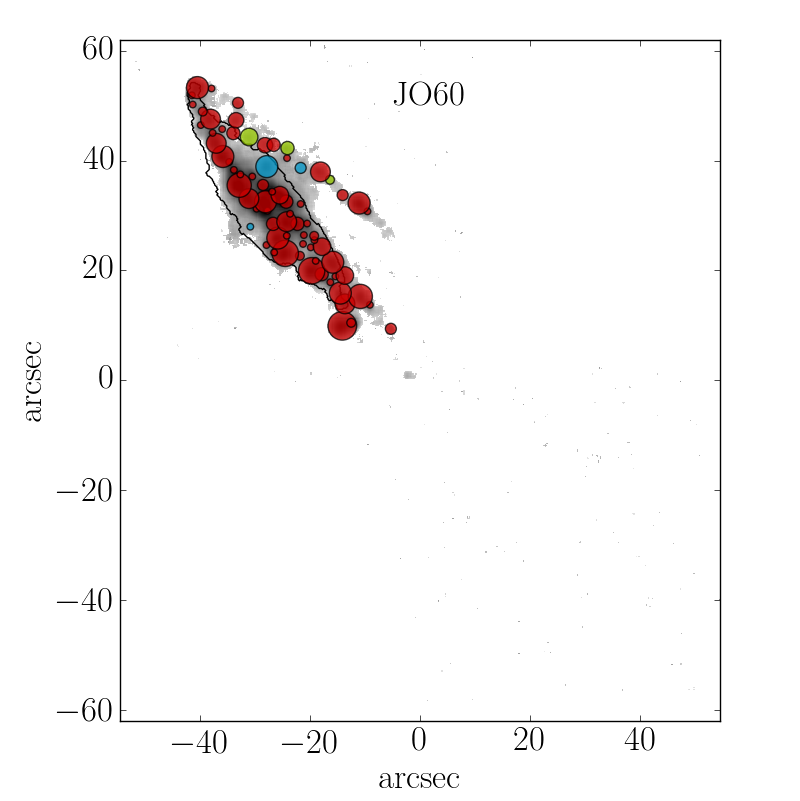}\includegraphics[width=2.4in]{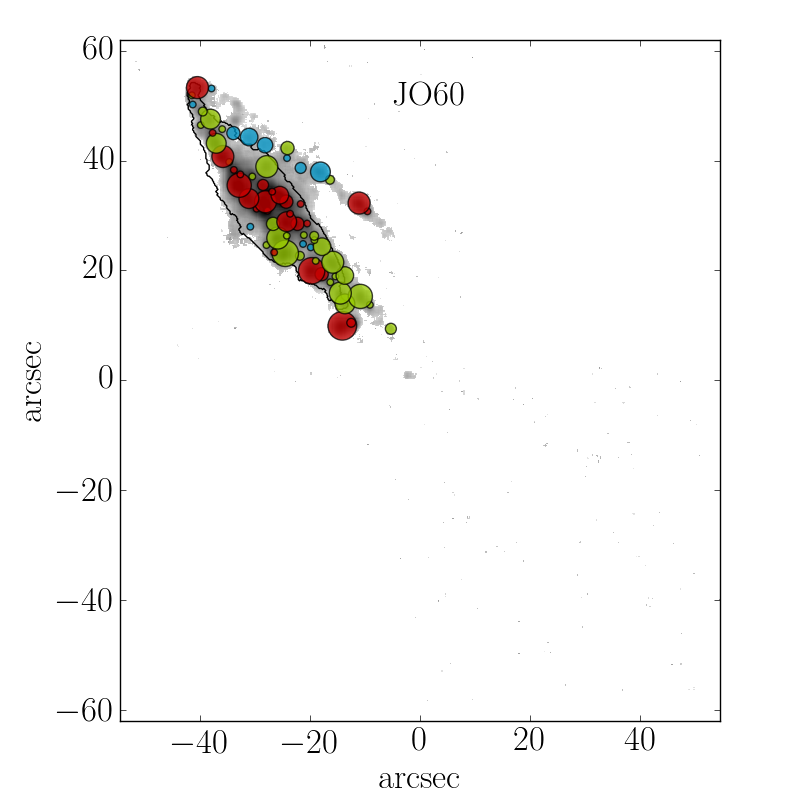}}
\contcaption{}
\end{figure*}

\begin{figure*}
\centerline{\includegraphics[width=2.4in]{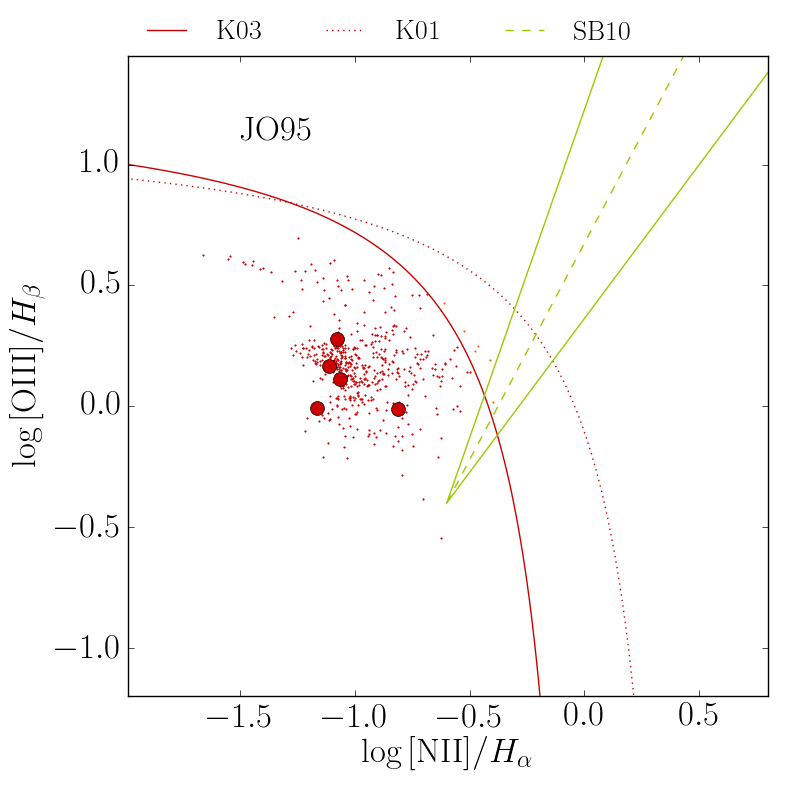}\includegraphics[width=2.4in]{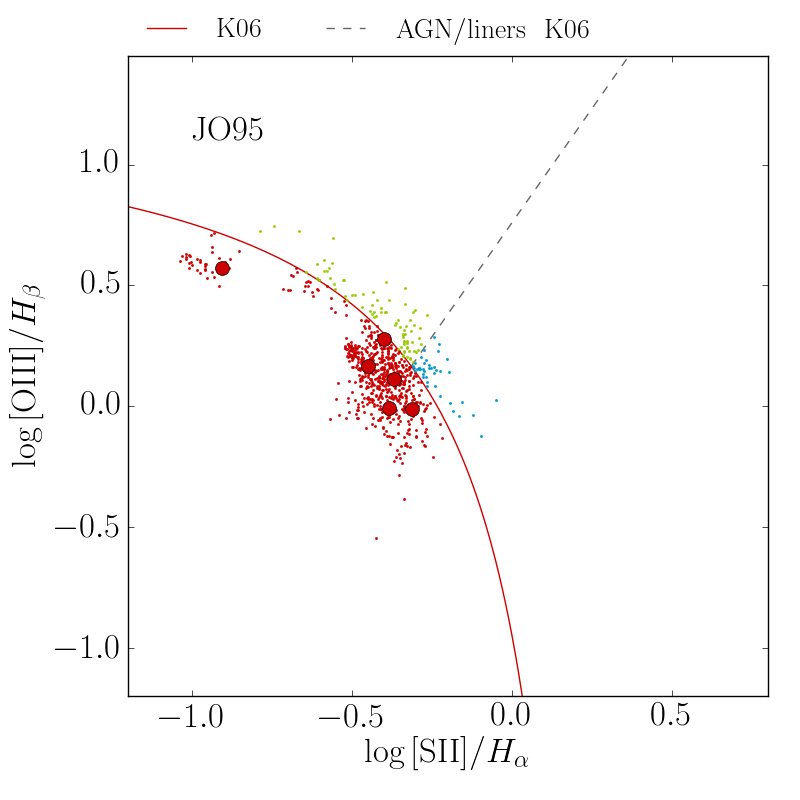}\includegraphics[width=2.4in]{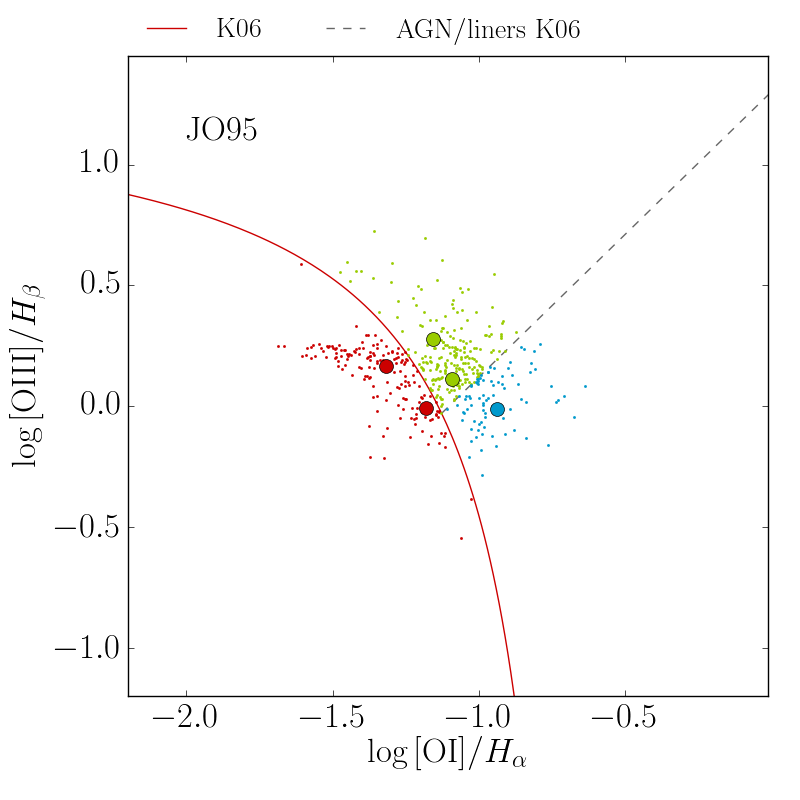}}
\centerline{\includegraphics[width=2.4in]{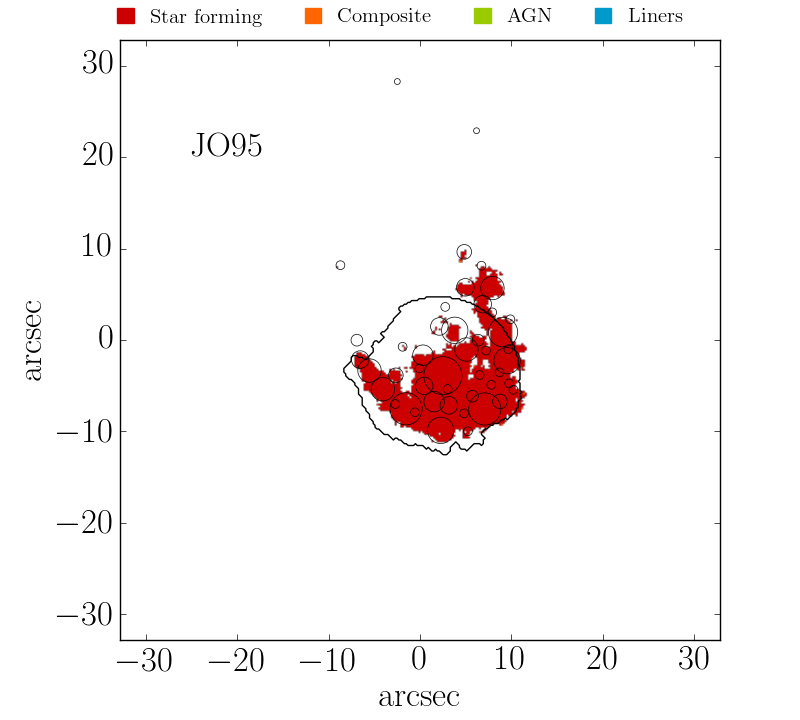}\includegraphics[width=2.4in]{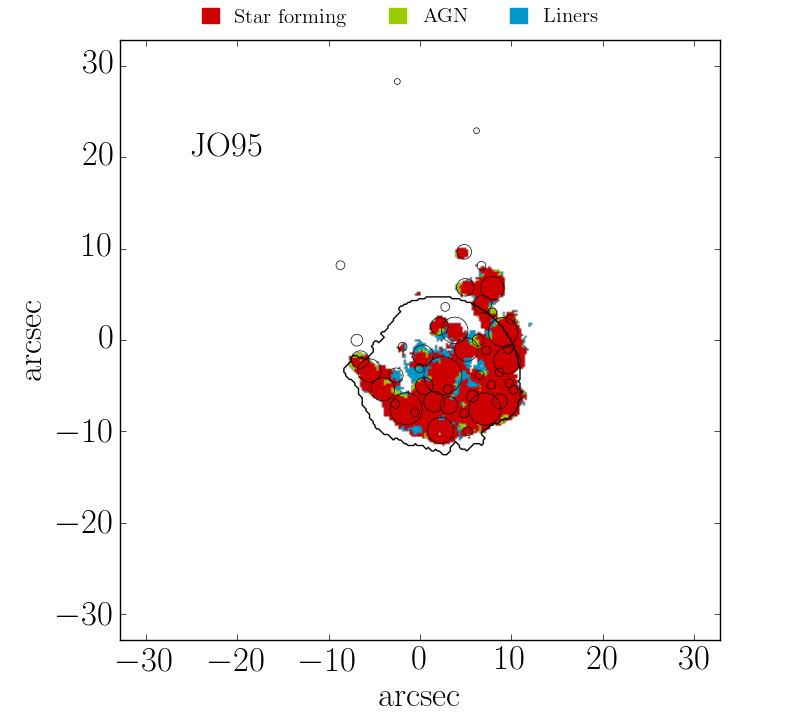}\includegraphics[width=2.4in]{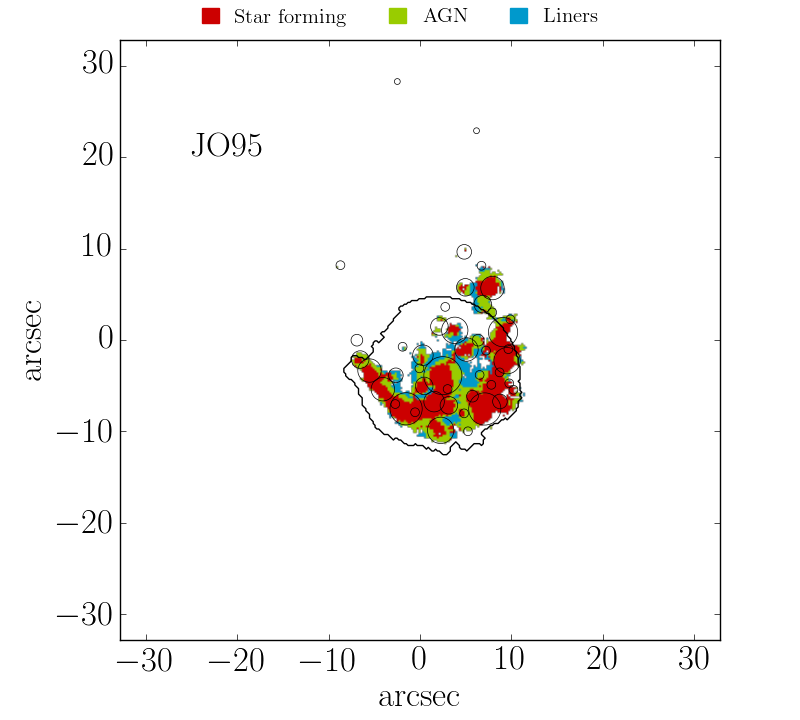}}
\centerline{\includegraphics[width=2.4in]{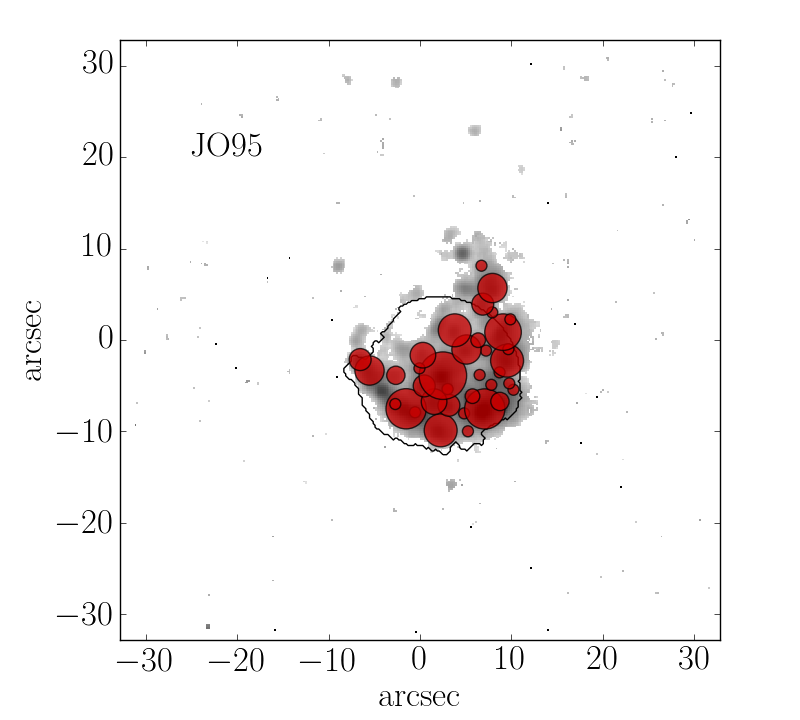}\includegraphics[width=2.4in]{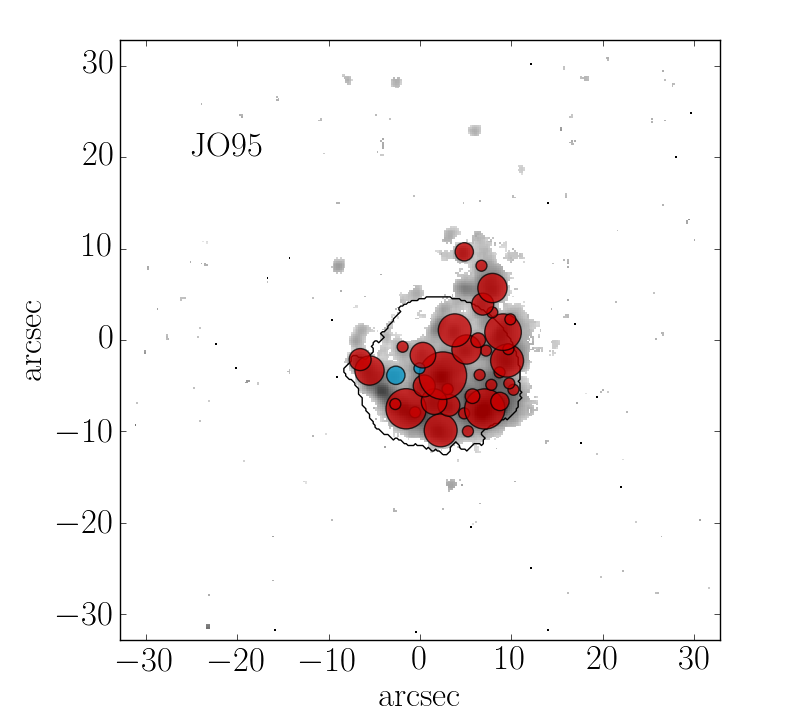}\includegraphics[width=2.4in]{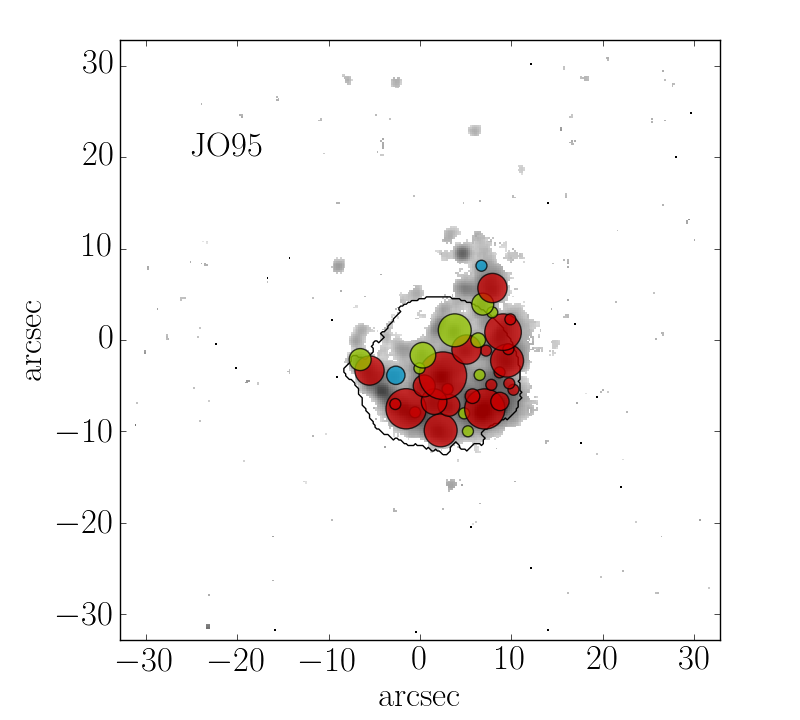}}
\contcaption{}
\end{figure*}

\begin{figure*}
\centerline{\hspace{5cm}\includegraphics[width=2.4in]{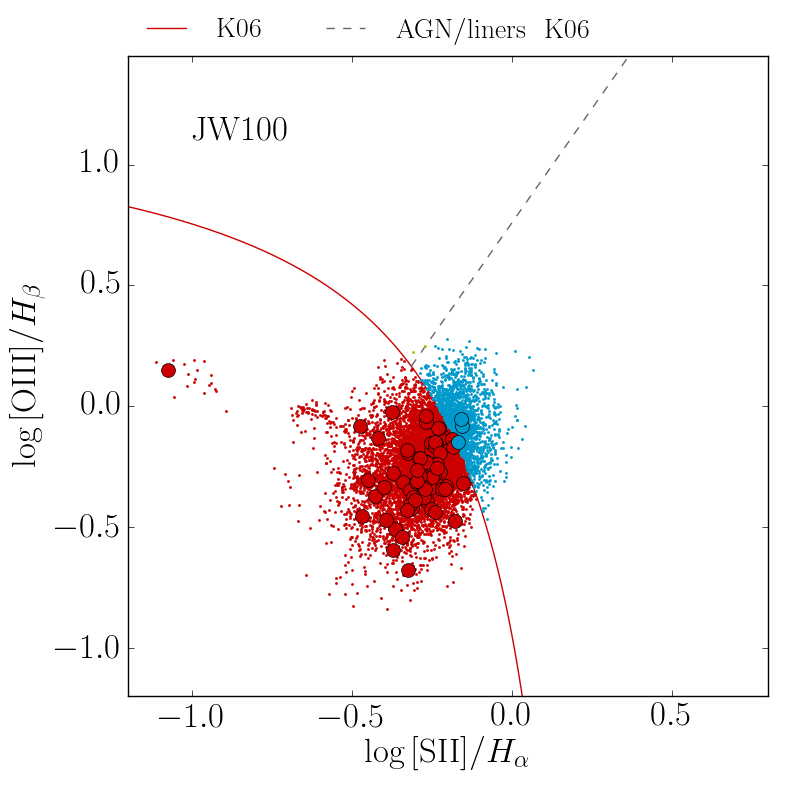}\includegraphics[width=2.4in]{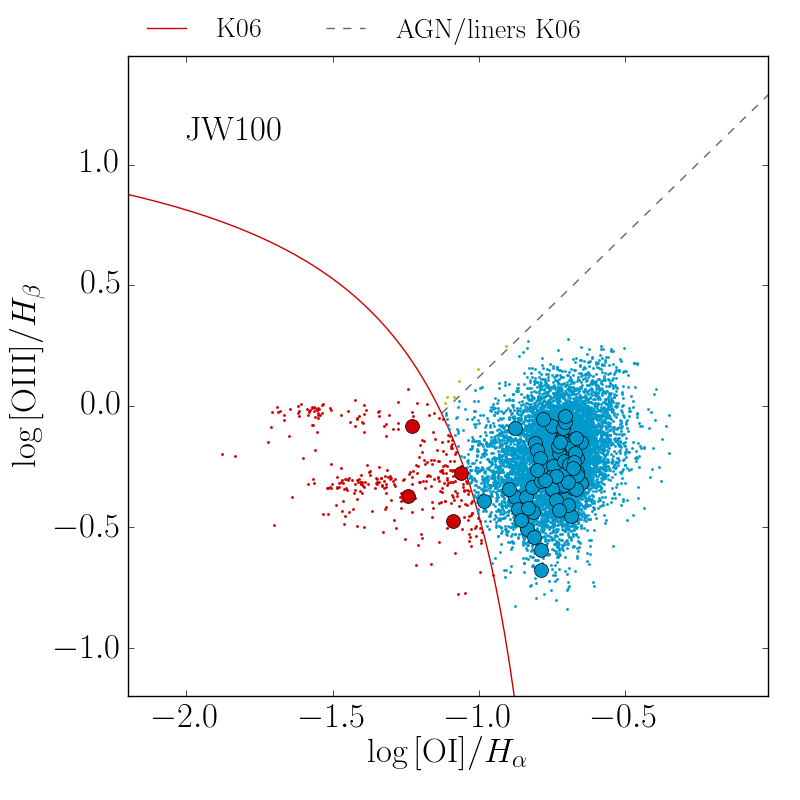}}
\centerline{\hspace{5cm}\includegraphics[width=2.4in]{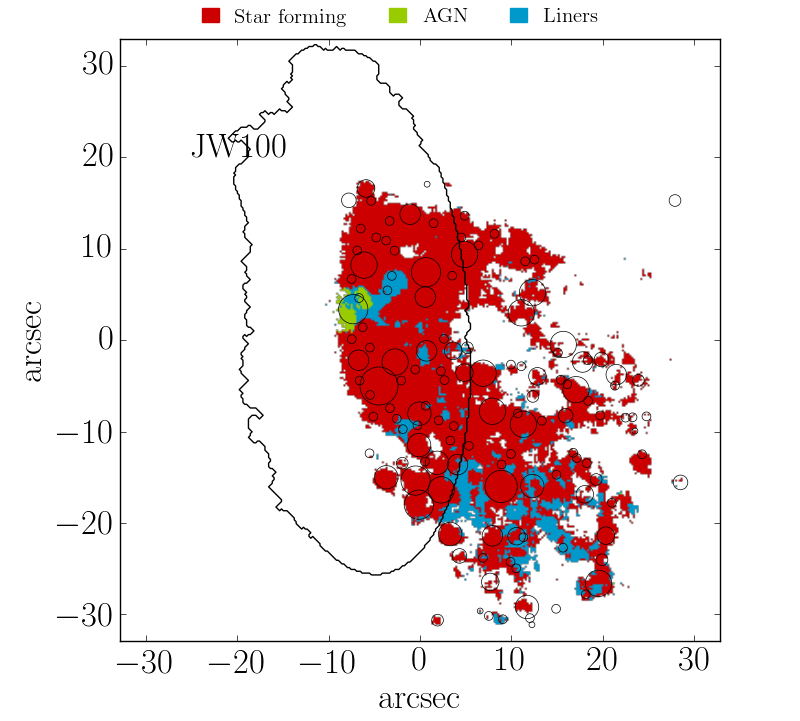}\includegraphics[width=2.4in]{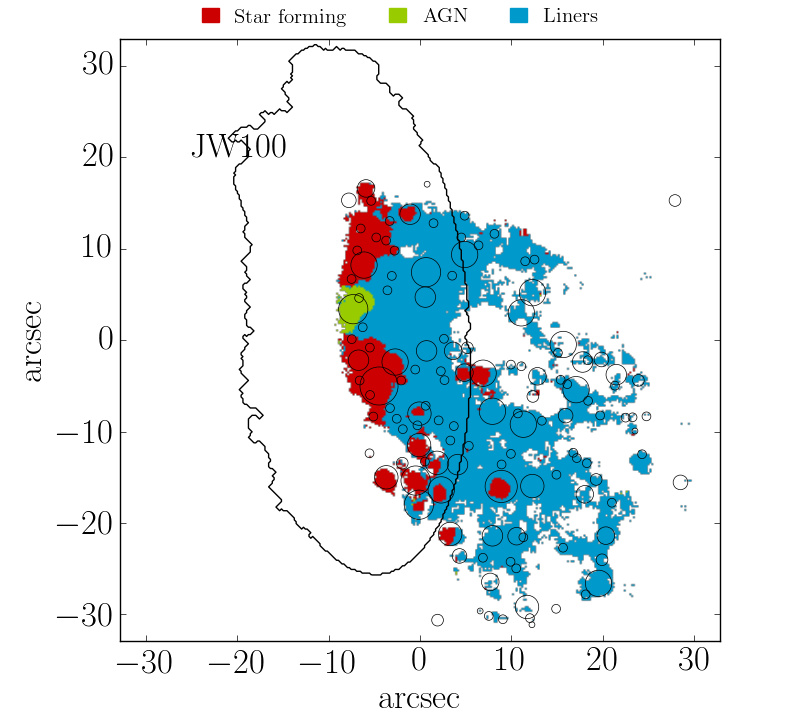}}
\centerline{\hspace{5cm}\includegraphics[width=2.4in]{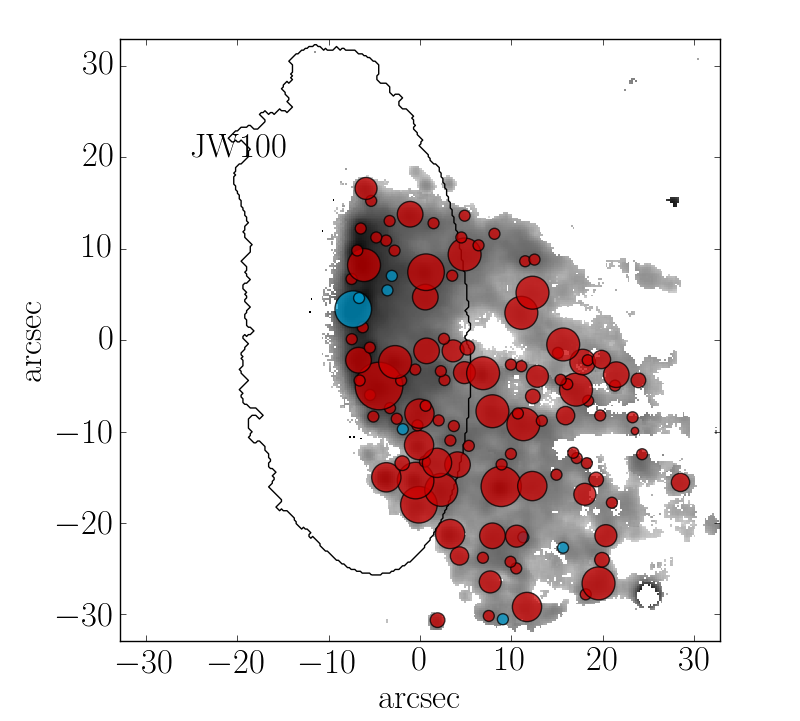}\includegraphics[width=2.4in]{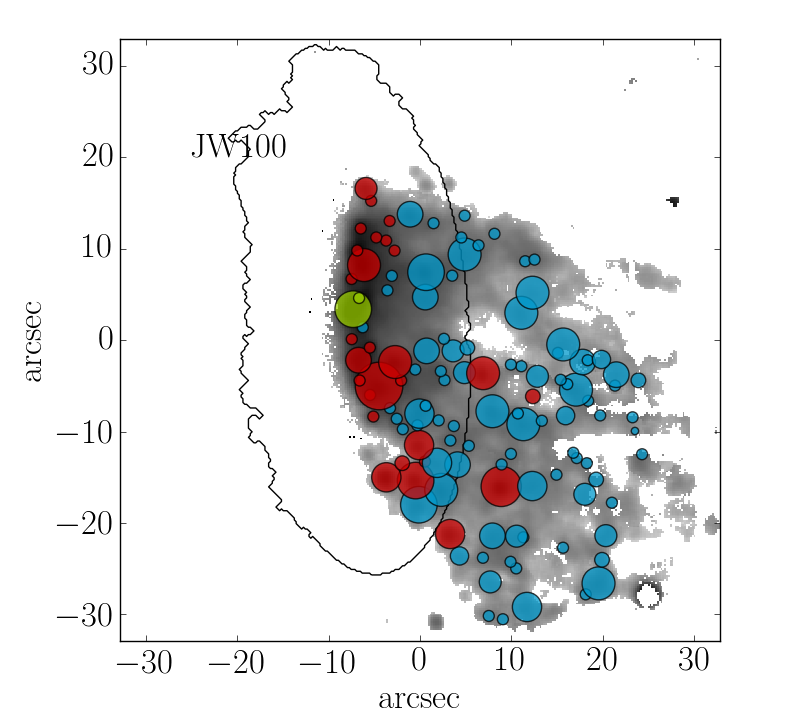}}
\contcaption{}
\end{figure*}

\begin{figure*}
\centerline{\includegraphics[width=2.4in]{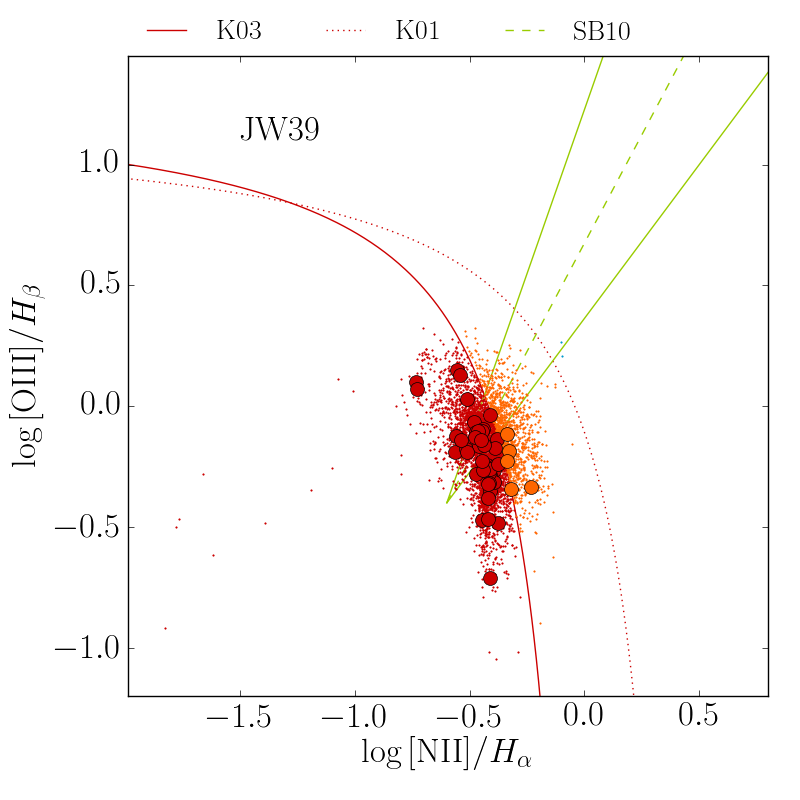}\includegraphics[width=2.4in]{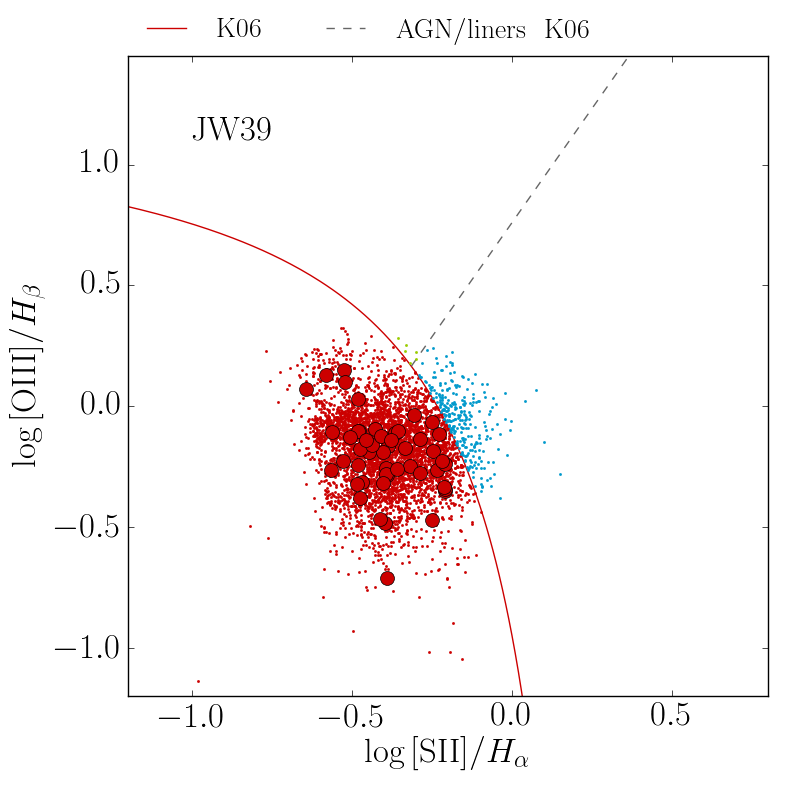}\includegraphics[width=2.4in]{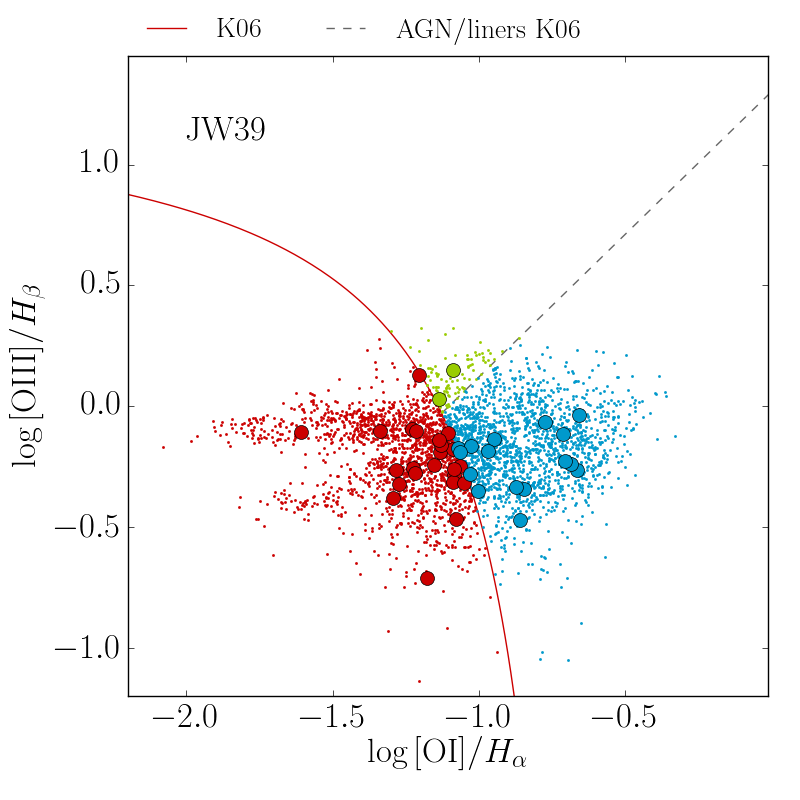}}
\centerline{\includegraphics[width=2.4in]{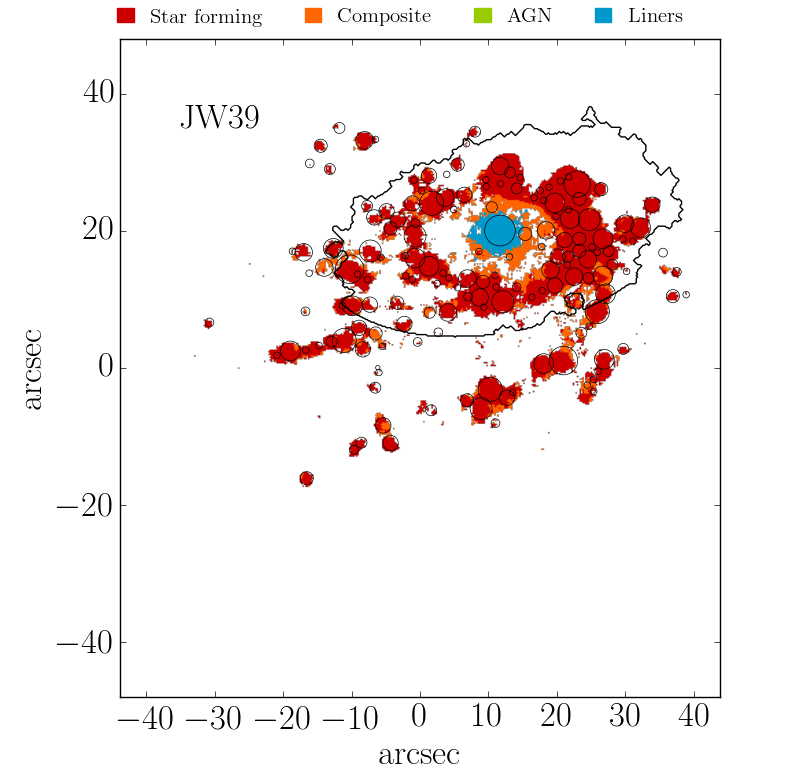}\includegraphics[width=2.4in]{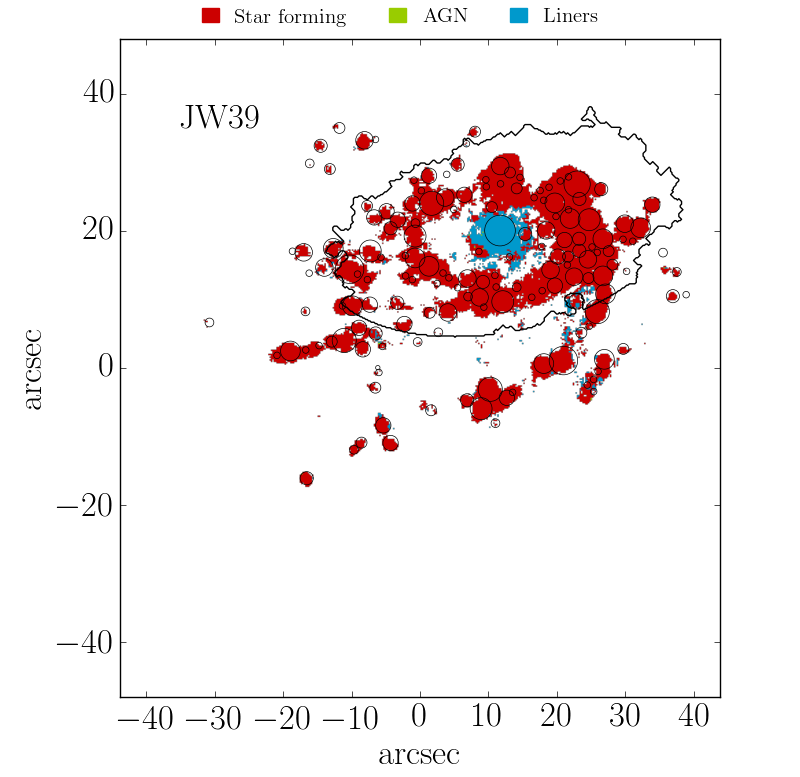}\includegraphics[width=2.4in]{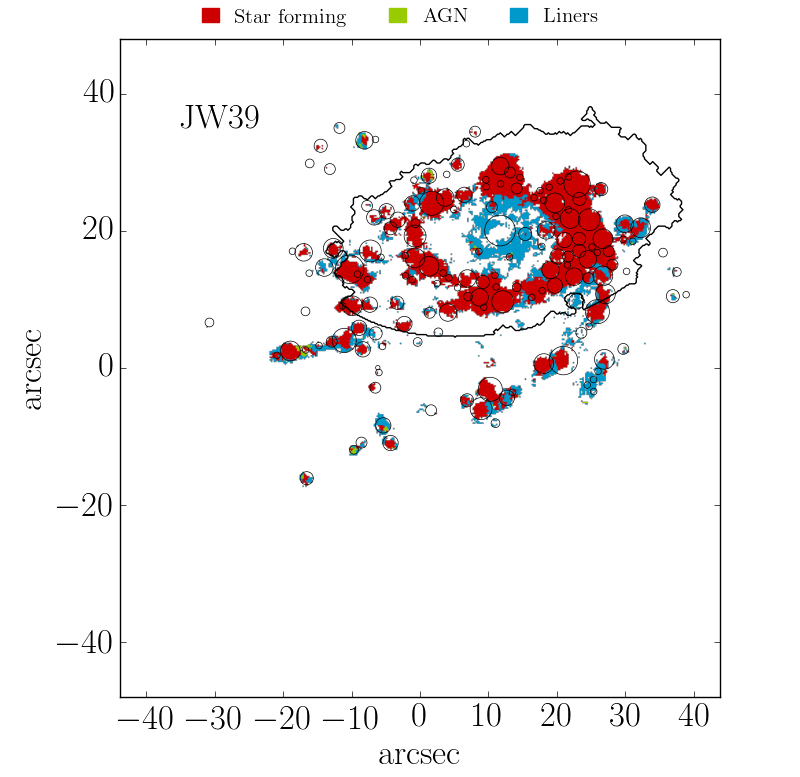}}
\centerline{\includegraphics[width=2.4in]{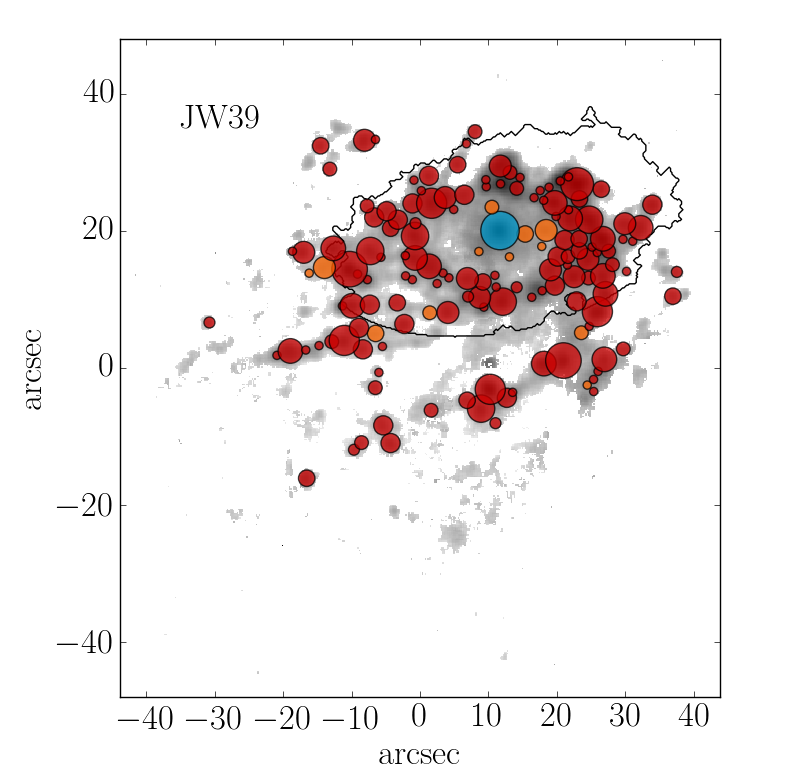}\includegraphics[width=2.4in]{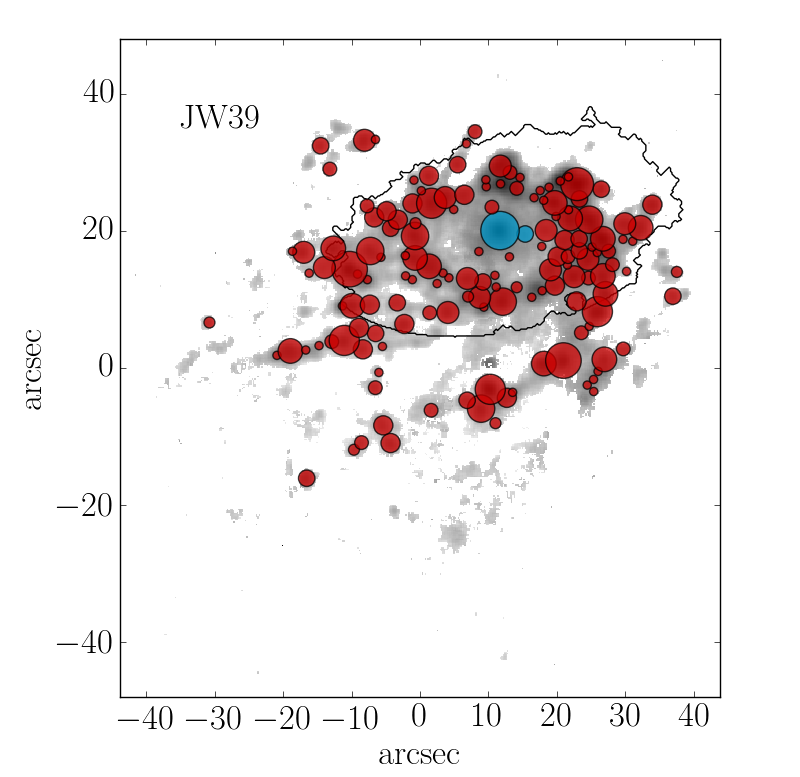}\includegraphics[width=2.4in]{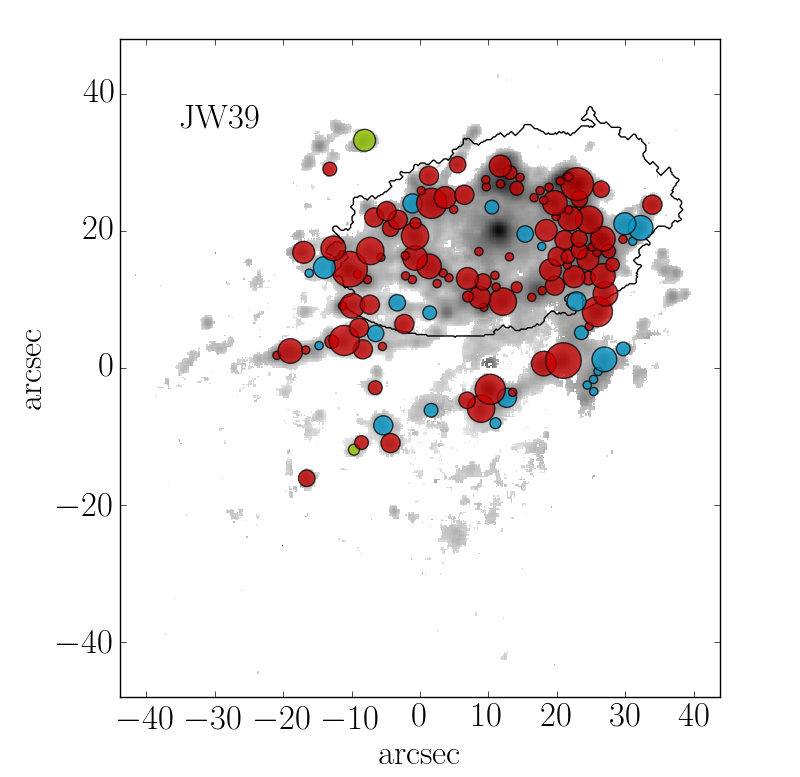}}
\contcaption{}
\end{figure*}

%\clearpage 

\subsection{In-situ star formation in the tails}

If the main source of ionization of gas in the tails are young massive
stars, as shown in the previous section, it is worth asking where such
stars are located.

In principle, the ionizing photons could originate from star formation
in the tails or in the disk. 
%In the former case, the
%location of the newly born stars would be

In the latter case, two possible situations could be envisaged:
either the ionizing photons could travel a long distance from the disk
before ionizing stripped neutral gas, or the gas itself could be
ionized within the disk, and then stripped to large distances.
Both of these situations are unrealistic, as discussed in \cite{Poggianti2017a} and below.

If the gas were ionized in the disk and then stripped, it should travel
at impossibly large speeds to reach large galactocentric
distances before recombining and decaying. In fact, for a gas density $n =10 \, \rm
cm ^{-3}$, the recombination time of hydrogen is about $10^4$ yr and
once recombined the decay time is negligible \citep{Osterbrock2006}. Thus, %once the ionizing photons have ionized the gas, 
the gas recombination lines will be visible only for this timescale from
the moment the gas was ionized. 
The recombination time goes linearly with the gas density and, as we will show in the next section, the gas density in the $\rm H\alpha$ 
clumps is often higher than $n =10 \, \rm  
cm ^{-3}$, implying recombination times even shorter than  
$10^4$ yr.  
The maximum distance from the disk at which we observe ionized gas
ranges in our galaxies between 20 and 100 kpc, meaning the gas should travel at speeds
between $2 \times 10^6$ and $10^7 \rm \, km \, s^{-1}$ to get there in $10^4$ yr.  
%In order to reach 20-100kpc away from the disk in such a short time,  
%the gas should travel at a speed of 
Even for a gas density 1000 times lower ($n =0.01 \, \rm
cm ^{-3}$), which is unjustified based on our measurements
and which would result in a recombination timescale of $10^7$ yr,
the gas should move with speeds of the order of 2000-10000 $km \,
s^{-1}$, which are much higher than the velocities at which these galaxies
are moving within the ICM.

The possibility that the ionizing photons formed in the disk manage
to escape to such large distances before encountering neutral gas to
ionize is also highly unlikely. Our JVLA data shows long tails of HI neutral gas
coexisting with the $\rm H\alpha$ tails (Ramatsoku et al. in prep.,
Deb et al. in prep.), and coexisting HI and $\rm H\alpha$ tails are
present in a few other jellyfish galaxies \citep[e.g.][]{Consolandi2017}.
%as it is the case in the other jellyfishes with multiwavelength data
%in the literature (e.g. ESO137-).

The in-situ formation of new stars in the tails is therefore the most
likely hypothesis, and is corroborated by other GASP results:

1) the stripping candidates in the \cite{Poggianti2016} atlas were
selected from B-band images for having unilateral debris material
which indeed turned out to be the brightest $\rm H\alpha$ clumps. This
B-band light %originates 
%is most likely 
stellar continuum from stars, with no significant contribution
from line emission in the observed band at these redshifts, hence the visibility of these clumps in the B-band
already points to the presence of blue stars in the tails.

In the same fashion, the FUV and NUV light of the O and B young stars in the clumps of the
tails was directly
observed with UVIT@ASTROSAT in JO201, one of the galaxies in our sample
\citep{George2018}, 
%and the UV light provides direct evidence for
%the presence of the young stars in the tails, 
where we found a remarkable
agreeement between the SFR of individual clumps derived from the FUV 
and from $\rm H\alpha$. GALEX data of several GASP jellyfishes, albeit
at much lower spatial resolution, support the same scenario.

2) In \cite{Moretti2018b} we presented APEX CO(2-1) data for four of the
galaxies presented in this paper and found large amounts of molecular
gas (several $10^9 M_{\odot}$) in the tails of these
galaxies. Molecular gas was also found in the
tails of 3 other jellyfishes in the literature \citep{Jachym2014,
Jachym2017, Verdugo2015}.
The CO observed in the tails is the smoking gun of the presence of
cold, molecular gas where new stars can be born.

3) The GASP MUSE spectra are fitted by our spectrophotometric code SINOPSIS
\citep{Fritz2017} that finds good fits to the spectra in the tails
for a vigorous ongoing and recent (past few $10^8$ yr) star
formation. These young stars can account both for the number of ionizing
photons required by the emission lines and for the observed continuum
level and (blue) shape.\footnote{We note that the continuum gas emission is not
  included in spectrophotometric models, and this remains a source
  of uncertainty.}

We conclude that in-situ star formation in the tails is the %major
cause of the photo-ionization of the star-formation-powered gas in the
tails.
The $\rm H\alpha$ clumps we identify in our galaxies are the obvious
sites for such star formation, and in the next section we show they are
%from here on we will interpret them as 
star forming clumps, possibly composed of smaller HII regions
and complexes which we cannot resolve at the 1 kpc resolution of the
MUSE data. 
%In the following, their properties are analyzed in detail.

\section{The physical characteristics of the $\rm H\alpha$ clumps}

%Discuss resolution issues and the fact that we are probably not
%sampling the scales of HII regions...IMPORTANT

We now focus on the ionization mechanism of the $\rm H\alpha$ clumps 
defined in \S3.1. 
%beginning with the analysis of their ionization mechanisms. 
%of Fig.~2. 
The emission-line flux ratios of individual clumps displayed in Fig.~2 were estimated 
from the MUSE spectra integrated within each knot. 
%as explained in \S3, and 
The ionization mechanism could be determined for all clumps 
from NI and SII DD, and for $\sim 85$\% of them from the OI DD. 
Since the cases in which the OI 
line is too weak to be measured are most likely to be SF regions given 
the location of the dividing lines in the DD, the 
remaining 15\% can be considered as star-forming. 

In total, there are 521 $\rm H\alpha$ clumps in the tails of our 
galaxies, and 1031 inside the stellar contours of the galaxy disks. 
%, both in the tails and in the disk. 
Similarly to the spaxel analysis, the NI DD analysis finds that the majority 
of the clumps (70\%) are powered by SF, and this fraction becomes 
98\% when considering SF+Composite emission, as can be appreciated 
inspecting the bottom panels of each galaxy in Fig.~2. Similar conclusions are 
reached from the SII DD, while according to the OI 
DD about 60\% of them 
%half of them  - sarebbe la meta', ma nel 60\% ho gia' contato quelli 
%per cui la riga OI non e' detettata 
are powered by SF. 
%\footnote{The OI-based LINER-like knots 
%are especially conspicuous in JO147 and JW100, in line with the 
%spaxel results.}
%45\% of them from the OI DD. 

%{\bf To do: exclude neighbours in JO147 throughout.}
%Yara, figure DD: overplot the blobs on the spaxel DD? Add blobs in 
%spaxel map to see the diffuse regions better; add legend 

%We will now turn to discuss the clump velocity dispersions and  
%other physical properties which
%.  All the physical properties of the $\rm H\alpha$ clumps discussed below 
%have been derived
%from the integrated MUSE spectrum within each clump, unless otherwise stated.

\subsection{Gas velocity dispersion and $\rm H\alpha$ luminosity
  function}

Fig.~3 shows the spaxel map of gas velocity dispersion for each galaxy as
measured from $\rm H\alpha$. Focusing only on the tails, there are regions of both
high and low velocity dispersion.
%\footnote{JW100 and JO201 have
%  non-negligible double components in the tail, here we show the
 % dominant component of the double Gaussian fit (Bellhouse et al. 2017).}
%\footnote{Since JW100 and JO201 have
%important double components in the tail, we show here the dominant
%component of the double Gaussian fit for JW100, while the JO201
%1-component plot shown is essentially unchanged with respect
%to the dominant component of the double Gaussian fits shown in
%Bellhouse et al. 2017, see this paper for details.}
The high velocity dispersion regions might be due to intrinsically
dynamically ``warm'' regions, or to the superposition along the line of sight
of various gaseous components at different velocities.
It is interesting that the two galaxies with the ``warmest'' tails,
JO147 and JW100 (and to a less extent JO201), are 
those for which the OI DD finds a strong LINER component. In these
tails the ``heated'' component of the gas is clearly more prominent than in the
others. 

%Yara: Fig.4 add AGN; separate line for only disk; 

The $\rm H\alpha$ clumps in the tails (see white circles in Fig.~3) 
%tend to 
generally correspond to the
regions with lowest gas velocity dispersion.  

Figure~4 presents the line ratios versus gas velocity dispersion of 
individual clumps in the tails. These plots show that 
high [NII]/$\rm H\alpha$, [SII]/$\rm H\alpha$ and [OI]/$\rm H\alpha$
ratios (corresponding to Composite/LINER-like emission) are found in 
those blobs with high gas velocity dispersion 
($>50-70 \, \rm km \, s^{-1}$), while clumps with low line ratios 
(star forming) have velocity dispersions typically below 
$50 \, \rm km \, s^{-1}$. These trends are very similar to those found 
by \cite{Rich2011} in two luminous infrared galaxies, and are 
consistent with high line ratios 
originating when the gas is ``heated'' by some other process than star formation. 
Thus, while clumps with low emission line ratios and low velocity 
dispersion have the typical characteristics of HII star-forming 
clumps, the fewer clumps with high ratios and high $\sigma$ are probably 
at least partly ``heated'' by other mechanisms such as thermal 
conduction from the ICM or turbulence due to the stripping motion, and/or are 
contaminated along the line of sight by the emission of more diffuse 
turbulent/heated gas.

The velocity dispersion
distribution of the clumps is presented in Fig.~5 (panel (a)),
where the black histogram indicates all the clumps in the tails of our sample galaxies.
We note that the mean error on the clump $\sigma$'s is about $\sim 4 \rm \, km \,
s^{-1}$ but uncertainties on the velocity dispersion measurements %begin to 
dominate for $\sigma \leq 17 \rm \, km \, s^{-1}$. This,
combined with the instrumental line width (46 $\rm km \, s^{-1}$ at 
$\rm H\alpha$, \S3), prevents us from measuring reliable clump
velocity dispersion values lower than $\sim 17 \rm \, km \, s^{-1}$. 

The median $\sigma$ of all clumps in the tail is $34.9 \, \rm km \, s^{-1}$
(first and third quartiles are Q1=23.6 and Q3=74.9).
As already clear from Fig.~4, the tail clumps powered by SF have lower velocity dispersions than
those ionized by Composite or LINER mechanisms (Fig.~5a): the median
$\sigma$ of SF-powered tail clumps is $27.2 (Q1=21.0, Q3=39.4) \, \rm
km \, s^{-1}$ according to the NII DD ($26.8 (21.2-38.3) \, \rm km \,
s^{-1}$ using the OI DD).\footnote{These medians should be 
lower limits but also close approximations of the true values for
the limitations on velocity dispersion measurements discussed in the text.}
In contrast, the median for the Composite
clumps is $105.9 (63.9-155.5) \, \rm km \, s^{-1}$, while for LINERs is 
$74.1 (61.5-75.1) \, \rm km \, s^{-1}$ and $87.4 (49.2-149.8) \, \rm km
s^{-1}$ for NII and OI DDs, respectively.\footnote{For completeness,
for AGN-powered clumps in the tails the median
is $35.5 (32.0-55.6) \, \rm km \, s^{-1}$ and $24.0 (19.2-32.4) \, \rm
km \, s^{-1}$.}
%ADD COMMENT ON SINGLE GALAXIES?

%Compare with Wilson 2011 sigma in our Galaxy NO

The SF-powered clumps in the tails are therefore kinematically quite cold
star-forming complexes, whose absorption- and dust-corrected 
$\rm H\alpha$ luminosity function is 
presented in Fig.~5b.
As before, we present the distribution for all the clumps in the tails (solid black line),
for SF-powered tail clumps (red solid NI DD, dashed OI DD), and for
all the clumps in the disks (dashed black histogram). 

The median clump luminosity in the tail is $4.5 \times 10^{38} \rm erg \, s^{-1}$, with the first
and third quartiles being $2.5 \times 10^{38} \rm erg \, s^{-1}$ and $1.5
\times 10^{39} \rm erg \, s^{-1}$.
The clump luminosities are typical of the so-called ``giant HII
regions'' ($L_{\rm H\alpha}=10^{37} - 10^{39} \rm erg \, s^{-1}$, ionized
by a few OB associations or massive stellar clusters, such as the
Carina Nebula in our Galaxy) and
``super giant HII regions''  ($L_{\rm H\alpha} > 10^{39} \rm erg \,
s^{-1}$, probably ionized by multiple star clusters or super star
clusters, with no analog in our Galaxy but observed in late-type
galaxies and interacting galaxies, such as 30 Doradus in the Large
Magellanic Cloud \citep{Lee2011}). These giant and super giant HII
complexes are conglomerates of many individual HII regions,
corresponding to high-density condensations interconnected by a more
diffuse medium \citep{Franco2000}.
Our spatial resolution is limited by the seeing ($\leq 1"=1$kpc at the
distance of our galaxies), therefore higher resolution IFU studies would be needed
to probe the scales of individual HII regions within these complexes. 
%and in principle
%the large sizes of the star-forming clumps 
%and their correspondingly high $\rm H\alpha$ luminosities might be
%affected by resolution effects. 
In \S5.5 we will discuss the issue of
clump sizes in more detail.
%..COMPARE WITH Bradley fig.3? or fig. 4 in Hodge 89, ... 

\begin{figure*}
\centerline{\includegraphics[width=3.5in]{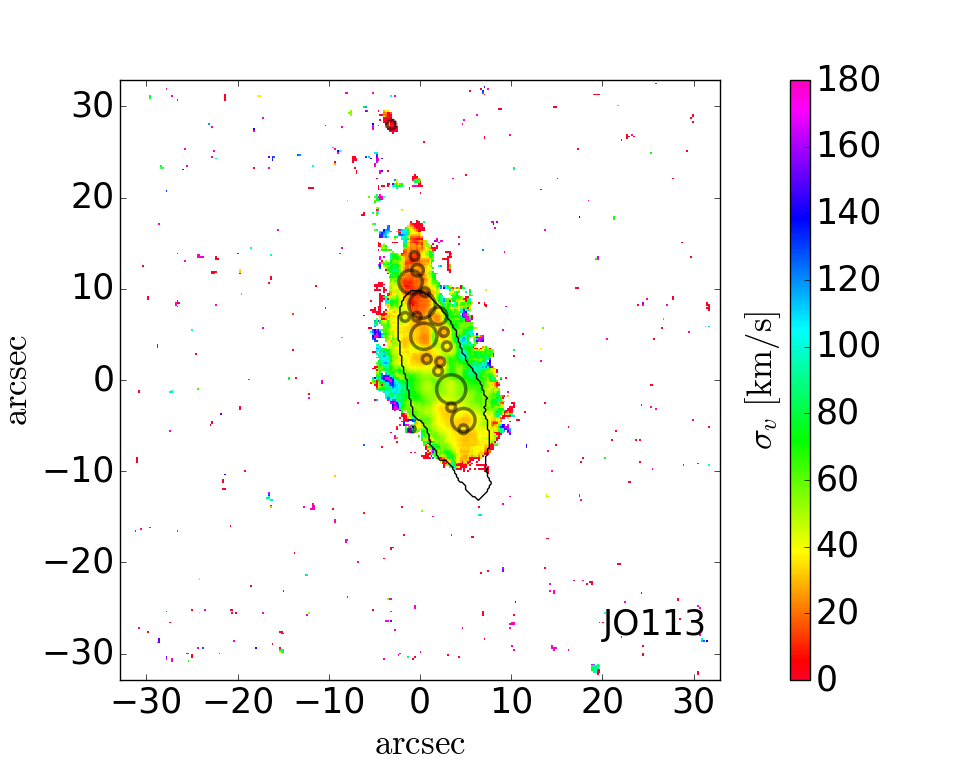}\includegraphics[width=3.5in]{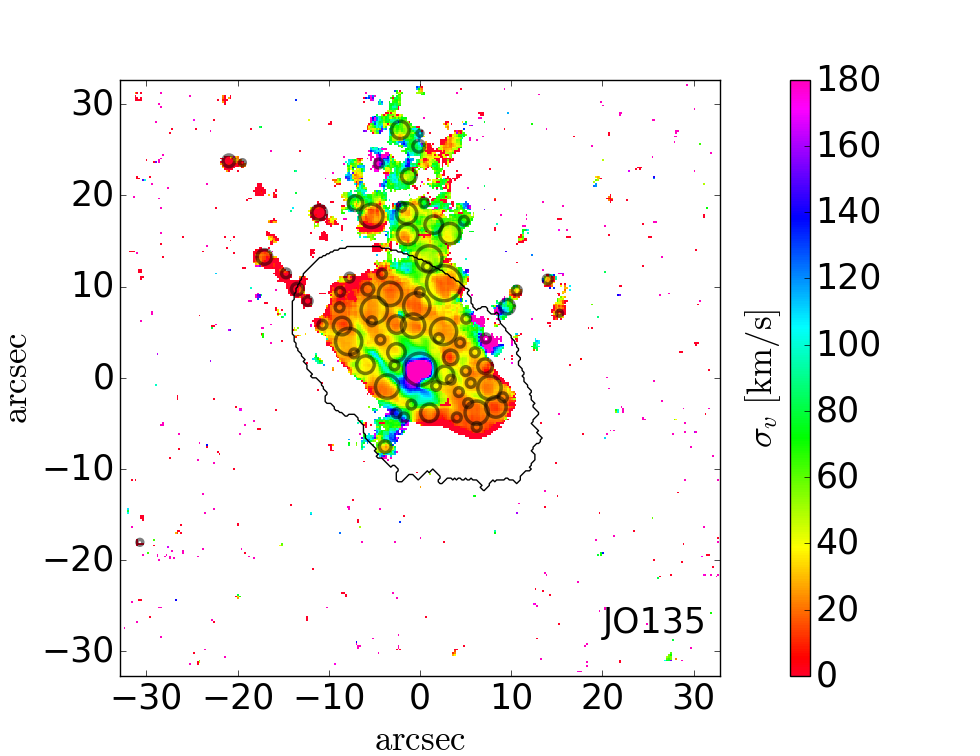}}
\centerline{\includegraphics[width=3.5in]{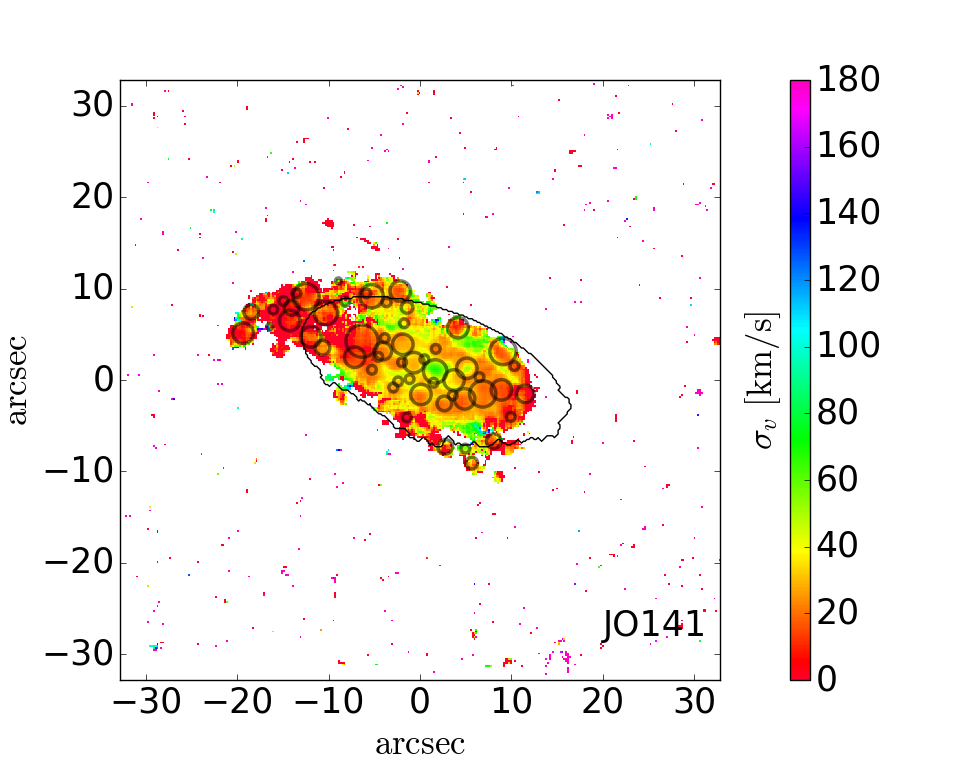}\includegraphics[width=3.5in]{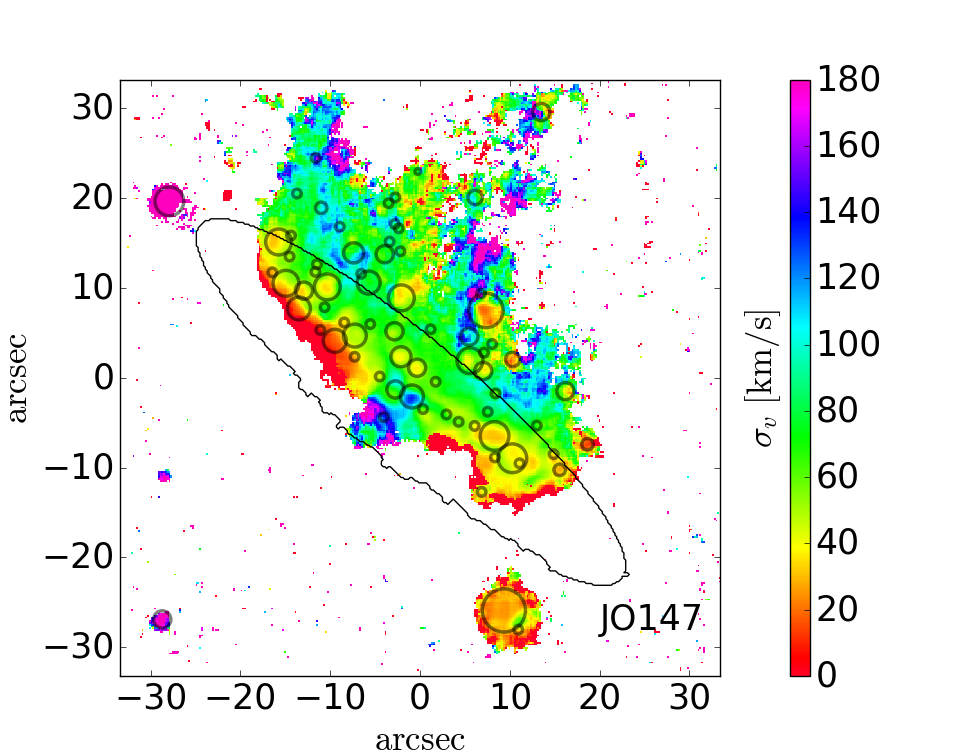}}
%\centerline{\hspace{2.1in}\includegraphics[width=3.5in]{plotsig_JO141_smo5_sn4_noflag.png}\includegraphics[width=3.5in]{plotsig_JO147_smo5_sn4_noflag.png}}
\centerline{\includegraphics[width=3.5in]{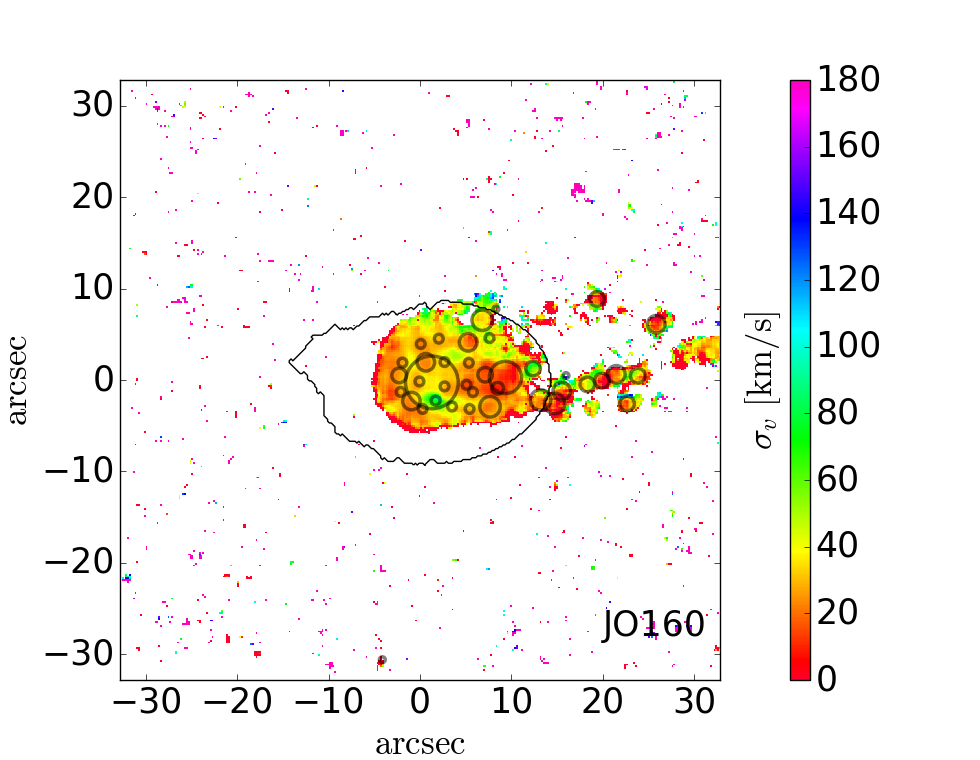}\includegraphics[width=3.5in]{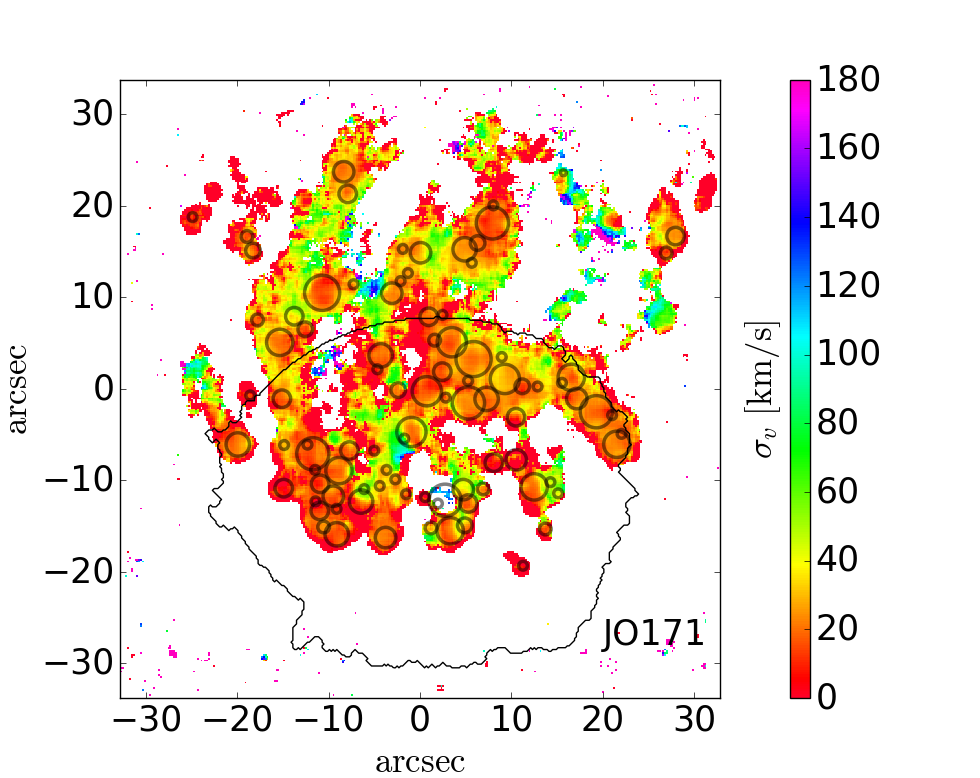}}
\caption{Gas velocity dispersion maps. Grey circles are the $\rm 
  H\alpha$ clumps.} % For JW100 we show the dispersion of the dominant component of the double Gaussian fits.}
\end{figure*}

\begin{figure*}
%\centerline{\hspace{2.1in}\includegraphics[width=3.5in]{plotsig_JO175_smo5_sn4_noflag.png}\includegraphics[width=3.5in]{plotsig_JO204_smo5_sn4_noflag.png}}
\centerline{\includegraphics[width=3.5in]{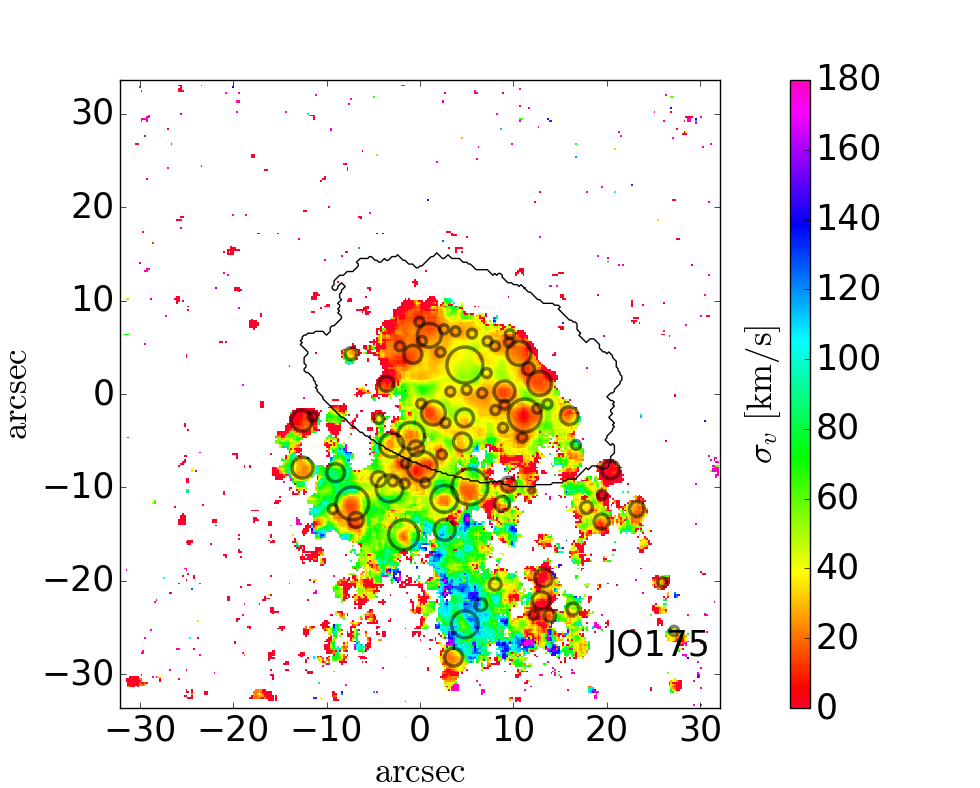}\includegraphics[width=3.5in]{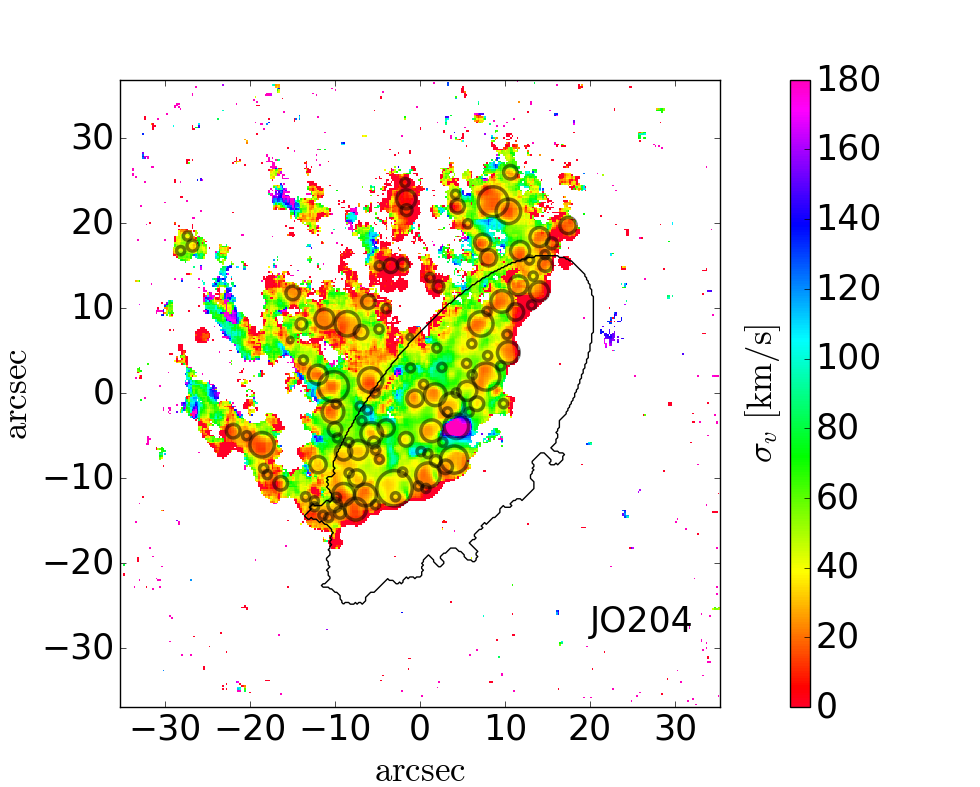}}
\centerline{\includegraphics[width=3.4in]{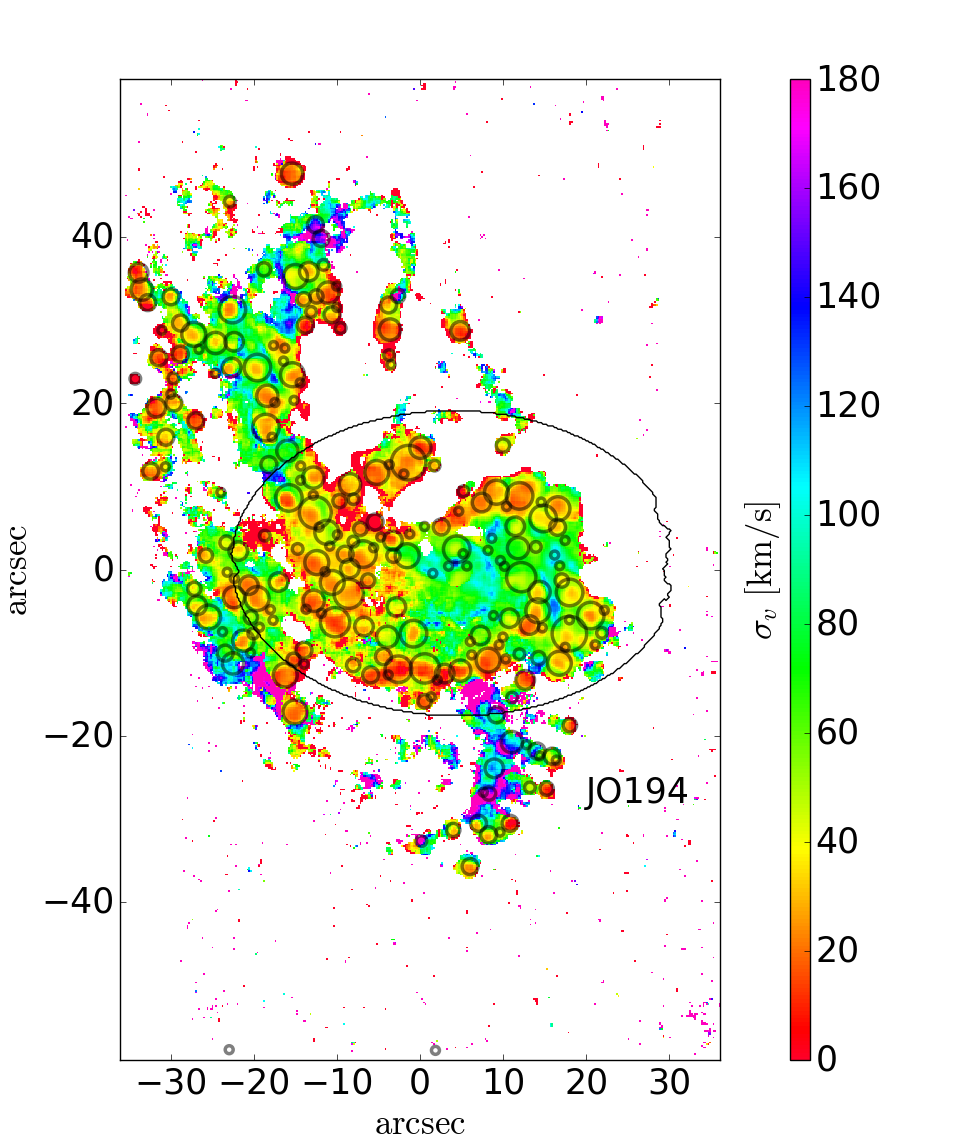}\includegraphics[width=3.8in]{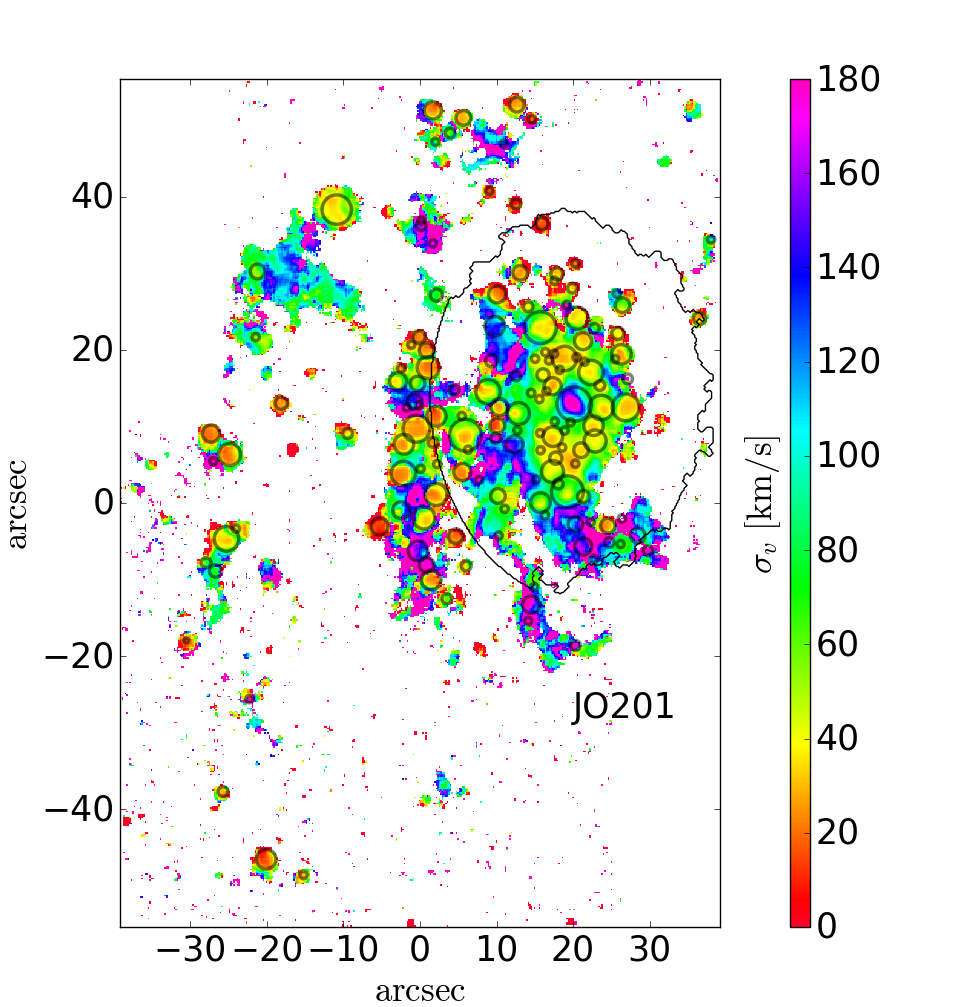}}
\centerline{\includegraphics[width=4.7in]{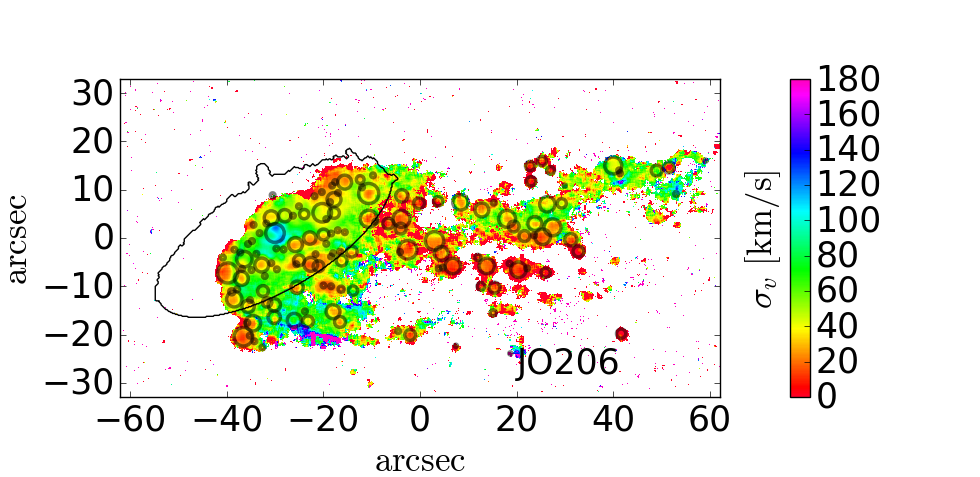}\hspace{-0.3cm}\includegraphics[width=3.5in]{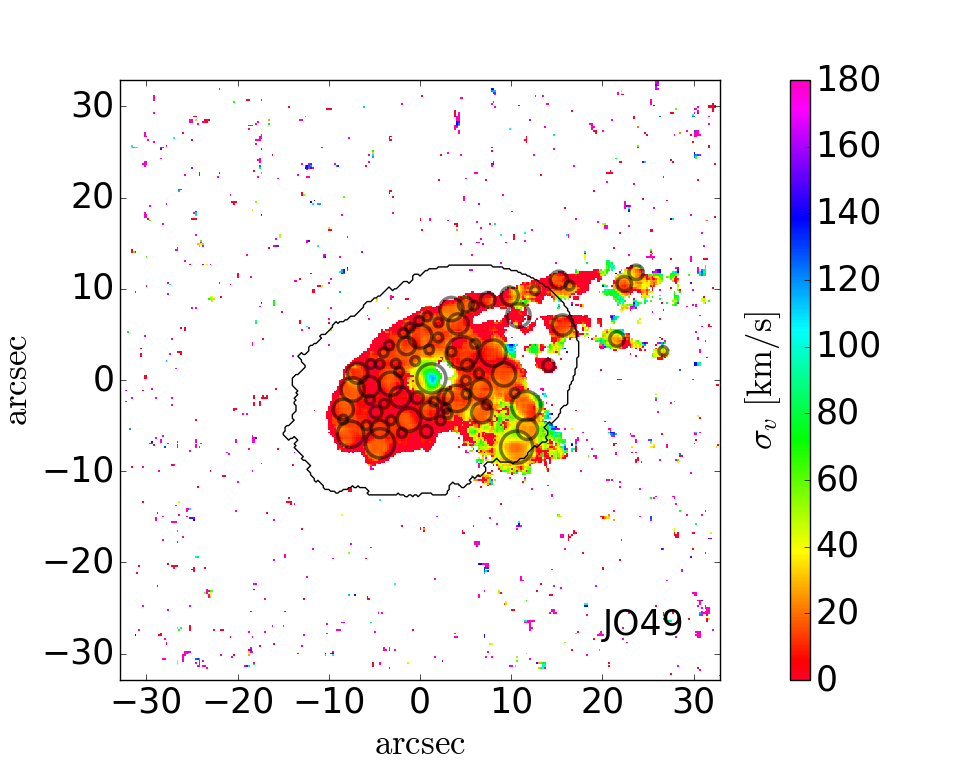}}
\contcaption{} % For JW100 we show the dispersion of the dominant component of the double Gaussian fits.}
\end{figure*}

\begin{figure*}
\centerline{\includegraphics[width=3.5in]{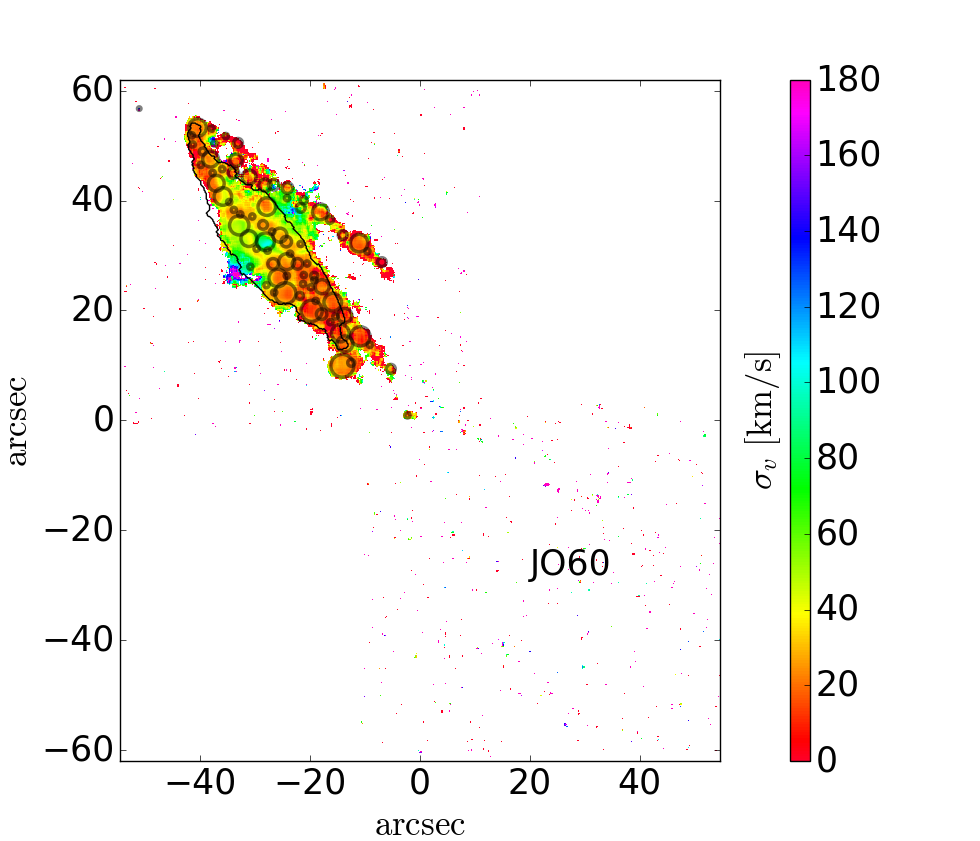}\includegraphics[width=3.5in]{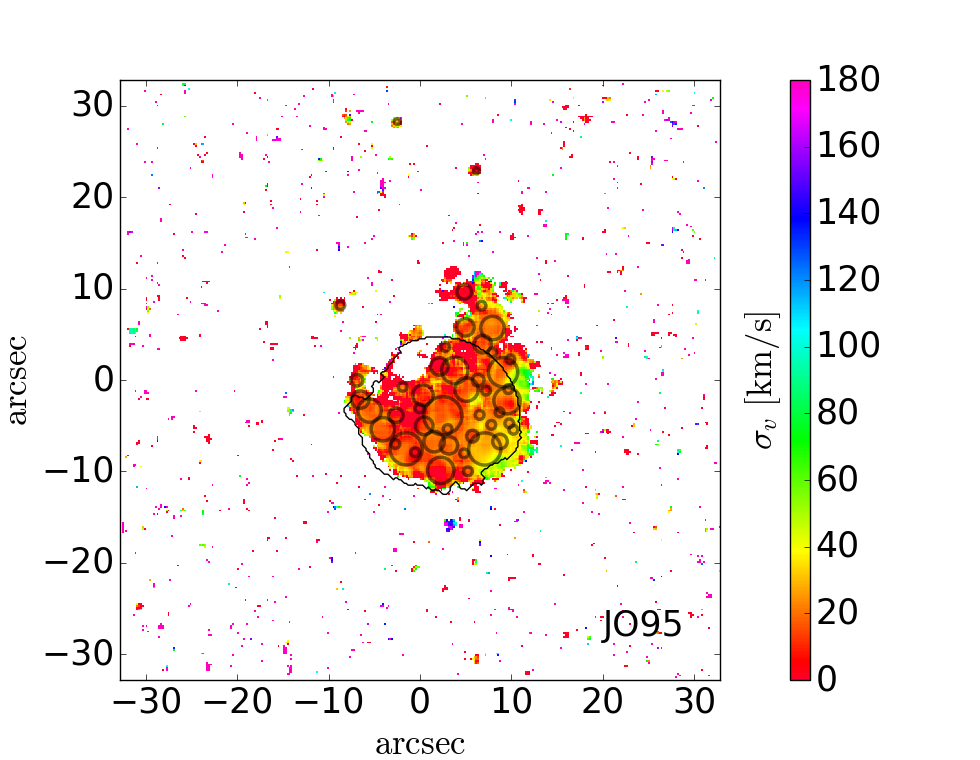}}
\centerline{\includegraphics[width=3.5in]{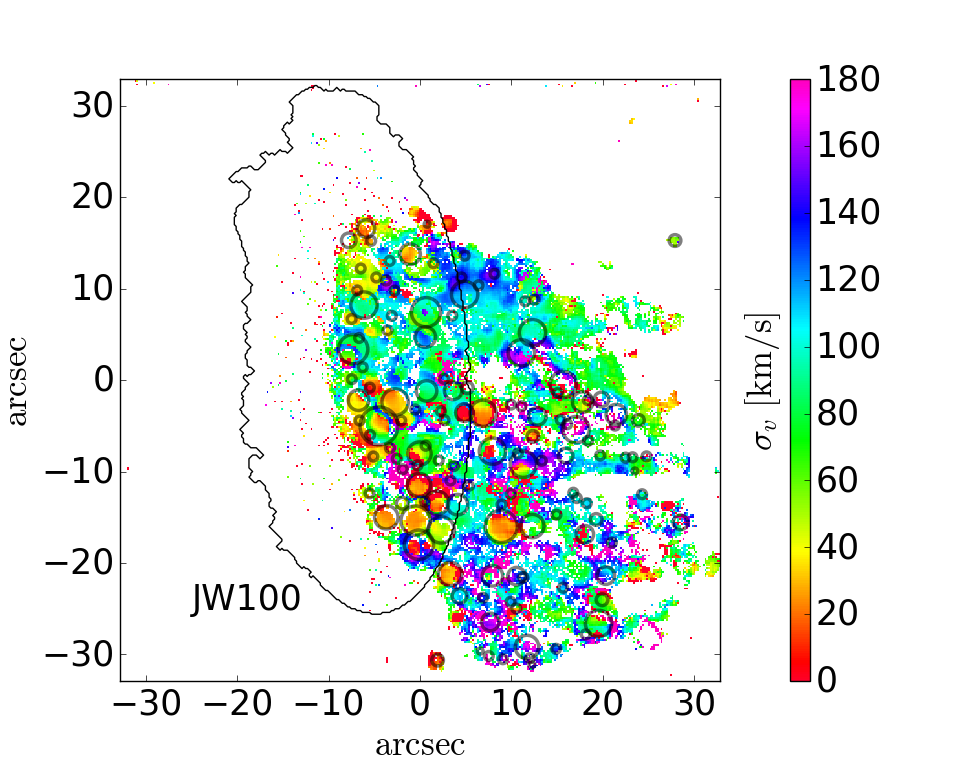}\includegraphics[width=3.5in]{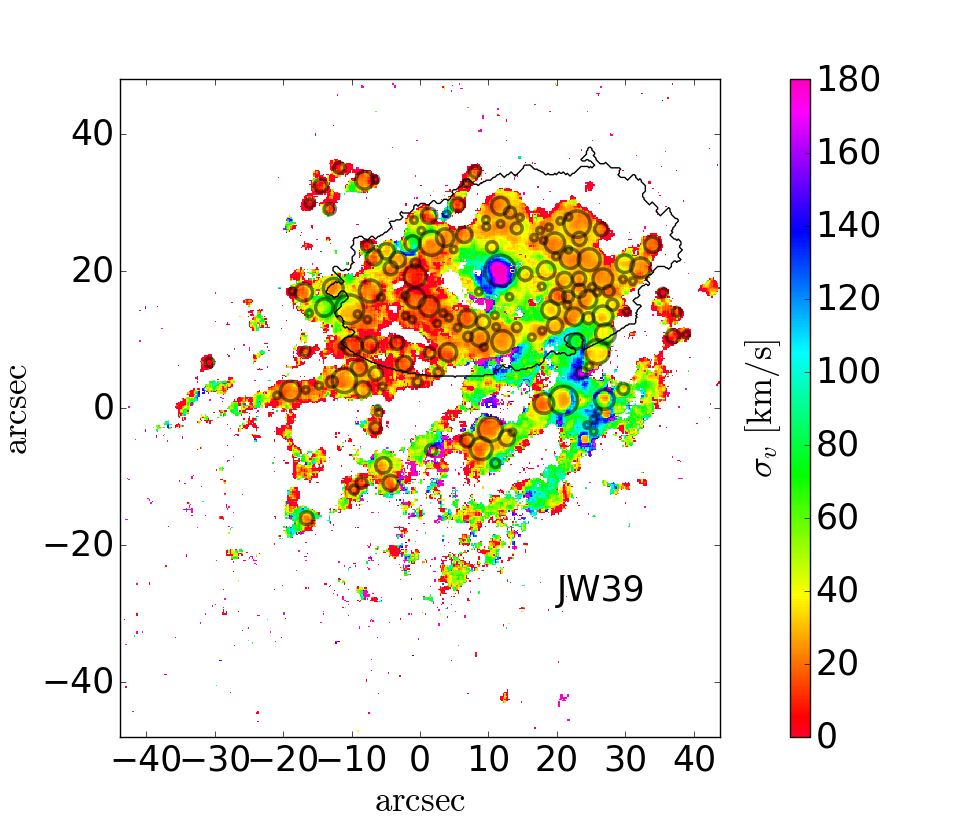}}
\contcaption{} % For JW100 we show the dispersion of the dominant component of the double Gaussian fits.}
\end{figure*}

\begin{figure*}
%\centerline{\hspace{2.1in}\includegraphics[width=3.1in]{JO113_ne_map_blobs.pdf}\includegraphics[width=3.1in]{JO135_ne_map_blobs.pdf}}
\centerline{\includegraphics[width=7.0in]{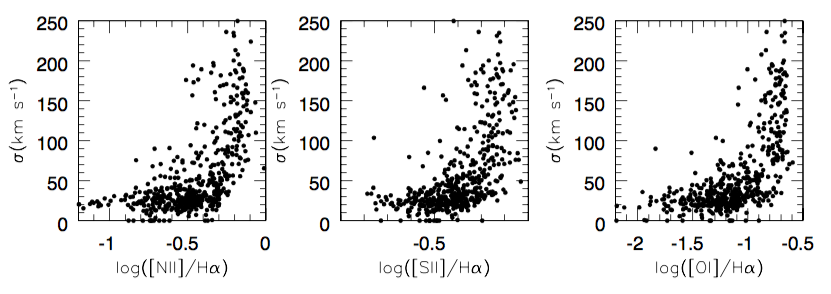}}
%\centerline{\hspace{2.1in}\includegraphics[width=3.1in]{JO141_ne_map_blobs.pdf}\includegraphics[width=3.1in]{JO147_ne_map_blobs.pdf}}
%\centerline{\hspace{2.1in}\includegraphics[width=3.1in]{JO160_ne_map_blobs.pdf}\includegraphics[width=3.1in]{JO171_ne_map_blobs.pdf}}
%\centerline{\hspace{2.1in}\includegraphics[width=3.1in]{JO175_ne_map_blobs.pdf}\includegraphics[width=3.1in]{JO204_ne_map_blobs.pdf}}
\caption{Plot of line ratio vs. gas velocity dispersion for 
  individual clumps in the tails.}
\end{figure*}

\begin{figure*}
\centerline{\includegraphics[width=2.7in]{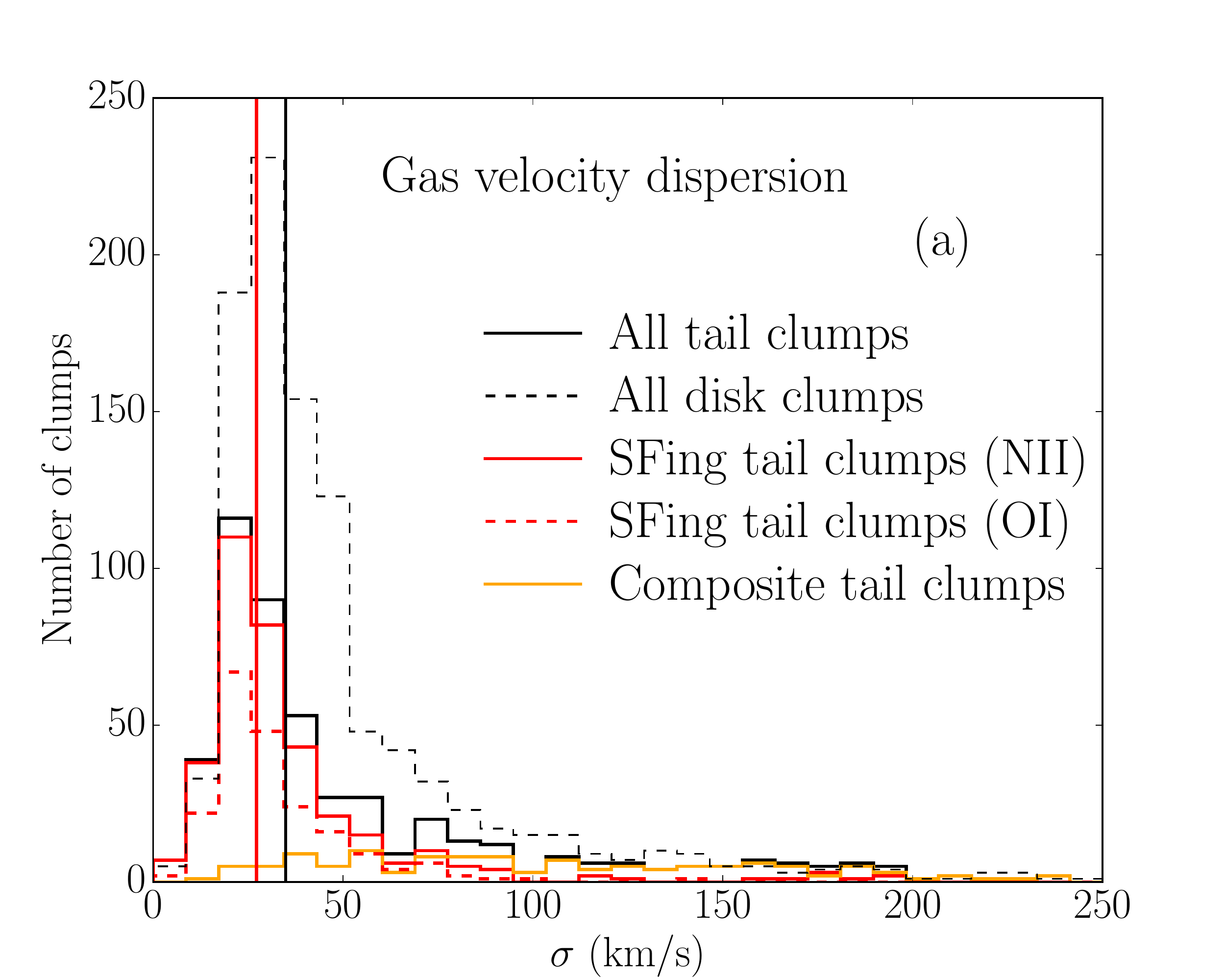}\includegraphics[width=2.7in]{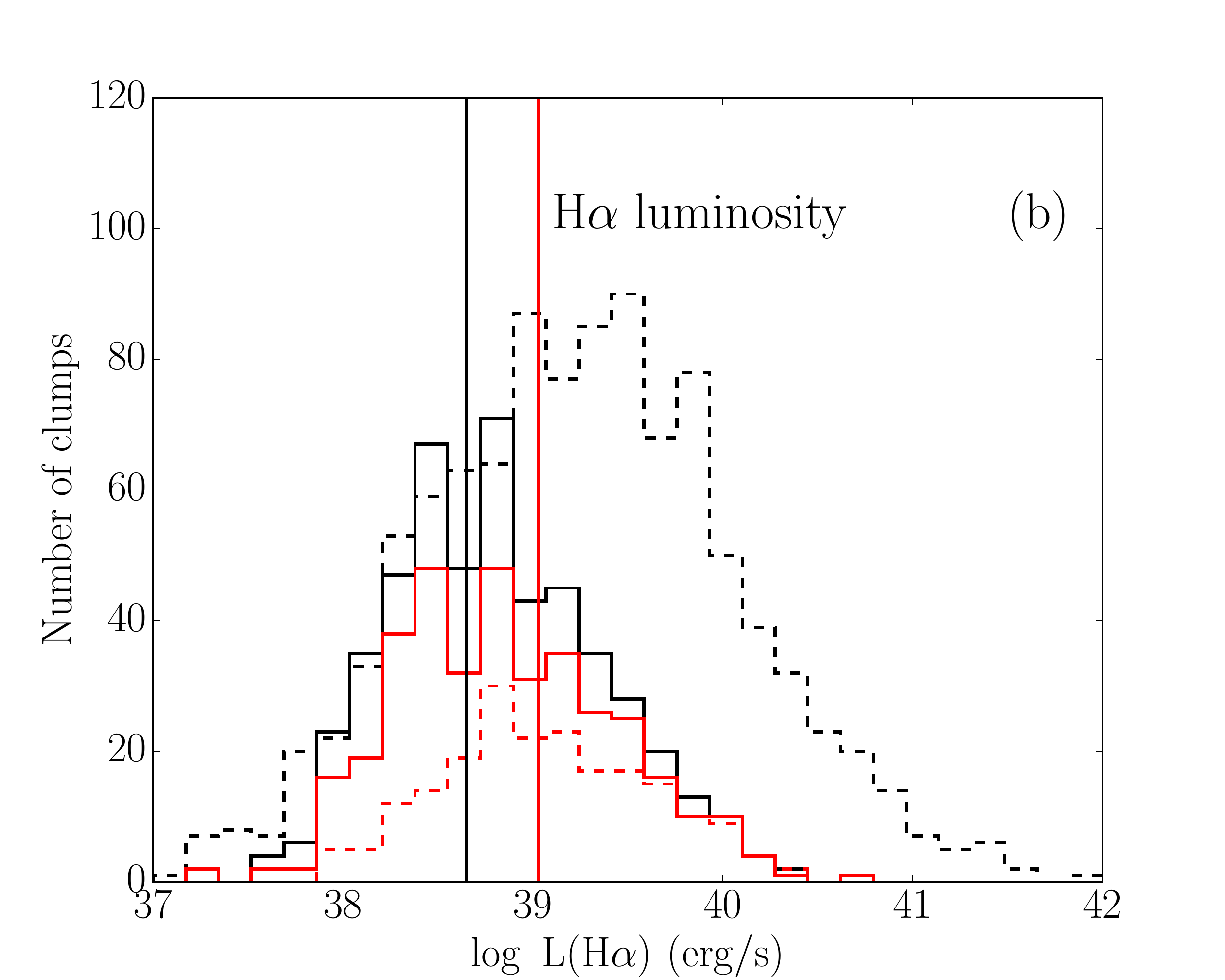}}
\centerline{\includegraphics[width=2.7in]{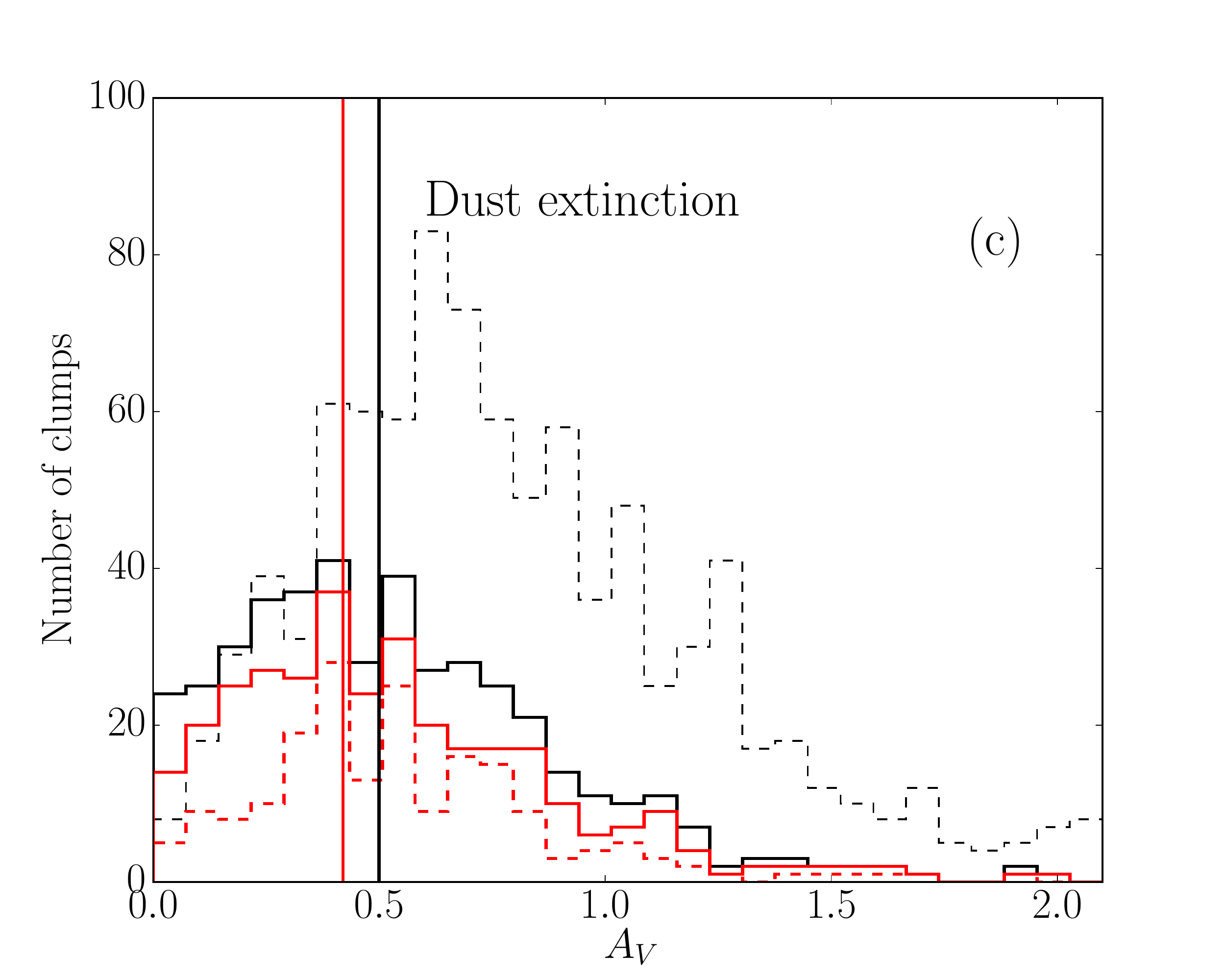}\includegraphics[width=2.7in]{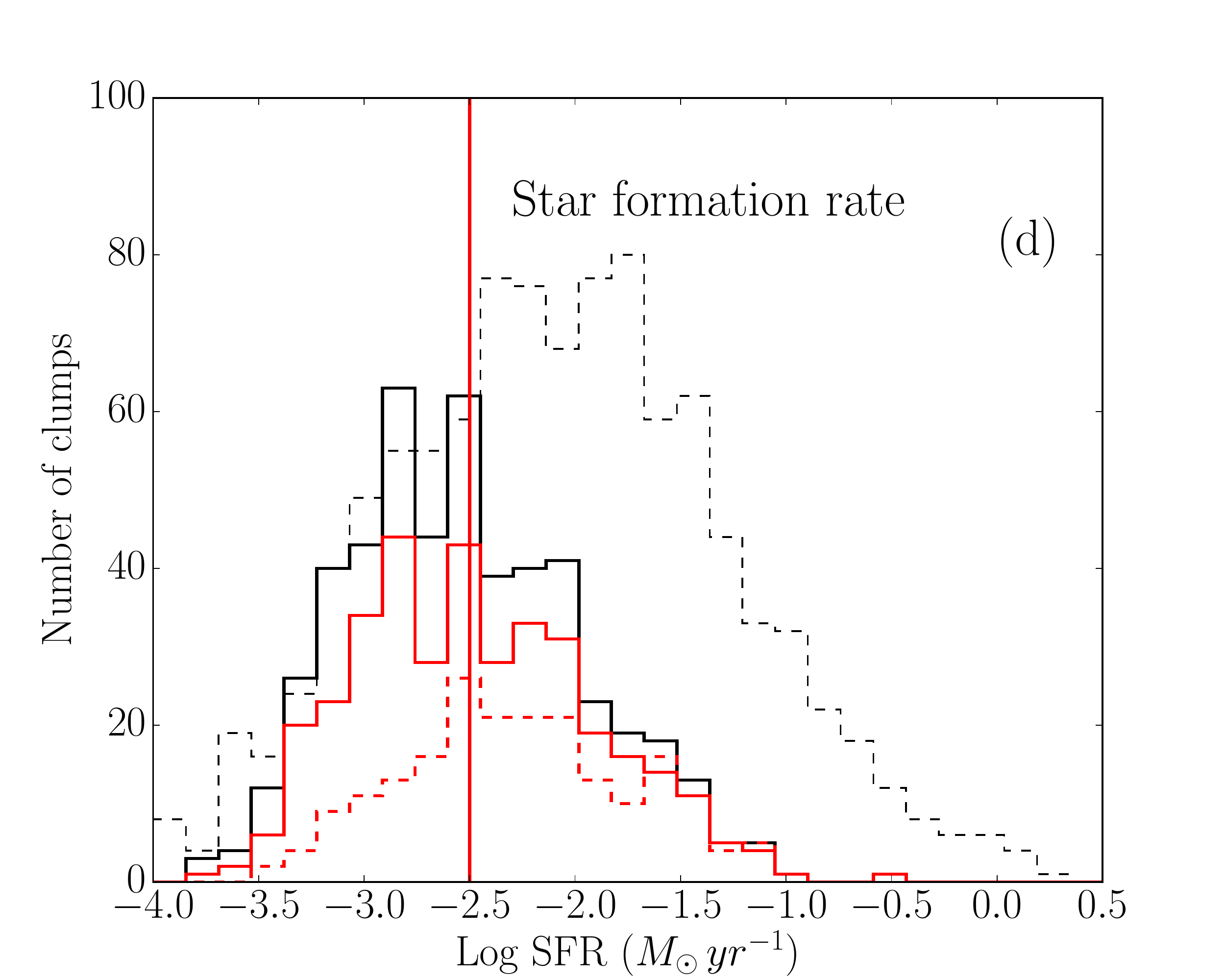}}
\centerline{\includegraphics[width=2.7in]{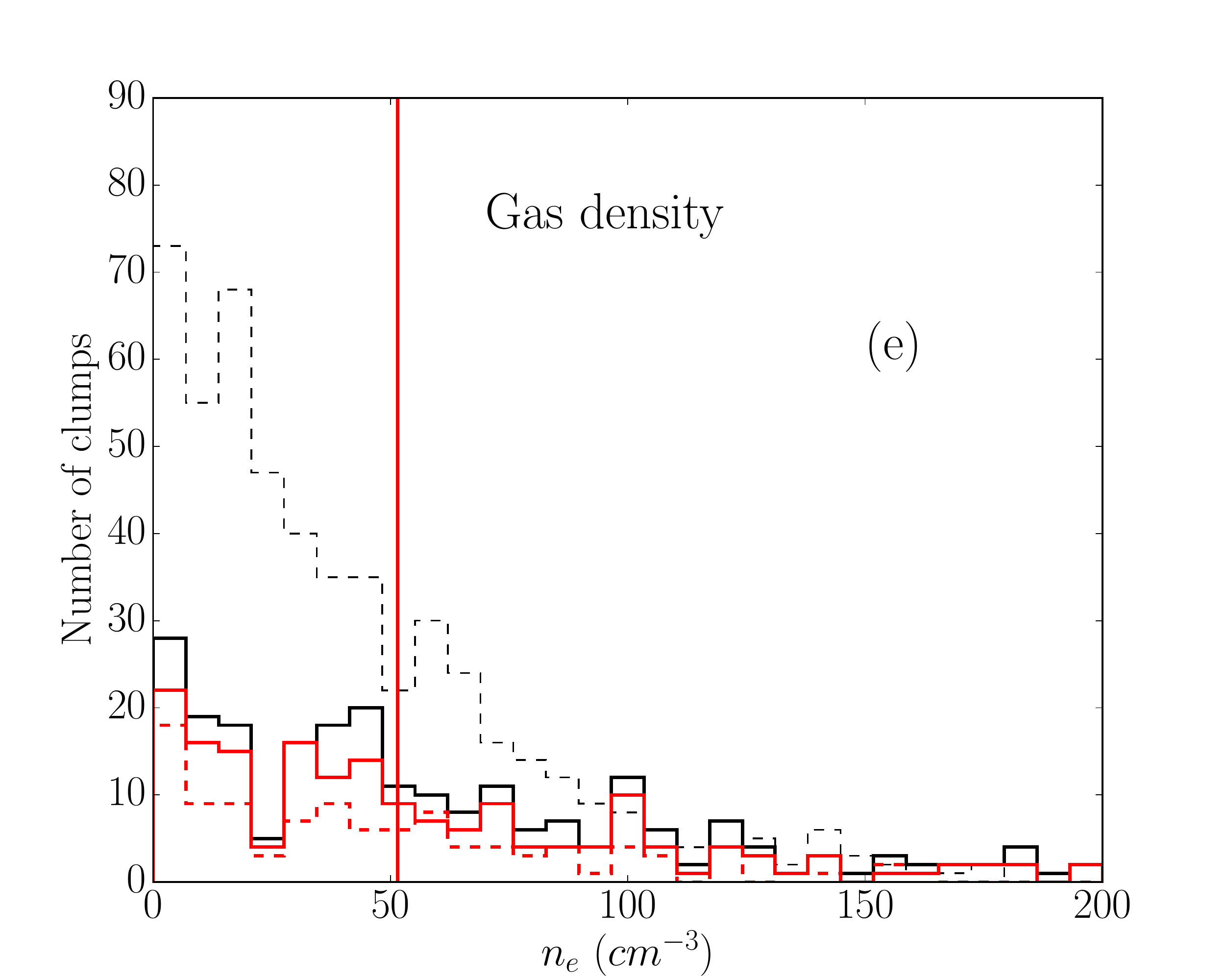}\includegraphics[width=2.7in]{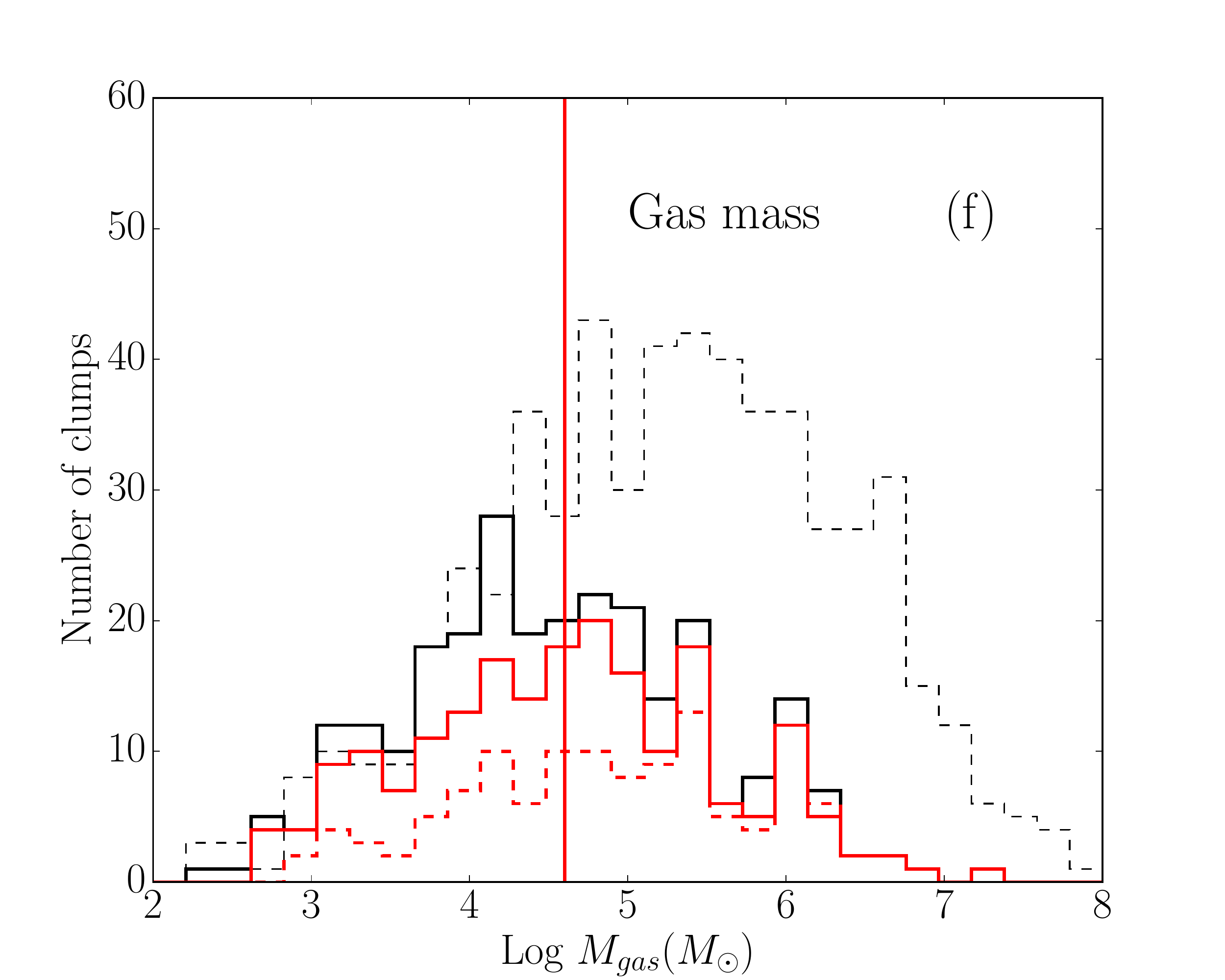}}
\centerline{\includegraphics[width=2.7in]{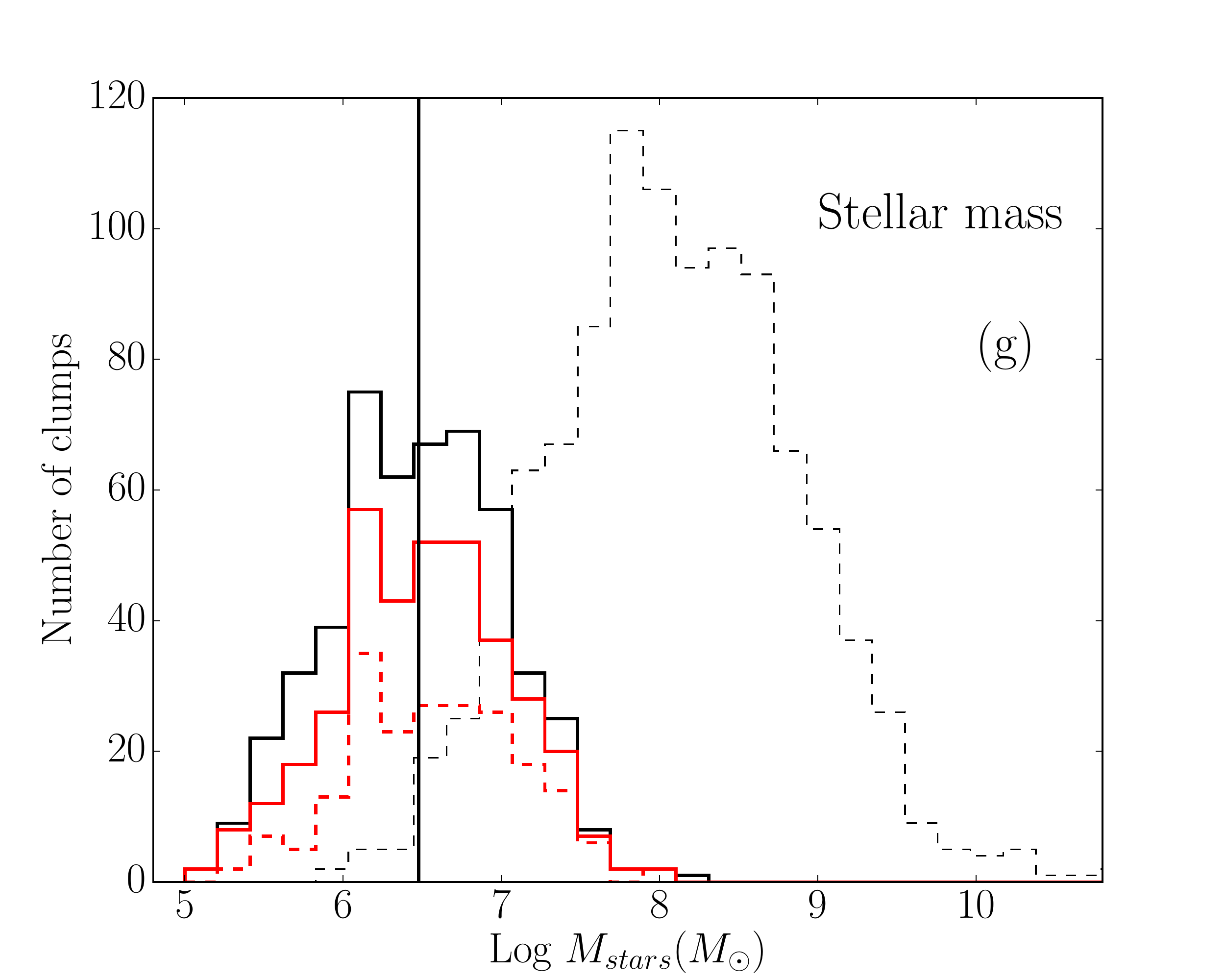}}
%\centerline{\includegraphics[width=2.7in]{Mstars_histo.pdf}\includegraphics[width=2.7in]{Tscale_histo.pdf}}
\caption{Clump physical properties. Panel a): Gaseous velocity dispersion distribution of clumps in the
  tails. Solid black: all clumps. Red solid: SF from NII DD. Red dashed:
  SF from OI DD. Orange: Composite from NII DD. Dashed black: clumps in
  the galaxy disks. Panel b): $\rm H\alpha$ clump luminosity
  distribution in tails (solid histograms) and disks (dashed histograms). 
Panel c): Dust extinction $A_V$. Panel d): SFR distribution. Panel e)
Gas density distribution. Panel f): Gas mass. Panel g): Stellar
mass. Vertical lines indicate median values.
 }
\end{figure*}

%Pb: galaxies with double components

\subsection{Dust extinction and star formation rates}

According to the Balmer decrement, the clumps in the tails are
extincted by non-negligible amounts of dust with a median $A_V=0.5
(0.27-0.74)$ mag, see Fig.~5c.  This
median value remains very similar if we select only SF-powered
clumps. Only 9\% of the tail clumps have an $A_V>1$ mag.  The
extinction in the tail clumps is lower on average than the one in the
clumps that are in the galaxy disks, whose median is 0.74 mag and where 
28\% of the clumps have an $A_V>1$ mag.

Having corrected the $\rm H\alpha$ emission for the extinction by
dust, we derive the SFR distribution of individual clumps (Fig.~5d).
While the median SFR of clumps within the disk is 0.012 $M_{\odot} \,
yr^{-1}$ (Q1=0.004, Q3=0.037), the one of the star-forming tail clumps is 0.003 (Q1-Q3=0.001-0.008) $M_{\odot} \,
yr^{-1}$ (0.005 using OI DD). Almost all the tail clumps have SFR lower than 
0.1 $M_{\odot} \, yr^{-1}$, while in the disk they can form up to 1
solar mass/year.
%.......Comment on IMF stochasticity......{\bf Jacopo, inputs?}

IMF stochasticity in regions of low SFR leads to underestimate on
average the SFR measured from the $\rm H\alpha$ emission. For
an observed $\rm H\alpha$ SFR of $\sim 10^{-3} \, M_{\odot} \, yr^{-1}$
this effect is about an order of magnitude assuming a Kroupa IMF and a
constant SFR over 500 Myr \citep{DaSilva2012, DaSilva2014}. A detailed analysis of the effects of stochasticity
on our SFR estimates is beyond the scope of this paper and will be
treated in a following work.
Until then, the SFR values quoted in this paper can be considered lower limits.

% Av and SFR DO NOT CORRELATE AT ALL! see plot SFR_vs_Av.pdf

\subsection{Gas and SFR densities}

It is interesting to estimate the ionized gas densities within the
clumps. 

The gas density can be measured only for about half of the clumps in
the tails (276 out of 521) because for the others the ratio of the two [SII] lines
falls out of the usable range (see
\S3). This is true also if we consider only clumps powered by SF
(e.g. 122/205 for OI DD). 
Thus, the values quoted below refer to the clumps for which the
\cite{Proxauf2014} calibration can be used, while an estimate for the
other half of the clumps cannot be obtained.
%For those clumps for which the density cannot be estimated, given the
%line ratios, $n_e$ might be
%are probably
%lower than the limit at which it can be measured, which is $n \sim 1.4 cm^{-3}$.
%Moreover, the $n_e$ estimates need to be taken with caution, because they 
%are computed under 
%the assumptions described in \S3, which might not be appropriate for 
%the physical conditions in the star-forming clumps of these galaxies. 

%The clump densities are shown in the maps of Fig.~?.
%SEE PLOT!! SO MANY WITH NO MEASUREMENT??
The median $n_e$ in tail star-forming clumps for which a measurement is
feasible is 51.5 $cm^{-3}$, with a large spread (see solid lines in Fig.~4e) with
Q1=21 and Q3=108 $cm^{-3}$.
The gas density distribution of clumps in the disk (where 
an estimate of $n_e$ cannot be obtained for 1/3 of the clumps) has a
different shape and is surprisingly skewed towards lower densities than the tail
distribution, though it covers a similar range of values, as can be
seen in Fig.~5e, with a median of 35.6$cm^{-3}$.

In Fig.~6 we show the density map of clumps in JO206, as an
illustrative case of the general trends. The clumps for which the
density cannot be estimated are grey circles, and the higher average
density in the tails compared to the disk is also evident.
However, given the incompleteness of the $n_e$ measurements especially
in the tails, the significant overlaps of different clumps especially
in the disk, and the spatial resolution limit which might result in
including into each clump also the lower density surroundings, a
definitive understanding of the differences between the disk and tail clump densities
must await higher resolution data.
What our analysis conclusively shows, however, is that dense gaseous clumps
%, with gas densities comparable
%to the typical ones found in the disks, 
are found in the tails. 

%WE COULD TRY using smaller radii (1/2*r) and remeasure the lines and
%the densities. Or show the spaxel density maps, with knots overlaid. TO-DO.

%SAY OVERLAP OF KNOTS , DISCUSS PROBLEM OF RESOLUTION

\begin{figure*}
%\centerline{\hspace{2.1in}\includegraphics[width=3.1in]{JO113_ne_map_blobs.pdf}\includegraphics[width=3.1in]{JO135_ne_map_blobs.pdf}}
\centerline{\includegraphics[width=5.1in]{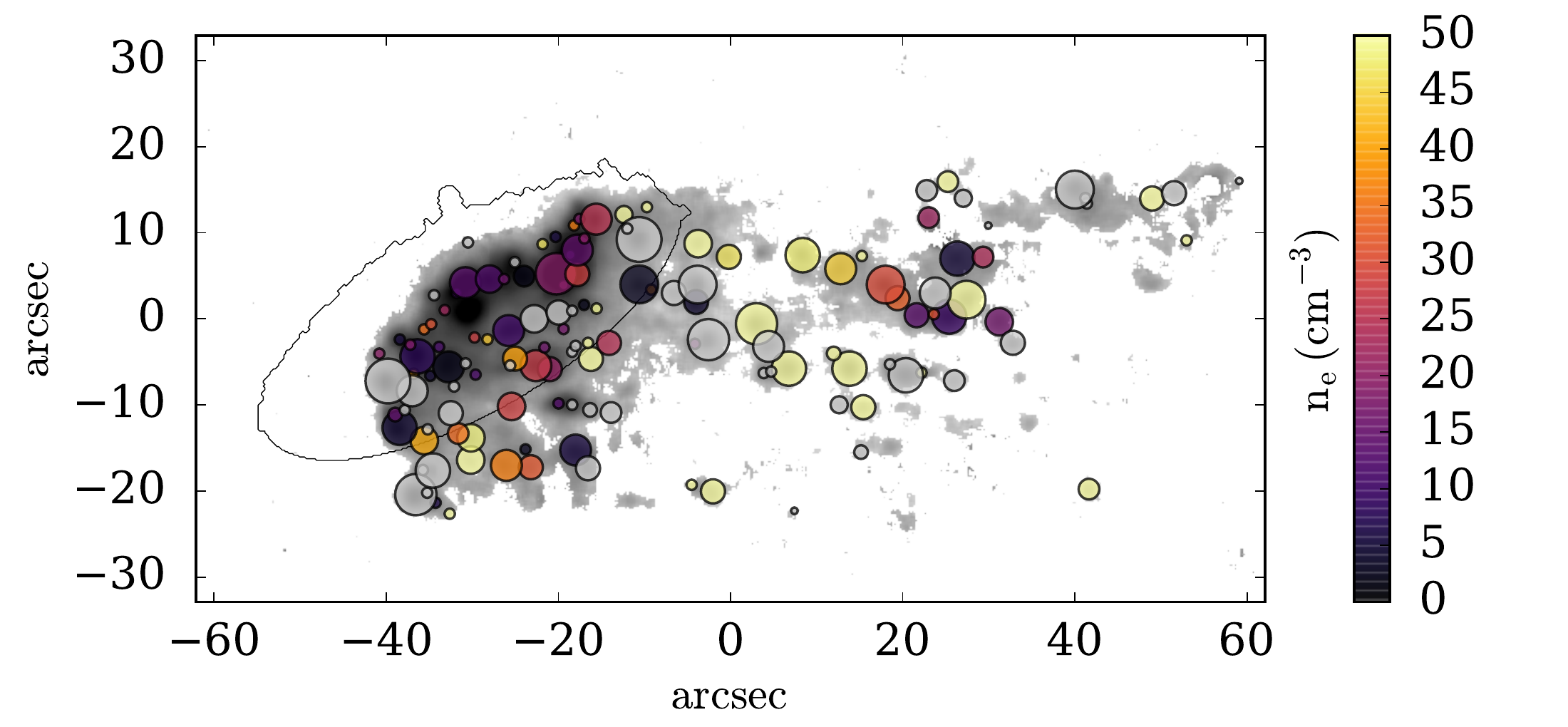}}
%\centerline{\hspace{2.1in}\includegraphics[width=3.1in]{JO141_ne_map_blobs.pdf}\includegraphics[width=3.1in]{JO147_ne_map_blobs.pdf}}
%\centerline{\hspace{2.1in}\includegraphics[width=3.1in]{JO160_ne_map_blobs.pdf}\includegraphics[width=3.1in]{JO171_ne_map_blobs.pdf}}
%\centerline{\hspace{2.1in}\includegraphics[width=3.1in]{JO175_ne_map_blobs.pdf}\includegraphics[width=3.1in]{JO204_ne_map_blobs.pdf}}
\caption{Gas density of the star-forming+composite clumps of an illustrative case,
  JO206. Grey-filled circles are the clumps for which 
 $n_e$ cannot be measured (see text). The underlying grey map shows the $\rm 
 H\alpha$ map.}
\end{figure*}

%\begin{figure*}
%\centerline{\hspace{2.1in}\includegraphics[width=2.3in]{JO194_ne_map_blobs.pdf}\includegraphics[width=2.7in]{JO201_ne_map_blobs.pdf}}
%\centerline{\hspace{2.1in}\includegraphics[width=3.1in]{JO206_ne_map_blobs.pdf}\includegraphics[width=3.1in]{JO49_ne_map_blobs.pdf}}
%\centerline{\hspace{2.1in}\includegraphics[width=2.5in]{JO60_ne_map_blobs.pdf}\includegraphics[width=3.1in]{JO95_ne_map_blobs.pdf}}
%\centerline{\hspace{2.1in}\includegraphics[width=2.8in]{JW100_ne_map_blobs.pdf}\includegraphics[width=2.8in]{JW39_ne_map_blobs.pdf}}
%\caption{}
%\end{figure*}

%Map Log ne spaxels (in and outrat) 
%Log ne istogramma spaxels 
%Log ne distribution blobs 

Turning to the SFR density (herafter SFRD) inside the clumps, this
cannot be simply estimated from the total SFR in the clump and the
clump area because of the likely overestimation of the clump sizes
(see \S5.5). 
The SFRD spaxel maps (Fig.~7) give us a view of the variation of the
SFRD from clump-to-clump, and between clumps and regions of diffuse
emission.
%{\bf To do: mask out AGN/LINER regions.}
The SFRD typically reaches logarithmic values between -2.5 and -1.5
$M_{\odot} \, yr^{-1} \, kpc^{-2}$
in the central spaxels of the brightest clumps in the tails, while
higher SFRD can be reached in the clumps inside the disks. 
%All these SFRD values should be regarded as hard lower limits to the 
%SFRD density reached within the knots, and higher spatial resolution 
%data would be required to probe the physical star formation scales. 

%can be estimated
%dividing the total SFR of each knot for its area. The median SFRD of
%knots in the tail is $log \, SFRD(\rm M_{\odot} \, yr^{-1} \,
%kpc^{-2})= -3$, ranging between logarithmic values of -3.8 and -1.9.
%As for the gas density, these values are probably influenced by the
%seeing that prevents us from resolving the star forming regions down
%to their physically meaningful scales, and from the fact that by
%construction brighter blobs have larger radii.

%istogramma SFRD knots ? not worth showing
%sigma vs SFRD(or Halpha) ? - no correlation
%sigma vs SF/blobs - no correlation
% SFRD non correla con Ne

%\clearpage

\begin{figure*}
\centerline{\includegraphics[width=2.0in]{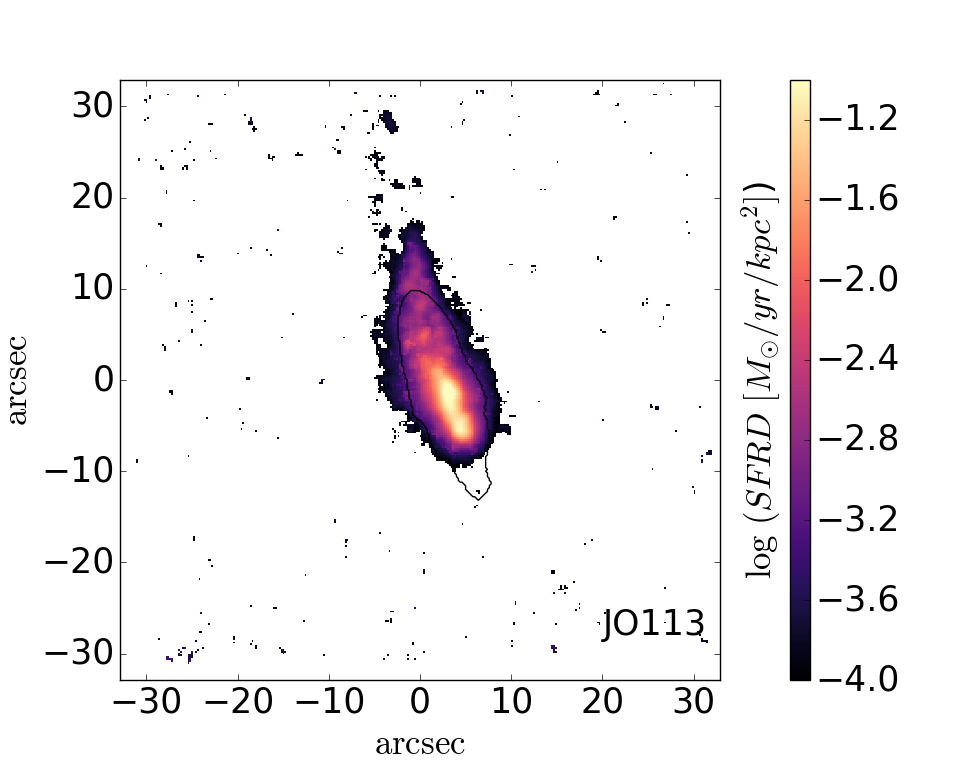}\includegraphics[width=2.0in]{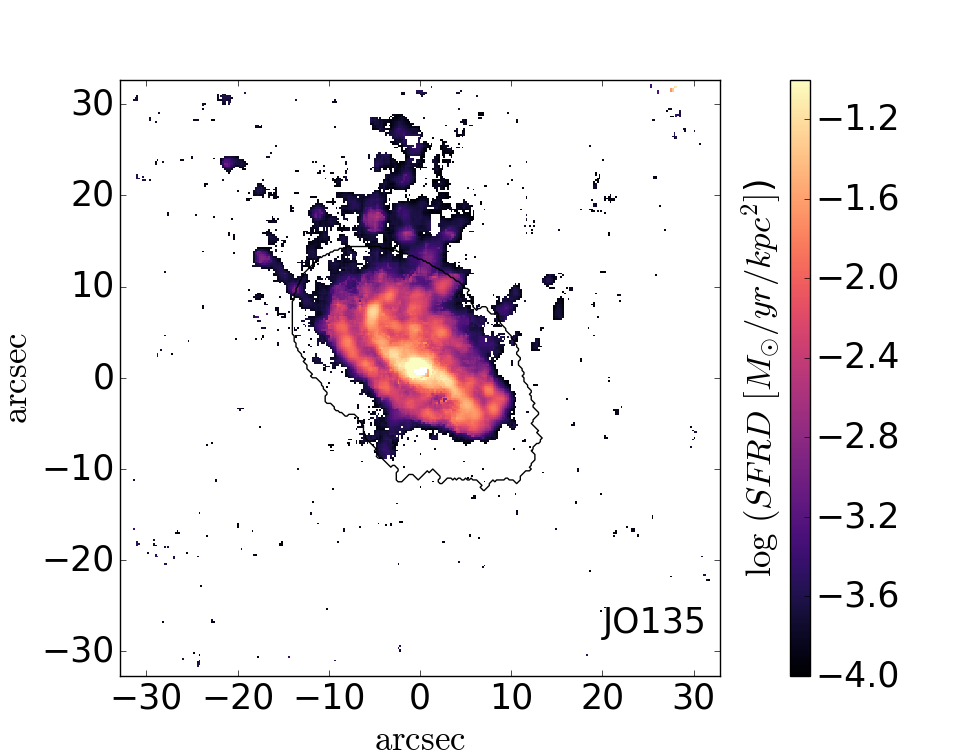}\includegraphics[width=2.0in]{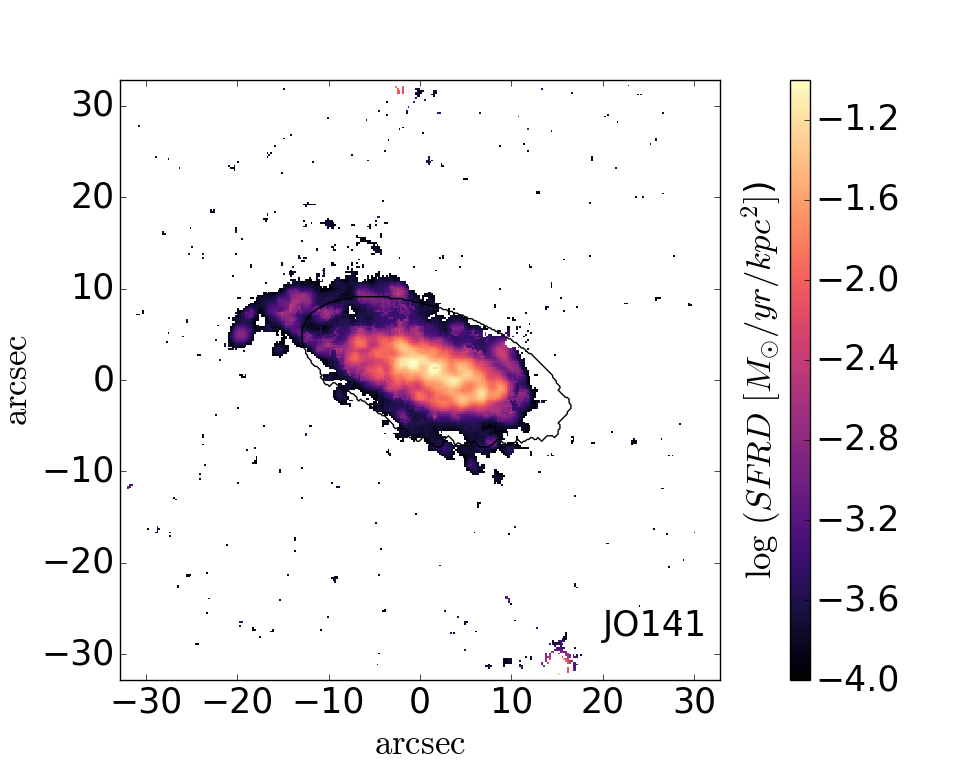}\includegraphics[width=2.0in]{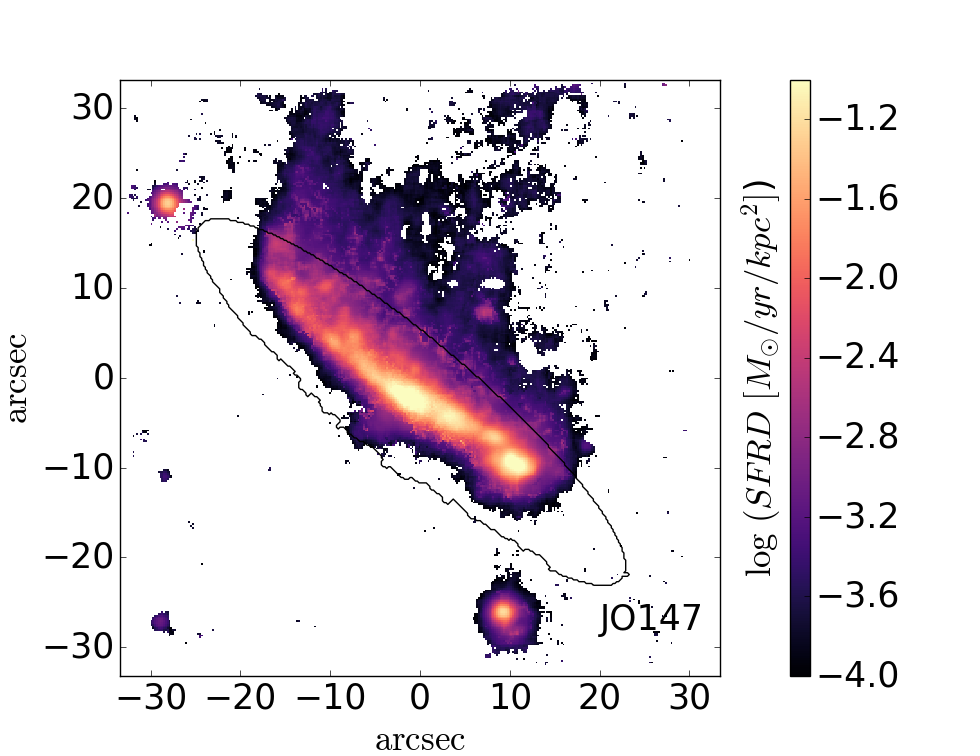}}
%\centerline{\hspace{2.1in}\includegraphics[width=2.0in]{plotsfr_map_JO141_Ha_smo5.png}\includegraphics[width=2.0in]{plotsfr_map_JO147_Ha_smo5.png}}
\centerline{\includegraphics[width=2.0in]{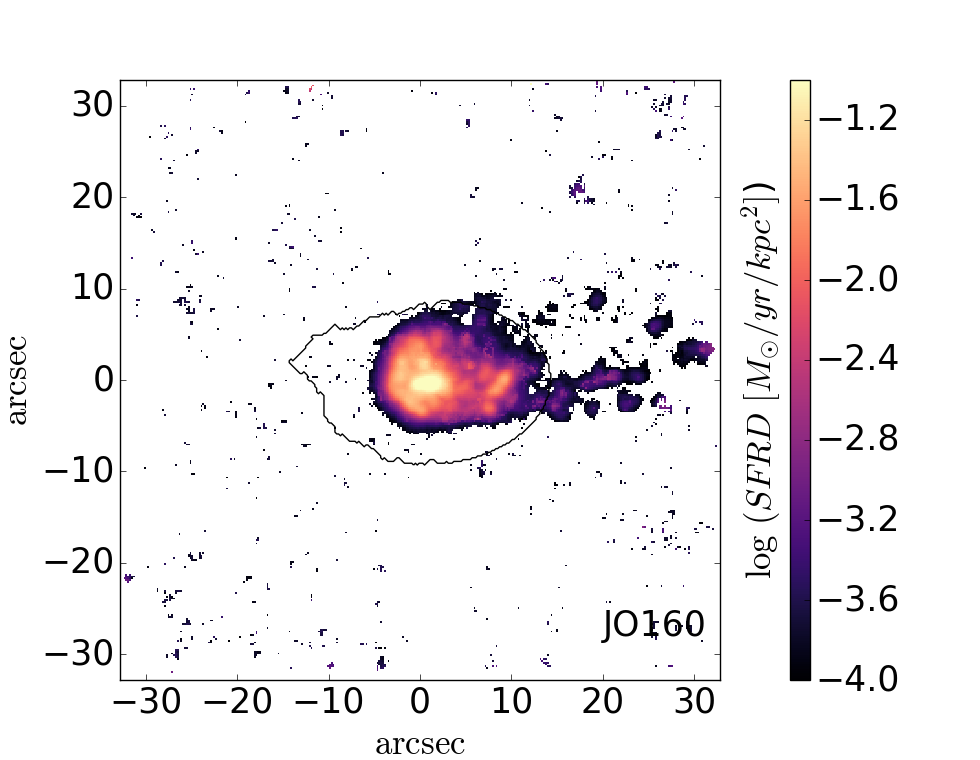}\includegraphics[width=2.0in]{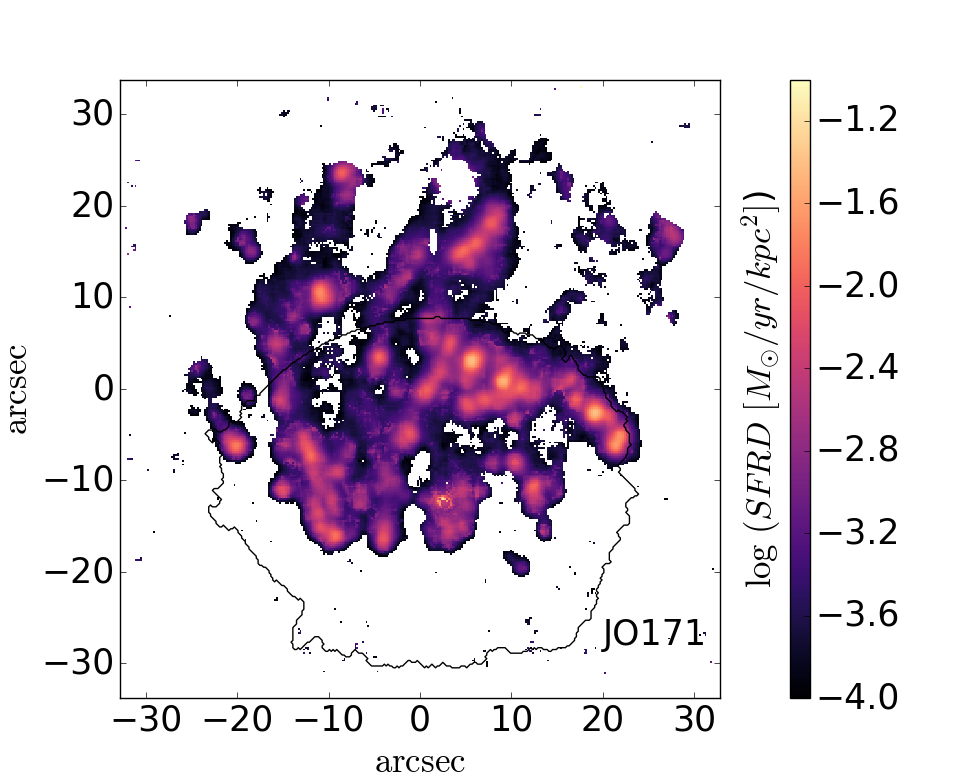}\includegraphics[width=2.0in]{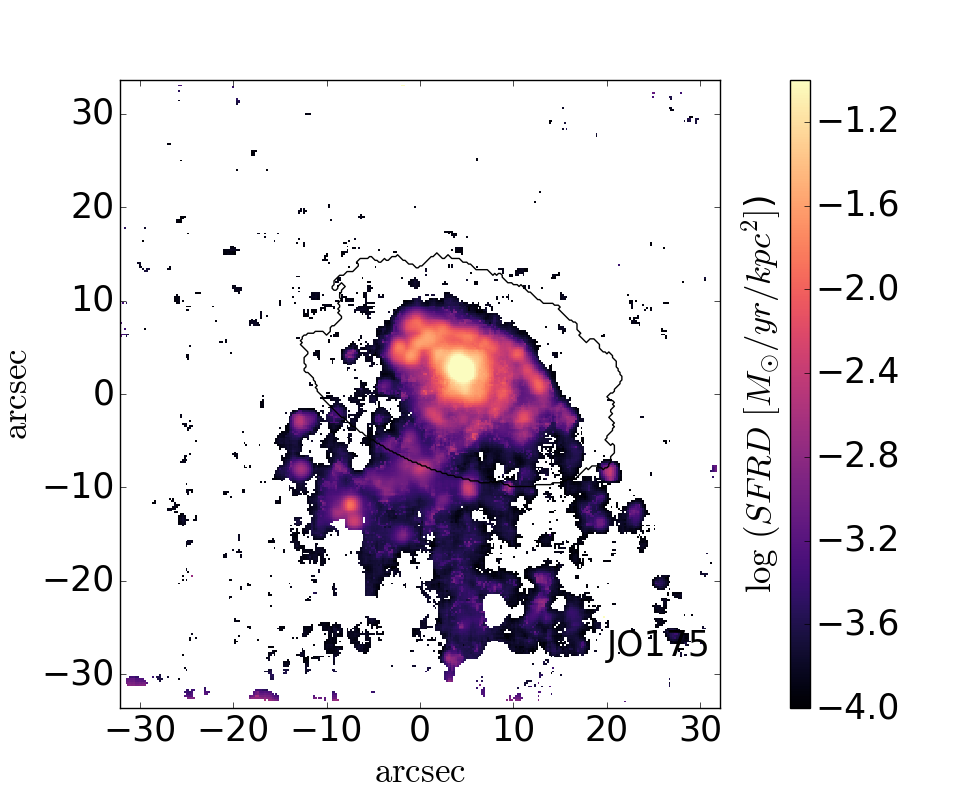}\includegraphics[width=2.0in]{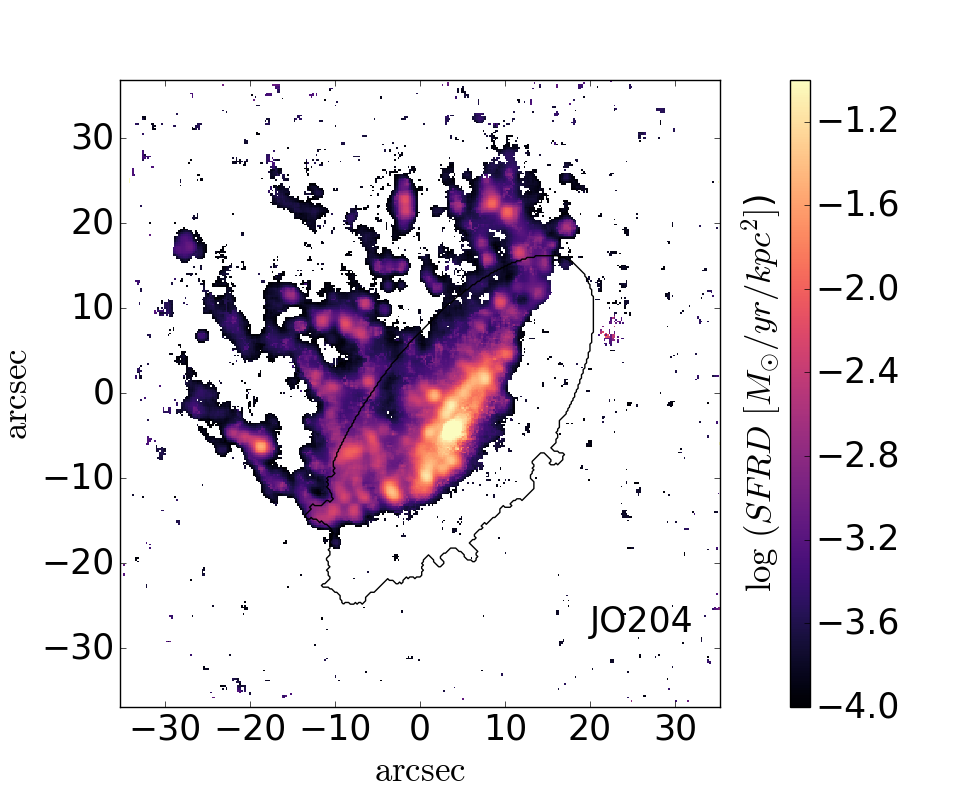}}
%\centerline{\hspace{2.1in}\includegraphics[width=2.0in]{plotsfr_map_JO175_Ha_smo5.png}\includegraphics[width=2.0in]{plotsfr_map_JO204_Ha_smo5.png}}
\centerline{\includegraphics[width=1.6in]{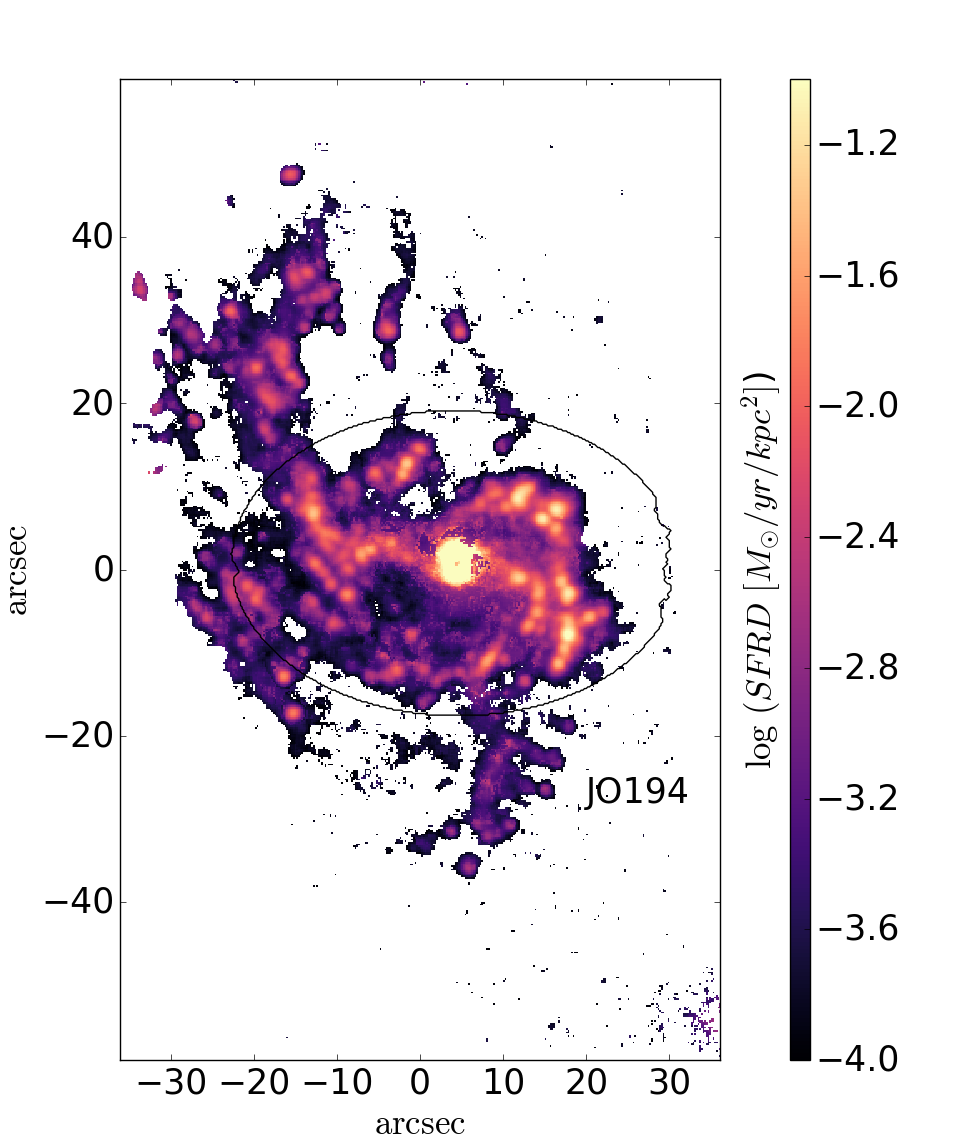}\includegraphics[width=1.8in]{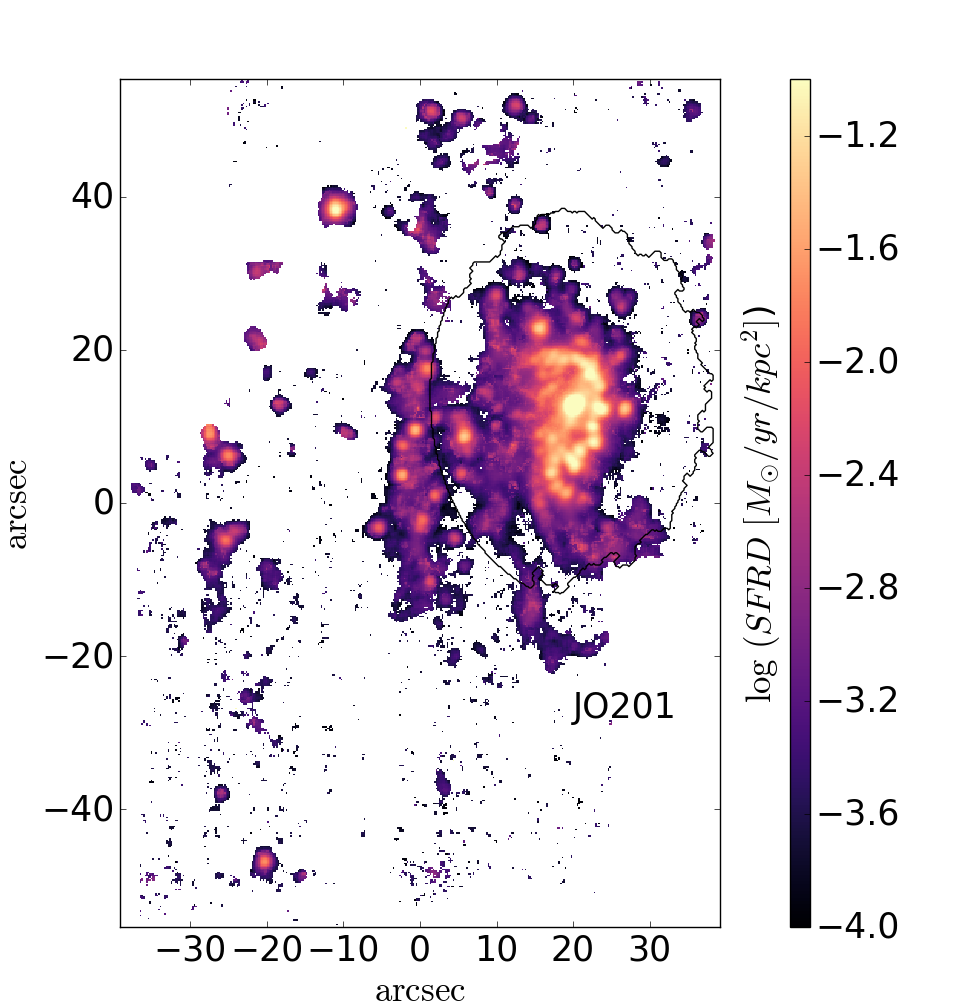}\includegraphics[width=2.5in]{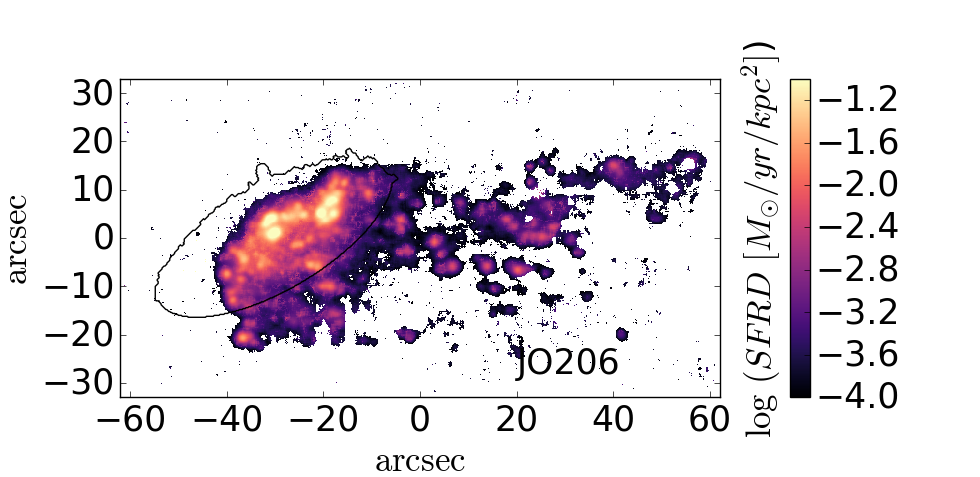}\includegraphics[width=2.0in]{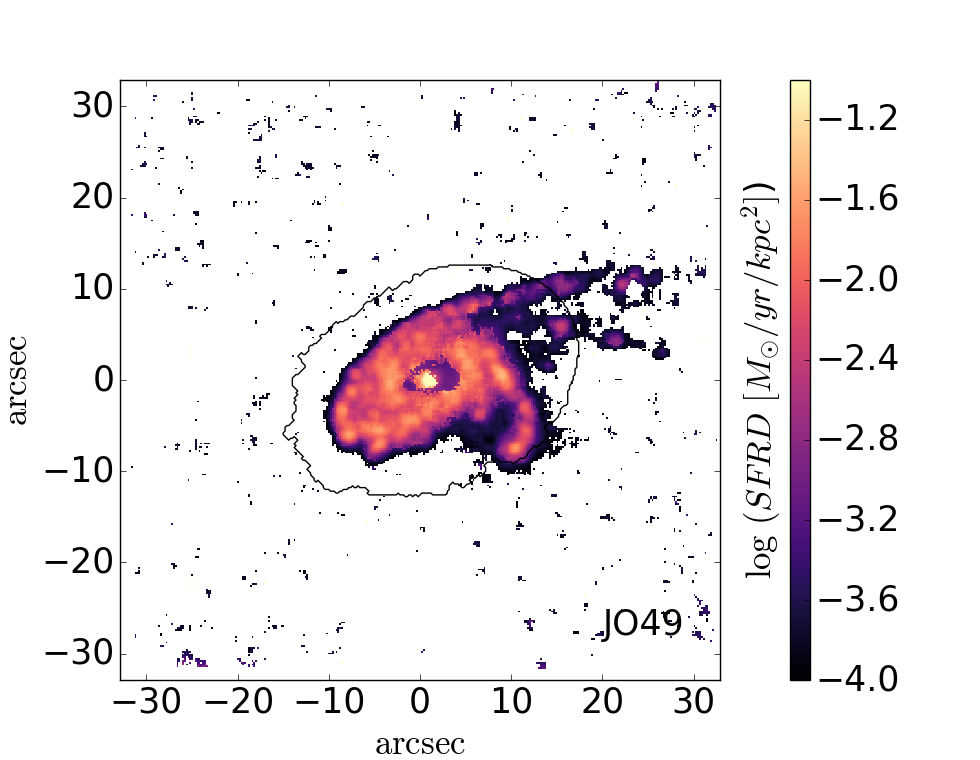}}
\centerline{\includegraphics[width=2.0in]{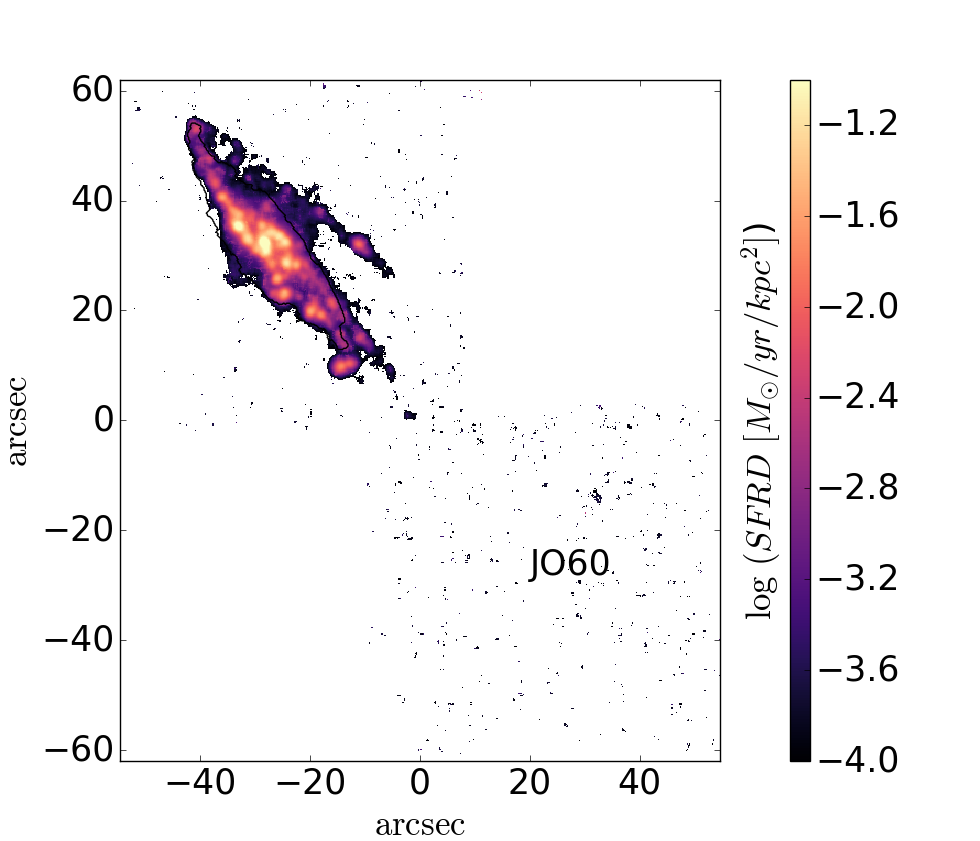}\includegraphics[width=2.0in]{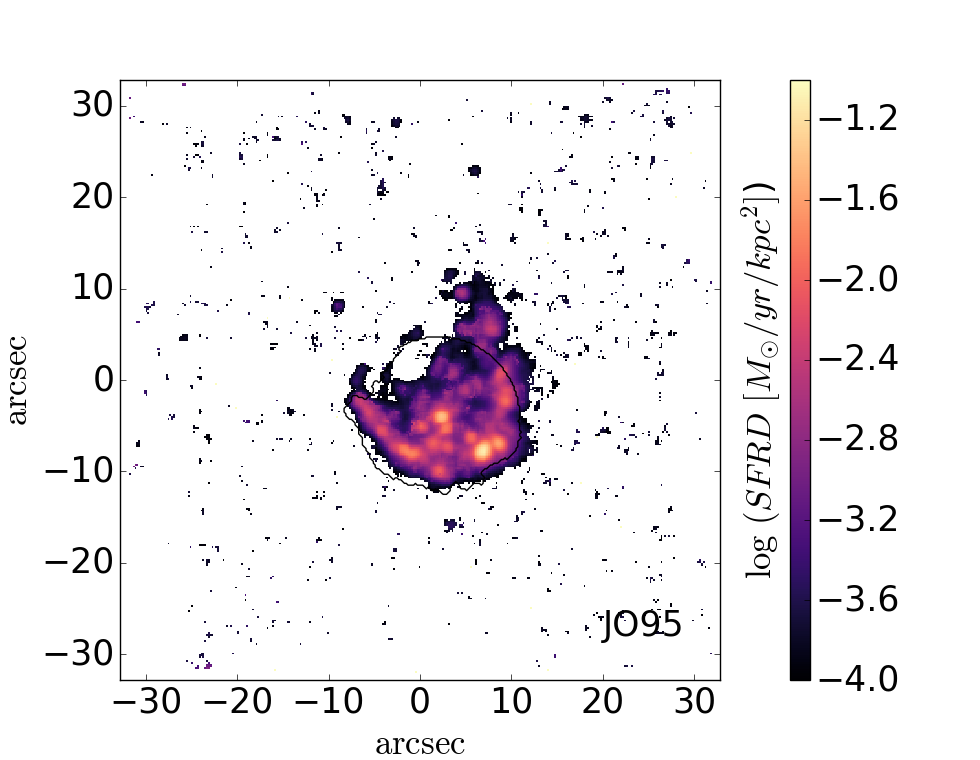}\includegraphics[width=2.0in]{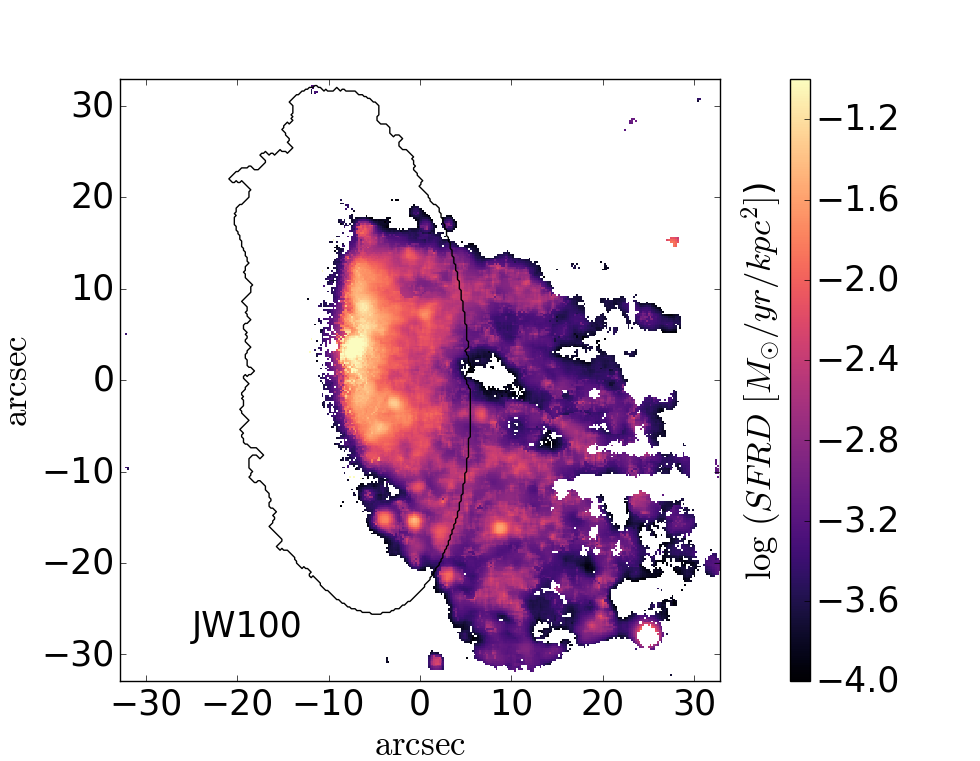}\includegraphics[width=2.0in]{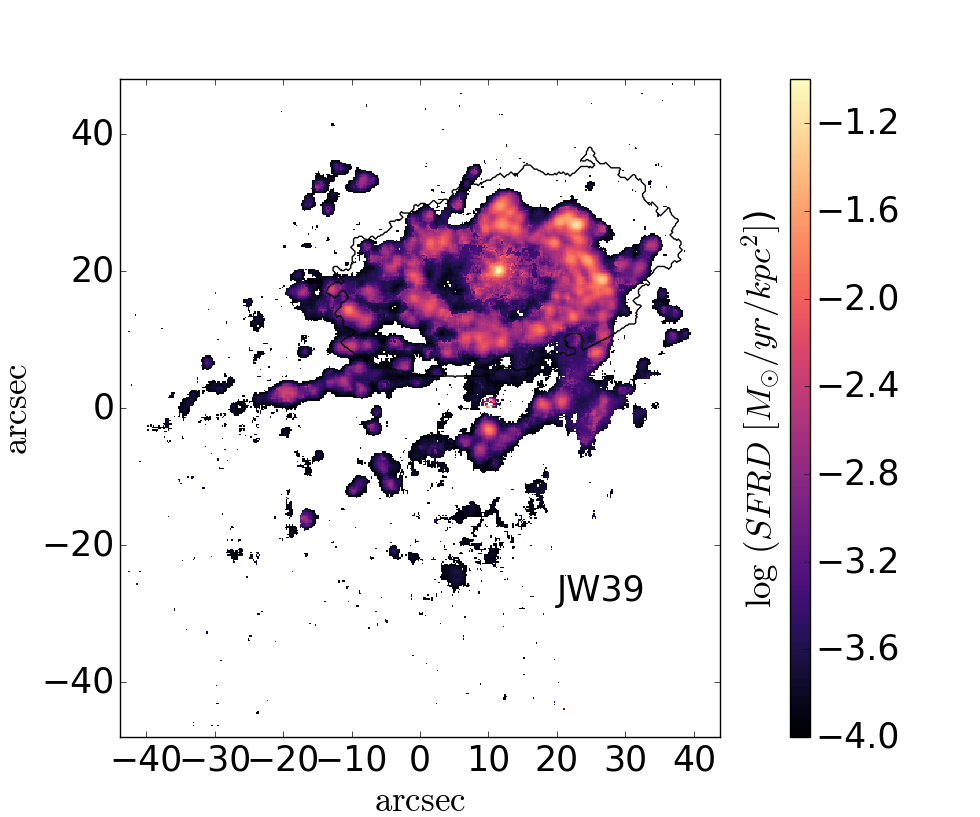}}
\caption{Star formation rate density maps.}
\end{figure*}

%\begin{figure*}
%\centerline{\hspace{2.1in}\includegraphics[width=2.3in]{plotsfr_map_JO194_Ha_smo5.pdf}\includegraphics[width=2.7in]{plotsfr_map_JO201_Ha_smo5.pdf}}
%\centerline{\hspace{2.1in}\includegraphics[width=2s.0in]{plotsfr_map_JO206_Ha_smo5.pdf}\includegraphics[width=2.0in]{plotsfr_map_JO49_Ha_smo5.pdf}}
%\centerline{\hspace{2.1in}\includegraphics[width=2.5in]{plotsfr_map_JO60_Ha_smo5.pdf}\includegraphics[width=2.0in]{plotsfr_map_JO95_Ha_smo5.pdf}}
%\centerline{\hspace{2.1in}\includegraphics[width=2.8in]{plotsfr_map_JW100_Ha_smo5.pdf}\includegraphics[width=2.8in]{plotsfr_map_JW39_Ha_smo5.pdf}}
%\caption{}
%\end{figure*}

\subsection{Gas and stellar masses}

The distribution of ionized gas masses of the clumps (Fig.~5f) shows that in the
tails they range between $\sim 10^3$ and $\sim 10^{7}
M_{\odot}$ (in the disk up to $10^8$), with a median for star-forming
clumps equal to $\sim 4 \times 10^4$ in the tails and $2 \times 10^5$
in the disk.\footnote{Since these values are computed from the gas
 density, the number of clumps for which the gas mass can be estimated
is the same as the number of clumps with $n_e$ measurements.}

%THERE IS A PROBLEM WITH THE STELLAR MASSES, WORK IN PROGRESS -- IGNORE
%PANEL IN FIG.23.
While inside the disk stellar
contours the clump stellar masses are derived from the fits that use SSPs of all ages
(because the disks contain also old stars in the line of sight of any HII
region), for the stellar mass estimates of clumps in the tails SINOPSIS
has been run placing an upper limit to
the age of the stellar populations ($6 \times 10^8$ yr), in order to
avoid having very low levels of unrealistically old stars in the tails, whose light
contribution is insignificant, but whose integrated stellar mass can
result in overestimating the stellar mass. We have tested that the
stellar masses of the tail clumps do not change appreciably varying the
upper age limit between a few $10^7$ and $10^9$ yr, therefore this
measurement can be considered stable. % for these realistic assumptions.
In the following, it is important to keep in mind that while the stellar masses of tail clumps are indeed the masses
of stars associated with the HII complexes observed, the
stellar masses of clumps in the disks are strongly influenced by the total stellar 
(old and young stars) mass density variations within the disks, and
cannot be considered representative of the population of young stars
in the HII complexes. In a sense, tail clump masses are ``true stellar
masses'' of the clumps, while disk clump masses are ``projected
stellar masses'' inflated by the underlying old stellar populations.

The stellar masses of the clumps in the tails range between $10^5$ and
$3 \times 10^{7} M_{\odot}$, with a median of $3 \times 10^6 M_{\odot}$ (Fig.~5g).
In the disks, the ``projected stellar masses'' (old+young stars) 
range between $3 \times 10^{6}- 3 \times 10^{9}$, with a median
$10^{8}$. Comparing the stellar masses of clumps in the tails and
disks is obviously not meaningful, for the reasons explained above,
but we do plot the ``projected stellar masses'' of disk clumps in
Fig.~5g for completeness. As for the SFR, the assumption of a standard \cite{Chabrier2003} IMF introduces 
a large uncertainty in the stellar mass estimates that should be kept
in mind.

The fate of the stars formed in the tails will be discussed in a
subsequent paper of this series discussing their contribution to the
intra-cluster light (Gullieuszik et al. in prep.).
Tail stars, especially those closer to the disk, may remain bound to the parent galaxy, and
fall back onto it contributing to the thick disk or the bulge
\citep{Abramson2011, Kapferer2009}. 
If unbound, they will remain an intracluster population of ``stripped
baryonic dwarfs'', as predicted by some simulations of ram-pressure stripping
\citep{Kapferer2008}.
If the tail stellar clumps have indeed stellar masses between $10^5$
and $10^7 M_{\odot}$, depending on their sizes they could resemble 
Ultra Compact Dwarf Galaxies (UCD) and Globular
Clusters (GC) for effective radii below 100pc, or Dwarf Spheroidals
(dSph) for sizes greater than 100pc \citep{Norris2014}, except that
they would be dark matter free.
In the next section we will present rough size estimates (Fig.~10,
median 160pc) 
%are core radii 
of the HII gaseous clumps, which are likely upper limits to the
sizes of the stellar clumps embedded within them. 
%We conclude that
Based on this, we suggest that
most likely the stellar clumps we see forming in the stripped tails
 may contribute to the abundant
population of UCDs and GCs observed in galaxy clusters \citep{Hilker1999, 
Drinkwater2000, Wittmann2016}, or even the
population of Ultra Diffuse Galaxies (UDGs) depending on their subsequent
interactions with the cluster potential and other galaxies, for
example via harassment \citep{Conselice2018}. Predicting
the metallicities and the subsequent dynamical history of these
stellar clumps within the cluster is
beyond the scope of this paper and will be pursued in future studies.

Finally, we compare the gas masses and stellar masses of tail clumps in Fig.~8
(green points), and notice a broad correlation, with
stellar masses typically one or two orders of
magnitude higher than the corresponding ionized gas
masses. A similar, but offsetted correlation exists between the gas
masses and the ``projected stellar mass'' of clumps in disks.
In the next section we will investigate the correlation between the SFR and
the gas and stellar masses of the clumps.

%Compare with Mstar distribution of Cava+ 2018 etc? La distribuzione di
%masse stellari clumps di Cava e' molto simile alla nostra dei dischi!

%Cf with Stephanie's notes

%{\bf STEPHANIE, in your notes I could not find any prediction for
%  clump gas masses or stellar masses in the tails, do they exist?
%The value you gave me in the notes 3e4Msun/kpc2 from TB2012 is a surface mass
%density I guess?}

\begin{figure}
\centerline{\includegraphics[width=5.0in]{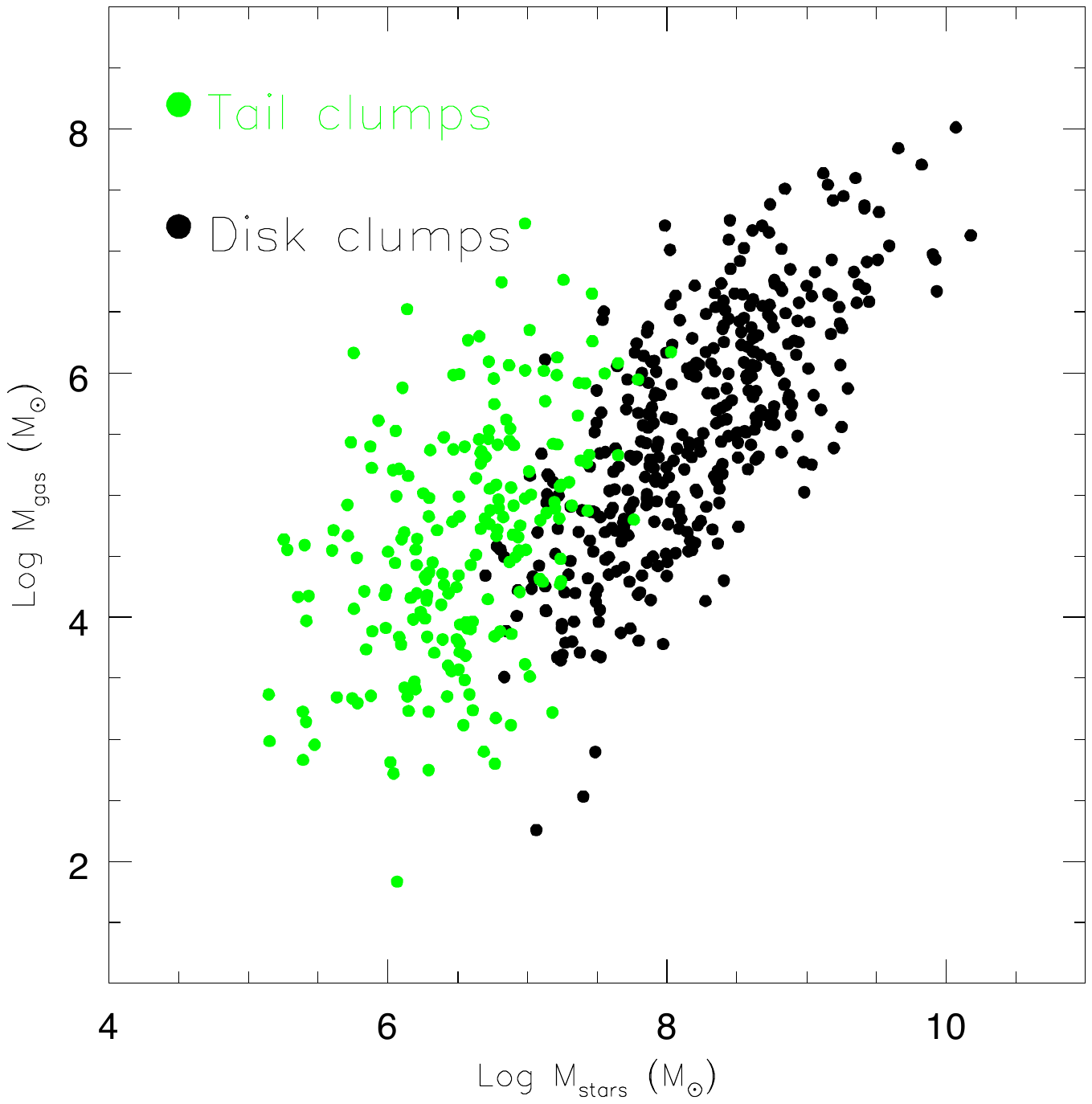}}
%\vspace{-10cm}
%\centerline{\includegraphics[width=5.0in]{wisnkioski.pdf}}
\vspace{-4cm}
\caption{Relation between ionized gas mass and stellar mass of 
  star-forming clumps in the tails (green) and ``projected stellar 
    masses''  of clumps in the disks (black, see text).}
\end{figure}

%% NON MOSTRO L'ISTOGRAMMA DEL TIMESCALE, NON HA MOLTO SENSO, CONTINUA
%% A VENIRE UN PO' TROPPO ALTO....

\subsection{Clump scaling relations}

%ANTICORRELATION Ne-Mgas !
%lack of correlations Ne - etc... 
%USE WISNIOSKI !!!! relations fig.6, and fig.7 etc etc!!!!! TO-DO 
%can use the scaling relations with sigma to estimate the rest!!!!!

Star-forming regions both at low- and high-z have been found to follow 
scaling relations linking their physical characteristics, such as 
sizes, gas velocity dispersion and $\rm H\alpha$ luminosity \citep{TerlevichMelnick1981, 
Larson1981, GallagherHunter1983, Monreal2007}. 

\cite{Wisnioski2012} showed that the gas velocity dispersions 
of star-forming clumps are 
unaffected by spatial resolution effects and that the $\rm H\alpha$ luminosities 
are quite insensitive to the chosen clump radius, being consistent when
measured within isophotal or core radii (see \S3.1).
Based on the robustness of the measurements of these two 
quantities, in Fig.~9 we inspect the $\sigma$-$L(\rm H\alpha)$ relation of our 
clumps, and contrast it with the \citeauthor{Wisnioski2012} relation (solid line) 
and their low-z data points (bottom panel in Fig.~9). 
%CLEAN THESE PLOTS FROM HIGH SIGMAS DUE TO DOUBLE COMPONENTS!!!! +
%CHANGE RELATION and WISNIOSKI POINTS TO OUR COSMOLOGICAL PARAMETERS -
%I checked, difference is negligible 
The paucity of points at $\sigma$ below $\sim 17 \, km \, s^{-1}$ 
in Fig.~9 is due to the intrinsic limit to our velocity dispersion measurements  
discussed previously.  Apart from this limit, it is remarkable that
both the clumps in the disk and those in the 
tails seem to broadly follow the scaling relations of normal star-forming 
galaxies at low-z, albeit with the presence in GASP galaxies of a subset of clumps
which seem to deviate from such relation due to their low $\rm H\alpha$
or, more likely, increased velocity dispersion. 

%\subsection{The real sizes of the star-forming clumps}

This result gives us confidence that normal-galaxy scaling relations 
might be used to have a very rough estimate of the true physical sizes 
of our star-forming clumps, below the limit imposed by the seeing. 
Adopting the relation between $\rm H\alpha$ luminosity 
and size for low redshift spirals, irregulars and starburst galaxies 
\citep[eqn. 6 in][]{Wisnioski2012}, we estimate the core radii 
of our clumps.
% and use it to derive the Jeans mass of the knots 
%following Wisnioski et al. (2012): $M_{Jeans}= pi^2 \sigma^2 
%r_{core}/3G$, where G is the gravitational constant. We note that the 
%$\sim 17 \, \rm km \, s^(-1)$ limit for the $\sigma$ measurement 
%might result in overestimating the Jeans mass in some of our knots. 
The distribution of estimated sizes %and corresponding Jeans masses 
is presented in Fig.~10. 
The typical expected core radii range from 100 to 400 pc in the tails 
(median 160 pc), and 
extend up to 600 pc in the disks (median 220 pc). These values are smaller than the 
1kpc corresponding to the seeing limit. For comparison, the median 
size in the \citeauthor{Wisnioski2012} sample is 150 pc. 
%The corresponding Jeans masses of the knots should range between $10^7$
%and $10^9 \, M_{\odot}$, with a median $\sim 10^8 \, M_{\odot}$ in the 
%tails and  $\sim 10^{8.2} \, M_{\odot}$ in the disks. 
%SOVRAPPORRE I VALORI A LOW-Z DI WISNIOSKI. 
%TAKE OUT JEANS MASSES??

%Expected radius from Wisnioski Lha-size relation vs Gianni's radius (relations.mac, diameter2:
%subctracting 1arcsec in quadrature) - but cannot be copmpared like 
%this, Gianni's radius is an isophotal radius, while they use an rcore 
%- must use as rcore the sigma of the Gaussian fittings our gas-only 
%halpha image, but below our seeing 
%TRUE CLUMP SIZES + JEANS MASS + RELATION JEANS MASS/LUM(HALPHA) see 
%relations.mac diameter2 
%SFRD vs Sigma* (fig 7 WISNIOSKI) 
%Cava18?
%CHECK SIZE-STELLAR MASS and STELLAR MASS-JEANS MASS 
%SHOW NICE CORRELATION SFR vs stellar (gas?) mass of the blobs --
%Mass surf den Sigma* distribution 

\begin{figure}
\centerline{\includegraphics[width=5.0in]{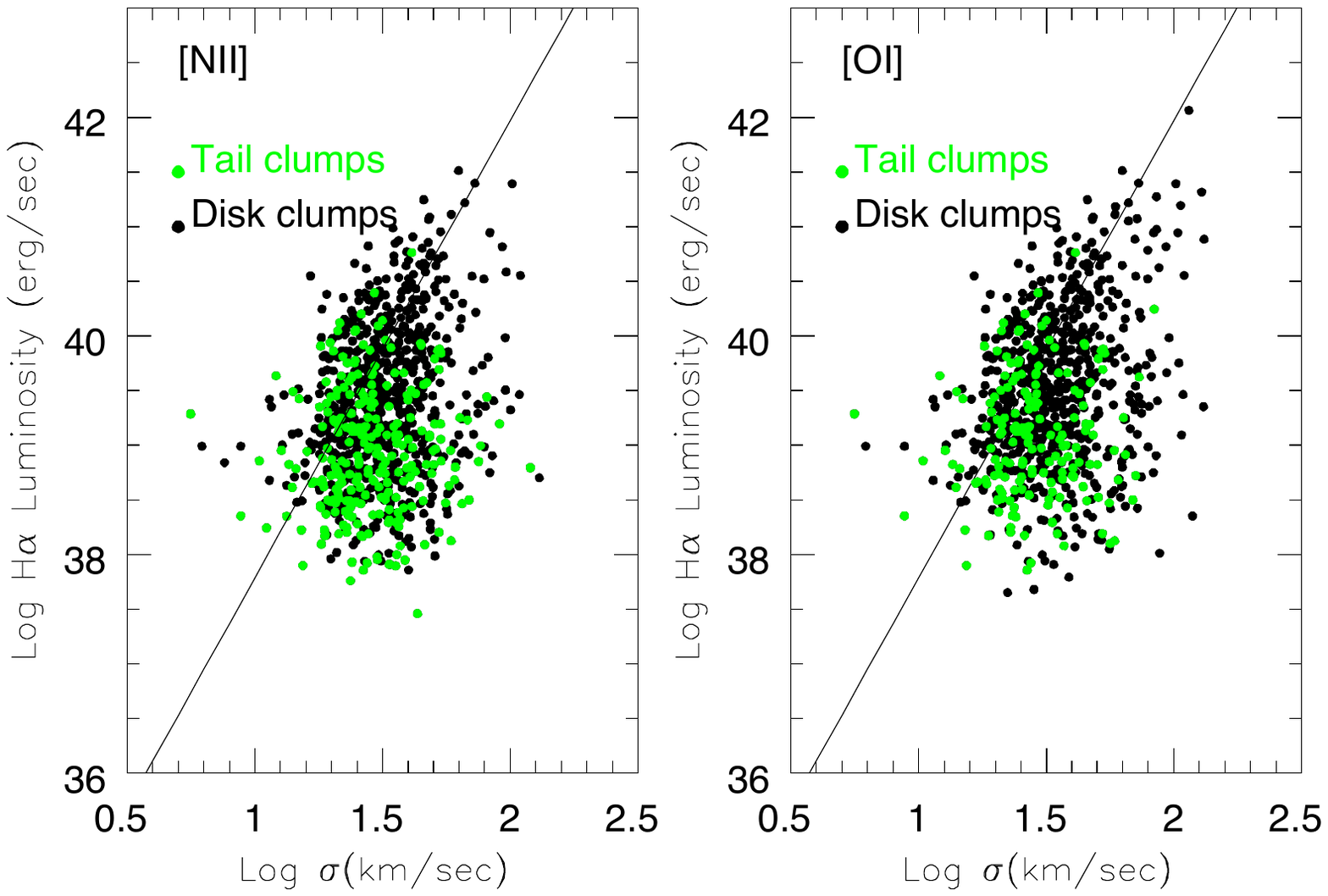}}
\vspace{-10cm}
\centerline{\includegraphics[width=5.0in]{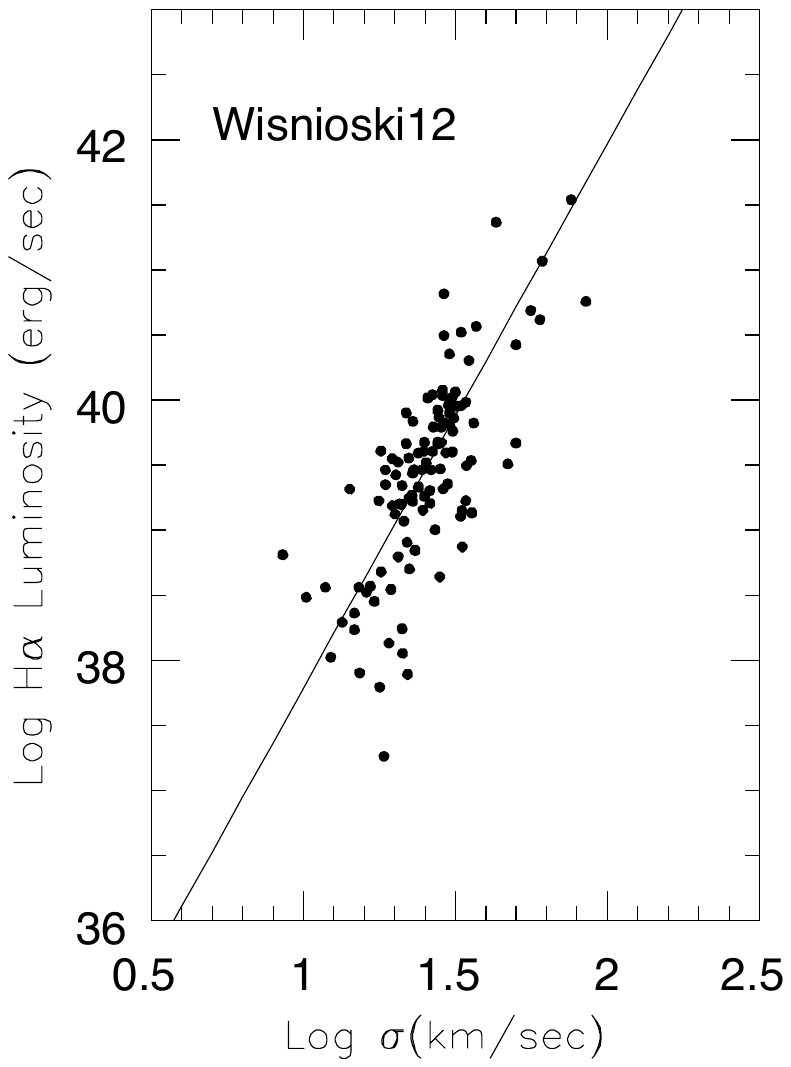}}
\vspace{-4cm}
\caption{Top. $\rm H\alpha$ luminosity versus velocity dispersion of the 
  star-forming clumps in the tails (green) and in the disks 
  (black). These are all clumps classified as star-forming by the NII 
  DD (top left panel) and the OI DD (top right panel). Solid lines represent 
  the scaling relation by \citet{Wisnioski2012}. Only points with
  reliable velocity dispersion estimates are plotted (errors on
  $\sigma$ $>0$ and excluding the worst error quartile).
% di fatto, ho tagliato a errore di sigma greater than 5 km/sec
Bottom. Same as 
  top, but plotted are HII regions in low-z galaxies from \citet{Wisnioski2012}. 
  The line is the same in all panels.}
\end{figure}

\begin{figure}
\centerline{\hspace{2cm}\includegraphics[width=4.5in]{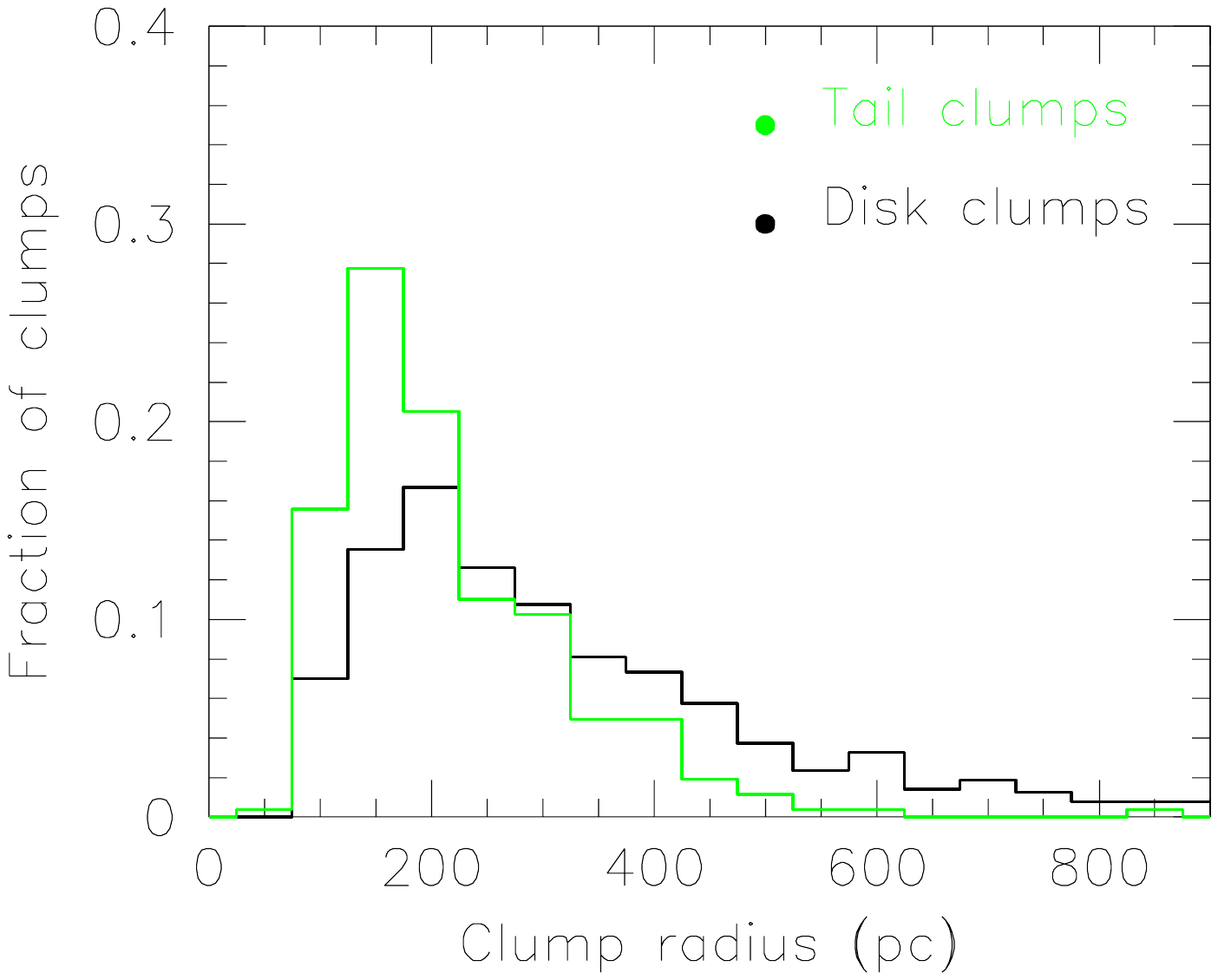}}
\vspace{-3.5cm}
\caption{Distribution of clump core radii inferred from the relation
  between $\rm H\alpha$ luminosity and size.
Green: tail clumps. Black: disk clumps.}
\end{figure}

Finally, we show that the GASP clumps follow very clear correlations
between clump SFR and clump stellar mass, and between SFR and gas mass
(Fig.~11). 
In the SFR-stellar mass relation, there is an offset between tail
clumps and disk clumps, which is due to the inflated ``projected
stellar mass'' of clumps in the disk. This offset disappears in the
SFR-gas mass relation, which is a tight sequence followed both in the
tails and in the disks: more massive gas clumps form a larger amount
of stars per unit time. Both the SFR and the ionized gas mass depend
linearly on the $\rm H\alpha$ luminosity (\S3), thus a correlation
is expected. However, the gas mass depends also on the gas density, therefore
the width of the correlation is related to the spread in gas density
from clump-to-clump. What is most striking is the fact that
star-forming clumps in the tails and in the disks follow the exact
same relation with a similar scatter, indicating that the range of
physical conditions is not too different.
%This shows once more that star-forming clumps
%formed in the stripped tails follow the same relations, hence the same
%physical laws, of clumps in galaxy disks.

\begin{figure*}
\centerline{\hspace{1cm}\includegraphics[width=5.0in]{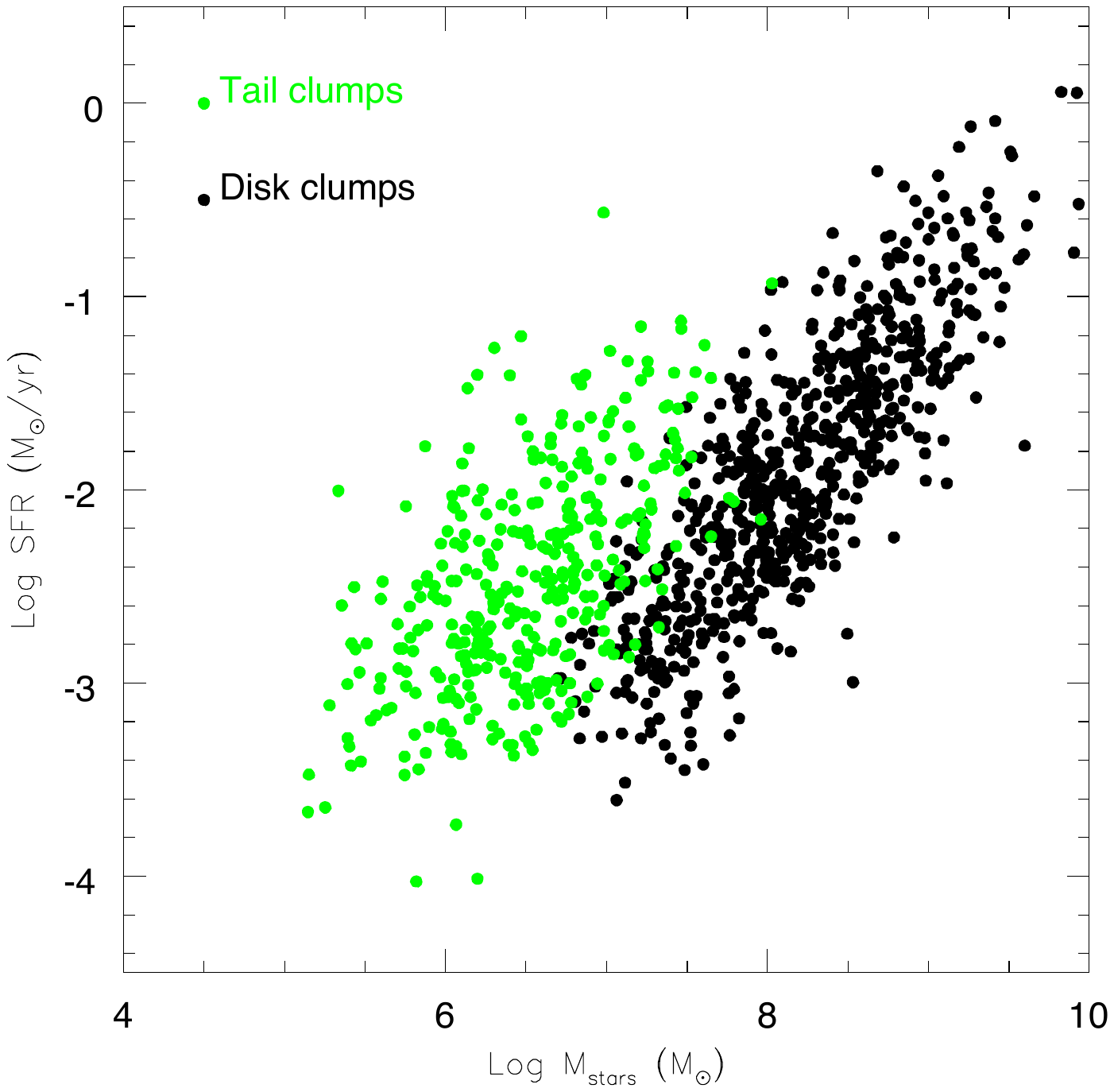}\hspace{-4cm}\includegraphics[width=5.0in]{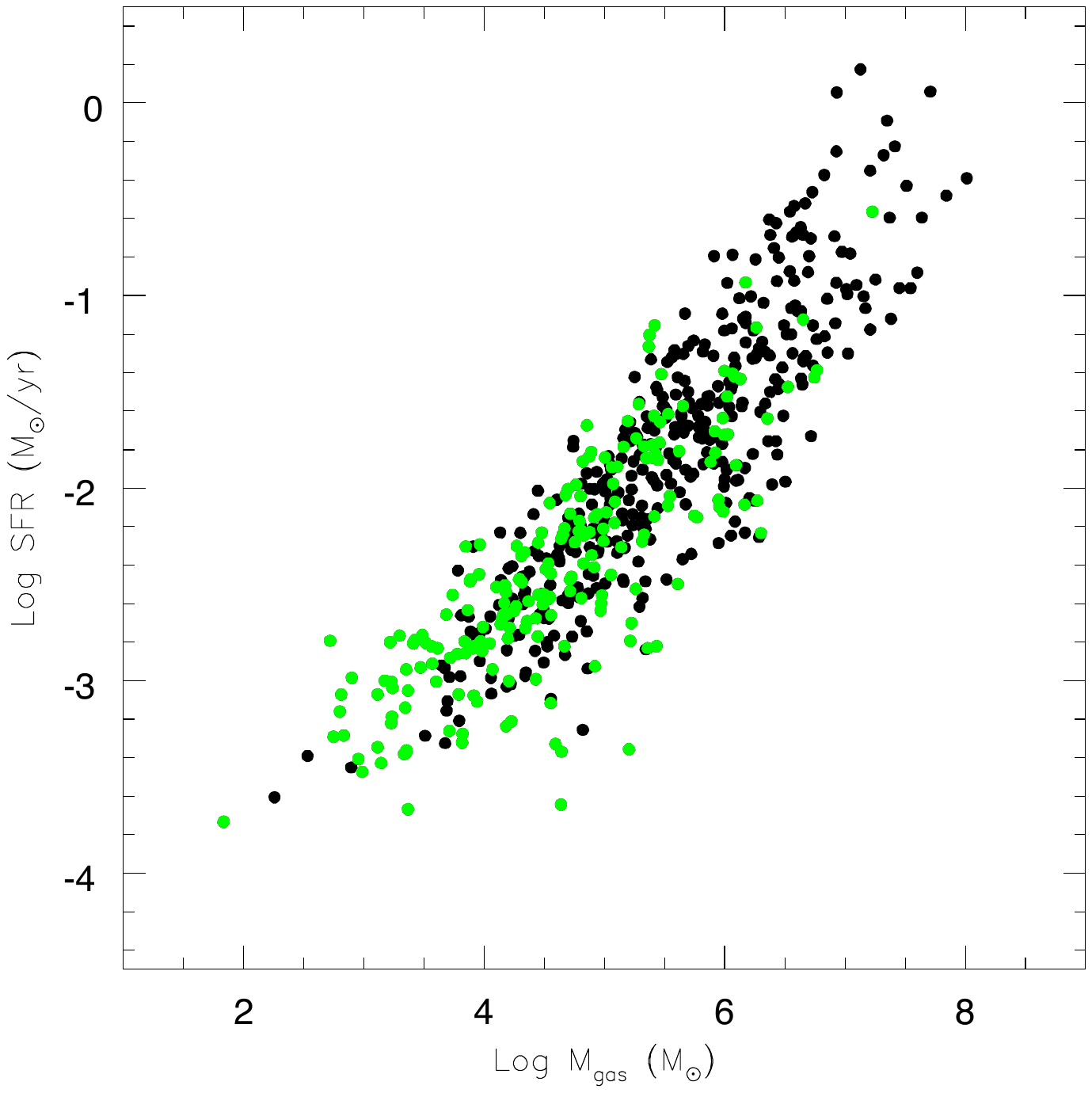}}
%\vspace{-10cm}
%\centerline{\includegraphics[width=5.0in]{wisnkioski.pdf}}
\vspace{-4cm}
\caption{SFR-stellar mass (left) and SFR-gas mass (right) relations of
clumps in the tails (green) and in the disks (black).}
\end{figure*}

\section{Diffuse emission}

%Are there tails without blobs? No 

The star-forming clumps are not the only location with $\rm H\alpha$
emission in the tails. Regions of diffuse emission with lower $\rm
H\alpha$ surface brightness are present in the interclump areas of
the tentacles of all our galaxies. In the following, we will name ``diffuse emission''
the $\rm H\alpha$ component that has not been assigned to any clump. 

%We can investigate what fraction of $\rm H\alpha$ luminosity in the tails originates from the
%diffuse component, and what is the powering mechanism for its ionization.

The diffuse component accounts for a significant fraction of
the $\rm H\alpha$ luminosities in the tails, on average 50\%, ranging
between 30 and 80\% (see Table~3).

There is a very strong anticorrelation between the fraction of $\rm
H\alpha$ luminosity that is in the diffuse component
and the total SFR in the tail (Fig.~12): the higher the SFR, the lower
the diffuse fraction. Hence, tails whose $\rm H\alpha$ emission is
dominated by clumps can reach much higher SFR levels, or, conversely,
tails with high levels of $\rm H\alpha$
emission/SFR are dominated by the star-forming clumps. 
%LEAVE FIGURE OR REMOVE?

\begin{figure}
\centerline{\hspace{1cm}\includegraphics[width=4.2in]{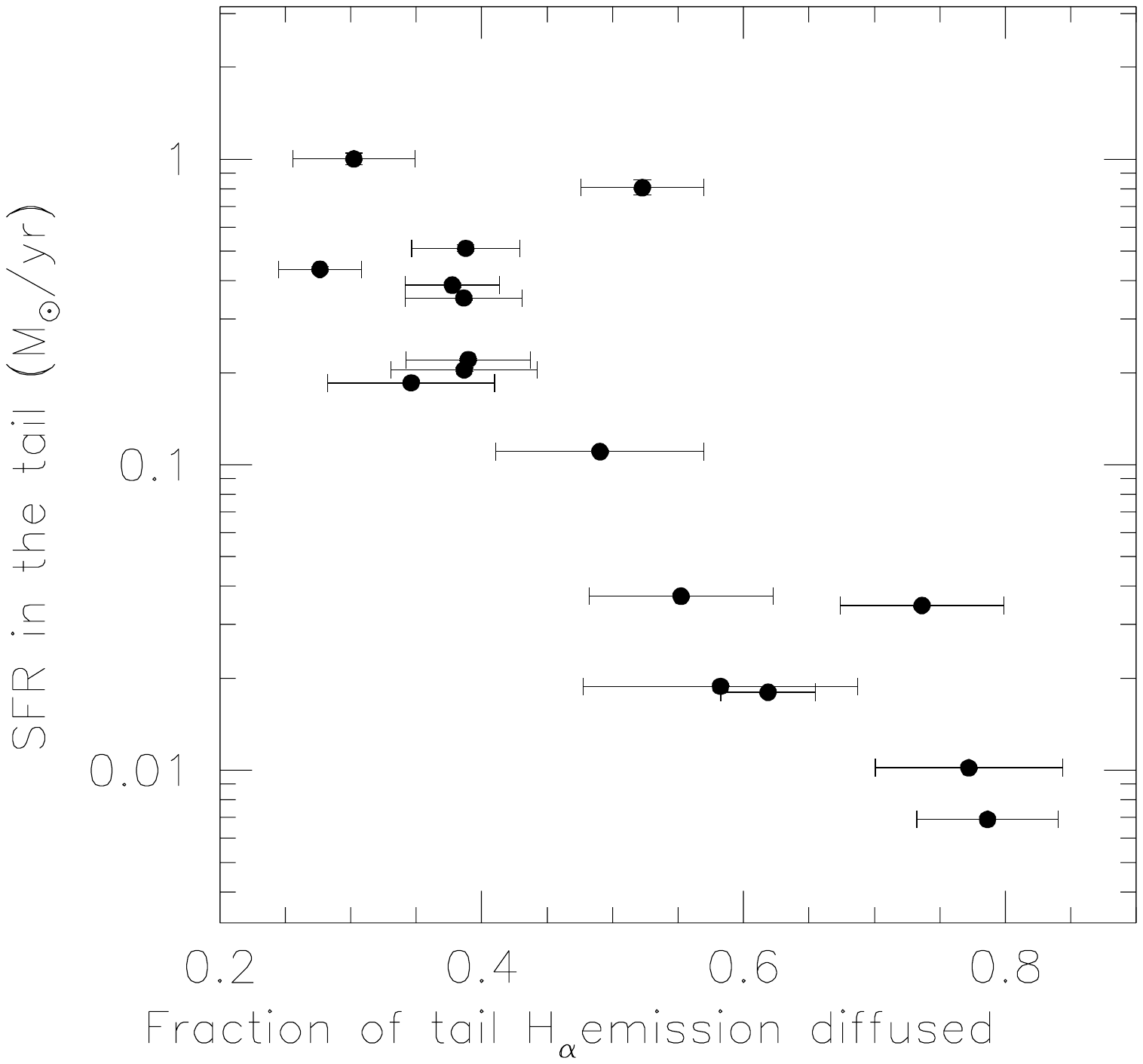}}
\vspace{-3cm}
\caption{Total SFR in the tail (Y axis, solar masses per year, clumps+diffuse) versus fraction of tail $\rm H\alpha$ emission 
that is in the diffuse component for the 16 galaxies in our sample.}
\end{figure}

In most of our galaxies, according to the NII DD
the dominant ionization mechanism of the diffuse emission in the tail
%in most of our galaxies
%the great majority of cases 
is star formation, or a combination of SF and
Composite, as summarized in Table~3. 
SF+Composite emission accounts for at least 98\% 
of the tail emission in all galaxies, except in
JO135 and JO204 that, as
discussed above, have AGN ionization cones extending in the tail, and
in JO147 where there is a 10\% LINER component.

Considering instead only the pure SF component (no Composite) in
  the NII DD, this dominates
($\geq 80\%$) the emission in the tails of 9 of our galaxies (JO113,
JO141, JO160, JO171, JO201, JO206, JO49, JO95, JW39). SF still
accounts for more than the half of the tail emission in JO175 and JO60
(66\% and 58\%, respectively), and only for about 20-30\% in JO135
(29\%) , JO204 (36\%) and JO194 (21\%).
Finally, as previously mentioned, JO147 and JW100 have tails dominated by a 
Composite emission, also in the diffuse component. 
 We note that the SII DD generally agrees closely with NII and
not with OI (see Fig.~2), but also that since the gas in the tails on average has 
lower metallicities than the one in the disk (e.g. Poggianti et 
al. 2017b, Gullieuszik et al. 2017), the NII DD could be affected by metallicity 
variations (Kewley et al. 2001).

As, although to a lesser extent, it was the case for the overall tail emission and for
the tail clumps (\S4.1 and \S5), especially for the diffuse emission
the OI DD indicates a much stronger contribution
from ionization mechanisms different from SF than the NII DD (see
Table~3). In addition to JW100 and JO147, whose NII and OI estimates agree
remarkably well indicating that the tail emission is dominated 
by Composite/''LINER-like'' processes, in all the other galaxies
the OI fraction of tail diffuse emission powered by SF ranges
  from 2\% to 63\%, with a median of  18\%. Four of our galaxies have a particularly strong SF 
component according to the OI DD ($\sim 40-60$\% in JO206, JO49, JW39
and JO194). Understanding the origin of the significant differences we observe from
one galaxy to another will be an important step in grasping the
physics at work, and will be the subject of future GASP investigations.

The OI DD is capable to highlight a contribution
from other physical processes which are particularly relevant for the origin
of the diffuse emission. Thermal conduction from the surrounding
hot ICM and turbulence are promising candidates in this respect.
It is also worth noting that the spaxels for which the OI DD classification can be derived are a subset
of the spaxels where NII DD can be used, due to the lower signal in the OI line, and where the OI is weak and a 
classification cannot be derived is most probably a SF-powered region. Thus, the OI DD might give a more 
biased view of the results in favor of non-SF mechanisms.
The differences between the NII and SII vs. OI results are probably
  indicating that the diffuse emission originates partly from star
  formation and partly from turbulence/thermal conduction, in a
  relative proportion which we cannot ascertain from the current analysis.
% However, all
%together our results seem to indicate that in most tails non-SF mechanisms are at
%the origin of a large fraction of the diffuse emission. }

Regardless of the DD employed, SF is present in tails, though is not sufficient to explain all the
ionized emission.
% some ionizing contribution
%  from SF is present in the tail diffuse emission, in some galaxies
%  more than others.
 Where are the stellar sources 
ionizing the gas that produces the diffuse emission? While in the case of the clumps there
are several lines of evidence demonstrating it is in-situ star formation
within the clumps, the diffuse tail emission might originate either
from low levels of in-situ star formation (a more widespread
population of lower luminosity HII regions), or from ionizing photons
escaping the HII regions within the clumps and going to ionize more
diffuse neutral gas located where no stars have managed to form.
This would resemble what happens in the disk of normal star-forming
galaxies, where the escape fraction is negligible over the whole
galaxy but is significant at the scales of single HII regions and
complexes \citep{OeyKennicutt1997, Wofford2013, Hernandez2018}. 
Given also the low density of the gas giving rise to the diffuse
component, leakage of ionizing photons from the tail clumps
is a plausible origin for the diffuse emission.

Under the assumption that the diffuse
gas is ionized locally by photons escaping the HII regions, the 
average escape fraction in a tail can be estimated from the MUSE
data, and it is equal to the \% of SFR in the tail that is in the diffuse
component (see Table~3), In fact, the escape fraction is the ratio
between the number of ionizing photons that escape the HII regions
(which is proportional to the SFR-powered $\rm H\alpha$ luminosity, hence the
SFR, in the diffuse component in the
tails) and the total number of ionizing photons (proportional to the
total SFR in the tails).
Using the NII DD, this escape fraction in our sample would be on average 18\% (median 15\%) and ranges between 6
and 46\%, in broad agreement with the values found in individual
star-forming regions  \citep{OeyKennicutt1998, Relano2002}.
This is probably an upper limit to the intrinsic escape fraction,
due both to the assumptions and the differences between the NII and OI
estimates.
This hypothesis cannot be confirmed based on the existing data,
%If this hypothesis is correct, these values should be confirmed 
and requires the comparison of our
$\rm H\alpha$ maps with high resolution, deep ultraviolet imaging as
might be obtained with e.g. HST.
% and our
%$\rm H\alpha$ maps.

\begin{table}
\scriptsize
\centering 
\caption{Diffuse versus clump emission and SFR. (1) Galaxy name; (2) Fraction
of tail $L_{\rm H\alpha}$ that is diffuse; (3) SFR in the tail
($M_{\odot} \, yr^{-1}$). (4)
Fraction of tail SFR that is diffuse; (5) \% of diffuse emission
powered by SF; (6) \% of diffuse emission powered by SF+Composite; (7)
Total SFR (disk+tail, $M_{\odot} \, yr^{-1}$); (8) \% of SFR that is
the tail. All the columns until column (8) included refer to the NII
DD. Column (9) presents the \% of tail diffuse emission powered by SF for OI DD.} % in the MUSE spectral range.} %\label{tab:decimal}
\begin{tabular}{lcccccccccccc}
\hline 
%$ID_{P16}$ & Diffuse  & SF \%        & SF+Comp \%  & $L_{\rm H\alpha}$ SF+C. &  SFR   & $L_{\rm H\alpha}$ SF+C. & Diffuse SFR& \% of tail SFR & Total $L_{\rm H\alpha}$ & Total & \% of SFR \\
%                & \%          & in diffuse & in diffuse        & in          tail                 & in tail & diffuse in tail & in tail & that is diffuse & SF+C. & SFR & in tail \\
(1) & (2) & (3) & (4) & (5) & (6) & (7) & (8) & (9) \\

\hline 
JO113&  0.74   &   0.035 &  0.44 & 0.98   &    1.00  & 1.70 &  0.020 &
0.04 \\
JO135&  0.58   &   0.019 &  0.06 & 0.29   &    0.49  & 1.94 &  0.010 &
0.15 \\
JO141&  0.55   &   0.037 &  0.10 & 0.98   &    1.00  & 2.51 &  0.015 &
-- \\
JO147&  0.39   &   0.204 &  0.22 & 0.03   &    0.91  & 4.64 &  0.044 &
0.04 \\
JO160&  0.79   &   0.007 &  0.23 & 0.99   &    1.00  & 1.94 &  0.004 &
0.12 \\
JO171&  0.39   &   0.351 &  0.22 & 0.81   &    1.00  & 1.63 &  0.216 &
0.19 \\
JO175&  0.49   &   0.110 &  0.19 & 0.66   &    1.00  & 2.48 &  0.044 &
0.08 \\
JO194&  0.28   &   0.436 &  0.13 & 0.21   &    1.00  & 8.31 &  0.052 &
0.63 \\
JO201&  0.30   &   1.002 &  0.11 & 0.80   &    1.00  & 6.06 &  0.165 &
0.14 \\
JO204&  0.39   &   0.221 &  0.13 & 0.36   &    0.71  & 1.68 &  0.130 &
0.18 \\
JO206&  0.39   &   0.511 &  0.19 & 0.90   &    1.00  & 5.32 &  0.100 &
0.39 \\
JO49  &  0.77   &   0.010 &  0.13 & 0.93   &    0.99  & 1.38 & 0.007 &
0.48 \\
JO60  &  0.35   &   0.185 &  0.08 & 0.58   &    0.99  & 4.47 & 0.041 &
0.02 \\ 
JO95  &  0.62   &   0.018 &  0.13 & 0.98   &    1.00  & 0.37 & 0.048 &
0.22 \\
JW100&  0.52   &  0.806  &  0.46  & 0.003 &    0.98  & 4.02 & 0.200 & 0.009 \\
JW39  &  0.38   &   0.387 &  0.15 & 0.78   &    1.00  & 3.31 & 0.117 &
0.55 \\
\hline 
\end{tabular}
%\tablecomments{(1) Paper I; (2) Paper V; (3) Paper IV; (4) Paper II;
%(5) Paper VI; (6) Merluzzi et al. (2013). Cluster redshifts from 
%Moretti et al. (2017) and Cava et al. (2009).}
\end{table}

%What fraction of Halpha is in blobs vs diffuse in the tails? 
%And in disk?

%Ionization mechanism of diffuse emission

\section{Star formation rates in the tails and the disks}

%We now investigate what fraction of the total SFR of these galaxies is
%hosted in the tails. 
The total SFR (disk+tails) of each of our galaxies is presented in Table~3 together
with the fraction of total SFR that is in the tail.

Our jellyfishes form in total 
%a few solar masses per year, 
between 1.5 and 6 $M_{\odot} \, yr^{-1}$, with 
two outliers being JO194 with $8.3 \, M_{\odot} \, yr^{-1}$ and 
JO95 with $\sim 0.4 M_{\odot} \, yr^{-1}$ (Table~3).
The location of GASP galaxies in the SFR-stellar mass diagram is the
subject of a separate paper \citep{Vulcani18c}.

The fraction of total SFR that takes place in the tail can vary
significantly, between less than 1\% and more than 20\% of the total
activity, as shown in Table~3.
The SFR occurring in the tail represents 10 to 20\% of the total in 6
of our galaxies (JO206, JW39, JO204, JO201, JW100 and JO171), which are those with the longest tails, as it
is reasonable to expect. In another 5 galaxies the tail SFR is about
4-5\% of the total (JO60, JO147, JO175, JO95, JO194), while for the
remaining 5 galaxies it is at the 1-2\% level (JO113, JO141, JO135,
JO49, JO160).
%a non-negligible
%fraction of the total ($\geq$5\%) in about half of the sample (JO171, JO194, JO201, JO204,
%JO206, JW100 and JW39) which are those with the longest tails, as it
%is reasonable to expect. In the other half of the sample (JO113, JO135, JO141, JO147,
%JO160, JO175, JO49, JO60, JO95) is typically at the level of a few percent.
%The dependence of the tail SFR on galaxy and cluster properties
%and the role of stars formed in the tails as contributors to the intracluster light 
%are presented in Gullieuszik et al. (in prep.).

Most of the SFR in the tails is concentrated in the clumps, typically
$>$80\%. \footnote{We note that the fraction of tail SFR that is in
  the clumps is equal to 1 minus column (4) in Table~3.} The only two galaxies with
significantly lower fractions ($\sim 55\%$) are JO113 and JW100, in
which the diffuse emission is more prominent.

\section{Discussion}

%\subsection{The fate of the star forming clumps in the tails}

%Boselli 1803.04177

%UCDs? (Wittmann's talk at Ringberg)
%dwarf galaxies? if dwarfs, what metallicities expected?
%Related to Marco's paper

%\% di SF che e' fuori 
%\% di SF fuori che e' in blobs and in diffuse (etc vedi abstract) 
% (in plot_bpt_N_diffuse.py) 
%escape fraction 
%our CO and UV studies
%strong selection bias:P16 

\subsection{The frequency of tail star formation and previous observational results}

Ongoing star formation is present 
%ubiquituous 
in the tails of all our sample galaxies.  
In a future work we investigate how the star formation in the tails
depends on the galaxy properties, the cluster properties and the galaxy
position in the velocity-clustercentric distance diagram using the
full GASP sample (Gullieuszik et al. in prep.).

%It is important to keep in mind that this does not necessarily imply 
%that star formation takes place in all the tails of ram pressure 
%stripped galaxies. 
Since the GASP targets were sourced from the \cite{Poggianti2016} 
atlas, they all have visible unilateral debris in the B-band 
images, therefore they might be expected to host SF in the tails by selection. 
Hence, our results do not rule out the possibility that stripping
can occur without extraplanar star formation.
%some gaseous ram pressure stripped tails do not form stars at all. 
Moreover, the sample presented in this paper consists of galaxies that are all
subject to strong ram pressure stripping, and with long tails of
extraplanar $\rm H\alpha$ emitting gas.
Galaxies in a more advanced stage of stripping 
(i.e. truncated gaseous disks with gas left only in the center, or 
fully stripped galaxies devoid of any gas) 
were excluded in this work, and require a separate analysis 
to assess whether they have had an $\rm H\alpha$ tail phase at some
point and whether
stellar-only extraplanar clumps are still visible. %, thus in the phase of peak-stripping. 

It is probable, as also expected from simulations, that SF in the tails
occurs only during the phase of peak-stripping, and when ram pressure
is sufficiently strong to produce significant gaseous tails.
Gaseous tails can be observed at various wavelengths, detecting gas in
different phases: a) with HI observations detecting stripped neutral
hydrogen, b) in X-ray detecting gas heated at the interface between the hot ICM and
the cold stripped ISM,  c) observing $\rm H\alpha$ emission with
narrow-band or IFU observations to probe the ionized/excited gas phase, and d) with
CO observations to study the molecular gas.
%X-ray tails seem to be rare, and none have been observed in the Virgo
%cluster (Sun et al. 2006, 2010).
How frequently tails of various gas phases coexist is still unknown,
as it is unknown how this frequency depends on the ICM conditions and
the galaxy properties. Studies of the HI \citep{Chung2007, Kenney2004, OosterlooVanGorkom2005, Crowl2005, Abramson2011} 
or X-ray tails \citep{SunVikhlinin2005, Sun2006, Sun2010} require
additional information to inform us about the ongoing or past tail SF,
that can come from $\rm H\alpha$ or UV data. 

The great majority of previous UV or $\rm H\alpha$ studies of stripped
tails find evidence for ongoing or recent SF in the tails.
The most solid evidence can be obtained with IFU or spectroscopic
studies, that allow us not only to detect $\rm H\alpha$ emission but
also to assess the ionization mechanism from multiple line
ratios. Before GASP there were three well studied such cases:

1) ESO137-001 in Abell 3627 (stellar mass $\sim 5-8 \times 10^9 M_{\odot}$)
was the first one which it was 
unambiguously shown to host star-forming HII regions in the cold stripped ISM by
\cite{Sun2010} using Gemini spectra \citep[see also][]{Sun2007}, as confirmed by MUSE
subsequent studies \citep{Fumagalli2014} that found these HII
regions form in low- velocity dispersion gas (25-50$\rm \, km \, s^{-1}$)
and have quite typical line ratios, densities, temperatures and
metallicities, suggesting they are formed in situ within the tails
\citep{Fossati2016}. Large amounts of molecular gas were found in
the tail of this galaxy by \cite{Jachym2014}.

b) UGC6697 ($10^{10} M_{\odot}$) in Abell 1367, a tidally interacting
system in which ram pressure stripping may have been enhanced by the
encounter. After several multiwavelength campaigns detecting radio
continuum, $\rm H\alpha$ and X-ray tails \citep{GavazziJaffe1987,
Gavazzi1984, Bothun1984, Gavazzi1989, SunVikhlinin2005}, it has been studied with
MUSE by \cite{Consolandi2017}  who found in the tail both diffuse emission and
compact knots of low velocity dispersion
with line ratios typical of HII regions, whose physical
properties do not differ from normal HII regions in galactic disks.

c) SOS 114372 ($7 \times 10^{10} M_{\odot})$ in the Shapley Abell 3558
cluster was studied using Wifes IFU data by \cite{Merluzzi2013} 
who found $\rm H\alpha$ knots and filaments in the one-sided
13kpc ionized gas tail, detecting a contribution from shock excitation
as well as star formation. We note that this galaxy is the GASP galaxy
JO147 presented in this paper, that was independently identified as a
gas stripping candidate by \cite{Poggianti2016}.

In addition, detailed GASP studies of five individual galaxies
showing star formation in the tails were published in \cite{Poggianti2017a, 
Bellhouse2017, Fritz2017, Gullieuszik2017, Moretti2018}, Bellhouse et al. in prep,  see
these papers for details.

In the absence of IFU or spectroscopic data, strong evidence for
ongoing SF in the tails can come from $\rm H\alpha$
narrow-band observations in combination with UV data. The latter, if
sufficiently deep, can reveal the stellar knots formed in the tails
and therefore confirm that the $\rm H\alpha$ emission is due to
ongoing SF, and viceversa $\rm H\alpha$ can confirm that the UV knots
belong to the galaxy and are not background sources. Well studied individual cases are:

d) NGC 4254 ($2.4 \times 10^{10} M_{\odot}$), in the Virgo cluster, in
which \cite{Boselli2018} has identified 60 GALEX candidate star
forming regions up to 20kpc outside of the disk, of which 30 have also
$\rm H\alpha$ emission. The 250kpc HI tail of this galaxy seems to be
driven by a recent gravitational encounter with another Virgo member,
and the knots in the tail are interpreted as coeval knots formed after
a single SF burst.

e) NGC4330 ($6 \times 10^9 M_{\odot}$) in Virgo, in which extraplanar UV
regions close to the disk were found by \cite{Abramson2011}, who
interpret them as extraplanar star formation in a galaxy undergoing
initial stripping, that has yet to reach peak stripping. Star-forming 
extraplanar regions are also visible in $\rm H\alpha$ imaging 
\citep{Abramson2011,Fossati2018}.

Moreover, another Virgo dwarf galaxy, VCC1249 ($1.2 \times 10^9
M_{\odot}$), interacting with a massive elliptical
companion, shows that the combination of tidal interaction and ram
pressure stripping led to the removal of HI gas from the disk and that
extraplanar HII regions were formed in situ in the gas removed
\citep{Arrigoni2012}.

To our knowledge, the only case for which it has been argued that no 
star formation occurs in a long tail with $\rm H\alpha$ emitting gas
is NGC4569 ($3 \times 10^{10} M_{\odot}$) in Virgo, which has an 80kpc
long $\rm H\alpha$ tail and is affected both by ram pressure and a
strong close interaction. \cite{Boselli2016} argue that no star
forming region is observed in the tail and conclude that other
mechanisms other than photoionization (such as shocks, heat conduction
or magnetohydrodynamic waves) are responsible for the emission.
%{\it Actually, clear Halpha knots in their fig.6!!! Should we write it?}

Two very well studied cases of galaxies in a ``post-stripping''
phase in which there is little star formation left but
 the stellar knots previously formed in the tails are still
strikingly visible are:

f) IC3418 ($4 \times 10^8 M_{\odot}$), a post-starburst passive galaxy in Virgo, with a 17kpc
tail of UV knots that were recognized as star-forming regions and
characterized by \cite{Hester2010, Fumagalli2011, Kenney2014}. IC3418 has little ionized gas left only in the outer tail.
%which is seen only in the outer part of the tail. 
This galaxy also has
a possible marginal detection of $\sim 10^6 M_{\odot}$ of molecular
gas in the disk, and only CO upper limits in the tail \citep{Jachym2013}.

g) RB199 ($8 \times 10^8 M_{\odot}$, \citealt{Sun2010}), another post-starburst disk but
in the Coma cluster, very similar to IC3418 in many ways (Yoshida et
al. 2008). This is a galaxy-galaxy merger remnant whose gas tail is
ascribed to ram pressure and whose 80kpc tail shows UV and $\rm
H\alpha$ bright knots forming stars.
  
Other strong evidence for star formation in stripped tails comes from
the works of \cite{Smith2010} and \cite{Yagi2007, Yagi2008, Yagi2010} in
the Coma cluster.
\cite{Smith2010} showed with GALEX data that 13
star-forming galaxies have tails with filaments and knots concluding that SF
occurs within the stripped gas by interaction with the cluster
environment, presenting also HST data for two of these that reveal
compact blue knots coincident with UV and $\rm H\alpha$ emission.
Yagi and collaborators mapped Coma with $\rm H\alpha$ imaging, and
found $\rm H\alpha$ clouds associated with 14 Coma members (6 of which
belong to the \citeauthor{Smith2010} sample). Some of these clouds are connected
with disk SF, some are clouds connected to a disk that is devoid of SF
(e.g. like IC3418 and RB199) and some others are totally detached
clouds. Extended ionized gas clouds, some of which are associated with galaxies,
some not, have also been detected in Abell 1367 by \cite{Gavazzi2001} and \cite{Yagi2017}.

Indirect support for the probable presence of SF in stripped tails comes from the
detection of a large amount of molecular gas in the tail of D100 ($2
\times 10^9 M_{\odot}$) in Coma \citep{Jachym2017} within the 60kpc
long $\rm H\alpha$+UV tail studied by \cite{Yagi2007, Yagi2010}, and \cite{Smith2010}.

Molecular gas is also present along the tail of NGC4388 \citep{Verdugo2015}, 
a Virgo galaxy in which in situ
star forming regions were found in the tail by \cite{Yagi2013} based on
photometric data and slit spectroscopy. This galaxy is an interesting
case for having both an HI tail and a faint X-ray tail \citep{OosterlooVanGorkom2005,
Boissier2012, Sun2010}. Some molecular gas was also
detected close to the giant elliptical M86 in Virgo, but it was
difficult to find a secure association with the neighboring galaxies
(including NGC4388) \citep{Dasyra2012}.

Finally, \cite{Cortese2007} studied two dwarf galaxies in two
clusters at z=0.2 with HST finding tails with bright knots and stellar
streams that the authors interpreted as star-forming knots
consistent with the formation of ultracompact dwarf galaxies.
At even higher redshifts, again HST revealed bright blue knots
consistent with star formation in the debris tails of jellyfishes in clusters \citep{Owen2006, Owers2012, Ebeling2014}.

We note that at odds with several of the well studied cases in Virgo, galaxies clearly  
interacting with a nearby companion have been excluded from our sample.  
Concerning the SF in the tails, our GASP results are very much in line with previous findings, and
extend them to a larger, homogeneously observed sample for which IFU
spectroscopy allows to study the ionization mechanisms in the tails and
the physical properties of the star forming clumps. 
They are formed in low velocity dispersion gas, as found by previous
studies \citep[e.g.,][]{Fumagalli2014, Fossati2016, Consolandi2017}.
The tail clumps in our sample extend to higher $\rm H\alpha$
luminosities and stellar masses than those identified by \cite{Boselli2018} 
in NGC 4254 ($L_{\rm H\alpha}=10^{37}-10^{38} \rm \, erg \,
s^{-1}$, $M=10^3-10^5 M_{\odot}$) and by \cite{Smith2010} in their two
galaxies with HST data ($M=10^4-10^5 M_{\odot}$), but similar $\rm
H\alpha$ luminosities to the clumps in ESO137-001 \citep{Sun2007}
and similar stellar
masses and SFRs to the
clumps in RB199 in Coma \citep[$M=10^6-10^7
M_{\odot}$]{Yoshida2008}.
Generally, the clumps presented in this paper extend to higher masses than
those in ram pressure stripped tails in Virgo \citep[see][and references therein]{Yagi2013}.
The characteristics of the clumps formed in the tails are expected to
depend on the ICM physical conditions. For example, the ICM in Virgo
is 10 times less dense and hot than in Coma, and higher density and
pressure can help the gas confinement and the SF in the
tails. On the other hand, our results demonstrate that tail SF does not occur
only in very massive clusters, but also in low mass clusters with
$\sigma = 500-600 \, \rm km \, s^{-1}$ where some of our most spectacular
cases are found (e.g. JO204, JO206, see Table~1).
We also note that, with the exception of SOS114372/JO147, no jellyfish
with a mass higher than
$3 \times 10^{10} M_{\sun}$ was studied in detail before
GASP. However, SF does not occur only in the tails of galaxies of a
given mass range, as demonstrated by the fact that
we detect it in galaxies with a wide range of stellar masses 
($3 \times 10^9-3 \times 10^{11} M_{\odot}$, Table~1), although the net
amount of tail SF depends on galaxy mass, as we discuss in Gullieuszik
et al. (in prep.).

%{\it Section too long? Shorten?}

%Add discussion on the fact that these are all undergoing STRONG RPS,
%peak-phase. Possibility that only in a given phase, and for strong 
%RPS, stars are formed in tails (cf also Boselli18). 
%Yara: how inclination and stripping stage affects the detectability of 
%tails and the analysis...Discussion of IMF stochasticity... 

\subsection{Expectations from hydrodynamical simulations}

A few simulations have examined star formation in ram pressure
stripped tails.  Most agree that star formation can take place
throughout the tail as at any time stripped gas will have a range of
densities and temperatures that will be accelerated at different rates
and have different collapse timescales (\citealt{Roediger2014, TonnesenBryan2012, Kapferer2009}; but \citealt{Steinhauser2016} find
no star formation in their tails).  

\cite{Kapferer2009} ran 12 simulations varying the ram pressure wind
velocity and density, and found that increasing the ram pressure
increases the SFR in the tail.  The SFR was more strongly affected by
the wind density than the wind velocity.  \cite{TonnesenBryan2012}
also argued that the SFR in the tail increased with increasing
surrounding ICM thermal pressure.  In \cite{Jaffe2018} we found
that the longest and most strongly star-forming tails tended to be
moving quickly near the center of clusters.  The simulations indicate
that it is the surrounding dense ICM that is driving the high star
formation rates in these tails.  The high velocities of the galaxies
towards the center of the cluster allow for gas to travel farther from
the galaxy before it collapses and forms stars.  

The average SFR over 500 Myr in the Kapferer tails ranged from 0.2-2
$M_{\odot} \, yr^{-1}$, depending on the ram pressure strength.  The SFR was much
lower in the \cite{TonnesenBryan2012} tail, peaking at about 0.06 $M_{\odot} \, yr^{-1}$.
As discussed in \cite{TonnesenBryan2012}, there are many
differences in the codes and initial conditions used in the
simulations that will affect the SFRs found in different works.  
The total SFR we measure in tails (column 3 of Table~3) ranges between
$\sim 0.01$ and 1$M_{\odot} \, yr^{-1}$, with a median of 0.20
$M_{\odot} \, yr^{-1}$ (Q1=0.03, Q3=0.44), in agreement with the
ranges spanned by the simulations.

\cite{TonnesenBryan2012} also considered the stellar clumps formed in
tails.  Using the Kennicutt's (1998a) relation to go from SFR to $\rm H\alpha$
luminosity, the authors found that the brightest clumps have an $\rm
H\alpha$ surface brightness 
$<3 \times 10^{38} \, \rm erg \, s^{-1} \, kpc^{-2}$, and only the few most massive stellar clumps have
mass surface densities of $3 \times 10^4 M_{\odot} \, \rm kpc^{-2}$ .  
%This puts the clumps in the simulated tail at the lower end of the observed distribution of clumps.  
A comparison with our observed values will need a careful assessment
of spatial resolution effects both in observations and simulations.
We will use simulations to examine how clump properties are related to 
the surrounding ICM in future work.  

%Explore Stephanie's hypothesis that: Simulation of RPS suggest that
%the observed long tails of ionized stripped gas are only possible to
%form under the most intense ram-pressure conditions. It is thus
%possible that only the massive galaxies were able to still have gas by
%the time they reached the vicinity of the cluster core and thus are
%the only ones displaying long bright tails. (from GASP IX).

%SEE ALSO PoggiantiGASPnotes.docx for stellar mass density, and for
%age gradients.

\section{Summary}

In this paper we have presented the analysis of the MUSE data of 16 cluster galaxies with
clear tails of ram pressure stripped gas from the GASP survey.
%, we have investigated the
%ionization mechanism of the tails using three different
%emission line ratio diagnostic diagrams. 
%We separate the tail emission in two components: emission in clumps,
%and diffuse emission. 
All galaxies present bright $\rm H\alpha$ clumps in their tails as  
well as interclump regions of more diffuse emission.  
We have found that:

1) The ionization mechanism of the tails has been investigated  using three different
emission line ratio diagnostic diagrams. Such mechanism could be
derived for all spaxels with sufficiently high S/N data, which on
average account for about 60\% of the total $\rm H\alpha$ luminosity in the tails.

According to the NII and SII diagnostic diagrams, the fraction of
$\rm H\alpha$ luminosity in the tails powered by SF ranges from galaxy
to galaxy between $\sim 70\%$ and 100\%, with a median fraction of
about 90\%. When using the OI DD, the SF-powered fraction ranges
between 87\% and 12\% depending on the galaxy, with a median of
64\%. The [OI] diagram is the most sensitive one to a contribution from
shocks, but in the stripped tails the extra [OI] emission might be due to thermal heating 
of the stripped gas by the hot ICM.

2) Timescale arguments rule out the possibility that the 
photons ionizing the gas in the tails originate from massive stars in the 
disks of these galaxies. Based on this, as well as on our molecular gas 
detections in the tails \citep{Moretti2018b} and our deep FUV and NUV 
imaging \citep{George2018}, we conclude that star formation occurs 
in situ in the tails, within the $\rm H\alpha$ clumps we 
identify. 

3) %In agreement with the spaxel-by spaxel analysis, 
The DD analysis of
the integrated spectra of $\rm H\alpha$ clumps finds that the vast
majority of clumps in the tails are powered by SF, or SF+Composite
emission, and have gaseous velocity dispersions $< 50 \, \rm km \,
s^{-1}$. Clumps with high emission line ratios have instead $\sigma > 50-70 \, \rm km \,
s^{-1}$, probably due to thermal conduction/turbulence effects
  and/or contamination by turbulent diffuse gas along the line of sight.

4) The  $\rm H\alpha$ star-forming clumps in the tails resemble giant and supergiant HII
regions and complexes, with a median gas velocity dispersion of 27
$\rm km \, s^{-1}$ and $\rm H\alpha$ luminosities in the range 
$10^{38}-10^{39} \rm \, erg \, s^{-1}$. 

5) We measure moderate values of
dust extinction in the clumps (median $A_V=0.5$ mag), which are
lower on average than those of clumps in the disks. The SFR of clumps in
the tails is also lower on average than in the disks (0.003
$M_{\odot} \, yr^{-1}$ versus 0.008 $M_{\odot} \, yr^{-1}$), but the SFR
estimates are subject to large uncertainties due to the unknown IMF and stochastic
effects in the IMF sampling at these low SFR values.

6) The density of the gas can be measured from the [SII] lines in the
MUSE spectra for about half of the
clumps. Measured values show a wide density distribution, between
20 $cm^{-3}$ and well above 100 $cm^{-3}$, with a median of $\sim 50
\, cm^{-3}$, indicating that dense gas clumps are
present in the tails.

7) The ionized gas mass of the clumps, whose median is $4\times 10^4
M_{\odot}$, broadly correlates with the stellar mass of the clumps (median $3\times 10^6
M_{\odot}$). On the basis of their stellar masses, we speculate that the
clumps formed in the tails of stripped gas  contribute to the large
population of UCDs/GCs/dSphs observed in nearby clusters, or even to the 
UDG population depending on their subsequent dynamical evolution.

8) The clump SFR correlates well with both the clump gas mass 
and stellar mass. Interestingly, the clumps in the tails follow the 
exact same correlation between SFR and gas mass of clumps in the 
disks. 

The star-forming clumps both in the tails and the disks
%behave normally also as far as other
%correlations are concerned, in particular 
roughly follow also the relation between gas
velocity dispersion and $\rm H\alpha$ luminosity of low redshift
star-forming galaxies. Assuming they also
share the $\rm H\alpha$ luminosity-size relation of clumps in the
disks of low-z spirals from the literature, we infer they should have
typical core radii between 100 and 400pc, with a median of 160pc in
the tails. These sizes are below our
spatial resolution limit and require higher resolution studies.

9) On average 50\% of the $\rm H\alpha$ luminosity in the tails is in
the form of diffuse emission. We find a strong anticorrelation between the
fraction of $\rm H\alpha$ tail emission that is
diffuse and the total SFR in the tail.
%: clump-dominated tails can reach much higher
%levels of in-situ SF.
% The higher the SFR in the tail, the
%lower the fraction of $\rm H\alpha$ tail emission that is
%diffuse. This shows that tails with high levels of SFR are
%dominated by the star-forming clumps.

The diffuse tail emission originates both from ionizing photons
  due to SF and from other mechanisms producing Composite or
  LINER-like emission line ratios.  The relative contribution of SF
  and non-SF processes varies significantly when
considering the NII and the OI DD, and from one galaxy to another.
%We hypothesize that the non-SF component is due to thermal conduction
%from the hot surrounding ICM or turbulence induced by ?????}

%For 9 of our galaxies, $>80$\% of the diffuse component is powered by
%SF. For other two galaxies this fraction is 60-70\%, for other three
%objects is 20-30\%, while the last two galaxies are dominated by
%Composite emission with a negligible SF contribution.

Under the hypothesis that the ionizing photons responsible for the SF
component of the diffuse emission originate within %, and escape from 
the star-forming clumps, we estimate an average escape fraction from
the clumps of 18\%, ranging between 6 and 46\% from galaxy to galaxy.

10) In most galaxies, the SFR in the tails represents only a small
fraction (a few percent) of the total SFR of the system
(tail+disk). In the 6 galaxies with the longest tails this fraction 
is much higher, between 10 and 20\%. Most of the SF in the tails is
concentrated in the clumps.

%The SFR in the tails and the disks of GASP galaxies are the subject of
%other papers in preparation. 

To summarize, we detect ongoing star formation in the tails of all
galaxies in our sample, that cover two orders of magnitude in galaxy stellar
masses (between a few times $10^9$ to a few times $10^{11} M_{\odot}$)
and are members of galaxy clusters with velocity dispersions between
$\sim 550$ and over a 1000 $\rm km \, s^{-1}$, from low mass to very
massive clusters.

In this work we have demonstrated that SF occurring in-situ in the
tails is a common phenomenon and started to unveil the physical properties of the
star forming clumps in the stripped tails of a statistically
significant sample of jellyfish galaxies. These tails are unique
laboratories to study the star formation process, in the absence of 
an underlying galaxy disk and in a fully gas-dominated regime.
%Among the several questions opened by this analysis,
Future works using the GASP MUSE data and its multiwavelength ongoing follow-ups 
(APEX, ALMA, JVLA and UVIT@ASTROSAT) will address some of the several
questions this work has opened.
%The galaxy-to-galaxy variations of the tail star formation properties
%Among the several questions opened by this analysis, 
%This analysis opens a number of questions, such as 
%which might
%help shedding some light on star formation processes in general

\section*{Acknowledgements}
%B.V. acknowledges the support from 
%an Australian Research Council Discovery Early Career Researcher Award
%(PD0028506). This work was co-funded under the Marie Curie Actions of the European Commission (FP7-COFUND).
We are grateful to the anonymous referee for her/his constructive and
detailed comments. We would like to thank Emily Wisnioski for providing us the data from
her paper. This work made use of the KUBEVIZ software which is publicly available at  
http://www.mpe.mpg.de/$\sim$dwilman/kubeviz/. We warmly thank Matteo Fossati and Dave Wilman for their help
with KUBEVIZ. We are grateful to Oleg Gnedin, Frank van den
Bosch, Elke Roediger, Bruce Elmegreen, Miroslava Dessauges-Zavadsky,
Sally Oey, Rob Kennicutt, Roberto and Elena Terlevich and all the
organizers and participants of the Ringberg 2017
conference on ``Galaxy evolution in groups and clusters at low
redshift: theory and observations'' 
(https://ringberg2017.wixsite.com/ap-ringberg2017)
and of the Cambridge meeting on ``The laws of star formation: from the
cosmic dawn to the present universe''
(https://www.ast.cam.ac.uk/meetings/2018/sf.law2018.cambridge)
for useful discussions.
We acknowledge financial support from PRIN-SKA 2017 (PI L. Hunt). 
Y. J. acknowledges support from CONICYT PAI (Concurso Nacional de Inserción
en la Academia 2017) No. 79170132.
Based on observations collected at the European Organisation for Astronomical Research in the Southern Hemisphere 
under ESO programme 196.B-0578. Based on observations taken with the 
AAOmega spectrograph on the AAT, and the OmegaCAM camera on the VLT.

%%%%%%%%%%%%%%%%%%%%%%%%%%%%%%%%%%%%%%%%%%%%%%%%%%

%%%%%%%%%%%%%%%%%%%% REFERENCES %%%%%%%%%%%%%%%%%%

% The best way to enter references is to use BibTeX:

%\bibliographystyle{mnras}
%\bibliography{example} % if your bibtex file is called example.bib

% Alternatively you could enter them by hand, like this:
% This method is tedious and prone to error if you have lots of references
%\begin{thebibliography}{99}
%\bibitem[\protect\citeauthoryear{Author}{2012}]{Author2012}
%Author A.~N., 2013, Journal of Improbable Astronomy, 1, 1
%\bibitem[\protect\citeauthoryear{Others}{2013}]{Others2013}
%Others S., 2012, Journal of Interesting Stuff, 17, 198
%\end{thebibliography}

\bibliographystyle{mnras}
\bibliography{gasp_all} % if your bibtex file is called example.bib

%%%%%%%%%%%%%%%%%%%%%%%%%%%%%%%%%%%%%%%%%%%%%%%%%%

%%%%%%%%%%%%%%%%% APPENDICES %%%%%%%%%%%%%%%%%%%%%

%\appendix

%\section{Some extra material}

%%%%%%%%%%%%%%%%%%%%%%%%%%%%%%%%%%%%%%%%%%%%%%%%%%

% Don't change these lines
\bsp	% typesetting comment
\label{lastpage}
\end{document}